
%
%
%
%
%
%
%
%
%
%
%
\input phys.tex
\twelvepoint\doublespace
%
\english
\hfuzz=100pt
\overfullrule=0pt
\hoffset = 1truecm
\chapters
\equchap
\equfull
\figchap
\tabchap
%
%
%
%
%
%
\message{Extended math symbols.}

\ifx\oldzeta\undefined    
  \let\oldzeta=\zeta    
  \def\zzeta{{\raise 2pt\hbox{$\oldzeta$}}} 
  \let\zeta=\zzeta    
\fi

\ifx\oldchi\undefined    
  \let\oldchi=\chi    
  \def\cchi{{\raise 2pt\hbox{$\oldchi$}}} 
  \let\chi=\cchi    
\fi


\def\square{\hbox{{$\sqcup$}\llap{$\sqcap$}}} 

\def\frac#1#2{{#1 \over #2}}

\def\half{\ifinner {\scriptstyle {1 \over 2}}
   \else {1 \over 2} \fi}

\def\bra#1{\langle#1\vert}  
\def\ket#1{\vert#1\rangle}  

\def\simge{\rlap{\raise 2pt \hbox{$>$}}{\lower 2pt \hbox{$\sim$}}}
\def\simle{\rlap{\raise 2pt \hbox{$<$}}{\lower 2pt \hbox{$\sim$}}}



\def\slashchar#1{\setbox0=\hbox{$#1$}  
   \dimen0=\wd0     
   \setbox1=\hbox{/} \dimen1=\wd1  
   \ifdim\dimen0>\dimen1   
      \rlap{\hbox to \dimen0{\hfil/\hfil}} 
      #1     
   \else     
      \rlap{\hbox to \dimen1{\hfil$#1$\hfil}} 
      /      
   \fi}      %


\def\vbig#1#2{{\vbigd@men=#2\divide\vbigd@men by 2%
\hbox{$\left#1\vbox to \vbigd@men{}\right.\n@space$}}}




\def\tr{\mathop{\rm tr}\nolimits} 
\def\Tr{\mathop{\rm Tr}\nolimits} 


\def\GeV{{\rm GeV}}   
\def\MeV{{\rm MeV}}   


%
%
\def\dk4{\int {d^4 k_E\over(2\pi)^4}}
\def\r2{<R^2>}

\def\ddmu{\partial_\mu}
\def\dumu{\partial^\mu}
\def\ddnu{\partial_\nu}
\def\dunu{\partial^\nu}
\def\prop{\int\limits_{1 / {\Lambda^2}} ^\infty}
\def\propchi{\int\limits_{1 / {(\Lambda^2 \chi^2)}} ^\infty}

\def\pti2{\int\limits_1 ^\infty}

\def\Sp{{\rm Sp}}
\def\Spto{{\rm Sp_{(to)} }}
\def\det{{\rm det}}
\def\Tr{ {\rm Tr}}
\def\ln{{\rm ln}}
\def\fm{{\rm fm}}
\def\MeV{{\rm MeV}}
\def\GeV{{\rm GeV}}

\def\NJL{Nambu--Jona-Lasinio-model\ }
\def\NJLM{Nambu--Jona-Lasinio-model\ }
\def\emes{E_{ mes}}
\def\eval{E_{ val}}
\def\esea{E_{ sea}}
\def\ebr{E_{ br}}
\def\etot{E_{ tot}}

\def\pirv{\vec\pi(\vec r)}
\def\sigr{\sigma(r)}
\def\pir{\pi(r)}

\def\gamf{\gamma_5}

\def\piv{\vec\pi}
\def\sld{\slashchar\partial}
\def\D{{\cal D}}
\def\C{{\cal C}}
\def\L{{\cal L}}
\def\Z{{\cal Z}}
\def\R{{\cal R}}
\def\T{{\cal T}}
\def\RR{{\cal R}}
\def\O{{\cal O}}
\def\B{{\cal B}}

\def\epsl{\epsilon_\lambda}
\def\epsn{\epsilon_\nu}

\def\epslv{\epsilon_{\lambda_V}}

\def\epsval{\epsilon_{ val}}
\def\roR{{r\over R}}

\def\ketl{{\ket{\lambda}}}
\def\ketlv{{\ket{\lambda_V}}}

\def\pd#1#2{\phi_{#2}^{\dag} (\vec #1)}
\def\ph#1#2{\phi_{#2} (\vec #1)}
\def\kon{\langle \bar{q} q \rangle_V}
\def\ddmu{\partial_\mu}
\def\dumu{\partial^\mu}
\def\ddnu{\partial_\nu}
\def\dunu{\partial^\nu}
\def\sld{\slashchar{\partial}}
\def\L{{\cal L}}
\def\A{{\cal A}}
\def\D{{\cal D}}

\def\V{{\cal V}}
\def\O{{\cal O}}

\def\K{{\cal K}}

\def\Sesp{S_{eff} (\sigma, \vec\pi)}

\def\Ssp#1{{S_{eff}} ^{#1} (\sigma, \vec\pi)}
\def\Se#1{{S_{eff}} ^{#1} }

\def\lpfi#1
 {{\cal L}_#1 (\psi , \bar{\psi} , {\cal F}) }

\def\qsch{{\tilde q}}
\def\kr#1{{#1 ^{\dag}}}
\def\id#1{\int{{d^4 #1}\over{(2\pi)^4}}}
\def\idk4{\int{{d^4 k_E}\over{(2\pi)^4}}}
\def\prop{\int\limits_{1 / {\Lambda^2}} ^\infty}

\def\pti2{\int\limits_1 ^\infty}
\def\el{\epsilon_\lambda}
\def\enu{\epsilon_\nu}
\def\eval{\epsilon_{val}}

\def\t#1{{{\tau^{#1}} \over 2}}

\def\str#1{#1 ^\prime}

\def\ln{{\rm ln}}
\def\det{{\rm det}}
\def\fm{{\rm fm}}
\def\signum{{\rm sign}}

\def\ax{A_\mu ^a}

\def\ges{{G_E}^{T=0} ({q}^2) }
\def\gev{{G_E}^{T=1} ({q}^2) }
\def\gpnq#1{g_{\pi NN} ({#1})}
\def\ga{g_A }
\def\gpn{g_{\pi NN}}

\def\R2{\langle R^2 \rangle}
\def\ms{\mu^{T=0}}
\def\mv{\mu^{T=1}}
\def\mp{\mu^{p}}
\def\mn{\mu^{n}}

\def\vap0{\bigg \vert _{\V = \A = \psi = \bar\psi =0} }
\def\erw#1{\langle #1 \rangle}
\def\so{\uparrow}
\def\su{\downarrow}

\def\romg#1{\uppercase\expandafter{\romannumeral #1}}
%

\def\dr3{\int {d^3 r}}

\def\kd0{\int {d k_0 \over (2\pi)} }

\def\det{\rm det}
\def\log{ {\rm log} }

\def\d4k{\int {d^4 k_E\over(2\pi)^4}}
\def\dk3{\int {d^3 k}}
\def\linie{\ \vrule height 14pt depth 7pt \ }
\def\xd4{\int {d^4 x}}
\def\dx3{\int {d^3 x}}
\def\dsl{\rlap{/}{\partial}}
\def\ksl{\rlap{/}{k}}
\def\ddmu{\partial_\mu}
\def\dumu{\partial^\mu}
\def\ddnu{\partial_\nu}
\def\dunu{\partial^\nu}

\def\ln{{\rm ln}}
\def\prop{\int\limits_{1 / {\Lambda^2}} ^\infty}

\def\pti2{\int\limits_1 ^\infty}

\def\dag{\dagger}

\def\beq{ $$ }
\def\eeq{  $$ }
\def\label{\EQN}
\def\Im{ {\rm Im} }
\def\Re{ {\rm Re} }


\titlepage
\topright{RUB-TPII-42/93}
\topright\thedate
\submit{Rep. Prog. Phys.}
\title{
\centerline{
BARYONS IN EFFECTIVE CHIRAL QUARK MODELS}
\centerline{
WITH POLARIZED DIRAC SEA}   }

\author{  \centerline{
        Th. Meissner $^{(b)*}$, E. Ruiz Arriola $^{(c)\$}$,
                                                             }
          \centerline{
        A. Blotz  $^{(a,b)\#}$ and K. Goeke  $^{(a)\dagger}$   }   }

\address{ \centerline{ ${(a)}$
         Ruhr-Universit\"at Bochum, Institut f\"ur
         theoretische  Physik  II}
         \centerline{
         D-44780 Bochum, Germany }       }

\address{ \centerline{  ${(b)}$
           Institute for Nuclear Theory, HN-12, University
            of Washington}
          \centerline{ Seattle, WA 98195, USA            } }

\address{ \centerline{  ${(c)}$
         Departamento de F\'{\i}sica Moderna,
         Universidad
         de Granada
                }
         \centerline{
         E-18071 Granada, Spain  }       }

\vfill
\lb
$^*$ Email:meissner@ben.npl.washington.edu             \lb
$^\#$ Email:andreeb@elektron.tp2.ruhr-uni-bochum.de     \lb
$^\dagger$ Email:goeke@hadron.tp2.ruhr-uni-bochum.de   \lb
$^\$ $ Email:earriola@ugr.es  \lb
\eject
\abstract{
In this paper concepts and numerical calculations are reviewed in which
baryons are considered as many body systems composed of three quarks
coupled to a polarized Dirac sea. The interaction is the one of
the Nambu--Jona-Lasinio model
and of local  and biquadratic form in a way which allows for
spontaneous chiral symmetry breaking. The models are treated explicitely
in the 0-boson and 1-quark loop approximation (Hartree mean field
approach), where the real part of the effective Euclidean action
is gauge invariantly
regularized.
In the  meson sector the parameters of the models, including a cut-off
for the Dirac sea, are fixed to PCAC, on-shell meson masses and
decay
constants. Furthermore the regularization schemes are chosen to get good
values for
vacuum condensates and  current masses leaving the constituent mass M
as only free parameter.
The review is focussed on the evaluation of baryon properties.
Assuming a spherically symmetric hedgehog ansatz for the quark fields
the solitonic solutions
of the system are obtained in a selfconsistent way by solving the
classical equations of motion.

After a semiclassical quantization procedure the solitons can be related
to the physical states of the spin $1/2$ and $3/2$ multiplets.
In this way observables and form factors can be calculated. Various
types of quark fields and quark-quark couplings are investigated and
reviewed.
These are in SU(2) given by  sigma and pion fields and in SU(3)
they are complemented by
kaon and eta-fields. In addition SU(2) vector mesons
of $\rho$, $A_1$ and $\omega$-type are considered as well. For these
models  relevant observables and form factors of the nucleons are
calculated. Some results on the other members of the baryon octet and
decuplet are reviewed as far as they are available and of theoretical
and experimental interest. For  scalar and pseudoscalar
couplings a clear picture emerges corresponding to localized valence
quarks coupled to moderately polarized sea quarks and being separated
from them by a finite energy gap. Using heat kernel and gradient
expansion techniques the models of the Nambu-Jona-Lasinio type can be
related to fully bosonized approaches of the Skyrme type, on the one
hand, and to
the chiral sigma model, representing valence quarks and dynamical
meson fields, on the other.     }

\endpage

%
%

\chap{Introduction}

%
\FIG\f11{Free Dirac spectrum for {\it current quarks} ({\it chiral symmetry
spontanously unbroken})}
\FIG\f12{Free Dirac spectrum for {\it constituent quarks} with mass $M$ ({\it
chiral symmetry spontanously broken})}
\FIG\f13{Dirac spectrum for the soliton. The bound valence
quarks $\epsilon_{val} >0$
polarize the Dirac sea weakly ({\it valence quark picture}).}
\FIG\f14{Dirac spectrum for the soliton. The bound valence quarks
$\epsilon_{val} <0$
polarize the Dirac sea strongly ({\it fully bosonized picture}).}
\TAB\tabmass{The current quark masses and corresponding charges
from  the  QCD vacuum  (Gasser and Leutwyler
1982) }

The present article deals with the description of nucleons and hyperons
by means of relativistic chiral effective models. The prominent
model discussed in detail is the Nambu-Jona-Lasinio (NJL) approach. It
incorporates certain  features of the Quantum Chromodynamics (QCD) and
is related to low energy strong interaction phenomena.

Quantum Chromodynamics  is generally considered to be the
proper theoretical framework for the description of structures and
reactions being dominated by the strong interaction. QCD is a SU(3)
colour gauge theory combining quarks of $N_f$ different flavours
(u,d,s,c,b,t)
as fermions and gluons (g) as gauge fields. Its Lagrangian reads
\beq  {\cal L}_{QCD} = \sum_\imath {\bar q}_\imath \left( i \gamma^\mu
          D_\mu
        -m_\imath
    \right)  q_\imath -{1\over 4} G_{\mu\nu}^a    G^{\mu\nu}_a
\eeq
with the covariant derivative
\beq D_\mu = \partial_\mu + i g A_\mu^a \half \lambda^a \eeq
and the field strength tensor
\beq    G_{\mu\nu}^a  = \partial_\mu A_\nu^a - \partial_\nu A_\mu^a
     - g f^{abc} A_\mu^b A_\nu^c  \eeq
Here $\lambda^a$ are the generators of $SU(3)_c$ (Gell-Mann matrices)
the $f_{abc}$
an the structure coefficients of $SU(3)$ and $m_j$ are the current
masses
of the quarks. The basic features of QCD are (Donoghue et al. 1992) the
following being all related to symmetry properties:

$\underline{\rm Universality}$: There is only one coupling
constant g for
all interactions among quarks and gluons. This is a direct consequence
of the local gauge invariance under the $SU(N_c)$-group, with
$N_c$ = number of colours.

$\underline{\rm Asymptotic\phantom{a}freedom}$:  At very high
energies quarks behave
as free particles. In this region the effective coupling constant
becomes  small and perturbation  theory may be applied. This
property has been observed  experimentally.

$\underline{\rm Confinement}$:  Up to now neither  free quarks  nor
gluons have
been observed outside the volume occupied by a hadron. This feature has
not fully been explained yet by theoretical reasoning. The idea is that
only colourless systems are stable.

$\underline{\rm Chiral\phantom{a}Symmetry}$: In the limit of
vanishing quark
masses, the QCD Lagrangian is invariant under the chiral  $SU(N_f)_R
\otimes~SU(N_f)_L$ group of global transformations
\beq  q(x) \rightarrow \exp{(i\alpha^a\lambda^a/2)} q(x) \eeq
\beq  q(x) \rightarrow \exp{(i\gamma_5\beta^a\lambda^a/2)} q(x) \eeq
where $a=1,\dots,N_f^2-1$, so that the corresponding conserved currents
read
\beq V_\mu^a(x) = {\bar q} \gamma_\mu \half \lambda^a q(x) \eeq
\beq A_\mu^a(x) = {\bar q} \gamma_\mu \gamma_5 \half \lambda^a q(x) \eeq
In the absence of strong interactions the quarks have a current mass of
electroweak origin which breaks chiral symmetry explicitly. That means
\beq \partial^\mu A_\mu^a=i {\bar q}(x)\{ {\hat m},\lambda^a \} q(x)
\eeq
where the current quark mass matrix ${\hat m}$ is defined as ${\hat
m}=diag(m_u,m_d,m_s,m_c,m_b,
m_t)$.
The attributed values are (Gasser and Leutwyler 1982)
given in Tab. 1.1.

The chiral symmetry is well realized with respect to up- and
down-quarks
because of their small current mass ($\simle 10\MeV$). The symmetry is
spontaneously broken resulting in a chiral condensate $<{\bar
u}u+{\bar d}d>^{1/3}=-(283\pm 31)\MeV$
(Gasser and Leutwyler 1992). This gives the up- and down-quarks a
dynamically generated constituent mass of about $350-450\MeV$. For
those
quarks apparently the spontaneous breaking of chiral symmetry dominates
over the explicit breaking caused by  finite current masses. For
heavy
quarks (c,b,t) the situation is rather opposite. The strange quark s
lies in an intermediate region and its relevance for low-energy
non-strange physics is at present under debate.

Finally let us mention that in the presence of external fields chiral
symmetry is also broken because of quantum corrections leading to the
chiral anomaly (Adler 1969, Bell and Jackiw 1969). Vector current
conservation implies an anomaly in the divergence of the axial current
(Bardeen 1069), which predicts successfully the decay rate for the
process $\pi^o\rightarrow~2\gamma$ if $N_c=3$.

$\underline{\rm Scale\phantom{a}Invariance}$:  In the chiral limit
($m_j=0$ for all
flavours) the QCD Lagrangian does not contain any dimensional
parameters. Thus it is invariant  under the scale transformation
$q(x)\rightarrow\lambda^{-3/2}q(\lambda x)$.  The
corresponding dilatational   current is not
conserved because of quantum corrections leading to the trace anomaly.

The QCD has been treated in a clean way only for high energy processes,
because there the asymptotic freedom (Politzer 1974, Gross und Wilczek
1973a,1973b)  facilitates      the solution tremendously
(perturbative regime). For
low energies, in particular the structure of hadrons, the gluon
self interaction $\sim~(gf^{abc}A_\mu^bA_\nu^c)^2$
creates non-linearities and causes
enormous problems (non-perturbative regime). Here the Lagrangian as
such has only been treated in a regularized way on a 4-dimensional
lattice in the form of lattice gauge theory (Wegner 1971, Wilson 1974).
Although these calculations are potentially exact they are practically
hampered  by technical and conceptual problems associated with the
choice of large current masses, fermion doubling, small size of
lattices, etc. In addition calculations of this sort require extreme
amounts of computer  time (Creutz 1983). Because of this dilemma
in the non-perturbative regime effective theories are very much in use.
They explicitely treat those degrees of freedom, which are relevant for
low energy structures, and ignore the others or parametrize them in
form
of coupling constants. This only makes sense if those low energy degrees
of freedom are clearly separated from the high energy ones. However,
such a question can  be decided only by an inspection of the low energy
phenomenology, which to some extend has been done already long before
the advent of QCD. Of course for such a limited theory one must not
expect that all properties of QCD, discussed above, are also properties
of the effective models. Hence the guideline for the effective models is
the necessity  to reproduce low energy phenomena  and to reflect
{\it some} of the QCD symmetry properties.

There are some basic phenomena of low energy hadronic physics which over
the many years of experience with effective models have been identified
as indispensible for the construction of an effective theory for baryon
and meson structure (Cheng and Li 1984). It is clear that the pion field
plays a dominant role in all hadron physics. This goes back to Yukawa's
original idea that nuclear forces are mediated by pions and corresponds
to the fact that basic feature of nuclear structure at low and
intermediate energies can successfully be explained in turn of nucleons
and mesons instead of their constituents quarks and gluons, as they are
identified at high energies. The pion is closely connected to the chiral
condensate. In fact it is the most prominent Goldstone-boson (Goldstone
1961, Nambu 1960) of the spontaneously broken chiral symmetry.

The pion decays via the axial current $A_\mu^a(x)$ and its decay
constant
$f_\pi=93\MeV$ is defined as (Cheng and Li 1984)
\beq <0\mid A_\mu^a(0) \mid \pi^b (p) > = i f_\pi p_\mu \delta^{ab}
\eeq
For hadrons  the axial current $A_\mu^a(x)$ is partially conserved due
to the
small but finite current masses of the up- and down-quarks. According to
the hypothesis of the partial conservation of the axial current (PCAC)
(Nambu
1960, Chou
1961, Gell-Mann and Levy
1960)
\beq \partial^\mu A_\mu^a (x) = f_\pi m_\pi^2 \pi^a (x) \eeq
with $m_\pi=139\MeV$ being the mass of the pion field $\pi^a$. The
iso-vector current $V_\mu^a(x)$ is conserved unless one explicitely
attributes  different current masses to up- and down-quarks.
Both
currents allow the formulation of chiral charges (Cheng and Li 1984)
\beq Q_V^a (x) = \int d^3x V_0^a (x) \eeq
\beq Q_A^a (x) = \int d^3x A_0^a (x) \eeq
and a corresponding charge algebra, which exists even if the currents
are not fully conserved:
$$ \eqalign{
        \bigl[ Q_V^a (x) , Q_V^b (x) \bigr] &= i f^{abc} Q_V^c (x) \cr
        \bigl[ Q_V^a (x) , Q_A^b (x) \bigr] &= i f^{abc} Q_A^c (x) \cr
        \bigl[ Q_A^a (x) , Q_A^b (x) \bigr] &= i f^{abc} Q_V^c (x)
    \cr} $$
The relevance of this commutator algebra lies in the fact that several
low energy theorems and sum rules involving strong and electro-weak
processes can be derived and confronted with experimental data. A
typical example is the Goldberger-Treiman relation (Goldberger and
Treiman 1958), which is experimentally fulfilled to a precision of
$7\%$:
\beq g_A M_N = f_\pi g_{\pi NN}  \eeq
Here $M_N=938\MeV$ represents the nucleon mass, $g_A=1.25$ is the axial
coupling constant of the neutron beta-decay and $g_{\pi NN}=13.6$ is the
pion-nucleon-nucleon coupling constant. In general the violation of the
chiral symmetry is less than $10\%$ and it is indeed after the isospin
symmetry the best known symmetry of strong interactions at low energies.
It is clear
that any effective theory of hadron has to incorporate the spontaneously
broken chiral symmetry and the corresponding {\it Goldstone} boson, i.e.
pions in $SU(2)$ and in addition the  kaons and the  eta
in
$SU(3)$.

Another fruitful idea in the study of electromagnetic properties of
hadrons is given by the vector meson dominance  model, which
assumes that
photons do not interact directly with hadrons but rather through the
virtual propagation of an intermediate massive vector meson state
(Sakurai 1960,1969).  In quantum field theory such an idea can be
formulated through the so-called current field identities
(Kroll et al. 1967)
  \beq J_\mu^{I=0} = {m_\omega^2\over g_{\omega NN} }
         \omega_\mu (x),\phantom{abcde}
       J_\mu^{I=1} = {m_\rho^2\over g_{\rho NN} }
         \rho_\mu^0 (x)        \eeq
for the isoscalar and isovector components of the electromagnetic
currents. Here the vector meson masses $m_\omega=783\MeV$ and $m_\rho =
770
\MeV$  appear together with the $\rho$  and $\omega$ exchange coupling
constants of
NN scattering. In most cases, the predictions of the {\it vector meson
dominance model} (VDM) have been
very satisfactory (see e.g. Gourdin 1974). The most famous example is
provided by the
prediction for the mean squared electromagnetic pion  radius
$$ \eqaligntag{
        <r^2>_\pi^{em} &=  {6 \over m_\rho^2}=0.39 {\rm fm^2}
&\EQ\radii
\cr}
$$
which compares reasonably well with the corresponding experimental
value of $(0.44\pm~0.01){\rm fm^2}$.

Finally, let us mention that the union of current algebra with
current-field identities leads to the field algebra whose most prominent
result is the KSFR relation (Kawarabayashi and Suzuki 1966, Riazuddin
and Fayazuddin 1967)
      \beq 2 g_{\rho\pi\pi}^2  f_\pi^ = m_\rho^22 \eeq
with $g_{\rho\pi\pi}=5.48$ being the decay constant of the strong
process
$\rho\rightarrow\pi\pi$.
The accuracy of this relation is $2\%$.

As soon as strange quarks are involved one has the Gell-Mann Okubo mass
formula (Gell-Mann, 1962; Okubo 1962), which deviates from the
experiment by less than
$3\%$:
\beq
\underbrace{ \phantom{1\over 1}
  M_\Sigma -M_N}_{254\MeV}  = \underbrace{{1\over 2} (M_\Xi -
M_N)}_{190\MeV}
+
\underbrace{{3\over
4}
(M_\Sigma
-    M_\Lambda )}_{58\MeV}  \eeq
and the Coleman-Glashow formula  (Coleman and Glashow, 1961)
\beq  \underbrace{M_n-M_p}_{1.3\MeV}
+ \underbrace{M_{\Xi^-}-M_{\Xi^0}}_{6.4\pm 0.6\MeV}
=
\underbrace{M_{\Sigma^-} -M_{\Sigma^+}}_{8\pm 0.8\MeV}
\eeq
which fits perfectly with experimental data and
which was originally attributed to the electromagnetic interactions
between the quarks.

An effective theory of low energy QCD degrees of freedom in the
baryonic sector can only be successful if it reproduces the above
phenomenological
features. Therefore we concentrate in this article on models showing
spontaneously broken $SU(N_f)_L\otimes SU(N_f)_R$ symmetry
with a possible explicit
breaking due to finite current masses. In some cases we will
also consider  scale invariance.  We consider in detail  the
Nambu-Jona-Lasinio model (Nambu and Jona-Lasinio 1961a, 1961b) and will
relate it to all others.  The theory will be explained for scalar and
pseudoscalar quark-quark interactions in the light quark sector (up,
down). Generalisations towards  vector interactions and/or SU(3)
will be
done in separate chapters. Thus the NJL-Lagrangian studied in detail is
\beq {\cal L}_{NJL} =
{\bar q} (i\gamma^\mu\partial_\mu - m_0 ) q +
      {G\over 2} \left[ ({\bar q}q)^2 + ({\bar q}i\gamma_5{\vec\tau}q)^2
\right]  \eeq
with q having an up- and down-component  and $m_0=diag(m_u,m_d)$. In
most
cases we will assume $m_u=m_d$. The NJL-model is distinguished because
it allowes to build conceptual bridges to most of the currently used
effectives models for baryon structure. In such a way these models and
also the NJL model can be better understood.

There are by now several attempts to relate the NJL to some low energy
limit of QCD. One has been given by Diakonov and Petrov (1986) assuming
the QCD vacuum to be an instanton liquid. By variational field theoretic
means they obtain after various plausible approximations the zero-boson
and one-quark-loop approximation to the NJL-Lagrangian.
This procedure is based on previous works by t'Hooft (1976a,1976b),
Callan, Dashen  and Gross (1978,1979a,1979b),
Carlitz and Cremer (1979) and Shifman,
Vainstein and Zakharov (1980a,1980b),  see also Shuryak (1986).
Ball (1987) on the other hand integrates out the gluon fields in some
certain approximation and obtaines in the end the NJL-Lagrangian in
$SU(N_f)_R\otimes SU(N_f)_L$
including vector mesonic couplings. Schaden et al. (1990) formulate QCD
in terms of path integrals over the field
tensor rather than the vector potential. At a certain level they perform
stationary  phase approximations and end up, after several
further
plausible arguments, at the NJL-model. Cahill, Roberts and others
(Cahill and Roberts, 1985; Roberts et al., 1988; Cahill, 1992) introduce
bilocal
meson and diquark fields into a functional form of QCD. The NJL model
appears as some pointlike  approximation. There are also steps in a
similar direction by Dhar and Wadia (1984, 1985), McKay and Munczek
(1985), Adrianov (1985) and Chanfray et al. (1991).
 One should note that
none  of these theories is really a clean cut derivation of NJL. For
example most of them are not able to say where in the chain  of
arguments and approximations on the way from QCD to NJL the confinement
has been lost, nor can they say how the NJL-model can be improved.

Actually the effective chiral models for baryon structure can roughly be
classified in three  groups: \lb
\item{i)} Chiral quark  models: Here the
system
is dominated by three valence quarks coupled to a dynamical  pion. In
die
chiral MIT-model (Chodos and Thorn, 1975) the quarks are confined by a
bag with infinite walls and chiral symmetry is preserved by a dynamical
pion field and the continuity  of the axial current at the
bag's surface. The linear versions  of these models can be related
to the
cloudy bag model (Theberge et al. 1980; Thomas AW 1983).  \lb
In the chiral $\sigma$-model (Gell-Mann and Levy
1960) we do
not have confining walls but a sigma  field with finite vacuum value.
Here the valence quarks are localized by binding forces exerted by the
sigma- and pion field (Birse and Banerjee 1984,1985, for a review see
e.g. Birse 1990 and references therein). The baryon number is carried by
the three valence
quarks. Altogether these models can be characterized by {\it valence
quarks and meson fields}.  \lb
\item{ii)} Skyrme models: In these models (Skyrme 1961, Adkins et
al
1983,
Zahed and Brown 1986, Holzwarth and Schwesinger 1987) one does not deal
with explicit quarks but solely with dynamical meson fields. They are
collectively quantized in order to get the proper quantum numbers of
e.g. a nucleon. The formalism is comparatively simple and hence these
models enjoy great popularity. Conceptually they are based  on the
large
$N_c$-expansion of the QCD ('t Hooft 1973, Witten 1979b). This would
require
an infinite set of boson fields which is in practice replaced by the
pion field in the simplest form, and by vector mesonic fields in more
recent versions (Meissner U G 1988). The baryon number is hidden in the
topology of the Goldstone fields. At the center of the baryon the
Goldstone fields must show a winding number, which is to be identified
with the  baryon number. These models can be characterized by {\it
meson fields and topology}.  \lb
\item{iii)}
 Nambu-Jona-Lasinio models: As described above one deals solely with
quarks without dynamic meson fields (Nambu and Jona-Lasinio 1961a,
1961b). Valence quarks appear in a natural way as bound, discrete and
localized states  (Kahana and Ripka 1984). Mesons enter the theory
only
as non-dynamic auxiliary fields  generated by the polarization of the
Dirac sea (Eguchi 1975, Kleinert 1976). In fact these  models can
be characterized by {\it valence quarks and sea quarks}.  \lb
\item{iv)}
    With respect to the above characterization there are some hybrid
models which combine certain features. This  is the chiral bag of the
Stony Brook group (Brown and Rho 1979). There the pion field outside the
bag shows some topological properties and carries hence a fractional
baryon number. The linear chiral sigma model of Kahana and Ripka (1984)
involves sea quarks and dynamic pion- and sigma fields and requires
renormalization: In fact these authors are the first to treat the
polarization of the Dirac sea in this context  (Ripka and Kahana
1987,
Soni 1987, Li et al. 1988). It shows, however, an instability
and hence has never been applied to nucleon properties.
Recently it was shown by Kahana and Ripka (1992) that the introduction
of vector mesons in this  model can cure the vacuum instability, though
no numerical calculations for nucleons were performed yet.

It is interesting to note, that the NJL-Model (valence- and sea-quarks)
can be directly related to the chiral bag models (valence-quarks and
mesons) and the Skyrme-models (topology and mesons).
In order to understand this, the figures 1.1-1.4 may be helpful. Fig.
1.1 shows schematically the single particles energies of the chirally
symmetric plane wave vacuum as it emerges from the NJL-model in the
0-boson and 1-quark loop approximation. We concentrate on up- and
down-quarks and hence a small gap of  $2m_0=m_u+m_d$ exists between
the
upper and lower continuum. In a plausible Fock-state picture the lower
continuum is occupied by $N_c=3$ quarks for each flavour in each single
particle level, as
indicated by crosses. The spontaneously broken chiral vacuum is shown in
Fig. 1.2. There the gap is 2M and the single particle states are again
of plane wave nature with the constituent mass M. The vacuum in both
cases is characterized by the baryon number $B=0$. The soliton for a
'small' constituent mass M is shown in Fig. 1.3. A single particle
state
from the positive continuum has entered the gap and has become a bound
state. It is explicitely occupied by $N_c=3$ valence quark and hence the
baryon number of the system has increased from $B=0$ to $B=1$. The
interaction of the continuum state with the valence quarks causes a
polarization of the single particle states of the positive and negative
continuum which is indicated by wavy lines. For  large constituent
masses
the bound state has moved further down, see Fig. 1.4, and approaches
the
Dirac sea. In such a case both continua are strongly polarized and the
Dirac sea incorporates another level and hence acquires the baryon
number $B=1$. The scenarios  of Fig. 1.3 and Fig. 1.4 are distinguished
by the values of the constituent mass M. As we will see for
sigma and  pionic
couplings in SU(2) as well as in SU(3)  the scenario of Fig. 1.3 is
the one which reproduces the relevant nucleon and hyperon observables.
If vector mesons are included the picture is not yet settled.

The formal techniques to relate the NJL model to the bag model and to
the Skyrme model are  the gradient expansion (Aitchison and Frazer
1984,
1985a and 1985b) the heat kernel expansion (Gasser and Leutwyler 1984,
Ebert und Reinhardt  1986), or the expansion of Chan (1985) all applied
to the fermion determinant of NJL. Actually these expansions are well
converging  for smoothly varying and largely extended boson fields
or
quark wave functions which is not always fulfilled for realistic
systems. Nevertheless, if one uses only a certain number of expansion
coefficients in the expansion of the effective action of the scenario of
Fig. 1.3,  the Lagrangian of Gell-Mann and Levi (1960) is obtained.
There the
baryon number is carried by the valence quarks and the polarization of
the Dirac sea creates dynamic meson fields whose possible topology does
not carry a baryon number. If one considers the scenario of Fig. 1.4
and terminates the expansion of the action in an appropriate way and if
one ignores certain destabilizing terms then a Skyrme type Lagrangian
(Skyrme 1961) is obtained. The highly polarized Dirac sea including the
additional valence level is described in terms of a dynamic pion field.
This exhibits a winding number to be identified with the baryon number.
One has to consider these relations between the models with some care:
First, the expansions are often badly converging. Second, a gradient
expansion of an observable of NJL does not necessarily agree with the
observable evaluated from the gradient expanded NJL Lagrangian.
Nevertheless the NJL model is extremely helpful in understanding the
meaning   of valence quarks and topology the concepts of which are the
basic features of all presently used   effective chiral models.

There are two other branches relevant in the context of the NJL-model,
one is the meson sector with baryon number $B=0$, the other is the
evaluation of properties of a hot and dense medium and the embedding of
a baryon in it. The meson sector has been the object of intensive
studies in the last years and several review articles have been written
on it (Vogl and Weise 1991, Klevansky 1992). We will discuss the meson
sector in this paper only as far it is needed to explain the appearance
of the chiral condensate and to adjust the parameters of NJL. The medium
calculations are not discussed in the present article at all. The
interesting point is that with increasing density and temperature of the
medium the chiral symmetry of the system is restored. The reader is
referred to   Hatsuda and Kunihiro (1985, 1987a, 1987b,
1988), Meissner U G (1989a, 1989b), Bernard V et al. (1986, 1987),
Jaminon
et al. (1989, 1992), Ruiz Arriola E et al. (1990), Christov C V
et al. (1990a,
1990b), Lutz and Weise (1991) and  Hatsuda and Kunihiro (1994).
\vfill\eject

%
%
%
\overfullrule=0pt

\def\op#1{\hat{#1}}
\def\bra#1{\langle #1 \,\vert}
\def\ket#1{\vert\, #1 \rangle}

\def\nl{\hfill\break}
\def\({\Bigl(}
\def\){\Bigr)}
\def\[{\Bigl[}
\def\]{\Bigr]}
\chap{The Nambu--Jona-Lasinio-Model With SU(2)-Flavor: \nl
Vacuum and Mesonic Sector}
\TAB\taband21{
               The cutoff $\Lambda$, the $\mu^2$ parameter, the
original NJL coupling G, the quark condensate ${\bar u}u$,
the current quark mass $m_0$ and the vacuum energy density for
the non-covariant (O(3)), covariant (O(4)), single (PT1) and double
(PT2) step proper time
and the Pauli-Villars (PV) regularization, defined in  Sect. 2 and
Sect.  6.}
\FIG\f21{ The cutoff $\Lambda$ for the
the non-covariant (O(3)), covariant (O(4)) and  single (PT)
step proper time regularization as a function of the constituent
quark mass. }
The purpose of this chapter is to present the \NJL
(NJL) (Nambu and Jona-Lasinio 1961) in the
simplest realistic case, i.e. with scalar and pseudoscalar couplings.
To begin with, we study the classical Lagrangian and its symmetries.
The bosonization procedure is presented by means of path integrals and
the stationary phase approximation is defined.
We extract the ultraviolet structure of the model and
present some different regularization schemes.
The gradient and heat kernel expansions are discussed as well as
the formalism for extracting vacuum and mesonic properties.
The parameters of the model are fixed in order to reproduce known
properties of the  mesonic sector.
Different ways of parameter fixings are introduced and the
corresponding numerical results are shown.

\sect{The Semibosonized NJL - Effective Action }

\subsec{Classical Lagrangian - Symmetries }

The simplest version of the \NJL
is
represented by the following Lagrangian
$$ {\cal L}_{NJL}= {\cal L}_{kin} + {\cal L}_{br} + {\cal L}_{int}
\EQN\njlsu2 $$
where the kinetic, interaction and chiral breaking mass terms are given by
$$ \eqalign{
& {\cal L}_{kin} = \bar{q}(x) i\slashchar\partial q(x) \cr
& {\cal L}_{br} = - \bar{q}(x) {\hat m} q(x) \cr
& {\cal L}_{int} = {G\over 2} \[ \(\bar{q}(x) q(x)\)^2 +
\(\bar{q} (x) i\gamma_5 {\vec \tau}q(x) \)^2 \]  \cr }
\EQN\e22 $$
respectively. Here $q(x)$ denotes a quark field with u and d flavors
and $N_c$ colors. The G is the coupling constant with dimensions of
length
squared. The $\hat m$ represents the current quark mass matrix given by
$$
{\hat m} = \left( \matrix{ m_u  &  0  \cr
                            0   & m_d } \right)=
                            m_1 {\bf 1} + {\tau_3} \Delta m
\EQN\e23 $$
where the average and difference masses   are defined as
$$ m_1 = {1\over 2} ( m_u + m_d ), \qquad
\Delta m = m_u - m_d
\EQN\e24 .  $$
Their attributed theoretical values at a scale of 1GeV
are (Gasser and Leutwyler 1982) given in \qutab{\tabmass}.
In the following we will assume the quark mass degeneracy $m_u = m_d
=m_1$, corresponding to  exact isospin symmetry.

At the classical level the total Lagrangian is invariant under the
global $U(1)$ transformation
$$ q(x) \to q'(x)= \exp (i\delta) q(x)
\EQN\e26 $$
The kinetic and interacting terms $\L_{kin}$ and $\L_{int}$
are also invariant under the $SU(2)_V
\otimes SU(2)_A$ global chiral group
$$ \eqalign{ & q(x) \to q'(x)= \exp (i \vec\tau\cdot\vec\alpha) q(x) \cr
             & q(x) \to q'(x)= \exp (i \gamma_5 \vec\tau\cdot\vec\beta)
q(x) \cr}
\EQN\e27 $$
These transformation properties generate the following Noether currents
(Cheng and Li 1984)
$$ \eqalign{
& B_\mu (x) = \bar q(x)\gamma_\mu q(x),
\qquad {\rm ( baryon \, current)} \cr
& \vec V_\mu (x) = \bar q(x) \gamma_\mu {\vec \tau\over 2} q(x),
\qquad {\rm ( vector \, current)} \cr
& \vec A_\mu (x) = \bar q(x) \gamma_\mu \gamma_5 {\vec \tau\over 2}
q(x), \qquad {\rm ( axial \, current)} \cr }
\EQN\e28 $$
whose divergences  on the classical level  are given by:
$$
\partial^\mu B_\mu = 0, \qquad  \partial^\mu \vec V_\mu = 0,
\qquad \partial^\mu \vec A_\mu = {i\over 2}
\bar q \{   {\hat m}, \vec\tau \} q
\EQN\e29 $$
corresponding to the conservation of the baryon number and isospin and
the partial conservation of the axial charge respectively.
Actually , even for ${\hat m}=0$ the axial current is not conserved
on the quantum level
in the presence of external vector and axial vector fields
reflecting the chiral anomaly (Wess and Zumino 1971, Wess 1972).
Finally let us mention that the total
Lagrangian shows a $SU(N_c)$ global invariance, leading to the
conservation of the color current.
Since the four fermion interaction in \qeq{\njlsu2} is of color singlet form,
this conservation does
not have dynamical implications.
However, there are other versions of the model which
involve color octet-octet interaction in the vector and axial channels
(Klimt et al. 1990 ; Takizawa et al. 1990) which are formally
equivalent to the Lagrangian \queq{\e22 } after a Fierz
rearrangement.
Those have been discussed at length in other reviews ( Vogl and Weise
1991; Klevansky 1992) and will not be considered here specifically.

\subsec{Collective Boson Fields}

Like the Fermi theory of weak interaction the Nambu--Jona-Lasinio model
is non renormalizable,
because the coupling constant of the 4-fermion interaction $G$ has the
dimension $ [G]= {\rm mass}^2$ (Itzykson and Zuber 1980, Cheng and Li 1984).
This means that with each increasing order in $G$ a new graph with
a higher degree of ultraviolet divergence appears.
In order to get a well defined theory
it is therefore necessary to
specify  how the infinities of the model have to be treated.

in which order the model has to be handled.

For our later purposes it will be convenient
to use a {\it semibosonized form} of the
4-fermion interaction \qeq{\njlsu2}.
The idea of {\it bosonization}
has first been formulated in solid state physics, where it is called
Hubbard-Stratononovich transformation (Negele and Orland 1987) and
has been applied to the present theory by Eguchi (1976), Kikkawa (1976)
and Kleinert (1978).
It consists in resummarizing the graphs of the 4-fermion
point interaction into a {\it quark-meson} interaction of the Yukawa type
by introducing collective scalar-isoscalar ($\sigma$) and
pseudoscalar-isovector ($\piv$) background fields, which carry the
 quantum numbers of the interaction channels $({\bar{q}} q)$ and
$({\bar{q}}i\vec\tau\gamma_5q)$,
respectively.
To this end we insert the functional identity:
$$
1=\int \D \sigma \D \vec\pi
 exp \biggl \{ -i\int d^4 x{{\mu^2}\over 2}
\Bigl [
\bigl ( \sigma + {g \over  \mu^2} (\bar{q}q - {m_0 \over G})\bigr )^2
+
\bigl ( \vec \pi    + {g \over  \mu^2} \bar{q} i
\vec\tau\,\gamma_5  q \bigr )^2
\Bigr ]
 \biggr \} \EQN\e210
$$
into the generating functional $Z_{NJL}$:
$$
Z_{NJL} = \int \D q
\D \bar q
e^{i \int d^4 x \Bigl \{
\bar q (i \sld - m_0) q + {G\over 2} \bigl [ (\bar q q)^2 +
(\bar q  i \gamma_5 \vec\tau q)^2 \bigr ] \Bigr \}  }
\EQN\e211
$$
and obtain:
$$
{\str {Z_{NJL}}} =\int \D \bar q  \D q \D \sigma \D \vec\pi
 e^{ i \int d^4 x {\str{{\cal L}_{NJL}}} (x)  }
 \EQN\e212
$$
with
$$
\eqalign{
{\str{{\cal L}_{NJL}}}&= {\bar{q}} iD q
- {\mu^2\over2}
(\sigma^2+\vec\pi^2) + {m_0\mu^2\over g} \sigma  \cr
iD &= i \sld -g(\sigma+i\vec\pi \vec\tau\,\gamma_5)   \cr  }
\EQN\e213
$$
and:
$$
G={g^2 / \mu^2}
\EQN\e214
$$
Here $g$ and $\mu$ are newly introduced parameters which arise due to
the present form of bosonization.   Their values will be fixed in sect.
2.6..

We want to stress that on this stage the fields $\sigma$ and $\piv$ are
non-dynamical collective background fields and no kinetic term
$\half \left [ \ddmu\sigma \dumu\sigma + \ddmu\piv \dumu\piv \right ]$
appears on the classical level.

 The semibosonized Lagrangian ${\str{\L}}_{NJL}$  remains invariant
under the
chiral transformation of  $SU(2)_R\otimes~SU(2)_L\otimes~U(1)_B$,
if
$(\sigma,\piv)$ transforms
under the $(\half,\half)$ representation of this group (Cheng and Li
1984). Hence we demand:
$$
\eqalign{
\sigma &\to \sigma + 2 \vec\beta \vec\pi \cr
\vec\pi &\to \vec\pi - 2 \vec\beta \sigma -
2 \vec\alpha  \times  \vec\pi  \cr  }
\EQN\e215
$$
\qeq{\e215} is consistent with the equations of motion for
${\cal L}^\prime$

\subsec{Grassmann-Integration - Effective Chiral Action}

The quark contribution ${\bar {q}} iD q$ in \qeq{\e213} is bilinear in the
quark fields $q$ and $\bar{q}$ so that the functional Grassmann integration
$\int \D \bar q \D q$ can be performed analytically. After
transformation to Euclidean space time (cf. appendix A) we get:
$$
{Z_{NJL}} ^{{\prime}{\prime}} = \int \D \sigma \D \vec\pi e^{- \Sesp}
\EQN\e216 $$
The {\it effective chiral action} $\Sesp$ contains the {\it fermion
determinant} $\Ssp F$
as well as an mesonic mass term $\Ssp {mes}$ and the
chiral breaking term $\Ssp {br}$:
$$
\eqalign{
\Sesp &= \Ssp F + \Ssp {mes} + \Ssp {br} \cr
\Ssp F &= -\ln\det (-iD) = -\Sp\ln (-iD) \cr
\Ssp {mes} &= {{\mu^2} \over 2} \int d^4 x_E (\sigma^2 + \vec\pi ^2) \cr
\Ssp {br} &= - {{m_0 \mu^2} \over g} \int d^4 x_E \sigma \cr}
\EQN\e217
$$
$\Sp$ denotes the total trace in functional as well as matrix space:
$$
\Sp \A = \int d^4 x \Tr_{\gamma \tau c} {\bra x} \A {\ket x} =
\int d^4 k \Tr_{\gamma \tau c} {\bra k} \A {\ket k}
\eeq

Generally the fermion determinant in Euclidean space time has a real as
well as an imaginary part, reading:
$$\Re \Ssp F = (-) \half \Sp \ln (D^{\dag} D)
\vert
\EQN\e217a
$$
and:
$$
\Im \Ssp F = (-) \half \Sp \ln \left( {D\over D^{\dag}} \right)
\EQN\e217b
$$
For time independent meson fields $\sigma$ and $\piv$ it will be shown in
chapter 3, that the imaginary part $\Im \Ssp F$ vanishes, because the $1$-
particle hamiltonian $h$ is hermitian. This changes if time-like
vector mesons (e.g. $\omega$ mesons)  are coupled to the Lagrangian
destroying the hermiticity of $h$. In some cases the imaginary part is
related to the chiral anomaly. Details will be discussed in chapter 7.
In addition imaginary parts of the action occur if the system is
considered in the rotating frame.

\sect{Constituent Quark Mass - Stationary Phase
Approximation}

The  main observation of Nambu and Jona-Lasinio (1961) in their
original
work was that the four fermion interaction generates a dynamical mass
for the fermions if the coupling constant is bigger than a certain
critical value. This was done in the canonical formalism using a
Bogoliubov-Valantin transformation from bare massless quarks to
constituent massive quarks. The idea is that if the interaction is
strong enough, the vacuum lowers its energy creating a
mass gap between the positive and
negative energy continua of the Dirac spectrum. Such a phenomenon is
called dynamical mass generation.  In the following sections we will
see explicitly how a dynamical mass  generation  takes place within the
path-integral approach to the NJL model.

\subsec{Spontaneously Broken Chiral Symmetry}

 From \qeq{\e213} one recognizes at once that for a finite vacuum expectation
value $\sigma_V$ of the $\sigma$-field the quarks acquire a finite
{\it dynamical} or {\it constituent mass} $M$ due to:
$$
M = {{d^2    {\str{{\L}_{NJL'}}}}\over { d     \bar q d    q}}
\Big\vert_{V }=
g \sigma_V
\EQN\e219
$$
which means that  chiral symmetry is {\it spontaneously broken}
(Cheng and Li 1984).
In order to have a translation and parity invariant vacuum one has to
demand that $\piv =0$ and
that $\sigma_V$ is independent of $x$.
The value of $\sigma_V$ will be determined later on.
Normal ordering with respect to this vacuum is performed by subtracting
the vacuum value
${S_{eff}} (\sigma_V, \piv_V)$ from the effective chiral action \qeq{\e217}

\subsec{Stationary Phase Approximation}

{}From now on we will assume {\it classical fields} for $\sigma$
and $\piv$ or in other words we
perform a stationary phase approximation in $0$th order with respect to
the stationary points of the effective action
$$
\eqalign{
{
{{\delta \Sesp} \over {\delta \sigma}}\linie}_{vac}   &= 0 \cr
{{{\delta \Sesp} \over {\delta \vec\pi}}\linie}_{vac}  &= 0 \cr}
\EQN\e220
$$
which is the {\it Schwinger-Dyson} or {\it gap equation}.
This means we have now specified the approximation in which the model will
be treated. It can be shown that the stationary phase approximation of the
semibosonized version is fully
equivalent to the Hartree approximation
in the original $4$-fermion version, which has been treated by a number of
authors using the Bethe-Salpeter formalism
(Bernard et al. 1984, Bernard 1986,  Ferstl.et al. 1986, Bernard
et al.1987,
Providencia et al. 1987, Bernard et al.1988, Klimt et al. 1990) and has been
reviewed by Vogl and Weise (1991) and Klevansky (1992).
Whereas in this chapter we restrict ourselves to the vacuum solutions
$\sigma_V$ and $\piv_V$ of \qeq{\e220}, the main aim of this article is
to study non-translational invariant solutions of this equation corresponding
to a system with finite baryon number,
which will be done in the next chapter.

\subsec{Vacuum Expectation Values - Quark Condensate}

In the following we need
the expectation value of the bilinear density
$\bar q (x)\K q(x)$
in the stationary phase approximation,
i.e. for classical $\sigma$ and $\piv$,
where $\K$ is an arbitrary matrix in spin,
isospin and color space.
This
can be expressed by:
$$
\eqalign{
\langle\bar q(x)\K q(x)\rangle &=
{ {\int \D q \D \bar q e^{- \int d^4 x_E \bar q(x) (-iD) q (x)}
[\bar q (x) \K q(x)] } \over
{\int \D q \D \bar q e^{- \int d^4 x_E \bar q(x) (-iD) q (x)} }  }  \cr
&= {\delta \over {\delta \kappa (x)}} \ln \int \D q \D \bar q
   e^{- \int d^4 x_E \bar q(x) (-iD - \kappa (x) \K ) q (x)}
   \Big\vert_{\kappa (x) =0}  \cr
&= {\delta \over {\delta \kappa (x)}}  \Sp \ln (-iD -\kappa (x) \K)
\vert_{\kappa (x) =0} \cr
&= {\bra {x_E}} \Tr_{\gamma\tau c} [ (iD) ^{-1} \K ] {\ket {x_E}}
\cr}
\EQN\e221
$$

One important special case is the expectation value
$\langle \bar q q \rangle$, the {\it quark condensate},
which is an order parameter characterizing the
strength of the spontaneous breakdown of chiral
symmetry, like it does $M$ or $\sigma_V$.
In the present model it can be easily determined from the general
expression \qeq{\e221} with $\K ={\cal I}$
by applying the Schwinger-Dyson equation \queq{\e220} to
\qeq{\e217}:
$$
\kon = {\bra{x_E}} \Tr [i D_V ]^{-1} {\ket {x_E}} = (-) {{\mu^2}
\over g} (\sigma_V - m_0)
\EQN\e222
$$

\sect{Divergences and Regularization}

\subsec{UV Divergences}

In order to extract the UV divergent terms in the fermion determinant $\Ssp F$
we write its real part with \qeq{\e219} in the form:
$$ \eqaligntag{   \Re
\Ssp F  &= (-) \half \Sp \ln D^{\dag} D \cr
&= (-) \half \Sp \ln [ -{\partial_E}^2 + M ^2 +
   ig \sld (\sigma + i \vec\tau
   \vec\pi \gamma_5) \cr &  + g^2 (\sigma^2 + \vec\pi^2 - {\sigma_V}^2
    )]
  &\EQ\e222a\cr}
$$

Introducing the abbreviations:
$$ \eqalign{
G &= (-{\partial_E}^2 + M^2 ) ^{-1}   \cr
V &=
ig \sld (\sigma + i \vec\tau
\vec\pi \gamma_5) + g^2 (\sigma^2 + \vec\pi^2 - {\sigma_V}^2 ) \cr}
\EQN\e223
$$
we obtain:
$$   \Re
\Ssp F = \half \Sp \ln G - \half \Sp \ln (1+GV)
\EQN\e224
$$
The first term $\half \Sp \ln$ is an infinite constant independent of
$\sigma$ and $\piv$ and vanishes after normal ordering.
The second term can be expanded in powers of
$(GV)$:
$$
\ln (1+GV) = \sum_{n=1}^{\infty}  (-)^{n+1} {1\over n} (GV)^n
\EQN\e225
$$
In momentum space the functional trace $\Sp$ gives an integral
$\id {k_E}$.
Therefore we see that all terms with
$n\geq 3$ are UV convergent.
The contributions $n=1$ and $n=2$ can be easily explicitly calculated
leading to (Eguchi 1976):
$$
\eqalign{   \Re
\Ssp F &= [(-) 4 N_c I_1 (M)] g^2 \int d^4 x_E (\sigma^2+\vec\pi^2)
\cr
&+ [2 N_c g^4 I_2 (M)]
\int d^4 x_E (\sigma^2 + \vec\pi ^2 - \sigma_V ^2)^2    \cr
&+ [(-) 4 N_c g^2 I_2 (M)] \int d^4 x_E \half
 [ (\partial^\mu \sigma) (\partial_\mu \sigma) +
   (\partial^\mu \vec\pi) (\partial_\mu \vec\pi) ]  \cr
&+ {\tilde S}_{eff} (\sigma,\vec\pi)  \cr}
\EQN\e226
$$
where the integral
$$
I_k (M) = \id {k_E}  {1\over \left( {k_E ^2 + M^2}\right)^k  }
\EQN\e227
$$
is quadratically  divergent for $k=1$ and logarithmically divergent for
$k=2$.    The
${\tilde S}_{eff} (\sigma,\vec\pi) $ contains only UV convergent terms and
is of higher order in the field amplitudes $V$ and their gradients
$\partial V$. A systematic expansion in $(V,\partial V)$ can be performed
by the gradient or heat kernel expansion and will be discussed in the
next section.

\subsec{Regularization Schemes}

The imaginary part  $\Im S_{eff}^F$ of the fermionic action in Euclidean
space is finite because it is in a fundamental way connected with the
anomaly structure of the theory. The real part $\Re S_{eff}^F$ is always
divergent corresponding to $I_1$ and $I_2$ being infinite.

The divergent integrals
$I_1$ and $I_2$ have to be
regularized {\it consistently} by applying a
regularization scheme to $\Ssp F$ with a finite UV cutoff $\Lambda$, which
enters as an additional parameter into our model.
This is not considered
to be a big problem since the model is believed to be an effective
low energy approximation of QCD, with the cutoff as the relevant scale
for low energy hadronic phenomena.
Physically, the cutoff
function models in some sense
the gluon cloud around a single quark and the correct cutoff function
should be derived ideally from QCD itself.
Because there exists no such derivation up to now one is forced to apply
general schemes which are used in the literature and fixes the cutoff
$\Lambda$ in order to describe the physics of the mesonic sector as well
as possible.
In practice there is no
reason to prefer some scheme in favor of the other provided some
constraints concerning symmetries are obeyed (Ball 1989).
The question of
different regularization schemes has been treated in detail by
Meissner Th
et al. (1990b) for the vacuum and by Blotz et al. (1990) and
Doering et al. (1992) in
the soliton sectors.
In this review we will mainly use two of them for
practical calculations: Proper time ( Schwinger 1951 ) and
Pauli-Villars (1949), as it is used in this article, both preserving
Lorentz-, chiral  and vector-gauge invariance (Ball 1989). Actually
both can be obtained from a generalized proper-time formalism (Ball
1989).

\subsec{Gauge Invariant Schemes}

The generalized proper-time regularization procedure
(Ball 1989 ) is based on
the following identity for the difference of two logarithms
$$ \log \alpha - \log \beta = \lim_{\Lambda\to\infty}
\int_0^\infty {d\tau \over \tau} \phi ( \tau, \Lambda ) ( e^{-\tau
\alpha} - e^{-\tau\beta} )
\EQN\e228 $$
with the additional condition $ \phi (\tau, \infty )=1$
and a properly chosen shape for $\phi (\tau , \Lambda)$ in the limit of
small $\tau$.
Using this
representation and assuming that the vacuum contribution is
subtracted, the generalized proper time regularized effective action
reads
$$ \log \det D^\dagger D - \log \det D_0^\dagger D_0 =
\int_0^\infty {d\tau \over \tau} \phi (\tau, \Lambda) \Bigl[
\tr e^{-\tau D^\dagger D} - \tr e^{-\tau D_0^\dagger D_0} \Bigr]
\EQN\e229 $$
The original {\it proper time} method of Schwinger corresponds to the
choice
$$ \phi (\tau)= \theta ( {1\over \Lambda^2} - \tau ) \EQN\e230 $$
with $\theta$ as a single step function \foot{See sect. 6 for the
double step proper time regularization, which offers some more degrees
of freedom.}.
The {\it Pauli-Villars} scheme is obtained as
$$ \phi (\tau)= 1 - (1 + \Lambda^2 \tau ) e^{-\tau \Lambda^2} \EQN\e231 $$
corresponding to usual mass subtractions in Feynman diagrams.

Some other schemes have been proposed, which however do not show the
invariances mentioned above and will not be used in this article.
Those are e.g. a relativistic four dimensional
sharp cutoff method
$$\Tr \log iD \to \Tr\{\theta(-\partial^2 -\Lambda^2 ) \log iD \}
\EQN\e231a
\eeq
and also a three dimensional sharp cutoff
$$\Tr \log iD \to \Tr\{\theta( -\nabla^2 -\Lambda^2 ) \log iD \}
\EQN\e231b    \eeq

\subsec{Dynamical Mesonic Terms}

As we can see the logarithmically divergent terms are proportional to
the kinetic mesonic part
$[ (\partial^\mu \sigma) (\partial_\mu \sigma) +
   (\partial^\mu \vec\pi) (\partial_\mu \vec\pi) ]$
and the mexican hat self interaction
$(\sigma^2 + \vec\pi^2 -\sigma_V ^2)^2$.
Those are exactly the dynamical mesonic terms appearing in the linear
chiral sigma model (Gell-Mann and Levi 1960).
We therefore have {\it generated} those parts in our model
from the {\it fermion determinant} $\Ssp F$.
The proportionality constants in front of them are dependent on the
cutoff $\Lambda$ and will therefore be determined by the physics
of the mesonic sector.

\sect{Gradient and Heat Kernel Expansion}

The effective action is a highly non-local functional in the $\sigma$
and $\vec \pi $ fields, i.e. it involves products of fields at all
space-time points. Therefore it is rather interesting to investigate its
behavior in some limiting cases. In this section we will study
the case of slowly varying fields, since then the effective action
reduces
to a local Lagrangian. We will also see how kinetic and interacting
contributions arise in this simple limit as a consequence of the
polarization of the Dirac sea. This allows also to study some
qualitative and quantitative features of the model. There are basically
two schemes: gradient and heat kernel expansion.

In essence, the gradient expansion is an expansion in the number of
Lorentz indices. The problem is merely technical and many methods have
been suggested (Aitchison and Frazer 1984, 1985a, 1985b). A very
elegant and powerful one has been proposed
by Chan (1985), so that it will be exposed here. The method has been
worked out assuming the cyclic property to be valid so that the
results are automatically vector gauge invariant. Chan (1985) has
applied it to second order elliptic operators up to the fourth order and
more recently Caro and Salcedo (1993) up to sixth order. Hence it
makes sense
to apply it to the real part of the regularized effective action  with
$$ D^\dagger D = -\partial^2 + i\slashchar\partial \hat \Sigma +
\hat\Sigma^\dagger \hat\Sigma
$$
with $\hat\Sigma=g(\sigma+i{\vec\tau}{\vec\pi}\gamma_5)$.
The main idea is to realize that the
integral representing a one loop Feynman diagram is invariant if all
external momenta are shifted by the same amount, so that one can as
well average over all possible shifts. In the proper-time scheme this
can be expressed as follows
$$ {\Re} S_{eff} =
\int_0^\infty {d\tau \over \tau} \phi (\tau, \Lambda)   \Sp
\exp \[ -\tau \( -\partial^2 + i\slashchar\partial \hat \Sigma +
\hat\Sigma^\dagger \hat\Sigma  \) \]
\eeq
$$={1\over \delta^{(4)} (0)} \int {d^4 k \over (2\pi)^4}
\int_0^\infty {d\tau \over \tau} \phi (\tau, \Lambda) \Sp
\exp \[ -\tau \( -(\partial+ik)^2 + i\slashchar\partial \hat \Sigma +
\hat\Sigma^\dagger \hat\Sigma  \) \]
\eeq
The next step is to expand the exponential in powers of the derivative
operator $\partial$ up to the desired order and to make use of the
cyclic property. This corresponds to a {\it derivative} or {\it gradient
expansion}. If one further expands around the vacuum configuration $\hat
\Sigma= M $ one gets the {\it heat kernel expansion}. In other words,
to get a given order of the derivative expansion one needs an infinite
number of terms from the heat kernel expansion. The final expressions
for the known orders are rather lengthy and will not be quoted here. For
our purpose it is sufficient and instructive to restrict ourselves to
the heat kernel expansion up to second order. The result is
(Kleinert 1976, Ebert and Reinhardt 1986)
$$ {\cal L} = N_cg^2 I_2(M) \tr_\tau\Bigl[
\partial_\mu{\hat\Sigma}^\dagger
\partial^\mu
{\hat \Sigma} + ( {\hat\Sigma}^\dagger {\hat\Sigma} -M^2 )^2 \Bigr]
 + ({\mu^2\over 4g^2}- 2N_cg^2I_{1}(M) ) tr_\tau({\hat\Sigma}^\dagger
{\hat\Sigma}
-M^2)
\EQN\deriv       $$
The $I_n$-integrals are given by \queq{\e227 }.
The important point of this expression is that even for slightly
space-time dependent fields the whole effect can be summarized in
a kinetic and interaction terms for the fields $\sigma$ and
$\vec\pi$. Thus although those did not appear explicitly in the original
lagrangian, they are indeed present due to the polarization of the Dirac
sea. In chapter 3 we will show how one can go beyond the limit of slowly
varying fields by computing the effective action in an exact manner.
The similarity of \qeq{\deriv } with the Gell-Mann--Levy (1960) sigma
model is
also very interesting and will be exploited later when fixing the
parameters. The effective Lagrangian gets even simpler if the so
called {\it chiral circle condition}
$$ \sigma^2 + \vec\pi^2 = f_\pi^2
\eeq
is imposed. Then one can use the parameterization
$$ \sigma + i\vec\tau \cdot \vec \pi = f_\pi U ; \qquad U=e^{ i \vec
\tau \vec \phi /f_\pi }
\eeq
with a SU(2) unitary matrix U  and $\vec\phi$ as  the non-linearly
transforming pion field (Coleman et al.~1969). The Lagrangian reads
then
$$ {\cal L} = N_cM^2 I_2(M) \tr_\tau \partial_\mu U \partial^\mu U^
\dagger
\EQN\ewbg $$
which resembles the Weinberg (1967) non-linear Lagrangian. The former
discussion carries along also for the currents. We just
quote the final result for the vector and axial currents for linear
chiral fields in the same approximation respectively
$$\eqalign{ & \vec V^\mu = \vec \pi \times  \partial^\mu \vec \pi
\cr
& \vec A^\mu = \sigma\partial^\mu \vec \pi - \vec\pi \partial^\mu
\sigma
\cr}
\eeq

\subsec{Gradient Expansion of the Imaginary Part - Wess-Zumino-Witten-Term}

The gradient expansion of the imaginary part of the fermion determinant
\qeq{\e217b} is more involved. The problem is, that the corresponding
term can not be written as $4$-dimensional space time integral over a
local Lagrangian. In order to obtain a closed analytical form for the gradient
expansion of $\Im \Se F$ one considers the change
$\delta_U \Im \Se F (U)$ under the variation $\delta U$
of the chiral field $U$ and
performs a gradient expansion of this expression. The actual calculation
(Dhar and Wadia 1984, Dhar et al. 1985, Diakonov et al.1988)
follows very closely the Goldstone-Wilczek expansion of the anomalous
baryon current as it is described in App. B.
The result is:
$$
\eqalign{
&i \delta_U \Im \Se F (U) \Big\vert_{grad} =
-{{i N_c}\over {48 \pi^2}} \int d^4 x \epsilon_{\alpha\beta\gamma\delta}
\cr &\Tr_{flavor} \left [
\left ( {U^{\dag}} \partial_\alpha U \right )
\left ( {U^{\dag}} \partial_\beta  U \right )
\left ( {U^{\dag}} \partial_\gamma U \right )
\left ( {U^{\dag}} \partial_\delta U \right )
\right ]        \delta U  \cr}
\eeq
which can be shown to be the variation of
$$
\eqalign{
&i \Im \Se F (U) \Big\vert_{grad} =
-{{i N_c}\over {240 \pi^2}} \int_{B_5}
d^5 x \epsilon_{\mu_1\mu_2\mu_3\mu_4\mu_5}
\cr &\Tr_{flavor} \left [
\left ( {U^{\dag}} \partial_{\mu_1} U \right )
\left ( {U^{\dag}} \partial_{\mu_2} U \right )
\left ( {U^{\dag}} \partial_{\mu_3} U \right )
\left ( {U^{\dag}} \partial_{\mu_4} U \right )
\left ( {U^{\dag}} \partial_{\mu_5} U \right )
\right ]  \cr}
\EQN\e2wzw
$$
where $B_5$ denotes a $5$-dimensional sphere, whose boundary is the
$4$-dimensional space-time ($\partial B_5 = R_4$).
The r.h.s. of \qeq{\e2wzw}
is the famous {\it Wess-Zumino-Witten term}, which has been discovered
in context of the integration of the chiral anomaly (Wess and Zumino 1971,
Wess 1972).
It turned out that it is indispensable to implement it in
any effective chiral mesonic model in order to get a proper description
of low energy mesonic phenomena (Witten 1983a).
In case of an $SU(2)$ chiral field $U$ it can be shown to vanish identically
as it does always, if $U$ is time independent. Further details
will be discussed in the context of  the $SU(3)$ NJL in chapter 6.

Goldstone and Wilczek (1981) have shown that in the lowest
non-vanishing
order of the derivative expansion  the baryon current is proportional
to the topological current
$$ < B^\mu (x) > =- {{1} \over 24\pi^2} \epsilon^{\mu\nu\alpha\beta}
\tr
\( U^\dagger \partial_\nu U U^\dagger \partial_\alpha U
U^\dagger \partial_\beta U \) + \cdots      \label\bartopo    $$
In the next chapter the relevance of such topological current in the NJL
model will be discussed.

\sect{Mesonic spectra from effective actions}

The calculation of on-shell mesonic two-point functions
within the NJL model has
been undertaken by several authors, both in a pure fermionic language
(Blin et al. 1988; Bernard et al. 1988; Bernard and Meissner UG 1988;
Klimt et al. 1990; Takizawa et al. 1990)  and in a bosonized version (
Alkofer and Zahed 1990; Jaminon et al. 1992).
As we shall see in Sect. 2.7. this is actually a natural step because
the parameters of the NJL model will be fixed by reproducing the meson
masses.
In most cases a
diagrammatic procedure has been considered within a Bethe-Salpeter
formalism in the ladder approximation. Similar results have been also
obtained in a
time-dependent Hartree-Fock approach (da Providencia 1987 ). The virtue
of the on-shell definition  method is that no approximation is
involved apart
from working in the leading order of the large $N_c$ expansion. As it
can be shown, the heat kernel and derivative expansions are low energy
approximations to the  complete on-shell  two-point function. In this
section we sketch
the calculation of mesonic spectra    in the bosonized version of the
model as it has been done by Jaminon et al. (1992).
This is based on the observation, that the relevant quadratic part
$S_{eff}^{F,2p}$
of
the effective action
\queq{\e222a } for SU(2) scalar fields  can be  always  written in the
form  (Jaminon et al. 1992)
$$ \eqaligntag{
      {S_{eff}}^{F,2p}(\sigma,\pi) = \half \int {d^4q\over (2\pi)^4 }
     \bigl( &\sigma(q) \sigma(-q) Z_\sigma(q) \left[ q^2 + m_\sigma^2(q)
     \right] +  \cr
    &\pi(q) \pi(-q) Z_\pi(q) \left[ q^2 + m_\pi^2(q) \right]
      \bigr)
     &\EQ\e2000\cr}      $$
Therefore one can define  mesonic masses from
$$ \eqaligntag{    {1 \over Z_\sigma (q^2) }
    { 1 \over \delta (p_1-p_2) }
   { {\delta^2{S_{eff}}^{F,2p}(\sigma,\pi)
    \over \delta \sigma(p_1) \delta \sigma (p_2)
      }\linie}_{p=p_1=-p_2,p_1^2=q^2}
        &=
  {\left (p^2 + m_\sigma^2 \right)\linie}_{p^2=q^2}
     \cr
     {1 \over Z_\pi (q^2) }
    { 1 \over \delta (p_1-p_2) }
   { {\delta^2{S_{eff}}^{F,2p}(\sigma,\pi)
    \over \delta \pi(p_1) \delta \pi(p_2) }\linie}_{p=p_1=-p_2,p^2=q^2}
   &=
   {\left (p^2 + m_\pi^2 \right)\linie}_{p^2=q^2}
    &\EQ\e2002\cr}       $$
which effectively corresponds to a rescaling of the fields according to
$$    \sigma^\prime = Z_\sigma^{1/2}(q^2)\  \sigma,\ \
      \pi^\prime = Z_\pi^{1/2}(q^2)\     \pi  \EQN\e2002a  $$
The on-shell definitions of the mesons can be found then by
evaluation of $Z_\sigma(-q^2=m_\sigma^2)$ and
$Z_\pi(-q^2=m_\pi^2)$ to obtain
$m_\pi^2=m_\pi^2(-q^2=m_\pi^2)$     and
$m_\sigma^2=m_\sigma^2(-q^2=m_\sigma^2)$.
where the Z-factors are given by the Feynman integrals of the form
(Jaminon and Ripka 1993)
$$  Z_\phi (q^2) = 4N_c g^2 \int_{reg}  {d^4k\over (2\pi)^4 }
               G(k+q/2) G(k-q/2)  \EQN\2003 $$
with the {\it propagator} $G(k)=1/(k^2+M^2)$.  From these expressions it
is clear that the corresponding off-shell definitions can be obtained by
setting $q^2=0$ in all the equations. These expressions then coincide of
course with the corresponding one, which one would obtain from a
gradient expansion of ${S_{eff}}^F$ from the very beginning.
Then the  approximated analogue of \queq{\e2000 }  is already given by
\queq{\e226 }. The field redefinitions become trivial in this case
and the mesonic spectra could be  read off directly from the
effective potential (Coleman 1985).
For definiteness we  present $Z_\phi (q^2)$
explicitly for the proper time regularization \qeq{\e230 }
\beq Z_\phi (q^2)={4N_cg^2\over 16\pi^2} \int dx \Gamma
(0,[M^2-x(1-x)q^2]/\Lambda^2)   \EQN\2003a2 $$
where we made use of  the incomplete $\Gamma$(n,x)-function
\beq   \Gamma(n,x):=\int_x^\infty dt\ t^{n-1} e^{-n},\ \ \ x\ge 0
\EQN\2003a1   \eeq

\sect{Fixing of the Parameters in the Mesonic Sector }

We have mentioned already that if the coupling constant $G$ is higher
than a
certain critical value,  the vacuum is unstable if the mass is
increased. This leads
eventually to the Goldstone phase. Assuming that this is the
case we will fix the parameters of the model. Two ways for parameter
fixing may be distinguished depending upon  the particular
approximation involved. We will refer to them for brevity as the
on-shell  and the off-shell  condition.
The off-shell fixing of the parameters is dicussed in detail
by Meissner Th et al. (1990b).

\subsec{Mesonic Properties    }

Looking at \qeqs{\e217,\e219 } we see that
we have $m_0$,$\mu^2,\Lambda,\sigma_V$ and the
constituent quark mass M as the {\it five} parameters of the model.
Thus one can impose e.g. the following {\it four}
conditions

\item{1)} The stationary phase condition $\delta
S_{eff}/\delta \sigma$ in the vacuum state
$${{\delta
S_{eff}\over\delta \sigma}\linie}_{vac}=\left(\mu^2-8N_cg^2
I_{1}(M) -{m_0 \mu^2 \over g \sigma_V} \right) \sigma_V = 0
\EQN\2003a$$   immediately determines $\mu^2$ in the broken phase
with $\sigma_V\ne~0$.

\item{2)} Fixing of the pion decay constant $f_\pi$   is done
by considering the expectation value of the axial current between
a pion state and the vacuum, which  in the semibosonized form
corresponds to consider
\beq  <0\mid A_\mu^a(x)\mid\pi^a>  = N \int {\cal D}{\bar q}
       {\cal D}{q} \left[ {\bar q}(x)\gamma_\mu\gamma_5\tau^a
       q(x)\right]
       e^{i\int d^4x {\cal L}_{NJL}^\prime}.   \EQN\2003b  $$
where N is the normalization constant.
This is a similiar like \qeq{\e221 } but now it has to be evaluated for
a plane-wave pion state and not the vacuum. In the
one-fermion loop approximation   we obtain then
$$  <0\mid A_\mu^a(x)\mid\pi^a> = \sigma_V Z_\pi^{1/2}(q^2) p_\mu
\pi^a(x)
. \EQN\2004 $$
Comparison with the known matrix element  for the pion-decay
from current algebra
(Cheng and Li  1984)
$$ <0\mid A_\mu^a(x)\mid\pi^a>=i p_\mu f_\pi(q^2=-m_\pi^2)
   \pi^a(x) \EQN\e2004a $$
gives immediately the  relation
$$  f_\pi(q^2) = \sigma_V Z_\pi^{1/2}(q^2)   \EQN\2004b  $$
for  $Z_\pi$ and $\sigma_V$.

\item{3)} From the definition of the pion mass in \qeq{\e2002 } one
obtains straightforwardly the explicit form
$$  m_\pi^2 = { m_0 \mu^2 \over g \sigma_V } {1 \over Z_\pi}
\EQN\2004c
$$

\item{4)}  Now one can either choose the normalization
of the pion wave function  \qeq{\e2002a }
according to
$$ Z_\pi(q^2=-m_\pi^2) = 1 \EQN\2004d $$
or one can  equivalently require the classical PCAC relation
(Goldberger and Treiman, 1958)
$\partial^\mu~A_\mu^a=~f_\pi~m_\pi^2\pi^a$, which in the present model
gives in analogy to
\qeq{\e29 } but now from \qeq{\e213 }:
$$  \partial^\mu A_\mu^a = {m_0 \mu^2 \over g} \pi^a  \EQN\e2004e $$
so that together  with  \qeq{\2004c } and \qeq{\2004b } again
\qeq{\2004d }
follows.

This fixes all the parameters except one.
Usually the constituent quark
mass $M$ is chosen as the  free parameter.
 Then from \qeq{\2004d }, which
determines the cutoff $\Lambda$ for a given constituent quark mass M,
and from
\qeq{\2004b },  one can  deduce
$\sigma_V=f_\pi$, so that \qeq{\2004c } inserted in
\qeq{\2003a } gives $\mu^2$. Using again \qeq{\2004c } gives the
current
quark mass $m_0$. So $M=gf_\pi$ is the only free parameter of the model.

The above description  refers to the on-shell definition of the
parameters, whereas the off-shell formulation corresponds to  setting
$q^2=0$ in \qeq{\2004d }.  Because the Goldstone bosons and
especially the pions have a small  mass compared to the constituent
quark mass, the difference between both methods is rather small.
However  this is no longer the case, when heavier mesons like
$\rho,\omega$ and
$A_1$ are taken into account.

Furthemore, using again   \qeq{\e222 }   and  \qeq{\2004c }, it follows
the famous Gell-Mann, Oakes and  Renner (1968) relation
\beq  <{\bar q}q> m_0 =- m_\pi^2  f_\pi^2 + {\cal O}(m_0) \EQN\gor $$

\subsec{Numerical Results}

As we have said already the off-shell definition  is only a low
momentum
approximation to the on-shell condition. For on-shell
pions
the external momenta are rather small  so that the
numerical deviations from the two methods  turn out to  be
less than $7\%$. The only important difference is that in the chiral
limit ($m_0=0$) the scalar meson
mass acquires a value $m_\sigma=2M$ in the gradient expansion, while
in the on-shell method the $\sigma$ meson lies in the quark-antiquark
continuum with a finite decay width into this channel. This is the first
manifestation of the lack of confinement within the model.

The dependence of $\Lambda$ against $M$ can be seen at
\qufig{\f21 }. 
In the relevant region of $M=350-450\MeV$, the numerical value of the
cutoff $\Lambda$ depends noticeably on the regularization scheme chosen.

{}From \qeq{\e222 } we can now evaluate the quark condensate and  compare
it with the generally accepted value  (Shuryak 1986)   of
\beq  <\half({\bar u}u+{\bar d}d)>^{1/3}=-(225\pm~24)\MeV
\EQN\qcon \eeq
The results for the vacuum sector, i.e. quark condensate, current quark
mass and the vacuum energy density  (de Grand et al. 1975) can be seen
at
\qutab{\taband21
}.
For these
observables, i.e. for the quark condensate and  the current quark mass
a plateau has been observed for constituent
masses above $400\MeV$.

\vfill\eject

%
%
\chap{Systems with Finite Baryon Numbers - Solitonic Solutions}
%
%
\FIG\f31{The bound state spectrum of the $1$-particle hamiltonian $h$ for an
exponential profile form $\theta (r) = - n \pi e^{-r \over R}$ with $n=1$.
$\epsl$ and $R$
are given in scaled units ($\epsl \over M$ and $R \cdot M$).
The  valence orbit $(0^+)$ comes down from the positive into the
negative spectrum (Kahana and Ripka 1984).
Furthermore a value for the thermochemical potential $\mu $ is shown.}
\FIG\f32{Sea energy $\esea$, valence energy $\eval$ and total energy
$\etot$ as well as the corresponding  baryon numbers $B_{sea}$,
$B_{val}$ and $B_{tot}=B_{sea}+B_{val}$, respectively,
for a linear profile form
$$
\theta (r) =   \cases{
- n \pi (1- {r\over R})  & if $r<R$ \cr
0                        & if $r>R$}
$$
with $n=1$ at a constituent mass of
$M=565 \MeV$. The total energy $\etot (R)$ shows a local minimum at $R = 0.8
\fm$.
$B_{val}$ and $B_{sea}$ both jump if $\eval$ gets negative ($R\approx 1.3
\fm$),
whereas $B_{tot} =1$ in any case. (Diakonov et al.1988, Meissner Th et
al. 1989) }
\FIG\f33{The mean field energy $E_{MF}$ of the selfconsistent solution in the
non-linear model
($m_\pi =0$) (full line) splitted in valence (dashed-dotted) and sea (dotted)
part in dependence
of the constituent quark mass $M$. (Meissner Th et al.1989)}
\FIG\f34{The quadratic isoscalar electric radius $\r2 = <R^2>_p + <R^2>_n$
of the selfconsistent solution in the non-linear model
($m_\pi =0$) (full line) splitted in valence (dashed-dotted) and sea (dotted)
part in dependence
of the constituent quark mass $M$. (Meissner Th et al.1989)}
\FIG\f35{The selfconsistent $\sigma$ field (normalized to $f_\pi$) in the
non-linear
model with $m_\pi =0$ for 3 different constituent quark masses $M$
(Meissner Th and Goeke 1991)
.}
\FIG\f36{The selfconsistent $\pi$ field (normalized to $f_\pi$) in the
non-linear
model with $m_\pi =0$ for 3 different constituent quark masses $M$
(Meissner Th and Goeke 1991)
.}
\FIG\f37{The baryon densities $b_0 (r)$ of the selfconsistent solutions in the
non-linear
model with $m_\pi =0$ for 2 different constituent quark masses $M$ (full
lines). The valence
contributions (dashed-dotted) are explicitly shown  as well
(Meissner Th and Goeke 1991). The upper two curves correspond to
$M=725\MeV$ and the lower two to $M=363\MeV$.}
%
%
%
%
%
\TAB\t31{The critical values of the constituent quark mass $M_{cr}$ and the
corresponding values for the coupling constant
$g_{cr} = {{M_{cr}}\over{f_\pi}}$ as well as the ratio $\lambda_{cr} =
{{\Lambda (M_{cr})}\over {M_{cr}}}$
for $m_\pi =0$ (chiral limit) and $m_\pi = 139 \MeV$. Solitonic solutions of
the non-linear model exist if
$ M > M_{cr}$ ($g > g_{cr}$, $\lambda < \lambda_{cr}$)
(Meissner Th and Goeke 1991).}
In this chapter we consider static (time independent) meson field
configurations.
First we show  the expression for the total energy
(sect. 3.1). A system with
baryon number $B=1$ is obtained by adding $N_c =3$ valence quarks, which is
formally  done
by introducing a thermochemical potential $\mu$. Furthermore  we
discuss the connection
between baryon number $B$ and topological winding number $n$ of the $\piv$
field
(sect.3.2).
The mean field equations of motion
for the non-linear model (meson fields restricted to the chiral circle) are
derived and solved
for a $B=1$ system (sect.3.3).
Systems with higher baryon and winding number are briefly discussed
(sect.3.4).
Finally we show that for the linear \NJL (off the chiral circle) no
solitonic
solution   exists but the system collapses to a zero energy and zero
size configuration (sect.3.5).

\sect{Static Meson Field Configurations - Energy and Static Expectation
Values}

\subsec{1-Particle Hamiltonian $h$ and Dirac Spectrum}

\noindent
In order to construct a system with finite baryon number $B$  one has to
consider meson
field configurations different from  the vacuum. Because we are
interested in static
properties of the baryon, we will restrict ourselves to
time independent meson fields $\sigma$ and $\piv$.
For those the Euclidean Dirac operator $iD$ can be separated into a trivial
time derivative
${\partial\over{\partial\tau}}$ as well as the time-independent 1-quark
hamiltonian $h$:

$$
-iD = \beta ({\partial\over{\partial\tau}} + h )
\eqn
$$
where
$$
h=
{{\vec\alpha \vec\nabla }\over i}
+ g \beta \left [
\sigma (\vec r) + i \vec\tau \vec\pi (\vec r) \gamf \right ]
\EQN\e32
$$
and in the vacuum:
$$h_V  =
{{\vec\alpha \vec\nabla }\over i} + \beta M
\eqn
$$
It is our aim to express the operator $\Sp \ln (-iD)$ through  the
eigenvalues of $h$. One should note that
$h$ is hermitian and traceless:
 $\Tr h =0$. Furthermore we make the {\it hedgehog ansatz} for $\sigma$ and
$\piv$:
$$
\eqalign
{\sigma (\vec r) &= \sigma (r) \cr
\pirv &= \hat r \pir }
\EQN\e34
$$
For chiral models which use a gradient or heat kernel expansion for the highly
non local
fermion determinant $\Sp \ln (-iD)$ like the Skyrme model (cf. sect.8.1)  or
the Gell-Mann--Levi
chiral sigma model (cf. sect.8.2) it has been proven (Ruiz-Arriola et
al. 1989) that the hedgehog shape \qeq{\e34}
is a necessary condition for the meson fields to be a solution of the time
independent classical
mean field equations of motion for a baryonic system. One has to admit that
there
exists up to now no such proof for the case if $\Sp \ln (-iD)$ is treated
exactly
as it is done in the present approach.

\noindent
Due to the hedgehog ansatz  \queq{\e34} the 1-particle
hamiltonian
$h$
\queq{\e32} commutes with the Grand
spin $G^2$ (where $\vec G = \vec J + \vec T$, $\vec J$:total spin, $\vec T$:
isospin), its
$z$-component $G_z$ as well as the parity $\Pi$ and the 4 observables
($h$,$G^2$,$G_z$,$\Pi$)
form a complete set of commuting operators. Therefore the eigenstates $\ketl$
of $h$:
$$
\eqalign{
h \ketl = \epsl \; &, \; \langle {\vec r} \vert \lambda \rangle =
\phi_\lambda (\vec r)  \cr
h_{V} \ketlv = \epslv \; &, \; \langle {\vec r} \vert \lambda_V \rangle =
\phi_{\lambda_V} (\vec r)  \cr} \EQN\e35
$$
can be characterized by the 4 quantum numbers $\epsl$,$G$,$G_3$ and $(-)^l$,
where $G_3$ is degenerated.
Following  Kahana and Ripka (1984) the eigenvalue problem
\qeq{\e35} can be solved numerically by putting the
system in a large but finite sphere with radius $D$ and appropriate boundary
conditions for the
radial part of the free eigenfunctions at $D$.
In doing so one obtains a basis consisting of discretized wave numbers $k_n$,
which belong to a given
grand spin $G$ and parity $(-)^l$.
For the numerical diagonalization of the matrix of $h$ one takes wave numbers
smaller than a given
numerical cutoff frequency $K_{max}$ into account.
$K_{max}$ and $D$ have to be chosen large enough so that the value of any
calculated observable does not
change any more by further increasing $K_{max}$ and $D$. It should be
emphasized that $K_{max}$ is a
purely  numerical cutoff and has nothing to do with the model intrinsic
physical UV cutoff $\Lambda$ introduced
and fixed in Chap. 2 in order to reproduce the pion decay constant
$f_\pi$.

\noindent
\qufig{\f31} shows the spectrum of $h$ for a meson profile {\it on the chiral
circle} (non-linear realization
of chiral symmetry) as example:
$$\eqalign{
\sigma (r) ^2 + \pi (r) ^2 &= \sigma_V ^2 + \vec\pi _V ^2 = f_\pi ^2 \cr
\sigr &= f_\pi \cos \theta (r)  \cr
\pir  &= f_\pi \sin \theta (r) \cr} \EQN\e36
$$
where the chiral angle $\theta (r)$ is parameterized by
$$
\theta (r) = - n \pi e^{ - \roR }
\EQN\e37
$$
with the topological winding number $n=1$ (Kahana et al.1984).
The parameter $R$    characterizes the size of the meson profile and in
a
way describes the magnitude of the
deviation between  the actual and the the vacuum field configuration
($R=0$). If $R$ is large enough
($R\cdot M \geq 0.5$) one finds bound orbitals both in the positive and the
negative spectrum. The positive
bound  state with  lowest single particle energy and quantum numbers
$G^P
=
0^+$,  will be called the valence orbit
${\ket \lambda} = {\ket {val}}$. Its single particle energy
decreases with increasing $R$, switches sign and  gets finally part of the
negative spectrum.  This level originates from the positive continuum
and gets bound and localized by interacting with the negative continuum
(Dirac sea). This simple non-linear mechanism creates the soliton.

\subsec{Static Energy}

\noindent
Because the meson fields are time independent, the trace over the Euclidean
time coordinate $\tau$ in the fermion
determinant can be performed trivially. Assuming anti-periodic boundary
conditions in the Euclidean time interval
$\left [ -{\T \over 2} , {\T \over 2 }\right ]$ the Euclidean time gradient
$\partial \over{\partial \tau}$ has the spectrum
$u_n = {{(2n +1)} \over  \T} \pi \, , \, n = 0,\pm 1,\pm 2, \dots$.
Therefore the expression for any functional $F$ of the form
$F \left [ {\partial\over{\partial\tau}} , f(\vec r) \right ]$
simplifies in the zero temperature limit ($\T \to \infty \, ,\, \Delta u_n \to
0$), which
leads to the the ground state of the system, to:
$$   \eqaligntag{
\Sp F_{\tau,\vec r} \left [ {\partial\over{\partial\tau}} , f(\vec r)
\right ]
&= \int d^4x_E\Tr  <x_E\mid F_{\tau,\vec r}[{\partial\over{\partial\tau}} ,
f({\vec
   r})]\mid x_E>   \cr
&= \T \int d^3x_E \int {du\over 2\pi} \Tr
              F_{\vec r} [-iu,f({\vec r})]
&\eq\cr}
$$
Especially if we take for $F$ the effective action $S_{eff}$ we find for the
imaginary part:
$$
\eqalign{
\Im S_{eff} (\sigma , \piv) &=
\Im {S_{eff}}^F (\sigma , \piv)   \cr  &=
(-) \sum_{\lambda} \int {{du}\over{2\pi}} \Im \ln (-iu + \epsl ) =
\sum_\lambda  \int {{du}\over{2\pi}} \arctan{{-u}\over{\epsl}} =0  \cr}
\EQN\e36a
$$
which is based on the fact that $h$ is hermitian and therefore $\epsl$ real.
As we have already mentioned this is not true if $\omega$-mesons are included
in the model (cf. chap. 7).
We therefore have:
$$
S_{eff} (\sigma,\piv) = \Re S_{eff} (\sigma,\piv) = \T \cdot E
\eqn
$$
where $E$ is the total static {\it energy} of the system.
The explicit expressions for the various parts of $E$ in terms of the
eigenvalues $\epsl$ read:
$$
E = \esea + \emes + \ebr
\EQADV\e310\SUBEQNBEGIN\e310a
$$
with:
$$
\eqaligntag{
\esea &= \left ( - {{N_c}\over 2} \right ) \left [ \sum_\lambda {\RR}_1 (\epsl
,\Lambda)
- \sum_{\lambda_V} {\RR}_1 (\epslv ,
\Lambda ) \right ]  \qquad \SUBEQ\e310b \cr
\emes &= {{\mu^2 } \over 2} \int d^3 r (\sigma ^2 + \piv ^2 - f_\pi ^2 ) \qquad
\SUBEQ\e310c \cr
 \ebr &= - m_\pi ^2 f_\pi \int d^3 r ( \sigma - f_\pi ) \qquad
\SUBEQ\e310d \cr}
$$
In case  of the proper-time regularization method, which we will
restrict ourselves on for explicitness in the
following, the regularization function ${\RR}_1$ is given by (Meissner Th
et al.1988):
$$
{\RR}_1 (\epsl , \Lambda) = (-) {\Lambda \over {\sqrt {4\pi}}} \int _1 ^\infty
ds s^{-3/2} e^{-s (\epsl / \Lambda)^2}
\EQN\e311
$$
A detailed and complete discussion of the various  regularization
schemes
can be found
in Blotz et al. (1990) and in Doering et al. (1992).
Furthermore it should be noted, that because of $\Tr h =0$
in the unregularized case ($\Lambda \to \infty$) \qeq{\e310\e310b} with
\qeq{\e311}
can be written as sum over the negative (occupied) states
$$
\esea =
\sum_{\lambda ,\, \epsl <0} \epsl -
\sum_{{\lambda_V} ,\, {\epslv} <0} {\epslv}
\eqn
$$
which corresponds to the naive picture  of the occupied negative states
(Dirac sea) (cf.\qufig{\f13}).
If a regularization is performed  $\tr~h=0$ is only approximately
true.
Since its numerical value in realistic cases of baryon number $B=1$ is
about
$50\MeV$
it actually does not matter if one runs over the whole spectrum or only
over the occupied states.

\subsec{Static Observables}

\noindent
Finally we consider the expectation value of the static observable
$ O := \int d^3 \vec r {q^{\dag}} ( \vec r) \O q (\vec r)$,
where $\O$ represents an arbitrary time independent operator in Dirac-
and isospin space. Because of
$O = {1\over \T} \int d^4 x {\bar q} (x) \beta \O q (x)$ we have
after Wick rotation into Euclidean space:
$$
\eqalign{
\langle O^E\rangle_{sea} &= {1\over \T}
{\int  {\D q \D {\bar q} \int d^4 x_E \left [ {\bar q} (x) \beta \O^E
q(x)
\right ] e^{ - \int d^4 x_E {\bar q} (x) (-iD) q (x)} } \over
  {\int\D q \D {\bar q} e^{ - \int d^4 x_E {\bar q} (x) (-iD) q (x)} } }
\cr
&= {1\over \T} {\partial \over {\partial \omega_E}} \Sp \ln (-iD -
\omega_E \beta \O^E)
\Big \vert_{\omega =0} \; - \; {vac. \; contr.}\cr}
\EQN\e311a
$$
Using the relation
$$
\lim_{k \to \infty} \int _{-k} ^{+k} {{du}\over{2\pi  i}} {1\over{u - ix}} =
\half {\rm sign} x
\EQN\e311b
$$
one gets in the unregularized case:
$$
\langle O^E \rangle_{sea} = {\half} N_c \sum
_\lambda ( - {\rm sign} \epsl ) O_\lambda^E \, - \, {vac. \; contr.}
\EQADV\e312\SUBEQNBEGIN\e312a
$$
with:
$$
O_\lambda^E :=
\int d^3 r {\phi^{\dag}}_\lambda (\vec r) \O^E \phi_\lambda (\vec r)
=
\int d^3 r {\bar \phi}_\lambda (\vec r) \beta \O \phi_\lambda (\vec r)
\SUBEQN\e312b
$$
In the regularized case one has to substitute  $[\Sp \ln ] \to [\Sp \ln
]_{reg}$, which gives after rotating back into Minkowski space:
$$
\langle O \rangle _{sea} = (+ N_c) \sum_\lambda {\RR}_2 (\epsl ,
\Lambda) \cdot O_\lambda \; - \; {vac. \; contr.}
\EQN\e313
$$
where  for the proper time method the explicit expression for the
regularization function reads:
$$ {\RR}_2 (\epsl , \Lambda) = (-) {1 \over {\sqrt {4\pi}}} \int _1 ^\infty ds
s^{-1/2} \left ( {\epsl \over\Lambda} \right )
e^{-s (\epsl / \Lambda)^2}
\EQN\e314
$$

\sect{Baryon- and Topological Winding Number - Chemical Potential -
Valencequarks}

\subsec{Baryon Number and Baryon Density}

\noindent
The baryon current $b_\mu$ is generally defined as:
$$
b_\mu (x) := {1\over {N_c}} {\bar q } (x) \gamma_\mu q (x)
\eqn
$$
Its 4-divergence vanishes $\ddmu b^\mu =0$ due to the $U(1)$-symmetry ($q
\mapsto e^{i\alpha} q$)
of the Lagrangian giving rise to a conserved charge,  the baryon number
$$
B = \int d^3 r b_0 (x)
\eqn
$$
Applying \qeqs{\e312,\e313} for $\O ={\cal I}$ we obtain for the
expectation value of $B$ in case of time
independent fields without regularization:
$$
\langle B \rangle = \half N_c \left [ \sum_\lambda ( - {\rm sign} \epsl ) \; -
\; \sum_{\lambda_V} (-{\rm sign} \epslv )
\right ]
\EQN\e317
$$
For the baryonic density $b_0 (\vec r)$ we find correspondingly:
$$
\langle b_0 (\vec r) \rangle = \half N_c
\left [
\sum_\lambda  ( - {\rm sign} \epsl ) \cdot
\phi_\lambda ^{\dag} (\vec r)
{\phi}_\lambda (\vec r) \; - \;
\sum_{\lambda_V}  ( - {\rm sign} \epslv ) \cdot
\phi_{\lambda_V} ^{\dag}  (\vec r)
\phi_{\lambda_V} (\vec r) \right ]
\EQN\e318
$$
Neither $\langle B \rangle$ nor $\langle b_0 (\vec r)\rangle $ are UV
divergent. For $\langle B \rangle$ this is clear
from \qeq{\e317} whereas for $\langle b_0 (\vec r) \rangle$ it follows from the
fact that the 2nd order
gradient expansion, which gives rise to the logarithmic divergence, vanishes
and the 4th order is already UV
convergent (cf. app. B).

Therefore there is no need for a regularization  of the baryon number.
Furthermore as we see at the end of the next section, this is
consistent
with a more refined derivation of the baryon number, which in that case
originates from the imaginary part of the Euclidean effective action. It
is therefore consistent to start with a proper time regularized real
part
of the effective action as it is done in Chap. 2 and treat the baryon
number  unregularized, as long as  one treats the imaginary
part unregularized. However, one should mention that recently by
Schlienz
et
al. (1993) also the imaginary part was regularized though this treatment
does not preserve the anomaly structure of the theory.

Furthermore  it should be noticed that due to the general method of
calculating static expectation values of
observables  as it has been described in the last section, our
expressions for $\langle B \rangle$ and
$\langle b_0 (\vec r)\rangle$  differ from the usual ones obtained in
Minkowski space, where the sum
$\sum_\lambda$is  performed only over the negative (occupied) states.
For the global quantity $\langle B
\rangle$ it is easy  to see that both cases lead to the same result,
whereas the local densities $\langle b_0(\vec r)\rangle$ may differ.

\subsec{Thermochemical Potential $\mu$; Separation of the Valence Part}

\noindent
In order  to constrain the baryon number of the system to a given value
(e.g. $\langle B \rangle =1$ for nucleons and
hyperons) we proceed the way known from
statistical mechanics (Huang 1987, Negele and Orland 1987),
and introduce the thermochemical potential $\mu$ as
a Lagrange multiplier into the generating functional $\Z$ which becomes
now the grand canonical sum of states $\Z (\mu)$
(Williams and Cahill 1983, Meissner Th et al.1990)
$$
\Z (\mu) = \int \D q \D {\bar q} e^{ - \int d^4 x_E {\bar q} ( -i D - \mu \beta
)q}
\eqn
$$
After integrating over the fermion fields we arrive at the fermionic part of
the grand canonical effective action
$$
S_{eff}^F (\sigma, \piv , \mu) = (-) \Sp \ln (- i D - \mu \beta)
\eqn
$$
The mesonic part remains unaffected.
After subtracting the vacuum contribution the  $S_{eff}^F$ can be
splitted in a natural way into 2 parts:
$$
\eqaligntag{
S_{eff}^F (\sigma, \piv , \mu) &-
S_{eff}^F (\sigma_V , \piv_V , \mu_V = 0) = \cr
\left [S_{eff}^F (\sigma , \piv , \mu) -
S_{eff}^F (\sigma , \piv , \mu =0)
\right ]
\; &+ \;
\left [
S_{eff}^F (\sigma,\piv,\mu =0) -
S_{eff}^F (\sigma_V , \piv_V , \mu =0) \right ]  \cr
= S_{eff}^{val} (\sigma ,\piv ,\mu) &+ S_{eff}^{sea}  (\sigma , \piv)
&\EQ\seffmu
\cr
}
$$
where:
$$
S_{eff}^{val} :=
\left [
S_{eff}^F (\sigma , \piv , \mu) -
S_{eff}^F (\sigma , \piv , \mu =0) \right ]
\eqn
$$
and
$$
S_{eff}^{sea} :=
\left [
S_{eff}^F (\sigma , \piv , \mu =0) -
S_{eff}^F (\sigma_V , \piv_V , \mu =0) \right ]
\eqn
$$
The $S_{eff}^{sea} (\sigma ,\piv)$ is of course nothing
but the fermionic contribution  (Dirac sea)
to the effective  chiral action $S_{eff}^F (\sigma , \piv) = \T \esea$
considered in the last section.  In general it needs regularization.
The {\it valence} contribution $S_{eff}^{val} (\sigma ,\piv , \mu)$
is finite and needs not to be regularized. It can be expressed in terms
of the eigenvalues
$\epsl$ by:
$$
S_{eff}^{val} (\sigma ,\piv ,\mu) = N_c \T \sum_{0 < \epsl <\mu} (\epsl
-\mu)
\EQN\e323
$$
where we have used the relation:
$$
\ln (-iu + \epsl - \mu) - \ln (-iu + \epsl) = \int_{\epsl} ^ {\epsl - \mu} dx
{1\over {-iu+x}}
\eqn
$$
as well as \qeq{\e311b}.
Similarly one can separate the grand canonical expectation value of any
observable
$O:= \int d^3 r {q^{\dag}} \O q $ into a valence and a sea part:
$$
\langle O \rangle (\mu) = \langle O \rangle _{val} (\mu) + \langle O \rangle
_{sea}
\eqn
$$
with
$$
\langle O \rangle _{val} (\mu) = \langle O \rangle (\mu) - \langle O \rangle
(\mu =0) =
N_c \sum_{0 < \epsl < \mu } O_\lambda
\eqn
$$
Especially for the baryon number $\langle B\rangle $, which is the
integral of the time component of the baryon current
$b_\mu^E=(1/N_c){\bar
q}\gamma_\mu^Eq$ in Euclidean space, one obtains with
setting ${\cal O}^E=\gamma_4$   in \qeq{\e311a } and using
$b_0^M=-ib_4^E$ (cf. app.A):
$$
\eqalign{
\langle B \rangle (\mu ) &= {-i\over{N_c \T}} \cdot
 {d\over d\omega_E} \Sp \log \left( -iD -\omega_E\beta{\cal O}^E
-\mu\beta \right) \cr
 &= {1\over {N_c}} \int_{-\infty} ^{+
 \infty}
{{du}\over{2\pi i}} \sum_\lambda {1\over{ u + i (\epsl - \mu)}} = \cr
&= (-) \half \sum_\lambda {\rm sign} (\epsl - \mu ) = B_{val} (\mu) + B_{sea}
\cr}
\EQN\e327
$$
with
$$
\langle B \rangle _{val} (\mu) = \sum_{0 < \epsl < \mu } 1
\eqn
$$
and
$$
\langle B \rangle _{sea} = (-) \half \sum_\lambda {\rm sign} \epsl
\eqn
$$
Analogous expressions hold for the baryon density.

\noindent
In any case one notices that the valence contribution is finite and that there
is no need for a
regularization of this part. It is also important to realize that the
thermochemical potential $\mu$
is a real number both in the Minkowski space and in the Euclidean space, which
causes finally simply
a shift in the eigenvalues $\epsl \to \epsl - \mu$. It is therefore
different from the treatment of
the time component of a four vector $\omega_\mu$, which has  to
be Wick rotated from Minkowski
into Euclidean space like the real time: $\omega_0 \to i \omega_4$, where
$\omega_0$ and
$\omega_4$ are both real (cf. App.A).
{}From that and from the antihermiticity of $\gamma_4$ it is clear from
\qeq{\e327 } that the baryon number $B(\mu)$ originates from the
{\it imaginary part} of the effective Euclidean action.  Therefore the
baryon number
is finite and need no regularization. This philosophy will be prosecuted
throughout this work.      Actually the imaginary part is not zero in
the present case due to the additional term:
$\omega_E\beta{\cal O}^E$

\noindent
Using \qeqs{\e327} and \queq{\e323} we can write the valence part of the grand
canonical
effective chiral action as:
$$
{1\over \T} S_{eff}^{val} = N_c \left [ \sum_{0< \epsl < \mu} \epsl - \mu
B_{val} (\mu) \right ]
\eqn
$$
As it is known from statistical mechanics the total energy of a system at zero
temperature
$\T \to \infty$ (ground state) is given by:
$$
E_{tot} (\mu) = {1\over \T} S_{eff} (\mu) + N_c \mu \langle B \rangle (\mu)
\eqn
$$
which leads finally to:
$$
\etot (\mu) = \esea  + \eval (\mu) + \emes + \ebr
\EQADV\e332\SUBEQNBEGIN\e332a
$$
with \qeq{\e310b} and
$$
\eval (\mu) = N _c \sum_{0< \epsl < \mu} \epsl = N_c \epsilon_{val}
\SUBEQN\e332b
$$

\subsec{Valence Picture and Bosonized Picture}

{}From \qufig{\f31} we can distinguish 3 different regions  concerning
the behaviour of the valence level $O^+$ and, associated to it,
the behavior of $E$ and $\langle B \rangle $ ($\emes = \ebr =0$),
which have been discussed already by  Meissner Th et al. (1990a).

\item{(A):}
$$
\eqalign{
&\epsval > \mu >  0    \cr
&\langle B \rangle (\mu) =0 \cr
&\eval (\mu) =0 \cr
&\etot = \esea \cr}
\eqn
$$
\item{(B):}
$$
\eqalign{
&\mu > \epsval >  0    \cr
&\langle B \rangle (\mu) = \langle B \rangle_{val} =1 \cr
&\eval (\mu) =
N_c \sum_{0 < \epsl < \mu} \epsl
= N_c \epsilon_{val}
\cr
&\etot = \esea +
N_c \sum_{0 < \epsl < \mu} \epsl
= \esea + N_c \epsilon_{val}
\cr}
\eqn
$$
\item{(C):}
$$
\eqalign{
&\epsval < 0    \cr
&\langle B \rangle (\mu) = \langle B \rangle_{sea} =1 \cr
&\eval (\mu) =0 \cr
&\etot = \esea \cr}
\eqn
$$

\noindent
Case (A) is uninteresting for our purpose, because the valence particle is part
of the positive continuum and $\langle B \rangle =0$.
The cases (B) and (C) give $\langle B \rangle =1$ as desired.
Hereby  the valence orbit is counted explicitly for $\etot (\mu)$
as well as for all other static observables if $\epsval >0$ (B). If $\epsval
<0$ the valence particle gets part of the
negative spectrum (Dirac sea) which now as itself carries baryon number
$\langle B \rangle (\mu) = \langle B \rangle_{sea} =1$ due to the fact that it
contains now 1 orbital more. For the
following it is convenient to introduce the notation:
$$
\eta_{val} = \cases{ 1 &, if $\epsval >0$ \cr
                     0 &, if $\epsval <0$ \cr}
\EQN\e336
$$
so that we can write for a system with $B=1$:
$$
\eqalign{
E_{B=1} &= N_c \eta_{val} \epsval + \esea + \emes + \ebr \cr
\langle O \rangle _{B=1} &= N_c \eta_{val} \langle O \rangle_{val} + \langle O
\rangle_{sea} \cr}
\EQN\e337
$$

\noindent
\qufig{\f32} shows $\eval$,$\esea$ and $\etot$ ($\emes =\ebr =0$) for a meson
profile with linear shape:
$$
\theta (r) = \cases { - n \pi (1 - \roR ) &, if $r < R$ \cr
                        0                &, if $ r \ge R $\cr}
\EQN\e339
$$
on the chiral circle with winding number $n=1$, $M=465 \MeV$ and $m_\pi
=0$ in dependence   of the size parameter $R$.
One clearly recognizes that $\etot (R)$ has a local minimum at $R \approx 0.8
\fm$ corresponding to a solitonic
solution with $B=1$. In the next section we will construct these solutions
selfconsistently by variation
of $\etot [\theta (r) ]$ with respect  to all degrees of freedom $\{
\theta (r) \}$.
On a first glance to \qeqs{\e336,\e337 } it appears as if observables
change discontinously
their value if the valence single particle energy changes sign. In fact
this is not true. The regularization function takes care of this
automatically and indeed one can show analytically and numerically that
no discontinuity and no kink appears
(see Fig. 3.2. as examples).

\subsec{Connection between Baryon Number $B$ and Topological Winding Number
$n$;}
\subsec{Gradient Expansion}

{}From \qufig{\f31} one can see that for very large  values of the
size parameter $R$ the valence orbital approaches
the states emerging from the negative continuum and at the
end cannot be distinguished any more  from them
(Kahana et al.1984, Kahana and Ripka 1984).
On the other hand we   know that for those large  profile sizes $R$
the gradients of the meson fields get small
$\partial_r\sigma\propto\partial_r\pi\propto {1\over R} $ so that the gradient
or heat kernel expansion, discussed in
sect.2.  should be valid.
The  gradient expansion of the baryon current has been performed by
Goldstone and Wilczek (1981) and is shown
in appendix B. Indeed  it turns out that the gradient expanded baryon
number of the Dirac sea $\langle B \rangle_{sea}$
is identical to the topological winding number $n$ of the meson profile
$$
\lim_{R \to \infty} \langle B \rangle _{sea} =n
\eqn
$$
which is $n=1$ in the case considered above.
This feature generally holds in the region (C). Therefore in the case for large
$R$ the valence particle
has got part of the negative spectrum and the baryon number coincides with the
topological winding number.
Similar considerations hold for higher winding numbers (Kahana et al.1984).
For sufficiently large $R$ one therefore gets close to the philosophy of the
{\it topological soliton models} like e.g. the
Skyrme model which contains no valence quarks but relates the baryon number of
the soliton purely to the topological
winding number of the Goldstone field.
Furthermore it turns out, that for large $R$ the energy of the Dirac sea $\esea
(R)$ approaches the kinetic energy of
the mesons, which is the leading order in the gradient expansion
(Meissner Th et al.1988):
$$
\lim_{R \to \infty}
\esea (R) = E_{kin} (R) = \half \int d^3 r \left [ (\vec\nabla \sigma)^2 +
(\vec\nabla \piv)^2 \right ] \propto R
\eqn$$
as long as this expansion is convergent or at least an asymptotic series in
$1\over R$ (Zuk and Adjali 1992).

\sect{Solitonic Solutions of the Nonlinear Model}

\subsec{Mean Field Equations}

\noindent
The equations of motion for a system with baryon number $B=1$
are given by the stationary points of the grand canonical effective
chiral action $S_{eff}(\mu,B=1)$. In practice the hedgehog ansatz
\qeq{\e34 } is used and the system is assumed to be constrained to the
chiral circle \queq{\e36 }. The restriction to the chiral circle is
quite essential and its origin will be discussed in sect. 3.5 and 8.4
(Sieber et al. 1992; Meissner Th et al. 1993; Weiss et al. 1993).
The stationary points of the
grand canonical effective chiral action $S_{eff}(\mu, B=1)$ (see
\qeq{\seffmu })   with respect to the chiral angle $\theta(r)$ are
then given by:
$$
\delta_{\{\theta (r) \}} S_{eff} (\mu, B=1) =0
\eqn
$$
which reduces in case of time independent meson fields to
the variation of the energy
$$
\delta_{\{\theta (r) \}} E (\mu, B=1) =0
\EQN\e342
$$
In order to perform the variation we start from \qeqs{\e337,\e310}
and use the spectral representation
$$
\epsl = {\bra {\lambda}} h {\ket{\lambda}} = \int d^3 r \phi _\lambda ^{\dag}
(\vec r) h \phi_\lambda (\vec r)
\eqn
$$
Because the variation $\delta \langle \lambda \vert \lambda \rangle = \delta
\int d^3 r \phi_\lambda ^{\dag} (\vec r)
\phi_\lambda (\vec r)$ vanishes we get for the variation of the single particle
energy $\epsl$:
$$
\delta_{\{\theta (r)\}} = g \int dr r^2 \left [ \left ( - \sin \theta (r)
\right ) \cdot s_\lambda (r) +
\left ( \cos \theta (r) \right ) \cdot p_\lambda (r) \right ] \delta \theta (r)
\eqn
$$
with
$$
\eqalign{
s_\lambda (r) &= \int d \Omega {\bar \phi}_\lambda (r,\Omega)
\phi_\lambda(r,\Omega) \cr
p_\lambda (r) &= \int d \Omega {\bar \phi}_\lambda (r,\Omega)  i \gamf
(\vec\tau {\hat r} ) \phi_\lambda(r,\Omega) \cr}
\EQN\e345
$$
which lead to the equations of motion:
$$
\eqalign{
&\sin \theta (r) \left [ N_c g \left ( S_0 (r) + \eta_{val} s_{val} (r) \right
)
- 4 \pi f_\pi m_\pi ^2 \right ] = \cr
&\cos \theta (r) \left [ N_c g \left ( P_0 (r) + \eta_{val} p_{val} (r) \right
) \right ] \cr}
\EQN\e346
$$
where:
$$
\eqalign{
S_0 (r) &= \sum_\lambda {\RR}_2 (\epsl, \Lambda) s_\lambda (r) \cr
P_0 (r) &= \sum_\lambda {\RR}_2 (\epsl, \Lambda) p_\lambda (r) \cr}
\EQN\e347
$$
and the regularization function ${\RR}_2 (\epsl , \Lambda)$ has been defined in
\qeq{\e314}.

\subsec{Numerical Treatment}

\noindent
Eq.\queq{\e346} has  been solved numerically by Reinhardt and Wuensch
(1988) as well as Meissner Th et al.(1989), who used
a standard selfconsistent procedure as it is known from Hartree and
Hartree-Fock calculations in atomic and nuclear physics.
One starts with a reasonably chosen profile $\theta_{0} (r)$ given e.g.
by \qeq{\e339 },
diagonalizes the $h$ as described in sect.3.1, calculates
$s_\lambda$,$p_\lambda$,$s_0$,$p_0$ from the eigenfunctions $\phi_\lambda (\vec
r)$ and the eigenvalues $\epsl$ from
\qeq{\e347} and obtains a new profile function $\theta_{1} (r)$ from the
equation of motion \queq{\e346}.
The procedure is iterated until a desired degree of selfconsistency is reached.
An equivalent method for solving \qeq{\e346} consists in parameterizing $\theta
(r)$ in terms of a parameter set
$\{ \alpha_k \, ,\, k = 0,1,2,,\dots\}$ covering the whole $r$-dependence of
$\theta (r)$ as
fully as possible and performing
a minimization of the total energy in the $\{ \alpha_n \}$-space.
First calculations (Diakonov et al.1988, Meissner Th et al.1988) used a
simple, one dimensional parameterization in terms
of the size $R$ of an   appropriate chosen meson profile  $\theta\left(
\roR\right)$ like it was done in
\qeqs{\e37,\e339} and minimized just with  respect to $R$ (cf.
\qufig{\f32}).
The  accuracy of this very rough ansatz was improved by increasing the
number of parameters to $k=3$ (Diakonov.et al.1989),
which already gives results close to those of the selfconsistent method. In a
very recent work $\theta(r)$ is calculated at
a given number $k = 10 -30$ meshpoints in a sufficiently large interval $[0,D]$
and a systematic search for the minimum
of $\etot$ in the $k-$dimensional space spanned by these meshpoints is
performed using an elaborate minimization
algorithm (Sieber.et al.1992).
In  other approaches (Diakonov.et al.1988, Adjali et al.1991,1992) the
energy of the Dirac sea $\esea [\theta]$
(\qeq{\e310\e310b}) is approximated by expanding $\Sp \ln D^{\dag} D$ ($iD = i
\sld - g (\sigma + i \vec\tau\piv \gamf)$) in terms
of the `perturbation' $g \sld (\sigma + i \vec\tau\piv\gamf)$
like it was done in sect.2.3.
The corresponding   expression is exact in the case of a very large
meson profile size as well as in the case of small
deviations from the vacuum configuration ($\sigma = \sigma_V =f_\pi$, $\piv
=0$). In doing so the numerical diagonalization
of $h$ is avoided and the variational problem reduces to the solution of
an integro-differential equation in $\theta$.

\noindent
Generally the variation is performed for a given value of the constituent quark
mass $M$ or equivalently the ratio
$\lambda = {\Lambda \over M}$, which is uniquely related to $M$ by
\queq{\2004d } and \queq{\2003a }.
Furthermore finite pion masses ($m_\pi =139 \MeV$) as well as the chiral limit
($m_\pi =0$) have been considered.
It turns out, that solitonic solutions of \qeq{\e346} exist if and only if $M$
exceeds an critical value $M>M_{cr}$ or
equivalently $\lambda < \lambda_{cr}$. The corresponding values are shown in
\qutab{\t31}. This cusp behavior is typical
for localized (solitonic solutions) of a system of coupled
non linear equations (see e.g. Lee 1981 and ref.
therein) and has been observed also in other
chiral quark meson models (Birse 1990 and ref. therein).

\subsec{Selfconsistent Fields; Mean Field Quantities}

\noindent
\qufig{\f33} and  \qufig{\f34} show the total ({\it mean field}) energy
$E_{MF} = \etot$ as well as the isoscalar electric quadratic
radius $\langle R^2  \rangle = 4 \pi \int dr r^4 b_0 (r)$ which is
nothing but the sum of corresponding proton and
neutron contributions,  in dependence of the constituent quark mass $M$
for $m_\pi =0$
(Meissner Th et al. 1989, Meissner Th and Goeke 1990).
Both quantities are divided into their
valence and sea   contributions, respectively. One should notice that
the mesonic self energy $\emes$ (\qeq{\e310\e310c}) vanishes
due to the chiral circle  constraint \qeq{\e36}.
Using $m_\pi =139 \MeV$ one finds that the chiral breaking term
$\ebr$ (\qeq{\e310\e310d}) is small and its influence
on the solitonic solutions is in general rather small of less than
$10-15\%$
(Meissner Th et al.1989, Meissner Th and Goeke 1991).
Apparently  it affects the form of the asymptotic behavior of of
the pion tail of the selfconsistent solution
$$
\lim_{r \to \infty} \pi (r) = A e^ {- m_\pi r}  {{1+ m_\pi r } \over  {r^2}}
\EQN\e348
$$
Thus it has some
influence on quantities which are sensitive to that form, especially if they
diverge in the chiral
limit $m_\pi \to 0$, like e.g. the magnetic polarizability and the $\Sigma$
commutator.
Actually \qeq{\e348} can be obtained by substituting the sea energy $\esea$
(\qeq{\e310\e310b}) in the variational principle
(\qeq{\e342})  through the gradient expanded expression. One has then to
solve  the resulting differential equation in the
coordinate space. It  contains formally an infinite number of
derivatives for which  in the asymptotic region ($r\to\infty$),
only the contribution from the kinetic term $\left( {{\partial
\pi}\over{\partial r}} \right )^2 $
survives.

\noindent
In \qufig{\f35} and \qufig{\f36} the meson profiles $\sigma (r)$ and $\pi (r)$
of the selfconsistent solutions of \qeq{\e346}
for three representative values of the constituent quark mass $M$ with
$m_\pi =0$ are presented. The corresponding
baryon densities are shown in \qufig{\f37}. As one can see the actual form of
$\sigma (r)$ and $\pi (r)$ as function of the
physical distance $r$ is quite independent of $M$.
If we denote  with ${R}$ some characteristic size of the profile, we
find that the condition
$$
\vert\partial_r\sigma\vert\approx\vert\partial_r\pi\vert\ll M \;
\Leftrightarrow \; 1 \ll M\cdot {R}
\EQN\e349
$$
for the validity of the gradient expansion is fulfilled if $M \geq 1000 \MeV$
and therefore in this case the baryon number
$\langle B \rangle _{sea}$ of   the Dirac sea and the topological
winding number $n$ of the pion field coincide
(cf.app. B).
Indeed as one can  see from \qufig{\f35} the valence energy becomes
negative $\eval <0$ if $M\ge 780 \MeV$.
Due to the discussion in sect.3.2 one therefore comes close to  the
philosophy of the topological soliton models if $M$ gets large
(cf.\qufig{\f14}),
whereas the valence quark picture holds for small $M$ (cf.\qufig{\f13}).
The important  point is the fact, that for a given $M$ the model can
{\it decide} between the two pictures {\it dynamically} by
the equations of motion \qeq{\e346} and therefore switch continuously between
them just by the variation of $M$. In order to make a definite
decision which picture is favored one has to calculate baryonic observables and
see for which values of $M$ they are
  described  properly. If one looks e.g. at  the isoscalar charge radius
$\langle R^2 \rangle = \langle R^2 \rangle_p + \langle R^2 \rangle_n$
(\qufig{\f36}) one recognizes at once that one is
restricted to rather small constituent quark masses ($M\approx 360 \MeV$) just
above the critical cusp $M_{cr}$ in order to reproduce the experimental value
of $0.63 \fm^2$.
The corresponding mean field energy lies at about $E_{MF} \approx 1200 \MeV$.
High values of $M$ corresponding to
the bosonized picture lead to a baryonic system whose extension is by far too
small.
The region $M \approx 350 \dots 450 \MeV$ is also consistent with that of
various non
relativistic constituent quark models (see.e.g.Bhaduri 1988 and ref.therein).
Furthermore one can see from \qufig{\f36} that the contribution of the valence
quarks is clearly the dominating one.
Counterexamples as e.g. the neutron squared charge radius or the
nucleon polarizability are found as well.
We will see in chapter 5 that this will be true for most of the nucleon
observables. If and how this feature changes if vector
mesons ($\omega$,$\vec\rho$,$\vec A_1$) are explicitly coupled will be
discussed in chapter 7.

\sect{Higher Winding and Baryon Numbers}

For the construction of the solutions  of the mean-field equation of
motion for $\theta (r)$,
one has assumed that $\theta(0)=-n\pi$,  with the
topological winding number $n=1$. There is,
however a priori no reason in this model, why $n$ should be fixed to
$1$, because the baryon number $B$ is carried by the valence
quarks, if one chooses an appropriate thermochemical potential
$\mu$. This is different from
the topological soliton models (e.g. the Skyrme model), where $B=n$
from the very beginning.
There is even no a priori reason why $n$ should be an integer number, if one
leaves mathematical arguments
like   continuity of the $\piv$ field in the origin or the convergence
of the gradient expansion for $B$ (cf. app. B) aside.
In contrast to to the Skyrme model, where $n \in Z$ is required in order to
obtain a finite energy, the total energy $\etot$ is
finite  for any value of $n$ in the present approach. Diakonov
et al.(1988,1989) and Berg et al.(1992) have investigated the
behavior of $\etot (n)$ as a function of $n$ for a fixed profile form. It
turned out, that the minimum of $\etot$ definitely
lies at the point $n=1$, which means in fact that this feature is also a
dynamical consequence of the equations of motion.
Unfortunately   the whole analysis is based purely on numerical
arguments and there is up to now no analytic proof of this fact.

\noindent
In addition Berg et al.(1992) studied systems with higher baryon numbers $B>1$.
They found that local minima corresponding to
solitonic solutions only exist if $n \le B$ and in any case these local minima
appear for integer winding number $n$.
For  example for the $B=2$ system they found a soliton in the $n=1$
sector with 2 valence orbitals ($0^+$,$0^-$) occupied if
$M > 370 \MeV$.
In the $n=2$ sector solitonic solutions exist if $M> 560 \MeV$. The main
difference between the two cases is the fact, that for
$n=2$ the $0^+$ as well as the $0^-$ orbital cross the zero line and get part
of the
negative spectrum whereas in the case of $n=1$
only the $0^+$ orbital comes down and the $0^-$ orbital remains in the positive
spectrum.
(cf.\qufig{\f31}).

\sect{Collapse in the Linear Model}

Up to now in  any case the meson fields $\sigma (r)$ and $\pi (r)$ have been
restricted to the chiral circle
$\sigr ^2 + \pir ^2 = f_\pi ^2$, known as {\it nonlinear realization of chiral
symmetry} (Weinberg 1968) .
Due to this constraint
the variation $\delta \etot =0$ is performed only with the chiral angle $\theta
(r)$ as degree of freedom.
The motivation to consider the action as an effective chiral theory of
the nucleon from an instanton liquid model of QCD
(Diakonov et al.1986)   even suggests that only these degrees of
freedom, namely the Goldstone ones, should be taken into
account. On the other side in spirit of the \NJL there is no reason at all why
the meson fields should be restricted to the
chiral circle and the constraint \qeq{\e36} seems to be an artificial one.
Sieber et al.(1992) performed a detailed analysis of the full linear model with
both $\sigma$ and $\pi$-degrees of freedom
and showed  that at least in case of the proper time regularization
scheme no localized solitonic solution with finite
energy exists  in this case but the system collapses into a
configuration with both size and energy being zero while
$B=1$ is maintained.
To this end they choose a parameterization for $\sigma (r)$ and $\pi
(r)$ in terms  of their size $R$ and their depths $U$
by:
$$
\eqalign{
\sigr &= f_\pi \left [ 1 + U f \left ( \roR \right ) \cos \theta \left ( \roR
\right ) \right ] \cr
\pir &=  f_\pi U f \left ( \roR \right ) \sin \theta \left ( \roR \right ) \cr
} \EQN\e350
$$
where $f(r)$ is a strictly monotonously decreasing function with $f(0) >0$,
$f(\infty) =0$ and $\theta (0) = -n\pi$.
Taking the limit $u \to \infty$, $R\to \infty$ but $UR^\alpha = const.$, the
collapse described above occurs if
$1< \alpha < {3\over 2}$.  This is due to the fact that in this limit
for $\alpha >1$ altogether  $n$ valence
orbitals cross the zero line and get part of the negative spectrum which means
that they are not taken into account
in the expression for the total energy any more ($\eta_{val} =0$ in
\qeqs{\e336,\e337}) and the baryon number
of the system gets  $\langle B \rangle = \langle B \rangle_{sea} =n$.
Furthermore it turns out that the Dirac sea approaches
its vacuum configuration,  because any deviation from this is suppressed by the
UV cutoff $\Lambda$.
The reason for the occurrence   of this behavior is the fact, that in
solving the quark loop of NJL without vector mesons  the sea
energy
$\esea$ gets regularized by
$\Lambda$ whereas the finite  baryon number $\langle B \rangle$ is not.
On the other hand $\emes$ vanishes for $R\to \infty$ as long as $\alpha <
{3\over 2}$, which proves the statement above.

\noindent
Earlier works (Meissner Th et al.1989, Reinhardt and Wuensch 1989,
Meissner Th and Goeke 1991), which claim the
  existence of solitonic solutions also in the linear model turned out
to be premature in this respect since too small values of  $K_{max}$
(cf. sect. 3.1.), where chosen in the basis for the diagonalization of
the single particle hamiltonian h.  This fact has been emphasized in the
mean time by various authors
(Watabe and Toki 1992, Sieber et al.1992, Kato et al.1993).

\noindent
In sect.8.4. it will be shown that a modified version (Meissner   Th
et al. 1993)  of the
\NJL, which in addition to the spontaneously broken chiral
symmetry also simulates the {\it anomalous breaking of scale invariance} in QCD
shows perfectly stable solitonic solutions
also without  the non linear constraint, while the properties of the
soliton as well as the relevant nucleon observables
stay nearly unchanged. Thus the restriction of $\sigma$ and $\pi$ to the chiral
circle is justified by a model
which implements the concept of trace anomaly in QCD.
Stability can also be achieved by
adding the $U_A(1)$ breaking t'Hooft term (Kato et al.
1993), at least for a certain range for t'Hooft coupling strength.

\vfill\eject

%
%
%
\chap{Restoration of Broken Symmetries - Collective Quantization}
%
%
\def\tu{{\tilde U}}
\def\bu{{\bar U}}
\def\tp#1{ [ {#1}^2 ]_{(2)}}

\def\op#1{ [ {#1} ]_{(1)}}
\def\opq#1{ [ {#1}^2 ]_{(1)}}
\def\qsch{{\tilde q}}
\FIG\f40{The rotational moment of inertia (right y-axis),
its valence and sea-quark  contribution, and the nucleon-delta-splitting
(left y-axis) is given in dependence of the constituent quark mass M.
The pion mass is chosen to be 139$\MeV$.  }
\FIG\f41{The mean field energy $E_{MF}$ (HEDGEHOG)
as well as the energies of nucleon $E_N$ (NUCLEON) and delta
$E_\Delta$
(DELTA) and the translational and rotational corrections
of the selfconsistent solution in the non-linear model
($m_\pi =139\MeV$) in dependence
of the constituent quark mass $M$
(Pobylitsa et al. 1992)
.}

As in nonrelativistic many particle physics we face the problem that the
hedgehog mean field solution of the classical equations of motion, which
we have constructed in the last chapter, breaks the rotational, isorotational
and the translational symmetry of the full theory. This means that the
eigenstates $\ket{\lambda}$ of the $1$-particle hamiltonian $h$ do not
carry good spin, isospin or momentum quantum numbers.
Because we want
to describe nucleonic systems, which have these quantum numbers, those
symmetries must be restored.
This is done by coupling the corresponding expectation value
($\langle \vec J \rangle$, $\langle \vec T \rangle$, $\langle \vec P \rangle$)
through a Lagrange multiplier to the effective chiral action $S_{eff}$
like
it was done in sect.3.2.in case of the baryon
number $B$ with the thermochemical potential $\mu$. This turns out to
be equivalent to consider the soliton in an (iso-)rotating or moving system,
called {\it cranking} (sect.4.1) or
{\it pushing} method (sect.4.3), respectively.
Because the grand spin
$\vec G = \vec J + \vec T$ is a good quantum number of the hedgehog single
particle state, spin and isospin degrees of freedom are coupled and it is
enough to consider one of them, e.g.. the isospin.
Whereas the spectrum of the momentum
operator $  {\vec P}$ is continuous, the isospin degrees of freedom have
to be quantized. For this we will use the semiclassical collective
quantization method (sect.4.2).
Zero point energies appear, because the expectation values of the
$2$-particle operators
${\vec T}^2$ and ${\vec P}^2$ do not vanish (sect.4.4). Finally we will be
able to write down the expressions for the masses of nucleon $N$ and delta
$\Delta$ in a system at rest and give numerical values (sect.4.4).

In this chapter we will throughout consider the non-linear version of
the \NJL, i.e. restrict $\sigma$ and $\piv$ to the chiral circle
$\sigma^2 + \piv^2 = \sigma_V ^2 = f_\pi ^2$ and use the notation:
$$\eqalign{
U &= \sigma + i \vec\tau \vec\pi  \cr
U_5 &= \sigma + i \gamma_5 \vec\tau \vec\pi  \cr}
\EQN\e41
$$

\sect{Iso-Rotational Motion: Cranking}

In order to restore the isorotational symmetry, i.e. to construct a system
with  good isospin quantum numbers (e.g. $N$,$\Delta$) we follow a
procedure
which was established in nonrelativistic many particle physics (see e.g.
Ring and Schuck 1980) and is known as {\it cranking} approach.
It has been used also quite successfully in case of the Skyrme model
(Adkins et al.1983) as well as in the chiral sigma model with valence quarks
(Cohen and Broniowski 1986). Furthermore one can generalize the method
to $SU(3)$-flavor and calculate hyperon properties, which will be done
in chapter 6.

\subsec{Adiabatic Isorotation}

The  main idea is to perform an adiabatic isorotation of the hedgehog
meson
fields with the angular frequency $\vec \Omega$:
$$
\eqalign{
\sigma (\vec x ) &\to {\tilde \sigma} (t, \vec x) = \sigma (\vec x) \cr
\pi^a (\vec x) &\to {\tilde \pi}^b (t, \vec x) = D^{ab} (t) \pi^a (\vec x) \cr
U_5 (\vec x) &\to \tu_5 (t,\vec x) = \RR^{\dag} (t) U_5 (\vec x) \RR (t) \cr}
\EQN\e410
$$
with the $SU(2)$-rotation matrix
$$
\RR (t) = e^{ i {{\vec\Omega\vec\tau} \over 2} t}
\EQN\e411
$$
and its $SO(3)$-representation $D$, defined by:
$$
\RR (t) \tau^a \RR^{\dag} (t) = \tau^b D^{ba} (t)
\EQN\e412
$$
For simplicity we will first stay in Minkowski space without
regularization
and valence quarks, i.e. we set the thermochemical potential $\mu$ to zero,
and comment on the general case later on.

The effective action in the rotating system is given by:
$$
\eqalign{
e^{i S_{eff} [\tu_5 ]}
&= \int \D \bar\qsch \D \qsch
e^{ i \int d^4 x \bar\qsch [i D ({\tilde U}_5)] \qsch } \cr
&= \int \D \bar q \D q
e^{ i \int d^4 x \bar q \left [ i D (U_5) +
\beta \vec \Omega {\vec\tau \over 2} \right ] q } =   \cr}
\EQN\e413
$$
with: ${\tilde{q}} = R^{\dag} (t) q$,
and therefore:
$$
S_{eff} [ \tu_5 (\Omega) ] = \Sp \ln \left [ iD + \beta \vec\Omega
{\vec\tau \over 2} \right ] = \Sp \ln \left [ i \partial_t -
\left ( h - \vec\Omega
{\vec\tau \over 2} \right ) \right]
\EQN\e414
$$
{}From this we see, that $\vec\Omega$ acts as a Lagrange multiplier
constraining
the isospin
$$
T^a = \int d^3 x \bar q (\vec x) \beta \t a q (\vec x)
\EQN\e415
$$
similar to the thermochemical potential $\mu$ in case of the baryon number
(sect.3.2).
The $1$-particle hamiltonian $h (\vec\Omega)$
in the rotating system can be read off from
\qeq{\e414} to:
$$ h (\vec \Omega) = h - \vec\Omega {\vec\tau \over 2}
\EQN\e416
$$

\subsec{Perturbative Cranking, Moment of Inertia}

We assume $\vec\Omega$ to be small, so that the problem can be treated
perturbatively.
We will comment on the validity
of this assumption later on.

The 1st order in $\Omega$ vanishes
$$
{{\delta S_{eff} (\Omega)}\over {\delta\Omega^a}}  \Big\vert_{\Omega =0} =
\langle T^a \rangle = 0
\EQN\e417
$$
because of hedgehog symmetry:
$$
\sum_{\lambda(G,G_z)}
{\bra{\lambda(G,G_z) }}
\t a
{\ket{\lambda(G,G_z) }}=0
\eqn
$$
In second order we get:
$$ S_{eff} (\vec\Omega) = \T E (\vec\Omega) = \T \left [
E_{MF} + \half \Omega^a \Theta^{ab} \Omega^b \right ]
\EQN\e418
$$
where the {\it moment of inertia} tensor $\Theta^{ab}$ is defined by:
$$
\Theta^{ab} = {1\over \T}
{{\delta^2 S_{eff} (\Omega)}\over {\delta\Omega^a \delta\Omega^b}}
\Big\vert_{\Omega =0} = {1\over \T}
\Sp  \left [ (i \partial_t - h)^{-1}
\t a
(i \partial_t - h)^{-1}
\t b
\right ]
\EQN\e419
$$
In terms of the $1$-particle eigenstates $\ket\lambda$ (cf.\qeq{\e35})
$\Theta^{ab}$ is given as a sum over all particle ($\epsn >0$) - hole
($\epsl <0$) matrixelements of the 'perturbation' $\vec\tau$:
$$ \Theta^{ab} = {{N_c}\over 2}
\sum_{{{\epsl <0}\atop {\epsn >0}}}
{   {{\bra\lambda} \tau^a {\ket\nu} {\bra\nu} \tau^b {\ket
\lambda}}\over
{\epsn - \epsl} }
\EQN\e420
$$
which is known as {\it Inglis formula} (Ring and Schuck 1980)

Of course it is necessary to regularize $S_{eff} (\Omega)$ and $\Theta$.
In case of the proper time regularization, which works in Euclidean space,
the expression for $\Theta^{ab}$ was derived by Reinhardt (1989):
$$
\eqalign{
\Theta_0 ^{ab}  &= { N_c \over 4} {1\over{\sqrt{4\pi}}} \prop ds
s^{-{3\over 2}} \sum_{\lambda\nu}
{\bra \lambda} \tau^a {\ket\nu} {\bra\nu} \tau^b {\ket \lambda}
{\cal R}_\Theta (\el,\enu,s) \cr
{\cal R}_\Theta (\el,\enu,s) &=
\left [
{ { e^{-s\enu^2} - e^{-s\el^2} } \over {\el^2 - \enu^2} }   ~ -~ s
{ { \el e^{-s\enu^2} + \enu e^{-s\el^2} } \over {\enu + \el} }
\right ]   \cr}
\EQN\e421
$$
Hereby  it is essential to perform the continuation of the rotation
frequency
$\vec\Omega$ into the  Euclidean space
($\vec\Omega\to\vec\Omega_E=i\vec\Omega$)
like
it is done for the time  component of a physical $4$-vector,  e.g. the
time as itself ($t\to\tau=it$) or the time component of a  $\rho$-
meson.
This means that both $\vec\Omega$ and $\vec\Omega_E$ are  real numbers
and the fermion
determinant in Euclidean space is: $\Sp \ln  ( - iD - i \beta
{\vec\Omega}_E
{\vec\tau \over 2} )$. Otherwise it turns out  that the regularized
moment of inertia for the vacuum configuration $U_5 = 1$
does not vanish, which has to be considered as unphysical (Reinhardt 1989).

The separation of the valence quark contribution by using the
grand canonical effective action in the rotating system
$S_{eff} (\mu, \Omega)$ with the thermochemical potential $\mu$
(cf. sect. 3.2.) is straightforward yielding:
$$
\Theta^{ab} \Big\vert_{val} =
{N_c \over 2} \sum_{\nu\ne val}
{ { {\bra {val}} \tau^a {\ket\nu} {\ket\nu} \tau^b {\ket {val}} } \over
{\enu - \epsilon_{val}}   }
\EQN\e421
$$
which is nothing but the standard 2nd order perturbation theory result
for the orbital $\ket{val}$ with the perturbed hamiltonian \qeq{\e416}.

On simple symmetry reasons $\Theta^{ab}$ is diagonal
and can be written as $\Theta^{ab} = \delta^{ab} \Theta$.
By using the $1$-particle eigenstates $\ketl$ and $\ket\nu$ of the
selfconsistent solitonic solution as they were obtained in sect.3.3,
Goeke et al.(1991) and Wakamatsu and Yoshiki (1991) calculated numerically
the moment of inertia, which will serve as the essential quantity for
obtaining the masses  of $N$ and $\Delta$ from the soliton mean field
energy
$E_{MF}$ (see \qufig{\f40 }).

\subsec{Observables in the Rotating System}

Generally the expectation value of an observable
$O = \int d^3 x q^{\dag} (\vec x) \O q (\vec x)$ in the rotating system
reads (in Minkowski space):
$$
\eqalign{
\langle {\tilde O} \rangle (\Omega) &= {1\over \T}
 { { \int \D \bar\qsch \D \qsch \int d^4 x
[ \bar\qsch \beta \O \qsch ]
e^{ i \int d^4 x \bar\qsch [i D ({\tilde U}_5)] \qsch }  }\over
{ \int \D \bar\qsch \D \qsch
e^{ i \int d^4 x \bar\qsch [-i D ({\tilde U}_5)] \qsch }  }}
\cr
&= {1\over \T} {\delta\over{\delta\omega}} \Sp \ln \left
[ iD (U_5) + \beta \vec\Omega {\vec\tau \over 2} +  \omega \RR \O \RR^{\dag}
 \right ] _{\omega =0} \cr}
\EQN\e423
$$
One should be aware that this expression differs from
the simpler one
$$\langle O \rangle (\Omega) =
{1\over \T} {\delta\over{\delta\omega}} \Sp \ln \left
[ iD (U_5) + \beta \vec\Omega {\vec\tau \over 2} +  \omega \O
 \right ] _{\omega =0}
\EQN\e424
$$
if $\O$ is an isovector by:
$$
\langle {\tilde O}^a \rangle (\Omega) = D^{ba} \langle O^b \rangle (\Omega)
\EQN\e425
$$

In  the perturbative cranking approach any $\langle {\tilde O} \rangle
(\Omega)$
is expanded up to 1st order in $\Omega$. If this order vanishes
generally only the
0th order is taken into account.
For the case
of $g_A$ and the isovector magnetic moment we will extend this
method to
higher orders and discuss the results shortly in sect. 5.1. and
5.2..

In the special case of the isospin operator \qeq{\e415} we have:
$$
\langle {\tilde T}^a \rangle (\Omega) = D^{ba} \langle T^b \rangle (\Omega)
= D^{ba} \Theta^{bc} \Omega^c = \Theta D^{ba} \Omega^b
\EQN\e426
$$
On the other side for the total spin:
$J^a =\int d^3 x \bar q(\vec x) \beta \left ({\sigma^a \over 2} +
\vec x \times {{\vec\nabla}\over i } \right )  q (\vec x) = G^a - T^a$
one finds:
$$
\langle {\tilde J}^a \rangle (\Omega)
= \langle J^a \rangle (\Omega)
= \langle G^a \rangle (\Omega) -
\langle T^a \rangle (\Omega) =
- \langle T^a \rangle (\Omega) =
- \Theta \Omega^a
\EQN\e427
$$
which is true because $[h,G^a ] =0$ and therefore $\langle G^a \rangle$
vanishes
up to 1st order in $\Omega$ as it is known  from standard quantum
mechanics
perturbation theory.
{}From \qeq{\e426} and \qeq{\e427} we read off the relation between spin
and isospin expectation values:
$$
D^{ba} \langle {\tilde J} ^b \rangle +
\langle {\tilde T} ^a \rangle
=0
\EQN\e428
$$
and furthermore
$$
\langle \vec{\tilde J}  \rangle^2
=
\langle \vec{\tilde T}  \rangle^2
\EQN\e429
$$

\sect{Collective Quantization of the Iso-Rotational Degrees of Freedom}

Up to now the expectation values
$\langle {\tilde T} \rangle (\Omega)$ and
$\langle {\tilde J} \rangle (\Omega)$
are continuous quantities. The {\it semiclassical collective quantization}
method, which was established by Adkins et al.(1983) in the Skyrme model,
can be used for quantizing the rotational and isorotational degrees of freedom.
The main idea is to express
$\langle {\tilde T} \rangle (\Omega)$,
$\langle {\tilde J} \rangle (\Omega)$
as well as the rotational contribution of $S_{eff} (\Omega)$, which can
be written as  $\int dt L_{rot}(\Omega)=\half \T \Omega^2\Theta$,
in terms of the components of the $SU(2)$ matrix $\RR (t)$:
$$
\RR = b^0 + i b^a \tau^a   \qquad{\rm with} \quad
\sum_{\alpha=0} ^4 {b^\alpha}^2 =1
\EQN\e430
$$
and in terms of its time derivatives $\dot b_0$,$\dot b_i$:
$$
\eqalign{
{\erw{{\tilde T}^a}} (\Omega)
&= \Theta D^{ba} (\Omega) \Omega^b = (-i) \Theta
\Tr_{\tau} \left [ {\kr \RR} {\dot \RR} \tau^a \right ]    \cr
{\erw{{\tilde J}^a}} (\Omega)
&= - \Theta  (\Omega) \Omega^a = (-i) \Theta
\Tr_{\tau} \left [ \RR  {\dot \RR}^{\dag} \tau^a \right ]    \cr
L _{rot} (B) &= \Theta \Tr_{\tau} [{\dot \RR} {\dot \RR}^{\dag} ] =
2 \Theta \sum_\alpha   [ {\dot b}^\alpha ]^2    \cr}
\EQN\e431
$$
and to consider $b^\alpha$ as space components on the 3-dimensional sphere
$S^3$.
The corresponding canonical conjugate momenta are given by:
$$
\pi^\alpha (b^\alpha)
= {{\partial L_{rot} (B)}\over {\partial {\dot b}^\alpha}}
= 4\Theta {\dot b}^\alpha
\EQN\e33
$$
The  quantization is performed in the standard way by substituting the
coordinates $b^\alpha$ and the conjugate momenta $\pi^\alpha$ through
the operators
$$
\eqalign{
b^\alpha &\to  {\hat b}^\alpha \cr
\pi^\alpha &\to  {\hat \pi}^\alpha  =
(-i) {\partial\over{\partial b^\alpha}} \cr }
\EQN\e434
$$
on which the following  commutator rules    are imposed
$$
[{\hat b}^\alpha , {\hat\pi}^\beta ] = (+i) \delta^{\alpha\beta}
\EQN\e435
$$
The ${\hat b}^\alpha$ and ${\hat \pi}^\alpha$ are considered to act on a
collective wave function $\ket{\Psi (b)}$
which is normalized on the
3-dimensional unit sphere $S^3$ with respect to the surface integration
measure $d \mu (b)$:
$$
\langle \Psi(b)  \vert
\Psi(b)  \rangle =
\int d \mu (b)
\langle \Psi (b) \vert
b \rangle \langle b \vert \Psi(b) \rangle =1
\EQN\e436
$$
The $d \mu (b)$ can be expressed by the angular representation of $S^3$
by:
$$
d \mu (b) = d\psi d\phi \sin \phi d\theta \sin^2 \theta
\eqn
$$
with
$$
\eqalign{
b_0 &= \cos \theta    \cr
b_1 &= \sin \theta \cos \phi   \cr
b_2 &= \sin \theta \sin \phi \cos \psi   \cr
b_3 &= \sin \theta \sin \phi \sin \psi   \cr  }
\eqn
$$
and
$\theta \in [0, 2\pi ]$,
$\phi   \in [0,  \pi ]$,
$\psi   \in [0,  \pi ]$.
The $\langle {\tilde T} \rangle (\Omega)$ and
$\langle {\tilde J} \rangle (\Omega)$
now go over into the isospin and spin operators
${\hat t}^a$ and ${\hat j}^a$, respectively, which are given by:
$$
\eqalign{
{\hat t}^a &=  {i\over 2} \left [ b^a {\partial\over{\partial b^0}} -
                                  b^0 {\partial\over{\partial b^a}} -
                  \epsilon^{abc}  b^b {\partial\over{\partial b^c}}
                   \right ] \cr
{\hat j}^k &=  {i\over 2} \left [ b^0 {\partial\over{\partial b^l}} -
                                  b^l {\partial\over{\partial b^0}} -
                  \epsilon^{klm}  b^l {\partial\over{\partial b^m}}
                   \right ] \cr               }
\EQN\e436
$$
and fulfill the desired commutation relations:
$$
\eqalign{
[{\hat t}^a , {\hat t}^b ] &= i \epsilon^{abc} {\hat t}^c \cr
[{\hat j}^k , {\hat j}^l ] &= i \epsilon^{klm} {\hat j}^m \cr  }
\EQN\e437
$$
The collective wave functions for proton ($p$) and neutron ($n$) with spin
$\uparrow$ and $\downarrow$ read:
$$
\eqalign{
\langle b{\ket {p\so}} = {i\over\pi} (b^1 + i b^2 )  \quad & \quad
\langle b{\ket {n\so}} = {1\over\pi} (b^0 - i b^3 )  \cr
\langle b{\ket {p\su}} = {1\over\pi} (b^0 + i b^3 )  \quad & \quad
\langle b{\ket {n\su}} =-{i\over\pi} (b^1 - i b^2 )  \cr  }
\EQN\e438
$$
whereas for the $\Delta$ we have e.g.:
$$  \langle b
{\ket{\Delta^{++},s_z ={3\over 2}}} = {{\sqrt 2}\over \pi} (b^1 + i b^2)^3
\EQN\e438a
$$
from which the other $\Delta$-states can be constructed by applying the
spin- and isospin lowering operators $j^-$ and $t^-$ respectively.

Finally by knowing those collective wave functions it is possible to
calculate any matrix element of an observable acting in the collective
space ($b$-space) between two collective wave functions.
The general recipe is the following: One uses \qeq{\e423} and expands the
expressions up to 1st non vanishing order in the rotational frequency
$\Omega^a$.
Then one replaces $\Omega^a$ by the operator ${{{\hat t}^a} \over \Theta}$.
This expression can now be sandwiched between the collective wave
functions.
Actually if terms higher than of first non-vanishing order in
$\Omega$ are taken into account this simple recipe must be modified due
to some time-ordering problems (see for details sect. 5.1. and
5.2.).

Important matrix elements, which we will need later on are e.g.:
$$
{\bra {p\so}} D^{b3} {\ket {p\so}} = \left (- {1\over 3} \right )
\delta^{b3}
\EQN\e439
$$
and of course:
$$
{\bra {p\so}} {\hat t}^3  {\ket {p\so}} = \half
\EQN\e440
$$
whereas for any collective state with quantum numbers $T$,$J$ we have:
$$
\eqalign{
{\bra {\Psi_{TJ}}}
{\hat t }^2
{\ket {\Psi_{TJ}}}
&=T \cdot (T+1)  \cr
{\bra {\Psi_{TJ}}}
{\hat j }^2
{\ket {\Psi_{TJ}}}
&=J \cdot (J+1)  \cr}
\EQN\e441
$$
Therefore we see with \qeq{\e429} that
for the collective quantized eigenstates the general constraint
$$
J  = T
\EQN\e442
$$
 holds.

\sect{Translational Motion: Pushing}

As it does for the (iso-)rotational symmetry the soliton breaks also the
translational invariance of the effective action. Because we want to
consider nucleonic systems with good linear momentum,
especially for momentum
zero, i.e. at rest, this symmetry has to be restored as well.
The basic
idea is the same as in the case of the isorotational motion: In order
to avoid approximately the complicated boosting procedure one couples
the corresponding operator, the linear momentum
$\vec P = {{\vec\nabla} \over i}$ through a Lagrange multiplier $\vec
v$
to the effective action $S_{eff} [U_5 (\vec x)]$.
By applying the generators
of the translation $e^{i \vec v \vec P t}$ it can be seen to be
equivalent to calculate the action for a field
$\bu_5 (\vec v)= U_5(\vec x - \vec v t)$ which moves with
velocity $\vec v$ ({\it pushing model}):
$$
\eqalign{
S_{eff} [\bu_5 (\vec v)] &=
\Sp \ln \left ( iD[\bu_5(\vec v)] \right ) =
\Sp \ln \left \{
e^{- i \vec v \vec P t}
 \left ( i D [U_5] \right )
e^{ i \vec v \vec P t} \right \} \cr &=
\Sp \ln \left ( iD[U_5]) - \vec v \vec P \right ) \cr}
\EQN\e445
$$
Assuming that $v$ is small, which we will also comment later on, we can treat
the problem perturbatively:
$$
S_{eff} (\vec v) = \T E(\vec v) = \T \left ( E_{MF}
+ \half v^2 M^* + \O (v^3) \right )
\EQN\e446
$$
Hereby we made use of the fact, that the 1st order in $\vec v$, which is
nothing but the expectation value of the linear momentum $\erw{\vec P}$
in the static configuration $U_5 (\vec x)$, vanishes due to
Ehrenfest's theorem already at the 1-particle level:
$$
{\bra{\lambda}} \vec P {\ket{\lambda}} =
{i\over 2}{\bra{\lambda}}[P^2,\vec x] {\ket{\lambda}} =
{i\over 2}{\bra{\lambda}}[h^2,\vec x] {\ket{\lambda}} = 0
\EQN\e447
$$
and because of \qeqs{\e312,\e313} also in general.
The inertial mass $M^*$ defined by:
$$
M^* = {1\over \T} {  {\delta^2 S_{eff}[\bu_5]} \over
{\delta v^2}}\Big\vert_{v=0}
\EQN\e448
$$
is generally given by the particle-hole matrix elements of $\vec P$ similar to
the Inglis formula \qeq{\e420}.
Other than for the isorotational motion in case of the translational motion
Lorentz invariance of the NJL Lagrangian   guarantees that the
{\it inertial mass}
$M^*$ is equal to the soliton mean field energy $E_{MF}$:
For small $v$, $S_{eff} [\bu_5 (v)]$ in \qeq{\e445} is the
nonrelativistic Galilei limit of the effective action, which arises by
{\it boosting} the soliton field in ${\hat {v}} = {{\vec v}\over v}$-
direction with the boost-velocity $\omega = {\rm arctanh (v)}$
(cf. app. C).
Indeed the soliton energy $E$ transforms under this boost like the
time component of a Lorentz $4$-vector (Betz and Goldflam 1983), i.e.the
energy $E(\omega)$ in the boosted system is given by:
$$
E (\omega) = \cosh (\omega) E_{MF} = E_{MF} + \half v^2 E_{MF} + \O (v^3)
\EQN\e449
$$
which after comparing with \qeq{\e446} shows the desired identity:
$$
M^* = E_{MF}
\EQN\e450
$$
An explicit proof of \qeq{\e449} for the unregularized action Minkowski
space using Poincare algebra as well as the selfconsistent mean field
equation of motion \queq{\e342} is given in appendix C.
Pobylitsa et al.(1992) have shown,  that this conclusion also holds, if
one considers a regularized theory with a finite cutoff in Euclidean
space as long as the regularization scheme is gauge-invariant.
This means, in other words, that the central
transformation identity
$$
\Sp \ln  \left \{ \B^{\dag} (\omega) \A \B (\omega) \right \} =
\ln \det \left \{ \B^{\dag} (\omega) \A \B (\omega) \right \} =
\ln \det \A  = \Sp \ln \A
\EQN\e451
$$
where $\A$ denotes an arbitrary operator and $\B (\omega)$ the generator
of the boost, has still to be valid, if $\Sp \ln = \ln \det$ gets
regularized.

\sect{Spurious Zero Point Energies - Masses of Nucleon $N$ and $\Delta$}

Spurious zero point energies for the translational as well as the
(iso-)rotational motion, which have to be subtracted
from the total energy, if a
semiclassical  quantization is performed, arise due to the fact, that
even at
mean field level the expectation values for the operators
${\vec P}^2$ and ${\vec T}^2$ are finite.

\subsec{Expectation Values of $1$- and $2$-Particle Operators}

Let us first have a look at the expectation values of
$1$- and $2$-particle operators
in general.
For this we consider the pure
$1$-particle operators $\op{O}$,$\op{O^2}$ and the
pure $2$-particle operator $\tp{O}$ defined by (Negele and Orland 1987):
$$
\eqalign{
\op{O} &= \int d^3 x {\hat{q}}^{\dag}_\alpha  (\vec x) \O_{\alpha\beta}
{\hat{q}}_{\beta} (\vec x ) \cr
\op{O^2} &= \int d^3 x {\hat{q}}^{\dag}_\alpha  (\vec x) (\O^2)_{\alpha\beta}
{\hat{q}}_{\beta} (\vec x ) \cr
\tp{O} &= \int d^3 x_1 d^3 x_2
{\hat{q}}^{\dag}_{\alpha_1} ({\vec x}_1)
{\hat{q}}^{\dag}_{\alpha_2} ({\vec x}_2)
\O_{{\alpha_1}{\beta_1}}
\O_{{\alpha_2}{\beta_2}}
{\hat{q}}_{{\beta_2}} ({\vec x}_2 )
{\hat{q}}_{{\beta_1}} ({\vec x}_1 )  \cr}
\EQN\e452
$$
where the $\alpha$'s and $\beta$'s denote some Dirac or isospin indices.
Using the anticommutator relations for the fermion operators
$\hat q$ and $\kr {\hat q}$
$$
\eqalign{
\{ {\hat{q}}_\alpha (\vec x),{\kr {\hat{q}}}_\beta (\vec y)\}
&=\delta^3 (\vec x - \vec y )\delta_{\alpha\beta}
\cr
\{{\hat{q}}_\alpha (\vec x ), {\hat{q}}_\beta (\vec y )\}&=0
\cr
\{{\kr {\hat{q}}}_\alpha (\vec x ),{\kr {\hat{q}}}_\beta (\vec y )\}&=0
\cr }
\EQN\e453
$$
we can decompose the squared $1$- particle operator
$\{ [ O ] _{(1)} \}^2$ into:
$$
\{ [ O ] _{(1)} \}^2  = \op{O^2} + \tp{O}
\EQN\e454
$$
Using the notation:
$$ e^{iS(\alpha)} := \Z (\alpha) = \int \D q \D {\bar{q}} e^{ i \int d^4 x
 {\bar{q}} (iD + \alpha (\gamma_0 \O)) q}
\EQN\e455
$$
the expectation value of the last summand in \qeq{\e454} can be written as:
$$
{\erw {\tp{O}}} =
{1\over {(i \T) ^2}} \left [ {1\over {\Z(\alpha)}}
{{\delta^2}\over{\delta\alpha^2}} \Z (\alpha)
\right ]_{\alpha =0}
= {\erw {\op{O}}}^2 + {1\over{i \T^2}}
{{\delta^2 S(\alpha) }\over{\delta\alpha^2}} \Big\vert_{\alpha =0}
\EQN\e456
$$
For static field configurations $S(\alpha)$ is proportional to $\T$ so that
the last term in \qeq{\e456} vanishes in the ground state
(zero temperature limit) $\T \to \infty$.
Therefore we finally obtain:
$$
\lim_{\T \to \infty} {\erw{\{[O]_{(1)} \}^2} } =
{\erw{\opq{O}}} + {\erw {\op{O}}}^2
\EQN\e457
$$

\subsec{Zero Modes of the (Iso-)Rotational Motion, Band-Head-Energy}

{}From \qeq{\e457} we obtain for the
expectation value of the operator
${\vec T}^2$ in the rotating system (cf.\qeqs{\e423,\e426}):
$$
\erw{ \{ [ {\tilde {\vec T}} ] _{(1)} \}^2  } (\Omega) =
\erw{ \opq{\tilde{\vec T}}} (\Omega) +
\left ( \erw{ \op{\tilde{\vec T}}} (\Omega) \right )^2
= \Theta^2 \Omega^2 +
\erw{ \opq{\tilde{\vec T}}} (\Omega)
\EQN\e457
$$
The expectation value of the $1$-particle operator
$\opq{\tilde{\vec T}} (\Omega)$ is independent of $\Omega$, because
$\opq{\vec T}$ commutes both with $h$ and $h(\Omega)$. Then  the
expectation
value can be shown to get no contribution from the Dirac sea and
hence simply reads:
$$
\erw{ \opq{\tilde{\vec T}}} (\Omega) =
\erw{ \opq{\vec T}}  =
N_c {\erw{B}} \half (\half +1) =
N_c {\erw{B}}_{val} \half (\half +1) = {9\over 4}
\EQN\e458
$$
If we consider $\Omega^2$ as a Lagrange multiplier for adjusting
the quantum number $T$, we have the condition:
$$
\erw{ \{ [ {\tilde {\vec T}} ] _{(1)} \}^2  } (\Omega) = T(T+1)
\EQN\e459
$$
and therefore find:
$$
\Theta^2  \Omega^2 = T ( T+1) -
\erw{ \opq{\vec T}}
\EQN\e460
$$
Inserting this equation into the expression for the energy in the rotating
system (cf.\qeq{\e418}):
$$
E(\Omega) = E_{MF} + \half \Theta \Omega^2
\EQN\e461
$$
one obtains for the energy of a system with isospin $T$ and spin $J$:
$$
E_{JT} = E_{MF} +
{ {T(T+1)}\over{2\Theta}} -
{{ \erw{ \opq{\vec T}}  }  \over {2\Theta}}
\EQN\e462
$$
which means that the {\it band head } term
${{ \erw{ \opq{\vec T}}  }  \over {2\Theta}}$ has to be subtracted
from the {\it cranked energy} $E_{MF}+{ {T(T+1)}\over{2\Theta}}$.
The expression \qeq{\e462 } is familiar from nuclear many body physics,
see  e.g. Ring and Schuck (1980) and Blaizot and Ripka (1988).

\subsec{Zero Modes of the Translational Motion, Center-of-Mass Energy}

Instead of a proper boosting
for the translational degrees of freedom
we adopt a purely non-relativistic treatment,
which consists in separating the whole motion into an collective as well as
an intrinsic one (Ring and Schuck 1980). Thereby we hope that
at least the magnitude of the effect is described reasonably.

For this let us look at a $N$-particle system interacting by a purely local
time- and velocity independent one-body force $V(\vec x)$.
We consider the static and the pushed total Hamiltonians:
$$
\eqalign{
H &= \sum_{k=1} ^N { { \vec P _{(k)} ^2 } \over {2 m_{(k)} }} +
V (\vec x _{(k)} )   \cr
H_v &= \sum_{k=1} ^N { { \vec P _{(k)} ^2 } \over {2 m_{(k)} }} +
V (\vec x _{(k)}  - \vec v t )   \cr}
\EQN\e463a
$$
respectively.
The corresponding solutions of the time-dependent Schr\"odinger equations
$$
\eqalign{
\partial_t \ket{\Psi (t)} &= H \ket{\Psi (t)} \cr
\partial_t \ket{\Psi_v (t)} &= H_v \ket{\Psi_v (t)} \cr}
\EQN\e463b $$
are connected by the unitary generators of the Galilei transformation
(see e.g.Fonda and Ghirardi 1972):
$$
\ket{\Psi_v (t)} =
e^{ i \half M v^2 t} e^{-i \vec P \vec v t} e^{ i M \vec R \vec v}
\ket{\Psi (t)}
\EQN\e463c
$$
with:
$$
\eqalign{
\vec P &= \sum_{k=1} ^N \vec p_{(k)}  \cr
     M &= \sum_{k=1} ^N      m_{(k)}  \cr
\vec R &= {1\over M} \sum_{k=1} ^N m_{(k)} \vec x_{(k)}  \cr}
\EQN\e463d
$$
Applying this transformation one can easily show that
the expectation
values of the operator
${\erw{\{[{\vec P}]_{(1)} \}^2} } =\sum_{ij} \vec p_{(i)} \vec p_{(j)} $,
in the
moving and in the rest system are related by:
$$
\left \langle \sum_{ij}
\tilde{\vec p}_{(i)} \tilde{\vec p}_{(j)} \right \rangle (v) =
\left \langle \sum_{ij} \vec p_{(i)} \vec p_{(j)} \right \rangle + M^2 v^2
\EQN\e464
$$
$\sum_{ij} \tilde{\vec p} _{(i)} \tilde{\vec p} _{(j)} $ describes the
{\it center of mass motion} and is also called the square of the
{\it collective} momentum $\vec\Pi_{coll} ^2 (v) =
\sum_{ij} \tilde{\vec p} _{(i)} \tilde{\vec p} _{(j)}$.
Using \qeq{\e457} as well as the fact, that ${\erw{\op{\vec P}}} =0$ in a
static system (Ehrenfest's theorem \qeq{\e447}) we obtain a relation, which
fixes $v^2$:
$$
\vec\Pi_{coll} ^2 (v) = M^2 v^2 + \erw{\opq{\vec P}}
\EQN\e465
$$
Substituting \qeq{\e465} into the expression for the energy of the soliton
in a moving system \qeq{\e449} one recognizes that the total energy
can be separated into a collective and an intrinsic part:
$$
E [\vec\Pi_{coll} ^2 (v)] = {{\vec\Pi_{coll} ^2}\over {2 E_{MF}}} +
\left [ E_{MF} - { {  \erw{\opq{\vec P}}}\over {2 E_{MF}}} \right ]
\EQN\e466
$$
where we have set $E_{MF}$ for the total mass $M$.
The collective term ${{\vec\Pi_{coll} ^2}\over {2E_{MF}}}$
is the kinetic center of mass energy and vanishes, if we transform to a
system with resting c.o.m. The relevant part is therefore the intrinsic
energy $E_{MF} - { {  \erw{\opq{\vec P}}}\over {2 E_{MF}}}$,
in which analogous
to \qeq{\e462} the spurious zero point energy
${ {  \erw{\opq{\vec P}}}\over {2 E_{MF}}}$ has got subtracted from
the classical mean field energy $E_{MF}$.

Summarizing we find finally for the total intrinsic energy
for a system with spin $J$ and isospin $T$ whose structure is known
from e.g. Ring and Schuck (1980) or Blaizot and Ripka (1988):
$$
E_{[JT,{\vec\Pi_{coll}}^2 =0] } =
E_{MF} +
{ {T(T+1)}\over{2\Theta}} -
{{ \erw{ \opq{\vec T}}  }  \over {2\Theta}} -
{ {  \erw{\opq{\vec P}}}\over {2 E_{MF}}}
\EQN\e468
$$

\subsec{General Aspects of the Zero Mode Treatment}

Aside from the fact, that we have handled the center of mass motion
purely nonrelativistically, our treatment of the spurious zero modes
goes clearly beyond the semiclassical
approach, which can be seen e.g. from the condition \qeq{\e460}
for the Lagrange multiplier $\Omega^2$, which is incompatible with
the one for $\Omega^a$ in \qeq{\e426} on a classical level since
\qeq{\e460 } allows imaginary omega.
One should also note that the 'correction' terms to $E_{MF}$, which as itself
is of
$\O (N_c)$
are of different order in
$N_c$, namely $\O ({1\over{N_c}}) $ for the centrifugal term
${ {T(T+1)}\over{2\Theta}}$ and $\O (N_c ^0)$ for
the band-head term
${{ \erw{ \opq{\vec T}}  }  \over {2\Theta}}$ and
the center of mass term
${{ \erw{ \opq{\vec P}}  }  \over {2M}}$, respectively.

In nonrelativistic many particle physics the form of \qeq{\e468} can be
obtained within certain approximation by using Peierls-Yoccoz projection
techniques (Ring and Schuck 1980) for the angular and linear momentum.
Those techniques have been successfully
applied in soliton models with valence quarks
(Birse and Banerjee 1984, Birse 1885, L\"ubeck et al.1986, Fiolhais et al.1988,
Fiolhais et al. 1991, Neuber and Goeke 1992, for a review cf. Birse
1990).
Unfortunately it is up to now not clear how to establish them in case
of the NJL soliton, because the regularized Dirac sea does not allow the
definition of a Fock state, which is necessary for applying projection
methods.

In case of the Skyrme model it turned out that considering RPA fluctuations
around the mean field solution, which breaks the (iso-)rotational and
translational symmetry of the full theory, allows a treatment of the
corresponding eigen modes (Moussallam and Kalafatis 1991, Holzwarth
1992),
analogous to the case of two dimensional soliton models (Rajaraman 1982).
Altogether these modes lower the classical soliton energy and
about $80\%$ of the lowering originates from   the rotational
and translational zero modes. Hence it is probably a good approximation
in \queq{\e468} to concentrate on the zero-modes of the baryon.

A full treatment of the RPA-modes in the present NJL model has not been
done by now since the Dirac sea complicates the formalism tremendously.
On the other hand, as long as there exists no consistent treatment of the
zero modes in our model, the method described above is as good as the
semiclassical quantization by itself and one can hope that
\qeq{\e468} describes at least roughly the mass of a particle at rest
($\erw{\vec\Pi_{coll}} =0$) with spin J and isospin T.

\subsec{Masses of $N$ and $\Delta$, Numerical Results}

{}From \qeq{\e468} we can read off the expressions for the total masses
of a nucleon $N$ ($J=T=\half$) as well as a $\Delta$ ($J=T={3\over 2}$),
respectively (Pobylitsa et al.1992):
$$
\eqalign{
M_N &= E_{MF} +
{3 \over{8\Theta}} -
{9\over{8\Theta}} -
{ {  \erw{\opq{\vec P}}}\over {2 E_{MF}}}  \cr
M_\Delta  &= E_{MF} + {15 \over{8\Theta}} - {9\over{8\Theta}} -
{ {  \erw{\opq{\vec P}}}\over {2 E_{MF}}}  \cr}
\EQN\e469
$$
For the numerical calculation we take the mean field energy of the
selfconsistent solutions (cf. sect. 3.3). The $\Theta$ as well as the
expectation
value $\erw{\opq{\vec P}}$ can be obtained  from the $1$-particle
eigenstates
$\ketl$ of this solution using \qeqs{\e421,\e313}.
The numerical results in dependence of the constituent quark mass $M$ are
shown in \qufig{\f40,\f41 } (Pobylitsa et al.1992).
In the relevant region $M\approx 400 \MeV$ it turns out that the rotational
zero point energy ${9\over{8\Theta}}$
lies around $100 \MeV$ whereas the translational zero point energy amounts
to about $300 \MeV$. This is the order of magnitude obtained also in
nonrelativistic quark models (see e.g.Bhaduri 1988 and ref.therein)
as well as in relativistic soliton models using Peierls-Yoccoz projection
techniques (Fiolhais et al.1991, Neuber and Goeke 1992).
On the other hand especially the value for the translational zero point
is quite high in comparison with the soliton mean field energy
$E_{MF} \approx 1200 \MeV$, so that the perturbative treatment in
$\Omega$ and $v$ appears more than questionable.
It is also interesting to note that the 'cranking' term
${ {T(T+1)}\over{2\Theta}}$ can be numerically of the same order of magnitude
as the band-head energy
${{ \erw{ \opq{\vec T}}  }  \over {2\Theta}}$, although they are of different
order in $N_c$.

Finally $M_N$ and $M_\Delta$ (\qufig{\f41}) are at $M \approx 400 \MeV$
with $M_N \approx 900 \MeV$
and $M_\Delta \approx 1100 \MeV$, respectively,
somewhat  close to their experimental
values ($M_N = 938 \MeV$, $M_\Delta = 1230 \MeV$).
Especially one notes that the nucleon becomes stable against decay into
$N_c =3$ free quarks as consequence of the subtraction of the spurious zero
point energies.

\vfill\eject

%
%
%
\chap{Nucleon Form Factors and Observables}
%
%
%
%
\FIG\f50{The quadratic radii  of the proton and the neutron
are given in dependence of the constituent quark mass M. The
experimental values are indicated.    }
\FIG\f51{The   proton electric form factor $G_E ^p (Q^2)$ for 4
different values of the constituent mass $M$
(Gorski et al.1992).}
\FIG\f52{The  neutron electric form factor $G_E ^n (Q^2)$ for 4
different values of the constituent mass $M$
(Gorski et al.1992).}
\FIG\f53{The  electric charge distribution of the proton for a
constituent mass of $M=465 \MeV$ (full line)
splitted in valence (short dashed) and sea (long-short dashed)
(Wakamatsu 1991, Gorski et al.1992).}
\FIG\f54{The  electric charge distribution of the neutron for a
constituent mass of $M=465 \MeV$ (full line)
splitted in valence (short dashed) and sea (long-short dashed)
(Wakamatsu 1991, Gorski et al.1992).}
\FIG\f55{The  proton magnetic form factor $G_M ^p (Q^2)$ for 4 different values
of the constituent mass $M$
(Christov et al.1993a).}
\FIG\f56{The neutron magnetic form factor $G_M ^n (Q^2)$ for 4
different values of the constituent mass $M$
(Christov et al.1993a).}
\FIG\f57{The magnetic moments for neutron and proton including
contributions up to first order in the rotational frequency $\Omega$
are presented in dependence of the constituent quark mass M. The
calculations are performed with $m_\pi=139\MeV$.}
\FIG\f58{The axial coupling constant $g_A$  including
contributions up to first order in the rotational frequency $\Omega$ are
presented in dependence of the constituent quark mass M. The
calculations are performed with $m_\pi=139\MeV$.}
%
%
\TAB\t52{The isoscalar $\mu^{T=0}$ and the isovektor $\mu^{T=1}$
magnetic moment
for 4 values of the constituent quark mass
$M$. The value in brackets denotes the contribution of the sea quarks.
(Wakamatsu and Yoshiki 1990).
The calculations are performed in the lowest non-vanishing order in the
the cranking frequency, i.e. up to  ${\cal O}(\Omega^{(1)})$ for the
isoscalar quantity and  up to  ${\cal O}(\Omega^{(0)})$
for the  isovector magnetic moment. }

\TAB\t54{The pion nucleon coupling constant $g_{\pi NN}$  on shell $q^2
=  m_\pi ^2$ and at $q^2 =0$  in the lowest order of the cranking
frequency $(\Omega^0)$ for
3 values  of the constituent quark mass $M$. For the calculation the
nucleon masses $M_N$ from Fig.4.2 have been used.
The experimental value of $g_{\pi NN} (q^2 =0)$ is extracted from the
Goldberger-Treiman relation $g_{\pi NN} (0) = {{M_N}\over{f_\pi}} g_A$.
(Meissner Th and Goeke 1991).}

We are now supplied with the basic techniques to calculate various
nucleon observables from the mean field solutions obtained in chapter 3.
We will consider especially the electromagnetic form factors of proton and
neutron including radii and magnetic moments (sect.5.1) as well as the
axial vector coupling constant $g_A$ and the axial form factor
$g_A (q^2)$ (sect.5.2).
Furthermore the pion nucleon coupling $g_{\pi NN}$ and the form factor
$g_{\pi NN} (q^2)$ are discussed and the Goldberger-Treiman relation
between $g_A$ and $g_{\pi NN}$ is exhibited (sect.5.3).
Finally we deal with the nucleon $\Sigma$-term (sect.5.4).
The spin of the proton that is carried by the quarks will be discussed
in the context of the $SU(3)$ soliton in chapter 6.

\sect{Electromagnetic Form Factors, Moments and Radii}

\subsec{Electromagnetic Current}

On a classical level the electromagnetic current
$j_\mu ^{em}$
arises as Noether current
of the electromagnetic $U(1)$ phase transformation:
$$
q \to e^{i{\hat Q} \alpha} q
\EQN\e51
$$
with the charge matrix
$$
{\hat Q} = \pmatrix{ {2\over 3} &               \cr
                         &  -{1\over 3}  \cr}
= {1\over 6} {\cal I} + {{\tau^3}\over 2}
\EQN\e52
$$
and is therefore given by:
$$
j_\mu ^{em} = {\bar{q}} \gamma_\mu {\hat Q} q
\EQN\e53
$$
It is the sum of an isoscalar and an isovector part.
Applying the general method described in sect.4.2. we find for the
sea part of the expectation value of $j_\mu ^{em}$  in the rotating
(cranked) soliton in Minkowski space (cf.\qeq{\e423}):
$$
\left\langle
j_\mu ^{em} (x)
\right\rangle = {1\over T}
{\delta\over{\delta A_\mu (x)}}
\Sp \ln \left [ iD + \beta {{\vec\omega\vec\tau}\over 2} + A_\nu (x)
\RR (\beta\gamma^\nu {\hat Q}) \RR^{\dag} \right ]_{A_\nu (x) =0}
\EQN\e54
$$
If one performs a Wick rotation to Euclidean space the time component
$A_0$ has
to be handled in the same way as the cranking frequency $\Omega$ (cf.
sect.4.2), i.e. the time component of a physical $4$-vector:
$A_0 \to A_4 = - i A_0$, where $A_4$ and $A_0$ are both real.
The valence part arises in the standard way by introducing a chemical potential
(cf. sect.3.2).

One should mention that
$\partial^\mu \left\langle j_\mu ^{em} (x) \right\rangle$ does not necessarily
vanish on the fermion 1-loop level, because of the regularization involved.
The same is true also for the divergence of the axial current (cf.\qeq{\e28}).
Apparently the result of the divergence depends on the
regularization scheme used and indeed, for momentum cutoff schemes the
Noether theorems are violated in the solitonic sector (D\"oring et al.1992).
On the other hand the proper-time method preserves those invariances.
This has been checked for $U(1)$, $SU(N_f)_V$ and $SU(N_f)_A$ transformations.

\subsec{Electric Form Factors of Proton and Neutron}

The electric form factor $G_E (q^2)$ is defined by the matrixelement of
the time component $j_0 ^{em}$ between nucleon states
$\ket{N_i (p_i ,s_i , t_i )}$ with $4$-momentum $p_i$, spin z-component
$s_i$ and isospin z-component $t_i$ (i=1,2) (Bernstein 1968)
$$
\bra{N_2 (p_2 ,s_2 , t_2 )} j_0 ^{em}(0) \ket{N_1 (p_1 ,s_1 , t_1 )}
= \bar{u} (p_2 , s_2 , t_2 )  \beta {\hat Q}
u(p_1 , s_1 , t_1 ) G_E (q^2)
\EQN\e55
$$
In \qeq{\e55}
$q = p_2 - p_1$ denotes the $4$-momentum transfer and
$$
u(p_i,s_i,t_i) =
\sqrt{ {E+M_N} \over {2M_N}} \pmatrix{
 1 \cr
 {{\vec\sigma \vec p_i} \over {E+M_N}  }  \cr }
\chi_{s_i} \chi_{t_i}
\EQN\e56
$$
the spinor of a free pointlike nucleon with mass $M_N$.
It is convenient to use the Breit system defined by:
$$
\eqalign{
&p_1 = (E,-{\vec q} /2)  \quad
p_2 = (E,+{\vec q} /2)  \quad
q=(0,{\vec q}) \cr
&E=\sqrt{M_N ^2 +{\vec q}^2 /4} \quad
 Q^2 = -q^2 ={\vec q}^2 \quad
 Q= \vert {\vec q} \vert   \cr}
\EQN\e57
$$
There \qeq{\e55} simplifies to:
$$
\bra{N_2 (p_2 ,s_2 , t_2 )} j_0 ^{em}(0) \ket{N_1 (p_1 ,s_1 , t_1 )}
= G_E (q^2)
\chi_{s_2}^{\dag}
\chi_{t_2}^{\dag}    {\hat Q}
\chi_{s_1}
\chi_{t_1}
\EQN\e55a
$$
The nucleon state
$\ket{N_i (p_i ,s_i , t_i )}$ is treated in the static approximation
and center of mass corrections are neglected. This means:
$$
 \ket{N ({-\vec q \over 2} , s_i , t_i )} =
 \ket{N ({\vec q \over 2} , s_i , t_i )} =
 \sqrt{(2 \pi^3) \delta (0) } \ket{s_i t_i}
\EQN\e58
$$
In our approach we take for $\ket{s_i t_i}$ the collective quantized nucleon
states $\ket{p \so}$, $\ket{p \su}$,
$\ket{n \so}$ or $\ket {n \su}$.

Like the electromagnetic current $j_\mu ^{em}$ itself $G_E (q^2)$ separates
into an isoscalar and an isovector part
$$
G_E (q^2) = \half \ges + {\hat{t}}_3  \gev
\EQN\e59
$$
It is easy to see that for the isoscalar part the first nonvanishing term
is of zeroth order in $\Omega$, whereas for the isovector part is the term
of first order in $\Omega$.   These two quantities are then related to
proton and neutron experimental currents by
$     \ges  =   G_E (q^2)_p  +  G_E (q^2)_n $
and $ \gev  =   G_E (q^2)_p  -  G_E (q^2)_n $.

{}From \qeq{\e55} one can extract $\ges$ and $\gev$ after some algebra
to (Gorski et al.1992):
$$
\eqalign{
\ges &= \int d^3 x e^{- i \vec q \vec x} {{N_c}\over 3} \left \{
| \ph x {val} |^2 - \half  \sum_\lambda {\rm sign} \epsl | \ph x \lambda
|^2
\right \}  \cr
\gev &= \int d^3 x
e^{- i \vec q \vec x}
{{N_c}\over {12\Theta}}
\int d^3 y \Biggl \{ - \sum_{\lambda \ne {val}}
{ {
[ \pd x {val} \vec\tau \ph x \lambda ]
[ \pd y \lambda \vec\tau \ph y {val} ] } \over {\epsilon_{val} - \epsl}}\cr
&- {1\over{4{\sqrt \pi}}} \int_{1/\Lambda^2} {{ds}\over{\sqrt s}}
\sum_{\lambda \ne \nu}
\left [
{{\epsl e^{-s \epsl^2} + \epsn e^{-s \epsn^2} } \over {\epsl + \epsn} }
+ {1\over s}
{{ e^{-s \epsl^2} - e^{-s \epsn^2} } \over {\epsl^2 - \epsn^2 } }
\right ] \cdot \cr &\cdot
[ \pd x \lambda  \vec\tau \ph x \nu ]
[ \pd y \nu  \vec\tau \ph y \lambda ]
\Biggr \} \cr }
\EQN\e510
$$
where $\Theta$ denotes the moment of inertia (cf. sect.4.1) and
$\phi_\lambda (\vec x) = \langle \vec x | \lambda \rangle$ the $1$-particle
eigenfunctions in $\vec x$-representation.

The results for the  proton and neutron form factor as well as the
corresponding
charge densities are   shown in \qufig{\f51-\f54} (Wakamatsu 1991,
Gorski et al.1992).
As one can see,
the proton form factor
$G_E ^p (q^2)$ is described very well in contrast to the
neutron form factor $G_E ^n(q^2)$.
Furthermore we recognize that the neutron charge density is dominated
by the
sea quarks at large distances $r$, which confirms the popular picture
that the long negative tail of the neutron charge distribution is made
by the pion cloud (Thomas 1983), which is connected in our model
with the polarization of the  Dirac sea by
gradient or heat kernel expansion of the fermion determinant.

The corresponding quadratic radii
$$
\langle R^2 \rangle_E =- 6 {{d G_E (q^2)}\over {d q^2} } \vert_{q^2 =0}
\EQN\e511
$$
are shown in \qufig{\f50}.
The fact that the $<R^2>_E^n$ is negative is due to the long negative
tail in the corresponding charge distribution. It originates in the
present model from the polarization of the Dirac sea, which in other
models corresponds to the pion cloud.
This negative tail
is obviously overstressed in the present model
$<R^2>_E^n=-0.21\fm^2$ compared to the experimental value of
$<R^2>_E^n=-0.12\fm^2$. One should compare these values with
the Skyrme model, which gives $-0.36\fm^2$ in the
scalar version (Braaten et al.1986a,1986b)
and $-0.24\fm^2$ with vector mesons (Kaiser et al.1987).

\subsec{Magnetic Moments and Form Factors}

The same considerations can be applied in case of the magnetic form factor
which is related to the space components of the electromagnetic current
$j_i ^{em}$ in the Breit frame through
$$
\bra{N_2 (p_2 ,s_2 , t_2 )} \vec j ^{em}(0) \ket{N_1 (p_1 ,s_1 , t_1 )}
=  i { {G_M (q^2)}\over {2M_N}}
\chi_{s_2}^{\dag}
\chi_{t_2}^{\dag}
(\vec\sigma \times \vec q)
\chi_{s_1}
\chi_{t_1}
\EQN\e512
$$
The main difference compared to $G_E (q^2)$ is the fact, that in case of
$G_M (q^2)$ the leading order of the isovector part is $\O (\Omega^0)$
whereas the leading order of the isoscalar part is $\O (\Omega^1)$.

At $q^2 =0$ one obtains the magnetic moments:
$$
\eqalign{
\ms &= \mp +\mn \cr
\mv &= \mp -\mn \cr}
\EQN\e513
$$
which have been calculated in the present model
by Wakamatsu and Yoshiki (1990).
The results are shown in \qutab{\t52} and compared with the experimental
values ($\mp = 2.79$, $\mn =-1.91$, $\mv = 4.70$ and $\ms = 0.88$).
One recognizes that for the relevant values of the constituent quark mass
$M \approx 400 \MeV$ the isovector part $\mv$ has a relatively large
sea quark contribution and its total value comes out too small.
The isoscalar part $\ms$ is clearly dominated by the valence quarks and
can be reproduced quite reasonable.

A calculation of the full $q^2$ dependence of $G_M (q^2)$ has been performed
recently (Gorski et al.1993)  and shows a good agreement with the
experimental
results  as far as the $q^2$ dependence is concerned
(\qufig{\f55,\f56}).

Similiar to the axial vector coupling constant, to
be discussed in sect. 5.2, actually the first non-vanishing term for the
isovector magnetic moment is of zeroth order in the rotational
frequency
$\Omega$. However there are important corrections in linear order
of $\Omega$, which are entirely due to the
time-ordering of operators  in non-local  theories
(Christov et al. 1993b).
Without going into mathematical details the
final results $\sim (\Omega^0,\Omega^1)$ are presented in
\qufig{\f57 } and compared with the experimental data (Christov et al.
1993b).  Apparantly the
corrections $\sim\Omega^1$ are quite important (Compare with the axial
vector coupling constant in \qufig{\f58 }).

Furthermore  one has to keep in mind that all calculations have
been performed
in zeroth order of the pushing velocity $v$, which means,
that any center of mass corrections to nucleon observables other than the
nucleon mass (cf. sect.4.4)
have been completely neglected.
On the other hand one knows from nonrelativistic quark models
(Bhaduri 1988 and ref.therein) as well as other effective quark models
(Betz and Goldflam 1983, Dethier et al.1983,
Fiebig and Hadjimichael 1984a,1984b,
Braaten et al.1986a,1986b, Luebeck et al.1986, Leech and Birse 1989,
Stern and Clement 1989, Fiolhais et al.1991, Neuber and Goeke 1992)
that those corrections might be noticeable.

\sect{Axial Vector Coupling Constant and Axial Form Factor}

The general form of the matrix element of the axial current
(cf.\qeq{\e28})
$$
A_\mu ^ a (x) = {\bar{q}} (x) \gamma_\mu \gamma_5 \tau^a q (x)
\EQN\e514
$$
between nucleon states (\qeq{\e58}) can be written as:
$$
\eqalign{
&\bra{N_2 (p_2 ,s_2 , t_2 )} \ax (0) \ket{N_1 (p_1 ,s_1 , t_1 )} \cr
= &\bar{u} (p_2 , s_2 , t_2 )
{\tau ^a \over 2} [\gamma_\mu \gamma_5 g_A (q^2)
+ q_\mu \gamma_5 h_A (q^2)]
u(p_1 , s_1 , t_1 ) \cr}
\EQN\e515
$$
where $u(p_i , s_i , t_i ) \, ,\, i=1,2$ denotes the free nucleon spinor
(cf.\qeq{\e56}).
$g_A (q^2)$ is the axial and $h_A (q^2)$ the polar form factor of the nucleon.
At zero momentum transfer $g_A(q^2)$ reduces to the axial vector coupling
constant $g_A = g_A (q^2 =0)$ given by:
$$
g_A = 2 {\bra{p\so}} \int d^3 x \langle {\tilde A} _{\mu =3} ^ {a=3} (x)
\rangle (\Omega) {\ket{p \so}}
\EQN\e516
$$

\subsec{The Axial Vector Current in Zeroth Rotational Order}

Using the method of collective quantization described in sects.4.1 and 4.2
we have in zeroth order $\Omega$ (cf.\qeqs{\e423,\e439}):
$$
g_A = 2 {1\over 3} \int d^3 x \left \langle q^{\dag} (\vec x)
{{\sigma_3 \tau_3}\over 2} q(\vec x) \right \rangle
\EQN\e517
$$
Furthermore the sea contribution of $g_A$ gets regularized due to
\qeqs{\e313,\e314} with $\O = {{\sigma_3 \tau_3}\over 2}$.

Using the mean field equations of motion \qeqs{\e342,\e346} it can be shown
that the PCAC relation
\qeq{\e29 } is maintained for the expectation value of the axial current,
i.e. it is valid also at the $1$-quark loop level to order
$\sim\Omega^{(0)}$ as long as the
regularization scheme applied respects chiral invariance (Meissner Th
and
Goeke 1991).
$$
\partial^\mu \langle A_\mu ^a (x) \rangle  = - m_\pi^2 f_\pi \pi^a (x)
\EQN\e518
$$
As we have mentioned in the last section this is true, if
one uses e.g. the  proper time regularization, but violated in case of
momentum cutoffs.

\qufig{\f58} shows the numerical results for various constituent masses $M$
calculated from \qeq{\e517} with the selfconsistent profiles.
In the relevant region $M\approx 400 \MeV$ one obtains in zeroth order
of $\Omega$ an axial vector coupling constant
$g_A\approx 0.8$,
which is too small compared to the experimental value of $g_A =1.23$
(Wakamatsu and Yoshiki 1990, Meissner Th and Goeke 1991).
The valence quark contribution is clearly dominating.

Applying \qeq{\e518} to \qeq{\e516} $g_A$ including the valence part can be
written as radial integral over the pion field:
$$
g_A = - {{8\pi}\over 9} m_\pi ^2 f_\pi \int dr r^3 \pi (r)
\EQN\e519
$$
This  sum rule turns out to be fulfilled numerically as well
(Meissner Th and Goeke 1991a).

The momentum dependence of the form factor $g_A (q^2)$ is usually parameterized
by a dipole form:
$$
{{\ga (q^2)} \over {\ga (0)} } = \biggl ( 1- {q^2 \over {M_A ^2}}
                                 \biggr ) ^{-2}
\EQN\e520
$$
where the most recent experiments determine the dipole mass to
$M_A =1.09 \pm 0.03 \GeV$ (Ahrens et al.1988).
Meissner Th and Goeke (1991a) obtained in the present model a value of
$M_A =0.95\GeV$, which is almost compatible with the experimental result.
Hence the $q^2$ dependence of the axial form factor is reproduced.

\subsec{The axial vector current with  first-order rotational
corrections  }

Recently it has been  shown, that due to the non-commutativity of
some collective operators  (Wakamatsu and Watabe
1993) and due to the explicit time-ordering of time-dependent
operators
(Christov et al.1993b, Blotz et al. 1993c), that
there are important contributions from the rotational corrections in the
first order of the frequency $\Omega\sim{1/N_c}$. Their origin lies in
the fact, that
the generators of the $SU(2)_{spin}$-group $J^c$ do not commute with
the D-functions $D^{ab}$. Furthermore one has to take care of the
time-dependence of the operators, because one has to refer to the
operator formalism (and not to the functional integral) when dealing
with the rotation matrices ${\cal R}(t)$
(Christov et al.1993b).
Therefore in the expansion of \queq{\e423 },
there emerges a term of the form  $\simeq [J^a,D^{3c}]$, which is
non-vanishing.

Actual calculations (For details see Christov et al. 1993b)
including these corrections improve the value for $g_A$,
such that
the total value of $g_A$ for $M\simeq~400MeV$
in NJL comes out to be $g_A=1.15\to~1.30$, which is quite
close to experiment. The formulas for these additional terms
are presented together with the corresponding terms in SU(3)
in chap. 6.
Detailed values for $g_A$ in SU(2) including these corrections
are presented in \qufig{\f58} and
the importance of the $\Omega^1$
corrections to the isovector magnetic moment has been shown in
\qufig{\f57}.

 There are two philosophies (Blotz et al. 1994). First:
One evaluates the pion and sigma field from the equation of motion
\qeq{\e346 }, which are given in ${\cal O}(\Omega^0)$. Then one
evaluates with this the axial vector current and a new pion field, both
in order ${\cal O}(\Omega^0+\Omega^1)$. This treatment is strictly
consistent with an expansion in the $1/N_c$ spirit as far as rotations
are concerned and it is consistent with the perturbative cranking
procedure. On the other hand PCAC is violated. If one uses the numbers
of Alkofer and Weigel (1993), one may obtain a rough feeling how far
PCAC is violated and for reasonable values of the constituent mass this
violation seems to be small. Second: One derives equations of motion for
the pion field and also for the axial vector current, both up to first
order in $\Omega$. In this treatment PCAC is strictly conserved. However
the treatment does not comply with a strict $1/N_c$ expansion and the
perturbative expansion of the cranking procedure. On the present level
of investigation one cannot prefer one procedure  to the other.

\sect{Pion Nucleon Coupling and Form Factor}

The pion nucleon form factor $g_{\pi NN} (q^2)$ is defined through the nucleon
matrix element of the pionic current:
$J_{\pi} ^a (x)$
$$
(\square + m_{\pi} ^2 ) \pi ^a  (x) = J_{\pi} ^a (x)
\EQN\e521
$$
as:
$$
\eqalign{
&\bra{N_2 (p_2 ,s_2 , t_2 )} J_{\pi} ^a (0) \ket{N_1 (p_1 ,s_1 , t_1 )} \cr
=&(-i) \bar{u} (p_2 , s_2 , t_2 ) \tau ^a \gamma _5 u(p_1 , s_1 , t_1 )
g_{\pi N N} (q^2) \cr}
\EQN\e522
$$
For static pion fields one has:
$ J_{\pi} ^a (\vec r) = (-\nabla^2 + m_\pi ^2 ) \pi^a (\vec r)$,
which remains  also valid in the cranking approximation because the 2nd
time derivative in \qeq{\e521} is of 2nd order in the cranking frequency
$\Omega$ and therefore neglected.
{}From \qeq{\e522}  by using \qeq{\e521} the $\gpnq Q$ ($Q^2 = -q^2$) can
be finally be extracted to:
$$
{{g_{\pi N N} (Q^2)}\over {2 M_N}} =
 - (m_\pi ^2 + Q^2 ) \cdot {{4 \pi} \over 3 }
\int dr r^3 \left ( j_1 (Qr) \over Qr \right ) \pi (r)
\EQN\e523
$$
Especially at $Q^2 = -q^2 =0$ one has:
$$
{{g_{\pi N N} (0)}\over {2 M_N}} =
 - m_\pi ^2  {{4 \pi} \over 9 }
\int dr r^3  \pi (r)
\EQN\e524
$$
Comparing \qeq{\e524} with \qeq{\e519} we recognize the famous
{\it Goldberger-Treiman relation}
(Goldberger and Treiman 1958, Cheng and Li 1984):
$$
\gpn (0) ={{M_N}\over{f_\pi}} g_A
\EQN\e525 $$
which arises as a consequence of the PCAC relation \qeq{\e518} and holds as
long as the regularization scheme respects chiral symmetry.

The above formula deals with spacelike $q^2 = - Q^2 <0$.
However, direct experimental
data for the $\pi -N$ interaction are only available for timelike $q^2$,
even on shell $q^2 = -Q^2 \ge m_\pi ^2$. The \qeq{\e523} can be analytically
continued to the full timelike region $q^2 =-Q^2 >0$ (Cohen 1986):
$$
\eqalign{
{{g_{\pi N N} (Q^2 <0)}\over {2 M_N}} &=
\Bigl [
 - (m_\pi ^2 + Q^2 ) \cdot {{4 \pi} \over 3 }
\int_0 ^R dr r^3 \left ( j_1 (Qr) \over Qr \right ) \pi (r) \Bigr ] \cr
 &+ \Bigl [
 - (m_\pi ^2 + Q^2 ) \cdot {{4 \pi} \over 3 } A e^{-m_\pi R} R
   \left ( j_1 (Qr) \over Qr \right ) \Bigr ]   \cr
 &+ \Bigl [
 -  {{4 \pi} \over 3 } A e^{-m_\pi R}  {1\over{\vert Q \vert}}
   \left ( m_\pi \sinh (\vert Q \vert R) +
   \vert Q \vert \cosh (\vert Q \vert R) \right )
   \Bigr ]   \cr}
\EQN\e526
$$
The A is given by the asymptotic form of the pion field in \qeq{\e348 }.
Especially for  on shell pions $q^2 = - Q^2 = m_\pi ^2$ one finds
$$
{{\gpnq {q^2 = m_\pi ^2}}\over {2M_N}} = {{-4\pi}\over 3} A
\EQN\e527
$$
with the experimental value $\gpnq {q^2 = m_\pi ^2} = 13.6$.
\qutab{\t54} shows the values for $\gpnq {q^2 = m_\pi ^2}$ and $\gpnq 0$
in our model, where the nucleon mass from \qeq{\e469} and \qufig{\f41} have
been used. Because of the Goldberger-Treiman relation \qeq{\e525} it is
clear that the discrepancy between the theoretical and the
experimental value of $\gpn = 13.6$ is
the same as in case of $g_A$ evaluated to $\Omega^{(0)}$ considered in
the last section.

In the spacelike region the $q^2$ dependence of the pion nucleon form factor
$\gpnq {q^2 <0}$ can be fitted to $NN$ scattering data
by using one boson exchange potentials (OBEP).
In a monopole parameterization one obtains
in such an approach
$$
{ {\gpnq{q^2 <0}} \over {\gpnq {q^2 = m_\pi ^2}} } =
{{\Lambda_{\pi NN} - m_\pi ^2} \over
 {\Lambda_{\pi NN} - q^2 }  }
\EQN\e528
$$
with a monopole mass of $\Lambda_{\pi NN} = 1530 \MeV$ (Machleidt et al.1987).
In our model we obtain a monopole cutoff of
$\Lambda_{\pi NN} = 790 \MeV$.
Also all other chiral models predict a much lower value than the OBEP,
e.g. the Skyrme model without vector mesons $580 \MeV$ (Cohen 1986), the
Skyrme model with vector mesons $850 \MeV$ (Kaiser et al.1987), the projected
linear chiral sigma model $690 \MeV$ (Alberto et al.1988).
The same order of magnitude for $\Lambda_{\pi NN}$ is needed in charge exchange
reactions (Esbensen and Lee 1985), in more recent estimates within meson
exchange models (Deister et al.1990, Janssen et al. 1993). A similiar
value of $\Lambda$ is also obtained by substituting
the experimental values for $\gpnq{q^2 = m_\pi ^2} =13.6$ and
$\gpn (0) ={{M_N}\over{f_\pi}} g_A =12.4$ from the Goldberger-Treiman relation
\queq{\e525} into \qeq{\e528} which then yields $\Lambda_{\pi NN} =468
\MeV$.
The discrepancy between the OBEP calculation and all the other approaches
might lie in the fact that for a correct description of the $NN$ interaction
other, more complicated processes than the $\pi NN$-vertex have to be taken
into account. A refitting of the $NN$-phase shifts by means of OBEP models
with different exchange processes yields also lower values for
$\Lambda_{\pi NN}$ (Holinde 1992).

\sect{Nucleon Sigma Term and Form Factor}

The nucleon sigma term $\Sigma_N$ is defined as analogon to the quark
condensate $\kon$ \queq{\e222} in the nucleon sector
$$
\Sigma_N = m_0 \int d^3 x  {\bra{N}} {\bar{q}} (\vec x) q (\vec x) {\ket{N}}
\EQN\e529
$$
As an isoscalar quantity the $\Sigma
_N$
stays uninfluenced by the cranking procedure in SU(2).
Because the  current mass $m_0$ couples to the Dirac operator like a
Lagrange
multiplier for  the condition $m_0 \bar{q} q$ (cf.\qeq{\e311a}) one can
write
$\Sigma_N$ in the convenient form:
$$
\Sigma_N = m_0 {{\partial E_{MF} ({\tilde{m}}) }\over
{\partial {\tilde{m}} }} \Big\vert_{{\tilde{m}} =0}
\EQN\e530
$$

The corresponding form factor is defined by:
$$
\sigma (q^2) {\bar{u}} (p_2 ,s_2 ,t_2) u (p_1 , s_1 ,t_1) =
m_0
\bra{N_2 (p_2 ,s_2 , t_2 )} {\bar{q}} (0) q(0) \ket{N_1 (p_1 ,s_1 , t_1 )}
\EQN\531
$$
which gives with \qeqs{\e56,\e58}:
$$
\sigma (Q^2 = -q^2) = \int d^3 r j_0 (Qr)
\langle {\bar{q}} (\vec r) q(\vec r) \rangle
\EQN\e532
$$
At the Cheng-Dashen point $q^2=2m_\pi^2$, the $\sigma (q^2)$ can be
determined
from $\pi-N$-scattering data to $\sigma (2 m_\pi ^2) \approx 60
\MeV$
(Gasser and Leutwyler 1982). From dispersion relations an extrapolation to
$q^2 =0$ has been performed, which gives
$\Delta_\sigma = \sigma (2 m_\pi ^2) - \sigma (0) = 15 \MeV$
(Gasser and Leutwyler 1991). Hence the present NJL model should evaluate
the sigma term $\Sigma_N$ of \qeq{\e529} to $45\pm~5\MeV$.
One should note however that
from chiral perturbation theory a much
smaller value $\Delta_\sigma \approx 5 \MeV$ is obtained
(Gasser et al.1988a,1988b).
In contrast to all other observables in our model $\Sigma_N$ turns out to be
somewhat  sensitive to the regularization scheme applied.
For the proper time scheme one gets
a value of $\Sigma_N = \sigma(0) \approx 35 \dots 40 \MeV$
(Meissner Th and Goeke 1991, Wakamatsu 1993),
which is  also obtained in the Pauli-Villars scheme (Schueren
et al.1992), whereas with an extension of the proper time method
$$
\int_{1/\Lambda^2} d\tau F(\tau) \to \int_0 ^{\infty} \phi (\tau) F(\tau)
\EQN\e533
$$
also higher values for $\Sigma_N$ can be obtained for an appropriate chosen
cutoff function $\phi(\tau)$ (Blotz et al. 1993b). Details for SU(3)
will be given in sect. 6.1..

The \qeq{\e532} can be analytically continued to the time like region
$q^2 = - Q^2  >0$. The scalar form factor has been calculated by
Schueren et al. (1992). One finds from the deviation from the
Cheng-Dashen point
(Schueren et al.1992)
$$
\Delta_\sigma = \sigma (2 m_\pi ^2) - \sigma (0) = 7 \MeV
\eqn
$$
similar to the value obtained from chiral perturbation theory
($\Delta_ \sigma \approx 5 \MeV$, Gasser et al. 1988a,1988b) but only
half as large as the
one by means of dispersion relations ($\Delta_ \sigma \approx 15
\MeV$, Gasser et al. 1991) and by a Bethe-Salpeter approach in the meson
exchange picture (Pearce et al. 1992).

Wakamatsu (1992b) has also calculated the isospin violation of the
$\bar{q} q$ content in the nucleon (Gottfried sum rule) and found a reasonable
agreement with the experimental result from NMC data.

\vfill\eject

%
%
%
\chap{The SU(3)-flavour NJL-model  }
\TAB\taband1{The  contribution
            of the valence and the sea part of
            moments of inertia for $M=391.5MeV$ and $M=418.5MeV$. }
\TAB\taband2{Values of $<Y,T,T_3\mid D_{88}\mid Y_R,J,-J_3>$ and
 of  $<Y,T,T_3\mid D_{8i}J_i\mid Y_R,J,-J_3>$.  }
\TAB\taband3{The deviation of the theoretical mass from the
experimental value is shown for the perturbative treatment with
a constituent quark mass
$M=390.6MeV$ and the Yabu-Ando method with a constituent quark mass
$M=418.5MeV$.
For two values of the strange current quark mass,
these deviations are  shown together  with the absolute
experimental mass.
The theoretical value of the $\Sigma^{*}$-mass
is adjusted to the experimental one.
}
\TAB\taband4{
The axial vector coupling constants $g_A^0$, $g_A^3$ and
$g_A^8$ for $M=423.5MeV$, given  in the  lowest order contribution
($\Omega^0$) and with linear rotational ($\Omega^1$)  and
additional strange quark mass ($m_s^1$)  corrections from the
effective action, evaluated for
wave functions, which  contain linear $m_s$
corrections. These are compared with 'experimental'  numbers from
recent  EMC and SMC measurements (EMC, Ashman et al. 1988,1989;
SMC, Adeva et al 1993)
}
\FIG\fig1{The deviation of the theoretical mass  from the
experimental one is shown for the hyperon spectrum and for the Yabu-Ando
treatment and a constituent quark mass of $M=419MeV$
(Blotz et al.1993b)
}
\FIG\fig2{The deviation of the theoretical mass from the experimental
one is shown for the $\Sigma$ and the $\Lambda$, comparing the
perturbative and the Yabu-Ando method for $M=419MeV$(Blotz et al.1993b).}

\FIG\fig3{
The hadronic part of the isospin splitting for the octet baryons
from our theory (Praszalowics et al. 1993) compared with the
experimental
ranges for these splittings according to Gasser and Leutwyler (1982). }

In the following chapter we will investigate the extension of the
SU(2)-NJL model to the larger symmetry group of SU(3) (Blotz et al.
1992,1993b,1993c,1993d; Weigel et al. 1992a,1992b). In the vacuum
sector, we extend in this way the number of Goldstone bosons to the
kaons and the eta. Especially these lightest mesons  are
often considered as the dominating    degrees of freedom  for the low
energy regime of the strong interactions. Because the masses of these
particles
are still much lower than the nucleon mass, the extension to this larger
symmetry group is a priori  sensible.

Historically the success of hadronic isospin symmetry SU(2) (Heisenberg
1932), which is based on the almost mass-equality between the up and
down quarks and the flavour independence of QCD, leads  to the discovery
of flavour SU(3) (Gell-Mann and Ne'eman 1964). It was found that,
assuming that u,d and s quarks  transform as the fundamental
representation of SU(3), the spin $1/2$ nucleons belong to the
8-dimensional representation and the spin $3/2$ particles to the
10-dimensional representation of the group.
{}From the experimental
observation of these  symmetrically arranged particles in the
multiplets
one is usually inclined to consider SU(3) as the more 'physical'
symmetry. That this can be supported within a selfconsistent chiral
model for mesons and baryons, is shown below.
After considering the modifications  of the vacuum sector due to SU(3),
which include the mixing of the $\eta$-$\eta^\prime$-system, the
$U_A(1)$-anomaly and  the Gell-Mann Okubo mass relation, the baryon
number one sector is described.

To this aim we will concentrate here on the quantization of rotational
zero modes (Adkins et al. 1983), which rests on the assumption that
SU(3) is indeed a good symmetry of the strong interaction. An
alternative treatment  considers the baryon as a bound state of a
heavy meson and the background field of the SU(2) soliton (Callan and
Klebanov 1985,1988; Callan et al. 1988; Weigel et al. 1993).
It is believed that this treatment gives an exact
answer for extreme heavy mesons, such as those from a SU(4)
representation, but it is still not clear whether it is more appropriate
for the intermediate energy scale of the kaon system of SU(3).

Therefore special attention is drawn here to
the quantization of the  generators of the full SU(3) group  and to
second
order corrections  in the strange current quark mass  for the splitting
of the multiplets. Finally the axial vector coupling constants $g_A^0$,
$g_A^3=g_A$ and $g_A^8$.
currently under refined experimental investigation, are presented with
some recently found corrections arising from  some
non-commutativity of generators
(Wakamatsu and Watabe
1993) and explicit time-reordering (Christov et al. 1993, Blotz et
al. 1993d).

\sect{The NJL with SU(3)-flavours: The vacuum sector }

The vacuum or mesonic sector of the SU(3)-NJL model is intensively
discussed in the review articles of Klevansky (1992) and Vogl and Weise
(1991). However they mainly used an operator formalism based on the pure
quark NJL Lagrangian. So we will summarize here the most important parts
within the functional formalism. This  has advantages in the
baryonic sector, which is our main concern.

\subsec{SU(3) Symmetry Breaking and the Redundant $U_A(1)$  }

The Nambu-Jona-Lasinio Lagrangian with scalar and pseudoscalar
couplings on the level of the four-fermion
interaction and with a SU(3) symmetry is conveniently written as
\beq
    {\cal L}_{NJL}={\bar q}(x) ( i \dsl - m)q(x) - {G\over 2}
\left[ ({\bar q}(x)\lambda^aq(x))^2 +
        ({\bar q}(x)i\gamma_5 \lambda^a q(x))^2  \right]
\label\l1  \eeq
where $m={\rm diag}(m_u,m_d,m_s)=m_1 {\bf 1 }
+m_2\lambda_3+m_3\lambda_8$ is
the current quark mass matrix. The $\lambda_a,a=0,\dots,8$ are the usual
Gell-Mann matrices
with $\lambda^0=\sqrt{({2\over 3})} {\bf 1}$. Under infinitesimal chiral
SU(3) transformations the quark fields transform as
\beq  q \rightarrow  (1 - i \half \lambda^a \alpha^a - i \half \lambda^a
\beta^a \gamma_5 ) q  \label{\l2} \eeq
\beq  {\bar q} \rightarrow {\bar q} (1 + i \half \lambda^a \alpha^a - i \half
\lambda^a
\beta^a \gamma_5 )   \label{\l3} \eeq
{}From the \qeqs{\l2,\l3 } it is clear that the singlets
$({\bar q}\lambda^0q)^2$ and  $({\bar q}i\gamma_5\lambda^0q)^2$
corresponding to U(3) chiral meson fields
have to be included in the Lagrangian in order to be invariant under the
chiral $SU(3)_R\otimes~SU(3)_L$ transformation.

This is in contrast to the SU(2) case, where we had the
freedom to choose the chiral fields from a SU(2) or U(2) representation.
In addition,
this classical Lagrangian \queq{\l1} is invariant under $U_A(1)\otimes
U_V(1)$ transformations, where the singlet and octet parts of the quark
bilinears ${\bar q}\lambda^a q$ and   ${\bar q}i\gamma_5\lambda^a q$
transform independently.

\subsec{Currents and Divergences - PCAC }

The Noether  currents of the classical Lagrangian from the infinitesimal
transformations
\queq{\l2,\l3 } are given by
\beq   V_\mu^a = -i{\partial {\cal L} \over \partial ( \partial_\mu q) }
\delta_V q = - {\bar q} \gamma_\mu \half\lambda^aq  \label\l4 \eeq
\beq   A_\mu^a = -i{\partial {\cal L} \over \partial ( \partial_\mu q) }
\delta_A q = - {\bar q} \gamma_\mu \gamma_5 \half\lambda^aq  \label\l5
\eeq
and their divergences from the Gell-Mann Levi equations are given by
\beq \partial_\mu V_\mu^a = {\partial {\cal L} \over\partial \alpha^a }
={\bar q} i \half \bigl[ \lambda^a, m \bigr] q  \label\l6 \eeq
\beq \partial_\mu A_\mu^a = {\partial {\cal L} \over\partial \beta^a }
={\bar q} i\gamma_5 \half \bigl\{ \lambda^a, m \bigr\} q  \label\l7 \eeq
Assuming isospin symmetry, i.e. $m_u=m_d={\bar m}$, one observes that
\beq  \partial_\mu V_\mu^a  = 0, \ \ \ {\rm for\ } a=1,2,3,0,8 \label\l8
\eeq
from which follows consistently the conservation of isospin and
hypercharge.
Furthermore the divergences of the axial currents are proportional to
the corresponding pseudoscalar quark bilinears, which will be identified
later with the corresponding composite meson fields. We have
\beq    \partial_\mu A_\mu^{a=1,2,3} = -{\bar q}i\gamma_5\lambda^a q
{\bar m} \label\l9 \eeq
\beq    \partial_\mu A_\mu^{a=4,5,6,7} = -{\bar q}i\gamma_5\lambda^a q
\half({\bar m}+m_s) \label\l10 \eeq
In matrix form the remaining two currents are given by
\beq \partial_\mu \mat{ A_\mu^0\cr A_\mu^8} =
  {\cal D} \mat{ v^0\cr v^8}   \label\l11 \eeq
with
\beq  {\cal D} = \mat{ &{2{\bar m} +m_s\over3}
&{\sqrt{2}\over 3}({\bar m} -m_s) \cr
&{\sqrt{2}\over 3} ({\bar m} -m_s)  &{{\bar m} +2m_s\over 3}  }
\label\l12 \eeq
and $v^a=-{\bar q}i\gamma_5\lambda^a q$.
So we   observe that the singlet and octet currents are mixed. However,
we can disentangle these terms by means of an orthogonal transformation
according to
\beq \partial_\mu \mat{ {\tilde A}_\mu^0\cr {\tilde A}_\mu^8} =
 {\tilde {\cal D}} \mat{ {\tilde v}^0\cr {\tilde v}^8}   \label\l13
\eeq
with
\beq  \mat{ {\tilde A}_\mu^0\cr {\tilde A}_\mu^8}
     = {\cal R} \mat{ A_\mu^0\cr A_\mu^8} , \ \ \
     \mat{ {\tilde v}^0\cr {\tilde v}^8}
     = {\cal R}  \mat{ v^0\cr v^8}   \label\l14 \eeq
Here $ {\tilde {\cal D}}= {\cal R} {\cal D} {\cal R}^{-1}$ is a diagonal
matrix and the rotation matrix is given by
\beq   {\cal R}(\theta) = \mat{ &\cos\theta &-\sin\theta \cr
        &\sin\theta  &\cos\theta  } . \label\l15 \eeq
The result of this diagonalization is an angle $\theta$, ${\rm tan}
2\theta=-2\sqrt{2}$ i.e.
$\theta\simeq-35.26$,
which is called the {\it ideal mixing} angle. We willl come back to this
point, when we consider the mesonic mass spectrum.

The eigenvalues of  $ {\tilde {\cal D}}$ are simply $\bar m$ and $m_s$.
This can be understood in the following way. Whereas the former states,
which we  denote  by $\eta_0$ and $\eta_8$ are group theoretically given
by $\eta_0\sim{\bar u}u +{\bar d}d+{\bar s}s$ and
$\eta_8\sim{\bar u}u +{\bar d}d-2{\bar s}s$, the rotated states are
given by ${\tilde \eta}_0=\eta\sim{\bar u}u +{\bar d}d$  or
${\tilde \eta}_8=\eta'\sim {\bar s}s$. This is reflected by the two
eigenvalues of ${\cal D}$, which are either strange or non-strange
masses. This will become clearer  when we consider the mesonic
two point functions.

\subsec{$U_A(1)$ Symmetry Breaking }

As a consequence of the $SU(3)_R\otimes SU(3)_L$ invariance in the
chiral
limit and assuming a spontaneously broken vacuum by a non-vanishing
expectation value of some scalar composite field, we immediatedly
get from Goldstone's theorem that according to the 9 broken
generators of the symmetry group there will be 9 massless Goldstone
bosons. In nature  however, the $\eta'$ with
$m_{\eta\prime}=957\MeV$ cannot be
regarded as a Goldstone boson, because its mass is similiar  to
two twice the constituent quark mass.
't Hooft (1976b) proposed a mechanism supported in QCD,
which breaks the redundant
$U_A(1)$ symmetry, but conserves the SU(3) symmetry. Such a term is
given by
\beq  {\cal L}_{det} = \kappa \left[ \det ( {\bar q}_i P_R q_j ) +
              \det ( {\bar q}_i P_L q_j ) \right]   \label\l16 \eeq
where $i=u,d,s$ and
  $P_{R/L} = \half(1\pm\gamma_5)$ are the right and left helicity
projection operators \foot{See Witten (1979a), Veneziano (1979) and
Alkofer and Zahed
(1990)  for a discussion
of a minimal $U_A(1)$ breaking expression that is free of flavour
mixing.}.
Under an infinitesimal chiral
transformation it transforms into
\beq
       {\cal L}_{det} = \kappa \bigl\{  \det (1-i\lambda^{a*}\beta^a )
        \det ( {\bar q}_i P_R q_j ) +
        \det (1+i\lambda^{a*}\beta^a ) \det ( {\bar q}_i P_L q_j )
        \bigr\}
        \label\l17              \eeq
This expression is obviously invariant if and only if the matrices
$\lambda_a$ are traceless and therefore
$\det (1+i\lambda^{a*}\beta^a)=1$. This is the case if the
transformation corresponding to $\beta_0$ is excluded from the full
axial group $U_A(3)$.
So one is left with a
$SU(3)_V\otimes SU(3)_A\otimes U(1)_V$ symmetric expression.

\subsec{Bosonization }

In order to bosonize expressions like ${\cal L}_{det}$, which are not
only quadratic in the fields but also of higher order, one has to modify
the Gaussian bosonization procedure of sect. 2.1. Therefore we introduce
according to Zahed and Brown (1986), Reinhardt and Alkofer (1988)
\beq  1 = \int {\cal D} S_a   {\cal D} P_a
        \delta (S_a-{\bar q}\half\lambda^aq)
        \delta (P_a-{\bar q}\half i\gamma_5\lambda^aq)  \eeq
\beq  = \int {\cal D} S_a   {\cal D} P_a
       \int {\cal D} \sigma_a   {\cal D} \pi_a
       \exp{i\int d^4x\sigma_a(S_a-{\bar q}\half\lambda^aq)
         +\pi_a (P_a-{\bar q}\half i\gamma_5\lambda^aq) }
          \label\l18  \eeq
In this way, any quark bilinear in ${\cal L}_{det}$ can be replaced by
the fields $S_a,P_a$. A stationary phase approximation for
$S_a,P_a$ and $\sigma_a,\pi_a$ immediatedly relate $S_a,P_a$ with
$\sigma_a$ and $\pi_a$, according to
\beq  {\partial S\over \partial S_a} = -4GS_a + \sigma_a + {\partial
       {\cal L}_{det} \over \partial S_a }=0   \label\l19 \eeq
\beq  {\partial S\over \partial P_a} = -4GP_a + \pi_a    + {\partial
       {\cal L}_{det} \over \partial P_a } =0  \label\l20 \eeq
Flavour mixing in this context  means that
in general $S_a=S_a[\sigma_b]$.
In the case of U(2) chiral
fields and the presence of the 't Hooft term it holds
that $S_a\sim\sigma_a$, whereas for  U(N), $N\ge3$, chiral fields
the 't Hooft term  leads to a mixing, such that the $S_a$ are in
general non-linear functions of the $\sigma_a$.

Because we  consider here the case of SU(3) symmetry and no ${\cal
L}_{det}$-term
we can use immediately  the stationary phase equation \queq{\l19,\l20}
in order to eliminate the $S_a,P_a$.
We  obtain for the Lagrangian after a  rescaling of the meson fields
$\phi_a\rightarrow g\phi_a$ the form
\beq  {\cal L}_{NJL} = {\bar q} \Bigl(-i\dsl + m + g (\sigma_a\lambda_a
    +  i\gamma_5\pi_a\lambda_a \bigr)q  +\half \mu^2 \bigl(
    \sigma_a\sigma_a + \pi_a\pi_a \bigr) \Bigr) \label\l21  \eeq
as one would obtain also from a Gaussian integral multiplicator
(Eguchi 1976).
After Grassmann integration over the quark fields  one obtains
finally the
effective action in terms of classical meson fields, corresponding to
a one-fermion loop approximation :
\beq  S_{eff} = -\Sp \log
        (-i\dsl + m + g (\sigma_a\lambda_a
      +  i\gamma_5\pi_a\lambda_a \bigr)
       +  \half \mu^2  \int d^4x   \bigl(
      \sigma_a\sigma_a + \pi_a\pi_a \bigr) \label\l22   \eeq
This is expression is in a sense formal, because we have to apply a
specific regularization   scheme. We will choose here the
double step proper time scheme, given by   \qeq{\e229 }
where
$\phi(\tau)=c\theta(1-1/\Lambda_1^2)+(1-c)\theta(1-1/\Lambda_2^2)$
contains now two more parameters. This additional freedom
will be  used later in the mesonic sector to fix the
non-strange current quark mass to the preferred value.

\sect{Fixing of the parameters }

As we have said already the expression \queq{\l22 } has to be
regularized and as we know from
sect. 2.2. it should be done from the very beginning in order to avoid
             discrepancies.
For pedagogic reasons the regularized form will not be written out
explicitly.
{}From the action \qeq{\l22},  which always can be cast into  the form
\beq  S_{eff}[\phi] = \int d^4x  V_{eff}[\phi] + \half Z[\phi]
\ddmu \phi \ddmu \phi + \dots ,  \label\l22a
\eeq
 one obtains  for the non-derivative part of \qeq{\l22a }
the effective potential
$V_{eff}$
\beq   V_{eff}(\sigma_a,\pi_a) = - \Tr \int_{reg} {d^4k\over (2\pi)^4 }
    \Bigl( \ksl  + m + g (\sigma_a\lambda_a    +
        i\gamma_5\pi_a\lambda_a)  \Bigr)   +
                    \half \mu^2 \bigl(
      \sigma_a\sigma_a + \pi_a\pi_a \bigr) \label\l23   \eeq
which differs for homogeneous fields from the effective action only by
the four dimensional space-time volume \foot{Compare with sect.
2.6., where the parameter fixing is done with  using
the effective action only.}.
Actually, spontaneous
breaking
of chiral symmetry takes place and the $\sigma_0$ and $\sigma_8$ are the
only candidates of fields, which can acquire a non-vanishing vacuum
expectation value. This follows from conservation of parity, strangeness
and isospin.

However it has been shown (Pagels 1975) that  the vacuum is
$SU(2)_R\otimes~SU(2)_L$  invariant in the chiral limit, if $\sigma_0$
{\it and}
$\sigma_8$ have non-zero vacuum expectation values.
{}From Goldstone's theorem (Goldstone 1961) we know that the number of
broken symmetries
of the Lagrangian is related to the number of massless Goldstone bosons.
Actually the physical world
knows 8 pseudoscalar light particles, namely  $\pi,K$ and $\eta$.
If we neglect the $U_A(1)$ breaking by instantons,
even the $\eta'$ can be regarded as Goldstone boson.
Since
other would-be  Goldstone bosons are not known experimentally, the
true
vacuum state of the NJL model, if it is physical, has to be SU(3)
invariant in the chiral limit in accordance with Goldstone's theorem.
Fortunately the NJL vacuum
is indeed characterized by a large vacuum
expectation value of the $\sigma_0$ and a vanishing one for
$\sigma_8$. This situation is changed for explicit symmetry breaking,
like
the presence of the 't Hooft determinant, where the $U_A(1)$  symmetry
is  broken, or current quark masses, where the $SU(3)_V$ symmetry is
broken.

Now from the actual form of the potential \queq{\l23} the non-trivial
stationary phase conditions are given by
\beq
   {{d
V_{eff}\over d\sigma_0}\linie}_{vac}={{d
V_{eff}\over d\sigma_8}\linie}_{vac}=0
. \label\l24  \eeq
 After some algebraic manipulations,  one is left
with
\beq
    \mu^2(1-{m_0\over M_u}) = 8N_cg^2 I_1(M_u)
   \label\l25 \eeq
\beq
    \mu^2(1-{m_s\over M_s})    = 8N_cg^2 I_1(M_s)
\label\l26 \eeq
where we have set
 $M_u=\sqrt{2\over 3} g \sigma_{0v} + m_1
+{1\over\sqrt{3}} (g \sigma_{8v} + m_3) $ and
$M_s=\sqrt{2\over 3} g \sigma_{0v} + m_1 - {2\over\sqrt{3}}(g
\sigma_{8v} + m_3)$ and
\beq  I_1(M_i) = \int_{reg} {d^4k\over{(2\pi)}^4}
      {1\over k^2 + M_i^2}       \label\l27 \eeq
\subsec{Mesonic Spectra }

The masses of the pseudoscalar and scalar meson nonets follow directly
from the second functional derivative of the effective action. This
gives an expression, which is up to a normalization factor $Z_{ab}$
proportional to the inverse of a bosonic propagator. So it follows from
general arguments that the analoque of \qeq{\e2002 } is
\beq
     {1\over Z_{ab}(q^2) }{1\over \delta^4(p_1-p_2) } {{\delta^2
      S_{eff}
     \over \delta \pi_a(p_1)    \delta \pi_a( p_2)
      }\linie}_{p_1=-p_2,p^2=-q^2}
      = {(p^2 + m_{ab}^2)\linie}_{p^2=-q^2}
       \label\l28 \eeq
As described in sect. 2.5. one can now either use
the normalization point $q^2=0$,
which correspond to a derivation of the masses from a gradient
expansion, or at the on-shell point $q^2=m_{ab}^2$, so that the free
propagator is defined with  its physical mass.

Defining
\beq  I_2(M_i,M_j,q^2) = \int_{reg}  {d^4k\over{(2\pi)}^4}
      {1\over k^2 + M_i^2}{1\over {(k+q)}^2+M_j^2 }
      \label\l29      \eeq
we obtain for the normalization factor
\beq
 Z_{ab} =4N_cg^2I_2(M_a,M_b,q^2=-m_{ab}^2)
\label\l30 \eeq
depending on the flavour  content of the given meson.
Therefore the corresponding fields have to be rescaled according to
$\phi^{a\prime}=Z_{aa}^{(1/2)}\phi^a$. This gives
\beq
   m_\pi^2 ={1\over Z_\pi} {\mu^2 m_0\over M_u}, \ \ \
    m_K^2 ={1\over Z_K} {\mu^2 \over 2}({m_0\over M_u} +{m_s\over M_s})
       + (M_u-M_s)^2     \label\l31         \eeq
For the $\eta_0$ and $\eta_8$, the situation is more involved, since we
have again mixing between them. However, similiar to the PCAC relations
\queq{\l9,\l10 }
above, we can remove the mixing by the same ideal mixing angle
$\theta=-35.26$, if  $q^2=0$. Then the masses of the new
$\eta$ and $\eta'$  are given by
\beq  m_\eta^2 = {1 \over Z_\eta} {\mu^2m_0\over M_u} \simeq m_\pi^2
         \label\l32  \eeq
\beq  m_{\eta\prime}^2 = {1 \over Z_{\eta\prime} } {\mu^2m_s\over M_s}
        \simeq
         2m_K^2-m_\pi^2
         \label\l33  \eeq
One can formally  assign masses  to the mixed $\eta_0$ and
$\eta_8$ states and gets
\beq  m_{\eta_0}^2 \simeq {1\over3} m_\pi^2 +{2\over 3}m_K^2 \label\l34
     \eeq
\beq  m_{\eta_8}^2 \simeq -{1\over3} m_\pi^2 +{4\over 3}m_K^2
      \label\l35
     \eeq
Within the NJL model
  it is clear that it is the $\eta_8$ particle,
which fulfills the Gell-Mann (1962) and  Okubo (1962) (GMO) relation
\beq  m_\pi^2 + 3m_{\eta_8}^2-4m_K^2 = 0 \label\l36 \eeq
In nature however it is the $\eta$, which should fulfill the
GMO relation.
This problem, however, can indeed be solved if one includes the 't Hooft
interaction, which has the
main effect
of pushing the $\eta_0$ mass, such that the   $\eta$ and $\eta'$
are driven back to their group theoretical states
$\eta_0$ and $\eta_8$ (Hatsuda and Kunihiro 1991).

\subsec{Decay Constants }

{}From the general form of axial vector matrix elements between
pseudoscalar mesons  and the vacuum
$<0\mid A_\mu^a\mid\pi^a(p)>=-ip_\mu
f_a   \pi^a(p)$, we immediately deduce the pion and kaon decay constants
as
\beq
f_\pi ={M_u\over g} Z_\pi^{1/2}, \ \ \
f_K ={M_s + M_u\over 2 g} Z_K^{1/2}  \label\l37  \eeq

\subsec{Chiral Perturbation for $m_s$ }

Because we will treat the baryonic sector perturbatively in
the
current quark masses, we will give here also the mass ratio for kaons
and pions perturbatively in first order  in $m_s$
\beq
    {m_K^2\over m_\pi^2} = {m_s +m_0\over 2 m_0}
     + {\cal O} \left( {m_o\over M_u} \right)
       + {\cal O} \left( {m_s\over M_s}  \right)
.\label\l38 \eeq
It was already noted by Hatsuda (1990) and Hatsuda and Kunihiro (1991),
that this relation
gets large corrections order by order in  perturbation theory.
However there are cancellation effects of the  non-linear
terms  in the denumerator and numerator of the exact expression for
$m_K^2/m_\pi^2$, so that the  exact result  almost coincides with the
approximate
relation \queq{\l38} (Schneider 1994). That means that although
perturbation theory for
the whole vacuum does not work, we are not {\it a priori }going into
severe troubles if we make use of \qeq\l38, which coincides with the
exact answer.
Later in the baryon
sector
we will examine the validity of perturbation theory for $m_s$ again.
So we summarize that
by  this equation the current quark mass ratio is fixed to
$m_s/m_0\sim 24.5$
for given
experimental mesonic data, $m_\pi=139MeV$ and $m_K=496MeV$.
Chosing  our regularization function $\phi (\tau)$ such as to
reproduce the most reasonable value of $m_0\sim 6MeV$, this corresponds
immediately to $m_s\sim 150MeV$, which is also a reasonable value
infered from examinations of hyperon spectra.
So the  quark
masses, that will be needed to fit the hyperon splittings in Sect. 6.3,  will
be
compared  with the ratio given by
\qeq\l38. Furthermore the kaon decay constant $f_K$ equals $f_\pi$ in
this  approximation   and for the  constituent quark masses one obtains
$M_u=M_d=M_s$.

\sect{Collective Quantization of the SU(3)-Soliton  }

In sect. 4.2. we have dealt already the quantization of rotational modes
in the case of SU(2) symmetry. Now in the case of the SU(3) group, the
procedure is a little bit more involved. This is a reflection of the
fact that SU(3) is now  a rank 2 group, whereas SU(2) has rank 1 and
that the configuration space for the SU(3) rotations is now restricted
due to some trivial embedding of the SU(2) isospin subgroup into SU(3)
(Mazur et al. 1984).
Especially this embedding, which was proposed first by Witten (1983b),
ensures in the end  that only the physical representations emerge as
the lowest possible ones (Balachandran et al. 1985, Chemtob 1985,
Mazur et al. 1984).

But first we concentrate on deriving the
left and right generators of the group. According to Witten, we make for
the SU(3) chiral field the following ansatz :
\beq   U(x)
          =\mat{&U_2(x) &0\cr&0&1} \label\s1   \eeq
where  $U_2(x)=(\sigma_{(2)}
 + i\gamma_5{\vec\pi}
 {\vec\tau})/f_\pi$ and the SU(2) sigma field $\sigma_{(2)}$ is defined
according to
$\sigma_{(2)}=\sigma_0\sqrt{2\over3}+{1\over\sqrt{3} }\sigma_8$.
Then the  $\sigma_0$  and $\sigma_8$ fields themselves are
constrained by $\sigma_0\sqrt{2\over3}-{2\over\sqrt{3}
}\sigma_8=f_\pi$,
as long as there is no  flavour mixing interaction present.
This gives the $\bf 1$ in the lower right corner of $U(x)$. However
this constraint is not really necessary and recently solitons were found
(Kato et al. 1993), where it is released.  In order to
define
quantization rules,  we  again introduce the time dependence by writing
\beq  U(x,t) = A(t) U(x) A^\dagger (t)  \label\s2      \eeq
where $A(t)$ is now a time dependent matrix of SU(3).
The theory remains invariant if one performs on $A(t)$ the symmetry
operations
\beq A(t) \rightarrow A(t) G_{spat}, \ \ G_{spat}\ \epsilon\ SU(2)
      \label\s3
\eeq
and
\beq A(t) \rightarrow G_{flav} A(t), \ \ G_{flav}\ \epsilon\ SU(3)
. \label\s4
\eeq
The right multiplication of $A(t)$ can be identified as spatial rotation
because any isospin rotation of the chiral field can be undone by a
suitable space rotation. This is due to the symmetric hedgehog ansatz
\quref{ }. Note that because $U(x)$ commutes with the hypercharge group
$U_Y(1)$, with generators $\exp{iq_8\lambda_8}$, the configuration
space for the generalized coordinates of the SU(3) matrix $A(t)$ is
restricted to
$SU(2)_I\otimes U_Y(1)$ (Balachandran et al. 1985).
However, following the work of Balachandran et al. (1985), Chemtob
(1985) and Praszalowicz (1985), one can regard  the  symmetry of
the collective coordinates as SU(3) and treat the freedom for
right multiplication $U_R(1)$ as a constraint for the states.

Then one can
perform the time dependent
rotation of the effective action as described in sect. 4.2 and obtain
\beq S_{eff}^{rot}  =  - \Sp \log \left(
  \partial_\tau + H + A^\dagger (t) \dot A (t) - i \gamma_4
  A^\dagger (t) m A(t)   \right) \label\s5      \eeq
with
\beq  H = -i\gamma_4 \left( i \partial_i \gamma_i - m U(x) \right)
    \label\s6
\eeq
Because of the antihermiticity of $A^\dagger(t)\dot A (t)$, any
expansion of $S_{eff}$ in odd powers of this quantity constitutes a
contribution of the imaginary part  of the effective Euclidean action
and is therefore a finite quantity. As in the case of the baryon number
in sect. 3.2, it is reasonable not to regularize these quantities. This
is because they  are  connected with topological  indices, which
otherwise are not exact integers. We will see this, when we consider
e.g. the right hypercharge in this model. To this aim, we write the
Maurer Cartan form  $A^\dagger(t)\dot  A(t)$ as
\beq   A^\dagger(t)\dot  A(t) = \dot q_\alpha A^\dagger
\partial_\alpha A = {i\over 2} \dot q_\alpha
C_{\alpha\phantom{a}}^{\phantom{a}A} \lambda_A = {i \over 2} \Omega_A
\lambda_A \label\s7  \eeq
where the $q_\alpha$ are the coordinates of SU(3) and the $
C_{\alpha\phantom{a}}^{\phantom{a}A}$ are the vielbeins, which fulfill
\beq  C_{\alpha\phantom{a}}^{\phantom{a}A}
      C^{\beta\phantom{a}}_{\phantom{a}A} = \delta_\alpha^\beta, \ \ \
\left(C_{\alpha\phantom{a}}^{\phantom{a}A} \right)^{-1}
  =  C^{\alpha\phantom{a}}_{\phantom{a}A}   \label\s8  \eeq
and the important Maurer Cartan identity, following from the definition
of the structure  constant $f_{ABC}$ of the $\lambda$-matrices,
\beq   C^{\delta\phantom{a}}_{\phantom{a}B}  \partial_\delta
       C^{\gamma\phantom{a}}_{\phantom{a}A}    -
       C^{\delta\phantom{a}}_{\phantom{a}A}    \partial_\delta
       C^{\gamma\phantom{a}}_{\phantom{a}B}  =
      - f_{BAE}
       C^{\gamma\phantom{a}}_{\phantom{a}E}  \label\s10  \eeq
The effective Lagrangian after this SU(3) rotation is given by  (in the
chiral limit):
\beq L_M^{rot} = \half \Omega_A I_{AB} \Omega_B - {N_c \over 2\sqrt{3}}
B(U) \Omega_8 = \half \dot q_\alpha g_{\alpha\beta} \dot q_\beta +
Z^\alpha   \dot q_\alpha    \label\s11 \eeq
where $B(U)$ is the baryon number of the system with chiral field U(x)
and the metric  \beq g_{\alpha\beta}=
C_{\alpha\phantom{a}}^{\phantom{a}A} I_{AB}
C_{\beta\phantom{a}}^{\phantom{a}B}, \ \ \ Z^\alpha = -
{N_c \over 2\sqrt{3}} C_{\alpha\phantom{a}}^{\phantom{a}8}
\label\s12 \eeq
Note that in contrast to SU(2), there is now a term linear in the
angular frequency $\Omega_8$. This is due to the imaginary part of the
effective Euclidean action, which is vanishing for SU(2).
As mentioned above such a term  poses a problem for the
quantization, because the corresponding generators are  constraint in
this case (see e.g. Balachandran et al. 1985 for a discussion of this).

Note that this Lagrangian also describes a particle moving in a monopole
gauge field (Jackiw 1983), given by the gauge-variant $Z^\alpha$.
As a result, it leads to a quantization condition for  electric and
magnetic
charge.  In the present model it is the Wess-Zumino term, which is the
leading order term of the
gradient expanded baryon current $B(U)$ \queq{\bartopo }, that plays the
role of
the monopole.
The consequence of this gauge variant term in the present quark model
is a quantization condition for the number of colors
(see App. D for details).

But let us proceed step by step.
{}From  \qeq\s11 we can define canonical momenta by
\beq  \pi_a = { \partial L \over \partial \dot q_a } \label\s13 \eeq
and right generators by  the symmetrized form
(Toyota 1987)
\beq R_a = - \half \{ \pi_\alpha, C^{\alpha\phantom{a}}_{\phantom{a}a}
           \}    .
\label\s14 \eeq
In the same way one can   define left generators by using
the form $\dot A(t) A^\dagger (t)$ with some vielbeins
$E_{\phantom{a}a}^{\alpha\phantom{a}}$ and
\beq L_a = - \half \{ \pi_\alpha, E^{\alpha\phantom{a}}_{\phantom{a}a}
           \}    .
\label\s15 \eeq
{}From the Maurer Cartan identity and imposing the canonical quantization
by
\beq \bigl[  \pi_\alpha, q^\beta \bigr] = -i \delta_\alpha^\beta, \ \ \
     \bigl[  \pi_\alpha, \pi_\beta \bigr] =  \bigl[  q_\alpha, q_\beta \bigr] =
 0
\label\s16 \eeq
one can derive
\beq  \bigl[ R_a, R_b \bigr] = -i f_{abc} R_c, \ \ \
       \bigl[ L_a, L_b \bigr] = i f_{abc} L_c  \label\s17 \eeq
and the action on $SU(3)$ transformation matrices as
\beq  \bigl[ R_a, A  \bigr] = -A \half\lambda_a, \ \ \
      \bigl[ L_a, A  \bigr] = - \half\lambda_a  A \label\s18  \eeq
%
%
Now we see from \qeq{\s17 } that the left generators $L_a$ obviously
obey
correct commutation relations for $SU(3)$ and can be identified with
left flavour rotations, because of the remarks at the beginning of this
paragraph and \qeq{\s18 }. The  generators $R_a$ can be related to
some right transformations according to \qeq{\s18 } and act on the
space of spatial rotations. Because the eighth component is connected
with the baryon number, the space is also called the {\it  spin-baryon
number}  space.

We set the baryon wave functions according to Salam and Strathdee (1982)
as
\beq  \Psi(A) = \eta \sqrt{n} <i\mid D^{(n)}(A)\mid j>  \label\s19 \eeq
where n is the dimension of the representation and $\eta$ is a suitable
phase factor. The basis states $<i\mid=<I,I_ 3,Y\mid$ and $\mid j>=\mid
J,J_3,Y_R>$ carry the flavour and spin quantum numbers. The Wigner
function $D^{(n)}(A)$ is a matrix of the adjoint representation of SU(3)
and is given by
\beq   D^{(n)}_{AB}(A) = \half \tr A^\dagger\lambda_AA\lambda_B
\label\s20 \eeq
Using the commutation relations \qeq{\s18 } one obtains the action of
the generators on the wave function
\beq  \bigl[ R_C, D_{AB} \bigr] = i f_{CDB} D_{AD}, \ \ \
      \bigl[ L_C, D_{AB} \bigr] = i f_{CAD} D_{DB}   \label\s21 \eeq
{}From these one can deduce the action of the generators $R_A,L_A$ on the
wave functions. Furthermore, one can identify left indices of $D_{AB}$
as
flavour quantum numbers, chosen as hypercharge $Y$ and isospin $I,I_3$.
The right indices therefore correspond to the right hypercharge $Y_R$
and spin $J,J_3$. Now we come back to our rotated Lagrangian
$L_M^{rot}$. Using the properties of the hedgehog ansatz, we find
\beq
   I_{AB} =
\cases{    I_1 \delta_{AB} &for A,B=1,2,3  \cr
                      I_2 \delta_{AB} &for A,B=4,5,6,7 \cr
                      0              &for A,B=8 \cr }  \label\s22 \eeq
where  $I_{AB}$ is given by
\beq I_{AB}  =   - {N_c\over 4}
             \int {d\omega\over 2\pi} {\rm   tr }\left[
       {1\over i\omega + H}\lambda_A  {1\over i\omega + H}\lambda_B
       \right] .  \label\s23 \eeq
and the separation into valence and sea-part is analoque to Sect. 4.2..
The right generators  get explicitly
\beq
   R_A=-{{\partial L_M}\over{\partial\Omega_A}}=\cases{
- I_1\Omega_A                                        ,&A=1,2,3\cr
- I_2\Omega_A                                        ,&A=4,..,7\cr
  {1\over2}\sqrt3 = {\sqrt{3}\over2} Y_R  &A=8.  }\label\s24  \eeq
so that we can go from L to the hamiltonian H by
\beq  H=-\sum_A R_A \Omega_A - L  \label\s25  \eeq
and obtain
\beq
   H_{coll}^{sym}= M_{cl}+  {1 \over 2 I_2} \sum_{A=1}^7 R_A^2 +
   ({1\over 2      I_1   } - {1 \over 2 I_2 } )
     \sum_{A=1}^3 R_A^2    \eeq
\beq   =   M_{cl}+  {1 \over 2 I_2} C_2(SU(3)_{R/L}) +
            ({1\over 2      I_1   } - {1 \over 2 I_2 } )
            C_2(SU(2)_{R})   \label\s26  \eeq
where the $C_2(SU(3)_{R/L})$ is the {\it Casimir operator} of SU(3) and
$C_2(SU(2)_{R})$ is the corresponding one for the right SU(2). The
eigenstates of these operators are the irreducible representations,
which are usually labelled by the quantum numbers p and q, according to
the rank 2 of SU(3). From the construction of states, which was done
e.g. by de Swart, it is clear that $Y_R=1$ restricts the possible
representations to those, which contain a basis state with $Y=1$.
Therefore only {\it triality zero} states  survive, for which $(p-q){\rm
mod\ }
3=0$. As lowest possible one these are the octet  $\{8\}=(p=1,q=1)$ and
the decuplet $\{10\}=(p=3,q=0)$. With formula \qeq{\s26 } at hand, we
can determine the masses for the center of the octet and decuplet
states. In order to remove the degeneracy of the states, within the
multiplet, we have to switch on a finite symmetry breaking via a finite
current quark mass. In this way, we obtain
\beq L_M^{m_s} = -{2\over 3} {m_s\over m_u + m_d} \Sigma \left(
           1 - D^{(8)}_{88} \right)
-{2m_s\over \sqrt{3}} K_{AB} D^{(8)}_{8A} \Omega_B \label\s27 \eeq
where $\Sigma$ is the $SU(2)$ sigma commutator, defined in sect. 2.1 and
$K_{AB}$ are the so called anomalous moments of inertia, given by
(compare with Park and Rho, 1988, in chiral bag models)
\beq
   K_{AB}  =  -i {N_c\over 4}
        \int {d\omega\over 2\pi} {\rm   tr }\left[
       {1\over i\omega + H}\lambda_A{1\over i\omega + H}
         \gamma_4 \lambda_B
       \right] .  \label\s28 \eeq
and obey  a similiar structure like the $I_{AB}$:
\beq
   K_{AB} = \cases{    K_1 \delta_{AB} &for A,B=1,2,3  \cr
                   K_2 \delta_{AB} &for A,B=4,5,6,7 \cr
                   0              &for A,B=8 \cr }  \label\s29 \eeq
Because they originate from the imaginary part of the effective
Euclidean action, they need no regularization. The quantization
condition changes to
\beq
   R_A=-{{\partial L_M}\over{\partial\Omega_A}}=\cases{
-(I_1\Omega_A-{2m_s\over\sqrt{3} }K_1 D_{8A} )       ,&A=1,2,3\cr
-(I_2\Omega_A -{2m_s\over\sqrt{3} }K_2 D_{8A} )      ,&A=4,..,7\cr
  {1\over2}\sqrt3,   &A=8.       }\label\s24  \eeq
so that one obtains for the symmetry breaking part of the collective
hamiltonian :
\beq   H_{coll}^{sb} = -
    m_s {K_2\over I_2} Y - {2 m_s \over \sqrt{3} } \left( {K_1\over
I_1}
    - {K_2 \over I_2} \right) \sum_{A=1}^3   D_{8A}^{(8)}(A) R_A +
\eeq
\beq
  {2\over 3}{m_s \over m_u +m_d}\Sigma \left( 1-  D_{88}^{(8)}(A)\right)
+  {N_c m_s \over 3}{K_2\over I_2}  D_{88}^{(8)}(A). \label\s30 \eeq
where we have used the relation $\sum_A D_{8A}R_A=L_8={\sqrt{3}\over
2}Y$. The different values for the moments of inertia $I_1,I_2,K_1$ and
$K_2$ from the selfconsistent solitonic solutions can be found in
\qutab{\taband1}.
for two values of the constituent quark mass M.

Now we are in the position to evaluate the mass splittings within the
hyperon multiplets. This will be done in the next section by evaluating
the expectation value of the collective hamiltonian in a given baryonic
state.

\sect{Static properties of SU(3) Baryons }

After writing down the collective hamiltonian, we can
evaluate it in different ways. First we can set baryon wave functions as
described in \qeq{\s19 }, which are eigenfunctions of the symmetric
hamiltonian $H_{coll}^{symm}$. All we have to do is  then to sandwich
the D-functions of the symmetry breaking part of the collective
hamiltonian $H_{coll}^{sb}$
between baryon wavefunctions (Adkins et al. 1983).
On the other hand, one can parametrize the
right generators by differential operators of the Euler angles of the
given representation (Yabu and Ando 1988, Park et al. 1991, Park and
Weigel 1992). Then it is
possible to evaluate the hamiltonian
between the exact wave functions, which can be perturbatively expressed
as a series in the symmetry breaking quark mass $m_s$.
After we have clarified the different procedures, we will comment on the
numerical evidence of the exact treatment.

\subsec{The Perturbative Treatment of the Collective Hamitonian }

Having collected the leading terms in view of a $1/N_c$
expansion for the collective hamiltonian, we have to sandwich this
operator between the wave-functions of the baryons. Because of the
symmetry breaking terms in $H_{coll}^{sb}$, the  functions
$D_{AB}^{(n)}(A)$ in octet and decuplet representation are not exact
eigenstates of the full hamiltonian. In principle, one has to write down
a perturbative expansion
\beq  \mid B > = \mid B,R >  + \sum_{N\ne R}
    { <B,N\mid H_{coll}^{sb} \mid B,R> \over E_R^B - E_N^B }
     \mid B,N> + \dots     \label\s31    \eeq
where R is the lowest representation in which the baryon B can be found
and the summation goes over all higher dimensional representations N,
which have non-vanishing overlap with the ground state in R. So it is
clear that the proton, e.g., is a sum of a proton in an octet
representation and corresponding states in $\bf{\bar 10}$ and $\bf
27$.
 Furthermore these states enters in \qeq{\s31 } linearly with the
symmetry breaking parameter $m_s$.
Because only the symmetry breaking terms in the
collective hamiltonian have a non-vanishing  overlap with these higher
representations, they appear in the mass for  the first time in  the
second order  of $m_s$.
 So  one can define a
perturbative
treatment of $m_s$ in first  order, in which
the first order $m_s$-correction of the wave-function \queq{\s31 }
do not appear.
The effects of these terms will be discussed in the next
section.

In order  to evaluate $H_{coll}^{sb}$ in the baryon states
$\mid B,R>$,
we need the integral over 3 D-functions. These integrals are well known
(Blotz et al. 1993b)
and
are summarized in \qutab{\taband2}
(cf. De Swart ,1963, for a general discussion).
As a result, we can  express the mass splittings in terms of the 2
quantities
\beq
    \Delta={2\over3}{m_s\over m_u+m_d} \Sigma + m_s \left(
     2 {K_2\over I_2} - 3 {K_1\over I_1} \right)
    \label\s32
\eeq
and
\beq
    \delta = m_s {K_1\over I_1} . \label\s33
\eeq
This is little bit astonishing, since we have {\it three} operators
$Y,D_{88}$
and $\sum_{A=1,2,3}D_{8A} R_A$. However due to a non-trivial group
theoretical property (Blotz et al. 1993b), the {\it two} quantities
$\Delta$ and
$\delta$ are
sufficient. Then we obtain for the splitting between the center of the
multiplets (experimentally $\simeq 230{\rm MeV}$):
\beq
  \Delta_{8-10} = { 3\over 2 I_1}
\label\s34
\eeq
which is solely given by the difference of the Casimir operators
$C_2(SU(3)_{R/L})$  and    $C_2(SU(2)_{R})$ in octet and decuplet
representation. One should note that this formula coincides with
$N-\Delta$ splitting   in SU(2) theory. This means that the same
expression has to be considered as a different  physical
quantity depending on the use of the SU(2) or SU(3) collective
quantization. Then we can write for the masses of the strange
baryons relative to the mass of the  $\Sigma^*$, which we fix for the
moment to the experimental value, as \foot{Semiclassical
quantization is known to yield always values for the masses which are by
several hundred \MeV\ too high. Remedies to this by means of band-head
and centre-of-mass corrections will be used below.}
\beq
    \eqaligntag{
\Delta m_N &=  -\frac{3}{10}\Delta-\delta -\Delta_{8-10}  &\EQADV\spl
\SUBEQBEGIN\spl1\cr
\Delta m_\Lambda &=  -\frac{1}{10}\Delta - \Delta_{8-10}
&\SUBEQ\spl2\cr
\Delta m_\Sigma &=  \frac{1}{10}\Delta  -  \Delta_{8-10}
&\SUBEQ\spl3\cr
\Delta m_\Xi &=  \frac{1}{5}\Delta+\delta - \Delta_{8-10}
&\SUBEQ\spl4\cr
\Delta m _\Delta &=  -\frac{1}{8}\Delta-\delta   &\SUBEQ\spl5\cr
\Delta m_{\Sigma^*}&=  0                          &\SUBEQ\spl6\cr
\Delta m_{\Xi^*}&=   \frac{1}{8}\Delta+\delta      &\SUBEQ\spl7\cr
\Delta m_\Omega &=   \frac{1}{4}\Delta+2\delta     &\SUBEQ\spl8\cr}
\eeq
The reason that we fixed the $\Sigma^*$ to the experimental value is
that the quantization always yields  too large values  for the masses
if one does not employ band head and center of mass corrections as
discussed below.

{}From the formulas \qeqs{\spl } above follow the mass relation of
Guadagnini (1984)
\beq
     m_{\Xi^*}-  m_{\Sigma^*} +   m_N ={1\over 8} (11m_\Lambda - 3
      m_\Sigma)  .       \label\s35
\eeq
It is interesting to note that it is obtained in
the pseudoscalar Skyrme model only by introducing the   hypercharge
operator and several coefficients (Guadagnini 1984) by hand. In our
approach this term arise naturally in the theory and
the coefficients in front of all of them are completly determined by the
selfconsistent soliton solution.

Furthermore one obtains  the Gell-Mann
Okubo relations
(Gell-Mann 1962, Okubo    1962)
\beq
   2 (m_N+m_{\Xi})=3m_\Lambda+m_\Sigma   \label\s36
\eeq
and
\beq
   m_\Omega-m_{\Xi^*}=m_{\Xi^*}-m_{\Sigma^*}=m_{\Sigma^*}-m_{\Delta}
\label\s36
\eeq
automatically in this approach by means of \qeq{\spl}.
One should note that these relations are mass sum rules which rely
basically on the fact, that the SU(3) flavour symmetry breaking part of
the strong interaction can be treated in 1st order perturbation theory
(Cheng and Li 1984).  Their validity shows therefore the consistency of
the
whole quantization procedure. On the other hand the values of the
hyperon masses themselves depend on the dynamics of the model which is
considered.
In \qutab{\taband1 }
 we have
collected the difference of the theoretical mass from the experimental
one  for
two values of the constituent quark mass M.
As one can see from the table for the value $m_s=150 MeV$, which
correspond
to \qeq{\l38 } for  $m_0=6.1{\rm MeV}$, the splitting prediction is
better than $50{\rm MeV}$, except for the $\Omega$, which drops out with
$80{\rm MeV}$.
However it is instructive to increase  $m_s$ to a value $m_s=200{\rm
MeV}$, i.e. this corresponds to enlarge $m_K$ to $\simeq~570\MeV$
(exp. $m_K=496\MeV$,  which
gives a very good
agreement with experiment up to
$20{\rm MeV}$ for all baryons under consideration.

\subsec{Comment on Validity of Chiral Expansion }

In Sect. 6.1, following \queq{\l38 }, we mentioned the failure of chiral
perturbation theory for
some vacuum parameters. Now we are in a position to answer this question
for the baryonic sector and it will be done here for
the case of the total masses as an example. Other observables  are
considered in the literature (Blotz et al. 1993a,
1993d).  These two  sectors need not to  have the same
behaviour, because the vacuum sector is dominated by different integral
equations  compared to  the soliton.
Actually we can write the classical soliton  mass in perturbation theory
for $m_s$ as
\beq  M_{cl} = M_{(0)} + m_s M_{(1)} +m_s^2 M_{(2)}
      + {\cal O} (m_s^3)             \label\s60
\eeq
with the result within the present model:
$$
\eqaligntag{
      M_{cl}&=  M_{(0)} +{m_s\over 4m_0} \Sigma -{2\over 9} m_s^2 N_0
        \cr
       &= 1250 + 426.9  -26.0 MeV  &\EQ\s61\cr}
$$
for the preferred values of  $M=418{\rm MeV}$, $\Sigma=56MeV$,
$N_0=0.668{\rm fm}$ and
$m_s=186{\rm MeV}$. So one can say that at least for the action
the validity of the chiral expansion  seems to be
more reliable for the baryon sector than for the vacuum
(Hatsuda 1990).

\subsec{The Yabu-Ando Diagonalization Method }

In contrast to the former method of evaluating the collective
hamiltonian between the {\it symmetric} eigenfunctions of the
Casimir operators, Yabu and Ando (1988) developed a method for treating
the baryonic wave functions to  all orders in  $m_s$.
The  idea is to the express  the rotation matrix A by its eight Euler
angles and derive also for the right generators $R_A$  an explicit
differential operator form in terms of these Euler angles.

If we adopt the same parametrization of the rotation matrix
A as discussed by Yabu and  Ando (1988), we can write
\beq   A = R(\alpha,\beta,\gamma) \exp{(-i\nu\lambda_4)}
      R(\alpha^\prime,\beta^\prime,\gamma^\prime)
     \exp{(-i\rho\lambda_8/\sqrt{3})} \label\s62    \eeq
where the    $R(\alpha,\beta,\gamma)$ is  the Euler-angle rotation
matrices of SU(2)-isospin.  Because of the hedgehog ansatz  for the
chiral field, isospin rotations and space rotations are intimately
connected in a way, that any isospin rotation can be undone by a
corresponding rotation in coordinate space. Therefore the
 $R(\alpha^\prime,\beta^\prime,\gamma^\prime)$  is the SU(2) spin
rotation matrix.  Now we would like to obtain an expression for the
right generators $R_a$ in terms of these eight Euler angles,
conveniently written as $\alpha_a$. Following the work of Nelson (1967)
and Park et al. (1991),
we make the ansatz
\beq R_a = i d_{ab} (\alpha) { \partial \over \partial \alpha_a }
       , \label\s63         \eeq
such that $R_a$ are  a linear differential operator of the
$\alpha_a$.
The unknown matrix $d_{ab}$   can be determined by inserting this
ansatz into the definition of the $R_a$
\beq  A R_b A^\dagger = A \lambda_b A^\dagger   \label\s64 \eeq
Now both sides of \queq{\s64 } can be explicitly worked out and give
with  \qeqs{\s62,\s63 }
\beq
      \lambda_a  E_{ac}(\alpha) d_{cb} (\alpha) =
        C_{ab}(\alpha)\lambda_a     \eeq
with the some known matrices $E_{ac}(\alpha)$ and $C_{ab}(\alpha)$.
Comparison yields $d_{ab}(\alpha)=E_{ac}^{-1}(\alpha)C_{cb}(\alpha)$.
Their rather lengthy forms are summarized by Park and Weigel (1992).
Then one
can directly act with these operators on the wave functions, which are
themselves D-functions and therefore also given in terms of A.

The result of this method can be seen in \qufig{\fig1}, where
the deviation
of the theoretical prediction from  the experimental one is shown in
dependence of the strange current quark mass. Although $m_s$  is
fixed to be $\simeq 150\MeV$ from the mesonic sector, it is instructive
to see that although the predictions for $m_s=150\MeV$ are already
good,
a slight increase to $m_s\simeq187\MeV$ leads to an almost perfect
agreement (see also \qutab{\taband3}). The difference to the
former
perturbative treatment is in addition illustrated in \qufig{\fig2}, where
the Yabu-Ando method is compared to the perturbative treatment for two
particles of the multiplets. As it is clear from these curves, the
Yabu-Ando method gives a significant contribution to the masses for
current masses above 100 $\MeV$.
Similiar calculations have been performed by  Weigel et al. (1992b).
The formalism
of these authors differs from the present one in the fact that
they do not follow stricly a perturbative expansion in $1/N_c$ and
$m_s$. Hence they also have different constituent quark masses for the
strange and non-strange quarks.  The resulting numbers are quite close
to those presented here.

Furthermore  Praszalowicz et al.
(1993a) made  a refined calculation in the lines of the preceding
sections with the inclusion of  isospin breaking current quark masses.
As a result they found that one can make a
prediction for the hadronic part of the isospin splitting within the
multiplets, which agrees with the experimental data (Gasser and
Leutwyler
1982) with a high accuracy.
In this work  it turned out  that the anomalous moments
of inertia play again a crucial role in order to predict the splittings
such as the neutron proton mass difference, which could  be explained
neither
in the former SU(2) pseudoscalar Skyrme model, where it vanishes, nor in
a U(2) or SU(3)
extension
(Jain et al.
1989) of the model.
In \qufig{\fig3} the hadronic part of the isospin splitting
within the octet is shown and compared with the experimental predictions
from Gasser and Leutwyler (1982). In addition it could be shown by
Praszalowicz et al. (1993) that the
complete agreement between theory
and experiment within the experimental error bars is achieved for that
$m_d-m_u$ difference, which follows from the mesonic sector from
the masses of the charged kaons and pions. This can be viewed as a
reflection of the fact, that perturbation theory works quite well for
the small isospin breaking parts.
In addition it was shown by Blotz et al. (1993e), that the recently
measured Gottfried sum, which was considered in the present model for
the case of SU(2) by Wakamatsu (1992b), comes out in the SU(3) version
quite close to the  value of the NMC measurements  (NMC, Amaudruz et al.
1991). In addition the  $\Sigma$-term comes out to be
$\Sigma\simeq~45\MeV$, which is again close to the experimental value of
Gasser et al. (1991)

\subsec{Total Masses and Zero-Point Energy Corrections }

By now we have discussed the mass splittings within the baryon octet and
decuplet and the splitting between the octet and decuplet. As mentioned
already the absolute energies of particles come out by several hundred
\MeV\ too high in the semiclassical approximation no matter which chiral
model is used.
Subtraction mechanism, as they are discussed by Pobylitsa et al.
(1992)
in  SU(2) and by Blotz et al. (1993b) in SU(3) (compare with Jain et
al.
1988b for a discussion in the Skyrme model) can in principle
cure the  situation.  In order to obtain reasonable values for the
absolute masses of the hyperons one has to subtract the spurious zero
mode energies similiar to the way described in sect. 4.4..
The corresponding terms of the rotational zero modes in SU(3)
and of the translational one read
$$ \eqaligntag{
       \Delta M^{rot}
      &={1\over 2I_2}  \langle  C_2(SU(3)) \rangle    +
       {1\over 2}\left( {1\over I_1}-{1\over I_2} \right)   \langle
       C_2(SU(2)) \rangle      \cr
      &={1\over I_2}{7\over 8} +
        {1\over I_1}{9\over 8}     \cr
     \Delta M_{transl} &=   { <[{\vec P}^2]_{(1)}> \over 2M_{cl} }
                                 &\EQ\zeros\cr}
$$
where we used the values of the Casimir operators for the fundamental
representation. These are  $\langle C_2(SU(3))\rangle=4N_c/3$
and $\langle C_2(SU(2))\rangle=3N_c/4$.
The translational zero mode subtraction coincide for SU(3) with the
SU(2) result $\Delta M^{transl}$ of \queq{\e469}.
Subtracting these terms from the classical soliton mass $M_{cl}$
gives  for the center of the octet and decuplet
\beq    M_8 = M_{cl} + {1\over I_2}{3\over 4} +{1\over I_1}
      {3\over 8} -\Delta M^{rot}-\Delta M^{transl} \EQN\subt1 \eeq
\beq  M_{10} = M_{cl} + {1\over I_2}{3\over 4} +{1\over I_1}
      {15\over 8}-\Delta M^{rot}-\Delta M^{transl}      \EQN\subt2
\eeq
If one performs these corrections to the classical energy
analogous to \queq{\e468} in the present case of SU(3)
(Blotz et al. 1993b) the absolute mass of the $\Sigma^*$-particle
changes  from   2262 MeV     to 1494  MeV (experimentally
$m_{\Sigma^*}=1385 MeV$). Thus with these corrections the masses of all
the hyperons of the octet and decuplet are reproduced with a constant
shift of $\simeq 100\MeV$, being in fact a quite impressive result.

\subsec{Expectation Value of Axial Currents }

In a similiar way to chap. 5 one can evaluate  the expectation
value
of the SU(3) axial vector current operator. However, as it was mentioned
in the SU(2) case already, there are important $1/N_c$ corrections
to  vector and axial vector  currents. These  are entirely due
to the fact, that the collective operators $\Omega$ and
$A^+\lambda^aA$ in general do not commute and have to be explicitly
time-ordered  (Christov et al. 1993b, Blotz et al. 1993d).
Without going into details we will nevertheless give here a shortcut
derivation. From the path-integral  one obtains for
${\hat A}_\mu^a(x),\
a=0,3,8$:
\beq
        <{\hat A}_\mu^a(x)>  =  {\delta\over\delta
           s(x)   }
        \bigl[
      \Spto \log \left( \partial_4 + H +
         i \Omega_E   - i \gamma_4 A^+m A    +
       i\   s(x)  \gamma_4 \gamma_\mu \gamma_5   A^+ \lambda^a A
       \right)
       \bigr]     \label\s65         \eeq
where $\Spto$ means now that the time dependent operators
$(A(t),\Omega)$  within
the trace have to be time-ordered, before performing the trace in time
direction. This however disagrees with the usual definition of the trace
mapping and its origin is the fact, that we are not going to perform
the path-integral over the rotation matrices $A(t)$ in the functional
sense (Blotz et al. 1993d, Christov et al. 1993b). Instead one usually
switches to the operator formalism, as it was done throughout this work
(cf. Dyakonov et al. 1988), which involves the evaluation of
time-ordered products of operators. Because of this, the trace in
\qeq{\s65 } has to be modified in  a way, that respects an explicit
time-ordering of the operators (cf. Blotz et al. 1993d).
To obtain a c-number value for the current, finally one has
to sandwich the operator  $<{\hat A}_\mu^a(x)>$
between suitable baryon wave-functions.

In addition to the SU(2) case,  \qeq{\s65 }
contains  the symmetry
breaking term from the strange current quark mass.
Therefore the expression for axial current up to the first order
in the rotational matrices  $i\Omega_E$ and the strange current quark
mass $m$ is given by
$$ \eqaligntag{
      <{\hat A}_\mu^a(x)>  &=            \Spto { 1 \over \partial_4 + H
               }
               i\gamma_4 \gamma_\mu \gamma_5 A^\dagger I_a A \cr
   &     -            \Spto  { 1 \over \partial_4 + H }
           i \gamma_4 \gamma_\mu \gamma_5 A^\dagger I_a A
            { 1 \over \partial_4 + H } \delta H
       + {\cal O} ( {\delta H}^2 )
     &\EQ\s66\cr}    $$
where the {\it perturbation } $\delta H$  is now  given by
\beq  \delta H =  i \Omega_E   - i \gamma_4 A^+m A
\label\s67
\eeq
Evaluating this yields for the $a=3,8$ part of the axial vector
coupling constant  (Blotz et al. 1993d)
$$ \eqaligntag{
      {\hat g}_A^a  &=             M_3 D_{a3}
         + {4 M_{44} \over I_2} d_{3bb} D_{ab} R_b - \cr
    &  { 2 i Q_{12} \over I_1} D_{a3} - { 2i Q_{45} \over I_2} D_{a3}
  \cr  &  +   {2 M_{83} \over  I_1} R_3 D_{a8} ( 1
           +{4m_s\over\sqrt{3}}
            {K_1\over I_1} D_{83} ) \cr
    &  +   { 4m_s\over \sqrt{3} } {N}_{83}  D_{a8}D_{83}
        + {8m_s\over \sqrt{3}} ( N_{44} - M_{44} {K_2\over I_2} )
       d_{3bb} D_{ab} D_{8b}     \cr
  &   -{4 m_s\over \sqrt{3} } N_{38} D_{a3} ( 1-D_{88} )
     &\EQ\s68\cr}               $$
and for the singlet part  (Blotz et al. 1993a)
\beq  {\hat g}_A^0 =  {2\sqrt{3} { M}_{83} \over I_1} R_3 - {4 m_s
     }
     D_{83}^{(8)} \left( {K_1\over I_1}{ M}_{83} - { N}_{83}
\right)       \label\s67
\eeq
There we used the following definitions for the various moments of
inertia. For the anomalous moment  $M_{bc}$, we find
\beq   M_{bc} = {N_c\over 4} \sum_{n,m}
          < n \mid \sigma_3 \lambda_b \mid m > < m \mid \lambda_c
         \mid n >
         {\cal R}_{\cal M} (E_n,E_m)
       \label\s68
\eeq
with  \beq
      {\cal R}_{\cal M} (E_n,E_m) =
     \half {   \signum (E_n-\mu) - \signum (E_m-\mu) \over
                  E_n -  E_m   }
           \label\s69
\eeq
and where the chemical potential $\mu$ lies always between the valence
level and positive continuum of states, in order to describe a baryon
number $B=1$ system.
For the
proper time regularized normal moments $N_{bc}=N_{bc,val}+N_{bc,sea}$
from the symmetry breaking, we find
\beq    N_{bc,val} = {N_c \over 2} \sum_n
       { < n \mid \sigma_3 \lambda_b \mid v > < v \mid \lambda_c
        \gamma_0
       \mid n > \over  E_n    -   E_v   }  \label\s70
\eeq
and
\beq   N_{bc,sea} = {N_c\over 4} \sum_{n,m}
           < n \mid \sigma_3 \lambda_b \mid m > < m \mid \lambda_c
          \gamma_0
         \mid n >
         {\cal R}_\beta   (E_n,E_m)       \label\s71
\eeq
with  \beq
      {\cal R}_\beta (E_n,E_m) =
      {1 \over 2 \sqrt{\pi}} \int_0^\infty {dt\over\sqrt{t}} \phi(t)
      \bigl[{ {E_n e^{-tE_n^2} - E_me^{-tE_m^2}
             \over  E_n      -      E_m   }
                       }  \bigr]
    \label\s72
\eeq
and for the {\it antisymmetric} moments $Q_{bc}$  from the explicit
time-reordering we define $Q_{bc}=Q_{bc,val}+Q_{bc,sea}$:
\beq   Q_{bc,val} = {N_c \over 2} \sum_{n}
        {  < m \mid \sigma_3 \lambda_b \mid n > < n \mid \lambda_c
         \mid m >  \over E_n - E_v } \signum E_n
       \label\s72a
\eeq
and
\beq   Q_{bc,sea} = {N_c \over 4} \sum_{n,m}
          < m \mid \sigma_3 \lambda_b \mid n > < n \mid \lambda_c
         \mid m >
        {\cal R}_{\cal Q} (E_n,E_m)
       \label\s73
\eeq
In the case that the regularization function $\phi(t)$ is given
by $\phi(t)=\sum_ic_i\theta(1-1/\Lambda_i^2)$ the
${\cal R}_{\cal Q}(E_n,E_m)$ takes the simple form
\beq
      {\cal R}_{\cal Q} (E_n,E_m) =
    c_i \int_0^1 {d \alpha\over 2\pi}{\alpha(E_n+E_m)-E_m\over
       \sqrt{\alpha(1-\alpha)} } { \exp{(
      -[\alpha E_n^2+(1-\alpha)E_m^2]/\Lambda_i^2)} \over
        \alpha E_n^2+(1-\alpha)E_m^2    }
\label\s73a
\eeq
In the infinite cutoff limit it again reduces to
\beq
      {\cal R}_{\cal Q} (E_n,E_m) =
     \half {      \signum (E_m-\mu) - \signum (E_n-\mu) \over
                E_m - E_n }
           \label\s74
\eeq
Note that the regularization function ${\cal R}_{\cal Q} (E_n,E_m)$ in
\qeq{\s73a } is now antisymmetric with respect to
the states m and n in contrast to ${\cal R}_{\cal M} (E_n,E_m)$
in \qeq{\s69 }. This is a reflection of
the fact that the matrix elements of \qeq{\s73 } are now antisymmetric
with respect to m and n.

\subsec{Numerical Results }

Similiar to the case of SU(2)
in Sect. 5.2., the terms with the {\it antisymmetric} $Q_{bc}$
expressions  also serve here as  large corrections to the lowest
order terms $(\Omega^0)$ and pushes especially the value of $g_A^3$,
which was $50\%$ too low  without these terms,
a little bit beyond its experimental value (see \qutab{\taband4}).
Together with the values of $g_A^8$ and $g_A^0$, which
has the interpretation of being the spin of the proton that  is carried
by the quarks,
the contributions
from the lowest order  $\Omega^0$ and from the symmetric
and antisymmetric rotational corrections are presented in
\qutab{\taband4} and
compared with the experimental numbers from recent EMC and SMC
experiments. One should note that $g_A^{(0)}$ should be interpreted as
the spin of the proton (normalized to unity), which is carried by the
quark.  As it is clear from these numbers, the antisymmetric
contributions from the time-reordering play a significant role.
At the present stage of the work the NJL numbers are very good. The
corresponding  expressions
emerge from the real part of the Euclidean effective
action, though they turn out to be finite and need no regularization.
Furthermore, and this is most important, they are not present
in the usual Skyrme model approach. So this is in a sense another
reflection
of the explicit quark degrees of freedom, which obviously cannot be
described
perturbatively in the chiral fields and which provides  therefore a
clear distinction between Skyrme type models and Nambu-Jona-Lasinio type
models.

\def\eeq{ $$ }
\vfill\eject

%
%

%
\overfullrule=0pt
\def\3{$\beta$}

\def\op#1{\hat{#1}}
\def\bra#1{\langle #1 \,\vert}
\def\ket#1{\vert\, #1 \rangle}

\def\nl{\hfill\break}
\def\({\Bigl(}
\def\){\Bigr)}
\def\[{\Bigl[}
\def\]{\Bigr]}
\def\MeV{\rm MeV}
\chap{The NJL-model with vector mesons }
\TAB\tabv1{
            A comparison of the mean field energies ( in MeV ),
the axial nucleon coupling constant $g_A $, the isoscalar mean squared
radius $<r^2>$ ( in fm$^2$ ) and the value of the polar field at the
origin $\Phi(0)={1\over f_\pi}\sqrt{\sigma^2 (0)+\pi (0)^2}$.
For a constituent quark mass $M=340\MeV$ the $\omega$ coupling constant
$g_\omega$ is increased up to the correct value $g_\omega=2.24$, which
is in accordance with the mesonic sector. The linear model is used for
the calculation.
}

\TAB\tabv2{
The same as Tab.7.1 in the linear and non linear version of the
\NJLM with $\sigma\pi\rho\omega$ and $A_1$ mesons  for different
constituent quark masses $M$. }
\FIG\figv1{
The selfconsistent vector  and axial vector fields $\omega,\rho,A_s$ and
$A_T$ for a constituent  quark mass $M=340MeV$. The parameters are fixed
according to the on-shell definition for the two-point functions, i.e.
$\Lambda=877\MeV$, $g_\rho=4.61$ and $g_\omega=2.24$.
                                          }
\FIG\figv2{
The polar field $A(r)={1\over f_\pi} \sqrt{\sigma^2(r)+\vec\pi^2(r)}$
(which equals one in the non linear model)
of the linear model is shown for
different constituent quark masses. }
\parskip0.5cm

\vskip0.5cm

In this chapter we consider the influence of vector couplings on the NJL
model. After giving some motivation, the bosonization and regularization
of the extended model are presented. The vacuum and mesonic sectors
are analyzed both assuming the mesons to be on-shell or off-shell.
Finally a system with baryon number one is constructed and its
solitonic solutions are analyzed.

\sect{ The Effective NJL Action with Vector Mesons}

\subsec{Why Vector Mesons ?}

Since their discovery (Nambu 1957, Frazer and Fulco 1959, 1960), vector
mesons have been intimately linked to the internal structure of the
nucleon. In
fact, the success of the vector dominance hypothesis (Sakurai 1960;
Gell-Mann and Zachariasen 1961) and its later field theoretical
realization through current-field identities (Kroll et al. 1967) in
many hadronic reactions is obvious  (for a review see
e.g. Gourdin 1974). In addition, vector mesons play a crucial role in low
energy nuclear physics, since they are responsible for the medium and
short range NN interaction (Machleidt et al. 1987). It seems natural to
ask how vector mesons may be implemented in a NJL type model.

Up to now, we have referred to the NJL model with scalar-isoscalar and
pseudoscalar-isovector couplings ($\sigma\pi$-version of NJL). The
introduction of other couplings makes
possible to describe a wider meson spectroscopy, since it is known that
the $\sigma\pi$ version of the model does not account for vector or
axial vector degrees of freedom, i.e. the corresponding correlation
functions do not posses poles.

If additional vector couplings are included, the NJL-model  indeed
incorporates in a natural way  important phenomenological principles
found long before
the advent of QCD: Sakurai's universality (Sakurai 1966, 1969) and
vector meson dominance realized by means of current-field identities
(Kroll et a. 1967). It also provides a relationship to effective
low energy Lagrangians with vector mesons (Kleinert 1978, Dhar et
al. 1984,1985; Ebert and Reinhardt 1986; Wakamatsu and Weise 1988)
both
in  massive Yang-Mills (Lee and Nieh 1967 ; Gasiorowicz and Geffen
1969;
Meissner UG 1988) or in hidden symmetry  (Bando et al. 1984, 1988)
representation. Furthermore, it includes the gauged Wess-Zumino term
(Witten 1983) with the vector mesons interpreted as internal gauge
fields (Dhar et al. 1984, 1985 ; Ebert
and Reinhardt 1986; Wakamatsu 1989). In addition, almost all attempts
claiming any parentage of the NJL model to QCD require the explicit
inclusion of vector mesons (Dhar et al. 1984,1985; Cahill and Roberts
1985; Schaden et al. 1990; Chanfray et al. 1991). An exception to this
is the instanton liquid model (Diakonov and Petrov 1986).

\subsec{Coupling of Vector Mesons - Bosonization}

The extended $SU(2)_R \otimes SU(2)_L \otimes U(1)_B$
classically invariant NJL-model
including scalar, pseudoscalar, vector and axial-vector couplings reads
$$ \eqalign{
 {\cal L}_{NJL} =& \bar{q} (i\slashchar\partial -m_0)q +
 {G_1\over 2} [(\bar{q} q)^2 + (\bar{q}i\gamma_5 {\vec \tau}q)^2] \cr &
 - {G_2\over 2} [ (\bar{q} \vec\tau \gamma_\mu q)^2 +
                  (\bar{q} \vec\tau \gamma_\mu \gamma_5 q)^2 ]
 - {G_3\over 2} (\bar{q} \gamma_\mu q)^2                          \cr }
\eqn$$
It is important to mention that the terms multiplying the coupling
constants $G_1$, $G_2$ and $G_3$ are themselves chirally invariant.
Thus at the classical level the conservations laws given  in chapter 2.
remain valid. Furthermore, since the coupling constants have dimensions
of inverse mass squared it is convenient to choose them as follows
$$ G_1 = {g_\pi^2 \over \mu^2 } \quad, \qquad
   G_2 = {g_\rho^2 \over 4 m_\rho^2 } \quad, \qquad
   G_3 = {g_\omega^2 \over m_\omega^2 }
\eqn $$
where $ g_\pi , g_\rho $ and $g_\omega$ are dimensionless coupling
constants and $ m_\rho $ and $ m_\omega$ are the vector meson masses.
The parameter $\mu $ has mass dimension and it will turn out to be
identical to the  $\mu$ introduced  in chap. 2.

Similarly to the $\sigma \pi$ case the corresponding generating
functional is made well defined by performing a Wick rotation. In order
to keep track of Lorentz invariance the time components of the vector
fields are rotated as well (cf. app. A)
$$ x^0 \to i x^4 , \qquad
\omega^0 (x_0 , \vec x) \to  i \omega^4 (x_4 , \vec x), \qquad
\vec \omega (x_0 , \vec x) \to  \vec \omega (x_4 , \vec x)
\eqn $$
Following the steps of chapter 2
we multiply in addition by the Gaussian factor
$$
\int D\omega_\mu D\vec \rho_\mu D\vec a_\mu
\exp \Bigl\{ - {1\over 2} \int dx \[ m_\omega^2 \omega_\mu^2 +
m_\rho^2 (\vec\rho_\mu^2 + \vec a_\mu^2) \] \Bigr\}
\eqn $$
An important point is that the Wick rotation makes the exponent in the
Gaussian factor to have a well defined negative sign. Similarly to
chapter 2 Euclidean indices will be understood unless otherwise stated.
The Gaussian factor is chirally symmetric if the fields are assumed to
transform as
$$\eqalign{ & \omega_\mu \to \omega_\mu \cr
& \vec \rho_\mu \to \vec \rho_\mu + \vec \alpha \times \vec \rho_\mu +
\vec \beta \times \vec a_\mu \cr
& \vec a_\mu \to \vec a_\mu + \vec \alpha \times \vec a_\mu +
\vec \beta \times \vec \rho_\mu \cr }
\eqn$$
under the global chiral $SU(2)_R \otimes SU(2)_L $ group. Performing the
shifts
$$ \eqalign{
& \omega_\mu \to \omega_\mu - {g_\omega \over m_\omega^2}
 \bar q \gamma_\mu q \cr
& \vec\rho_\mu \to \vec\rho_\mu - {g_\rho \over m_\rho^2 }
\bar q \gamma_\mu q{\vec \tau\over 2} q \cr
& \vec a_\mu \to \vec a_\mu - {g_\rho \over m_\rho^2}
\bar q \gamma_\mu \gamma_5  q{\vec \tau\over 2} q \cr }
\eqn $$
one gets the semibosonized Lagrangian
 $$ \eqalign{& {\cal L}^{eff} (x) =  \bar q \(-i\slashchar\partial
     +g_\pi \left(\sigma +i\gamma_5 \vec\tau \vec \pi \right)
 + g_\rho \left(\vec{\slashchar\rho} +\vec{\slashchar a} \gamma_5
\right) {{\vec \tau } \over 2}-g_\omega \slashchar\omega + \bar m_0 \) q
\cr & \qquad \qquad +{\mu^2 \over 2} (\sigma^2 + \vec\pi^2)
 +{m_\rho^2 \over 2} (\vec \rho_\mu^2 + \vec a_\mu^2)
 +{m_\omega^2 \over 2 } \omega_\mu^2  \cr }
\eqn $$
At the classical level, the quark part of this Lagrangian is invariant
under {\it local} chiral transformations. We will see later that this
local symmetry is broken by the regularization thus leading to a chiral
anomaly. After integration of the quarks the extended  action
reads
 $$ \eqalign{& S_{eff}[\sigma,\pi,\omega,\rho, a] =
 - \Sp  \log \(-i\slashchar\partial
     +g_\pi \left(\sigma +i\gamma_5 \vec\tau \vec \pi \right)
 + g_\rho \left(\vec{\slashchar\rho} +\vec{\slashchar a} \gamma_5
\right) {{\vec \tau } \over 2}  - g_\omega
\slashchar\omega + \bar m_0 \)  \cr &
\qquad  +{\mu^2 \over 2} \int d^4 x (\sigma^2 + \vec\pi^2)
 +{m_\rho^2 \over 2 }\int d^4 x (\vec \rho_\mu^2 + \vec a_\mu^2)
 +{m_\omega^2 \over 2 }\int d^4 x \omega_\mu^2  \cr }
\eqn $$
Again we treat  the theory in the stationary phase approximation.
The corresponding classical equations of motion give the current field
identities:
$$ J_\mu^B = {m_\omega^2 \over g_\omega} \omega_\mu, \qquad
\vec J_\mu^V = {m_\rho^2 \over g_\rho} \vec \rho_\mu, \qquad
\vec J_\mu^A = {m_\rho^2 \over g_\rho} \vec a_\mu
\EQN\cfi $$
with $\vec J_\mu^V, J_\mu^B$ and $J_\mu^A$ being the baryon-, vector-
and axial vector currents
\beq \vec J_\mu^V  ={\bar q}\gamma_\mu\half\vec\tau q,\ \
     \vec J_\mu^A  ={\bar q}\gamma_\mu\gamma_5\half\vec\tau q,\ \
     J_\mu^B =  {\bar q} \gamma_\mu q  \EQN\curen  $$
The parameters $m_\rho, g_\rho, m_\omega, g_\omega$ will be
considered
later. Due to the Wick rotation the effective action is a complex
number which can be separated into real and imaginary parts
$$ S = {\rm Re} S + i {\rm Im} S   \quad ,\quad
   {\rm Re} S = {1\over 2} (S + S^\dagger)  \quad ,\quad
   {\rm Im} S = {1\over 2i} (S - S^\dagger)
\eqn$$
On the basis of the combined symmetry operation $Q=G\gamma_5$ with $G$
the usual G-parity (charge conjugation plus rotation of 180 degrees
around the y isospin axis) it has been shown (Doering et al. 1992)
that the real part of the action is an even function of the $\omega$
field whereas the imaginary part is an odd function. In particular,
if the $\omega$ field vanishes the Euclidean action becomes a real
number. Notice that this argument does not apply anymore if one
considers SU(3) flavour.

\subsec{Regularization of the Effective Action - Chiral Anomaly}

As we have said the quark contribution to the semibosonized Lagrangian
is formally invariant under chiral local transformations. On the other
hand, a regularization has to be introduced to make the effective action
finite. However, there is no regularization which preserves both vector
and axial gauge symmetries simultaneously. Thus there appears a chiral
anomaly (Adler 1969; Bell and Jackiw 1969). The subject of
chiral anomalies is rather involved and we refer the reader to the
recent review of Ball (1989) for a detailed discussion. (For a more
elementary level see e.g. Petersen 1985). We
just mention here that there is some mathematical ambiguity as to which
symmetry should be conserved and which should be destroyed. This freedom
depends on the particular physical application. The fact that QCD is
the fundamental theory underlying the NJL model suggests shifting the
anomaly to the pure axial sector, i.e. to make use of a vector gauge
invariant regularization.

Dhar et al. (1984,1985) have proven that the proper prescription to
achieve vector current conservation is to regularize the real part and
not to regularize the imaginary part, so that the effective action
becomes
$$ S_{eff} = -{1\over 2} \int_0^\infty {d\tau\over \tau}
\phi(\tau, \Lambda)\Sp e^{-D^\dagger D} + {1 \over 2}\(\Sp \log
(iD)-
\Sp \log (-iD^\dagger) \) + {\rm mass\ terms }
\eqn $$
Then the real part of the one loop
contribution is both vector and axial gauge invariant while the
imaginary part exhibits an axial anomaly. This prescription leads to the
original  Bardeen's form of the chiral anomaly (Bardeen 1969),
however with the vector and axial fields interpreted as dynamical
degrees of freedom.

In the case of the NJL model with dynamical vector mesons one gets
the following anomalous Ward identities  (Wess and Zumino 1971, Wess
1972), which are generalizations of
the Noether theorem in quantum field theory, in Minkowski space
$$ \eqalign{
&  \partial^\mu J^B_\mu = 0 \cr
&  \partial^\mu\vec J^V_\mu=0 \cr
&  \partial^\mu\vec J^A_\mu=2i m_0 \bar q \gamma_5 {\vec \tau\over 2}
q + {N_c \over (4\pi)^2} g_\omega g_\rho \epsilon_{\mu\nu\alpha\beta}
\partial^\mu \omega^\nu \Bigl({1\over 2} \partial^\alpha \vec
\rho^\beta
+ g_\rho (\vec\rho^\alpha\times\vec\rho^\beta-\vec a^\alpha
\times\vec a^\beta)\Bigr)\cr}
\EQN\ward$$
As we see, the anomalous contribution to the divergence of the axial
current survives, even in the chiral limit $m_0 =0$ and in the absence
of external fields. This is called an internal anomaly which does not
even disappear if vector mesons are integrated out. In fact it has been
proven by Wakamatsu (1989), Ruiz-Arriola and Salcedo (1993a), that this
anomaly breaks some low energy theorems of QCD.

\sect{ Fixing of the Parameters in the Vector Mesonic Sector }

As it has been done in previous cases, the parameters have to be
fixed by looking at different mesonic properties. As input parameters
we use the pion weak decay constant $f_\pi=93 MeV$ and the meson masses
$m_\pi =139MeV , m_\rho=770 MeV$ and $m_\omega=783 MeV$.

The main new point in the meson sector is the occurrence of a mixing
between the pion and the axial meson very similar to the one found in
the early
massive Yang-Mills approach (Lee and Nieh 1967). This causes after
redefinition of the physical axial field a finite renormalization of
the pion kinetic energy (Kleinert 1978; Ebert and Reinhardt 1986). As a
consequence the corresponding cutoff increases
with respect to the case without vector mesons. In addition, the axial
mass acquires a substantial contribution from a partial Higgs mechanism.
For simplicity we will work on the chiral circle.

\subsec{Real Part -  Massive Yang-Mills and Hidden Symmetry Approach }

In the second order heat kernel approximation the real part of
the effective action reduces to the following expression (Ebert and
Reinhardt 1986; Wakamatsu and Weise 1988)
$$
\eqalign{  {\cal L}_{\rm real} \to {\cal L}_{MYM} &=
  I_2 M^2 \tr \[ D_\mu U^\dagger D^\mu U + (U^\dagger + U) \]  \cr &
  -{1\over 6} I_2 \Bigl\{ g_\rho^2 \tr \Bigl[ (V_{\mu\nu}^R)^2
+ (V_{\mu\nu}^L)^2 \Bigr] + g_\omega^2 \omega_{\mu\nu}^2 \Bigr\} \cr &
\qquad \qquad +{m_\rho^2 \over 2} \tr [ (V_\mu^R)^2 + (V_\mu^L)^2 ]
+ {m_\omega^2 \over 2 } \omega_\mu^2  \cr }
\EQN\mym $$
where the regularization dependent integral $I_2 (M)$ has been
defined in \qeq{\e227 }.
The constituent quark mass $M=g_\pi~f_\pi$
has been also introduced. The covariant derivatives and field strength
tensors read
 $$
 \eqalign{ &D_\mu U = D_\mu^L U - U D_\mu^R =
\partial_\mu U - i g_\rho (V_\mu^L U - U V_\mu^R) \cr
& V_{\mu\nu}^{R,L}=\partial_\mu V_\nu^{R,L}- \partial_\nu V_\mu^{R,L} -
ig_\rho [ V_\mu^{R,L} , V_\nu^{R,L} ] \cr }
\eqn$$
where
$$\eqalign{ & U = {1\over f_\pi}(\sigma + i\vec \tau \cdot \vec \pi)\cr
&V_\mu^R = { 1 \over 2} (\vec \rho_\mu + \vec a_\mu)
\cdot \vec \tau \cr
&V_\mu^L = { 1 \over 2} (\vec \rho_\mu - \vec a_\mu)
\cdot \vec \tau \cr}
\EQN\fields $$
The former expression for $\cal L$ can be identified with the old
massive Yang-Mills Lagrangian (Lee and Nieh 1967) if one demands
 $$ {1\over g_\rho^2} = {1\over 4 g_\omega^2} = {2\over 3} I_2
\eqn$$
and also that $m_\rho $ and $m_\omega$ are the physical vector meson
masses. Also it has been shown (Ball 1987; Wakamatsu and Weise 1988)
that if one performs a field dependent chiral rotation the hidden
symmetry Lagrangian of Bando et al. (1984, 1988)  is obtained. As we
can see at \queq{\mym} the chiral symmetry of the Lagrangian suggests
that the chiral partners
$\rho$ and $a$ have the same mass. However, due to
the spontaneous breaking of chiral symmetry there appears a term
of the form $ M \vec a_\mu \cdot \partial^\mu \vec \pi $
which would imply an unphysical decay process. This problem may be
solved by introducing a new physical axial field $\vec a_\mu'  =
\vec a_\mu + \xi \partial_\mu \vec \pi $ and fixing the parameter $\xi$
in a way that the mixing for the new field disappears. This feature is
also present in the hidden symmetry approach. In any case the solution
to the $A-\pi$ mixing leads to the following algebraic conditions
(Ebert and Reinhardt 1986; Wakamatsu and Weise 1988)
 $$ f_\pi^2 = {m_\rho^2 \over m_\rho^2 + 6M^2 }
  4 M^2 I_2 \quad ,
 \quad g_\rho^2 ={M^2\over f_\pi^2} {6m_\rho^2 \over m_\rho^2 + 6M^2}
 \quad ,\quad  g_\omega^2 = {1\over 4} g_\rho^2
\EQN\heatvec $$
and the axial meson mass is given by $m_A^2 = m_\rho^2  + 6 M^2 $.
After this the parameters may be fixed in the way described below. For a
given constituent quark
mass $M$ the cutoff is adjusted to reproduce the pion decay constant
$f_\pi= 93$ MeV and the vector meson mass $m_\rho= 770$ MeV. Then the
rest of the parameters as for instance $g_\rho$ and $m_A$ are determined
uniquely. It is interesting to note that in the limit $m_\rho \to \infty
$ the above conditions become identical to those of sect. 1.6. For
$m_\rho= \sqrt{6}M$ and if a  constituent quark mass
of
$M=315 MeV$ is chosen, the KSFR relation $2g_\rho^2 f_\pi^2=m_\rho^2$
(1967) and the Weinberg sum rule $m_A^2=2m_\rho^2$ (Weinberg 1967)
are fulfilled simultaneously. Let us finally mention that the fact
that the massive Yang-Mills Lagrangian and further relations are
obtained even for a finite cutoff indeed requires the continuation of
vector fields into Euclidean space  as it is done in this review.

\subsec{Imaginary Part - Gauged Wess-Zumino Term}

As we have already mentioned  vector mesons generate an imaginary
part for
the Euclidean action. This has been computed in the low momentum limit
(Dhar et al. 1984, 1985; Ebert and Reinhardt 1986), reproducing the
vector gauged Wess-Zumino term (Wess and Zumino 1971; Witten 1983)
after rotation to Minkowski space
$$\eqalign{ & {\cal L}_{im}= {\cal L}_{GWZ} =
{N_c \over 24\pi^2} g_\omega \omega_\mu \epsilon^{\mu\nu\alpha\beta}
tr \Bigl\{ \Bigl(U^\dagger \partial_\nu U U^\dagger \partial_\alpha U
U^\dagger \partial_\beta U\Bigr) \cr &
+ 3 i g_\rho \partial_\nu \Bigl( \partial_\alpha U U^\dagger V_\beta^L
-   U^\dagger \partial_\alpha U V_\beta^R + i g_\rho [ U^\dagger
V_\alpha^L U V_\beta^R - V_\alpha^L V_\beta^R ] \Bigr) \cr }
\EQN\gaugwz
$$
which is $SU(2)_V \otimes U(1)_B $ gauge invariant. It must be said that
this part of the Lagrangian saturates the anomalous Ward identity
\queq{\ward}. Hence higher order terms ought to be chirally gauge
invariant. Another interesting point is that although such an effective
Lagrangian preserves vector gauge invariance it breaks global chiral
symmetry and hence breaks some low energy theorems such as the amplitude
for the decay $\gamma \to 3\pi $ (Wakamatsu 1989, Ruiz Arriola and
Salcedo 1993a).
The main reason can
be found in the $a-\pi$ mixing which produces systematic corrections in
any vertex with external pions. This is a clear drawback of the model,
also present in the topological soliton model. Nonetheless it reproduces
the correct result for the neutral pion decay via intermediate neutral
vector mesons $\pi^0\to\omega^0\rho^0\to\gamma\gamma$ and the
experimental value for the strong decay $\omega\to 3\pi $ (Gomm et
al. 1984; Kaymakcalan et al. 1985).

\subsec{Meson Propagators}

The calculation of on-shell meson propagators proceeds similarly as
sketched in sect. 2.6. for the scalar-pseudoscalar case and has been
treated in detail in several works. Here we will follow Jaminon et
al.~(1992).
As in the heat kernel expansion, there appears an $a-\pi $
mixing term, which can be diagonalized after a proper redefinition of
the axial field. In summary, the following conditions are obtained
$$
 f_\pi^2 = M^2 {4 N_c F(-m_\pi^2) \over
           1 +{\left(2M\over m_\rho \right)}^2
           {F(-m_\pi^2)\over S(-m_\rho^2)} }
 \quad , \quad g_\rho^2 = {1 \over N_c S(-m_\rho^2)}
 \quad , \quad g_\omega^2 = {1 \over 4N_c S(-m_\omega^2)}
\EQN\propvec $$
with the proper time regularized functions $F(q^2)$ and $S(q^2)$
$$
F(q^2) = {1\over 16\pi^2} \int_{-1}^1 {du\over2}
\int_{1\over \Lambda^2}^\infty {d\tau \over \tau}
e^{-[M^2 +{1\over4} (1-u^2)q^2]\tau}
\eqn $$
$$
S(q^2) = {1\over 16\pi^2} \int_{-1}^1 {du\over2} (1-u^2)
\int_{1\over \Lambda^2}^\infty {d\tau \over \tau}
e^{-[M^2 +{1\over4} (1-u^2)q^2]\tau}
\eqn $$
It should be mentioned that the approximation $ S(-m_\rho^ 2) \sim
S(0)= {2\over 3} I_2 $ and $ F(-m_\pi^2) \sim F(0)=I_2 $ corresponds
to the heat kernel
approximation as can be seen by comparison of formulas \queq{\propvec}
with \queq{\heatvec}.  Moreover, in the limit of
$\rho-\omega$ degeneracy the relation $ g_\rho =  2g_\omega$ holds.
Jaminon et al. (1992) have found that for constituent quark masses lower
than $M=385$ MeV the $\rho$ meson is no longer a bound state.
This is a direct consequence of the absence of confinement in the model.
For these masses various remedies have been proposed. Takizawa et
al. (1991) suggest to look for the complex solutions of the
Bethe-Salpeter equation, where the corresponding imaginary part
represents the decay width into free quark-antiquark pairs. Jaminon et
al. (1992) propose rather to take advantage of the approximate linear
behaviour of the inverse propagator in the time-like region and to use
a linear extrapolation into the space like region. For constituent
quark masses around 500 MeV they find the approximate formula
$$ m_A^2 \sim m_\rho^2 + 3.1 M^2  \eqn$$
which differs from the corresponding result in the heat kernel
approximation \queq{\heatvec}. Finally, it has been found
(Schueren et al. 1993) that these results do not depend strongly on
the particular regularization scheme employed.

\subsec{Numerical Results}

Vector mesons do influence the vacuum properties since their masses
and the $A-\pi$ mixing enter explicitly the determination of the
cutoff. In general the cutoff increases as compared to the case
without vector mesons. This
change is more dramatic if the parameters are fixed off-shell rather
than on-shell (Schueren et al. 1993). The  interesting result is
that only in the latter method the results for the vacuum parameters
seem to be reasonable and fairly independent of the regularization.
For illustration we quote the results for $M=350$ MeV and in the
proper time regularization
$$
<\bar q q> =-(271 \MeV)^3  \qquad   m_1 = \half(m_u+m_d)= 8.2 \MeV
$$
The effect of vector mesons on pure pionic properties, i.e. pionic
radii and threshold parameters for $\pi\pi$ scattering, has been
investigated (Ruiz Arriola 1991a; Schueren et al. 1993). A very
interesting point is that vector mesons generate in a natural way an
axial coupling constant for the constituent quarks $g_A^Q $ lower than
one as given by the expression (Vogl et al. 1990; Ruiz Arriola 1991a)
$$ g_A^Q = 1 - {g_\rho^2 f_\pi^2 \over m_\rho^2} \eqn $$  due to
intermediate axial vector meson contributions to the axial current even
in the leading order of the large $N_c$ expansion. A value smaller than
one has been also predicted in somehow different
approaches and models (Peris 1991, 1992; Weinberg 1992; Blotz and
Goeke 1992) as subleading large $N_c$ corrections.

The main result found by Schueren et al. (1993) is that vector mesons
do not change noticeably the calculated pionic properties, at least if
the $\rho$ meson mass is fixed to its experimental value. As in the
$(\sigma,\pi)$ case a strong dependence of the pion scalar and
isovector radii on the constituent quark mass is found. In fact the
precise numbers do not differ very much if
vector mesons
are introduced. In particular, fitting the pionic radii would require a
quark mass of 250 MeV. As we have said, for this value of the mass the
$\rho$ and $\omega$ mesons lie in the continuum, and as it will be
seen below no solitons have been found.

More recently it has been found that a $g_A^Q\ne 1$ in the \NJL
breaks the proper QCD anomalous structure. This can be only obtained for
$g_A^Q=1$ and correspondingly $g_\rho=0$ (Ruiz Arriola and Salcedo
1993). The problem arises whether  vector mesons can be described in a
NJL type model without breaking the QCD anomaly.

\sect{ Solitonic Solutions and Nucleon Observables }

The solitonic sector of the NJL model has been studied with quark
couplings corresponding to $\rho$-mesons
(Alkofer
and Reinhardt 1990), $\rho$ and $A$ (Doering et al. 1992 ; Alkofer et
al. 1992), $\omega$ (Schueren et al. 1992a; Watabe and Toki 1992;
Alkofer et al. 1993) and $\rho$, $A$ and $\omega$ mesons (Doering et
al. 1993 ; Schueren et al. 1993 ; Ruiz Arriola et al. 1993; Zueckert
et al. 1993). The  main difficulty arising in the
calculations including  the $\omega$-meson is the fact that the
Euclidean static energy is
complex, due to the $\omega$ meson. In this section we will present a
way from the complex valued Euclidean action to the description
of selfconsistent solitonic solutions for hedgehog field configurations
(Schueren et al. 1992a; Doering et al. 1993 ; Ruiz Arriola et al.
1993). At the end of this section we comment on other approaches
(Watabe and Toki 1992 ; Alkofer et al. 1993)

\subsec{Statement of the Problem}

To understand the nature of the problem related to the Wick rotation let
$H$ and $H^\dagger $ denote the single particle Dirac
hamiltonian and its hermitean conjugate given by
$$ \eqalign{
 H &= h +i \left[- g_\omega \omega_4 + g_\rho {\vec\tau \over 2}
                 (\vec\rho_4 + \vec A_4 \gamma_5) \right]
\cr
 H^\dagger &= h -i \left[- g_\omega \omega_4 + g_\rho {\vec\tau \over 2}
                 (\vec\rho_4 + \vec A_4 \gamma_5) \right]
\cr} \eqn$$
respectively. The hermitean part $h$ of the hamiltonians $H$ and
$H^\dagger$ has the form
$$
 h = -i \alpha_i  \nabla_i
 +\beta g_\pi (\sigma + i \gamma_5 \vec\tau \cdot \vec\pi)
 +\alpha_i \left(-g_\omega \omega_i + g_\rho {\vec\tau \over 2}
 (\vec\rho_i + \vec A_i \gamma_5) \right)
\eqn $$
Due to the Wick rotation $H$ is in general non-normal, i.e.
$[H,H^\dagger] \ne 0$, and therefore $H$ and $H^\dagger$ may not be
diagonalized simultaneously. For time independent configurations the
real part of the action reads after regularization
$$ {\rm Re} S_f = -{T \over 2} N_c \int_{-\infty}^\infty {d\nu \over 2\pi}
\int_0^\infty {d\tau \over \tau} \phi (\tau, \Lambda)
 \tr e^{(i\nu + H)(-i\nu + H^\dagger)}
\eqn $$
and the imaginary part
$$ {\rm Im} S_f =  {T \over 2} N_c \int_{-\infty}^\infty {d\nu \over
2\pi}
\Bigl\{ \tr \log (i\nu + H) + \tr \log (-i\nu + H^\dagger)\Bigr\}
\eqn $$
Now one would expect to proceed similiar to perturbation theory,
i.e.
one should evaluate
the trace, compute the $\nu$-integral and
rotate back to Minkowski space. Unfortunately, since the one particle
Hamiltonian is non-normal, the $\nu$ integration in the real part cannot
be done analytically. Thus, the eigenvalues in the exponent should be
computed for any value of the variable $\nu$. After that, the analytical
continuation to Minkowski space should be done numerically. Needless to
say that such an approach is not feasible in practice.

\subsec{Total Energy of the B=1 Soliton}

The calculation of the functional trace for baryonic systems is
conceptually involved and has been discussed in full detail by Schueren
et al. (1992b,1993). The interesting feature is that the
analytical behaviour of the spectrum is such that
the rotation back to Minkowski space can be performed by solving the
eigenvalue problems
$$ H^ {\pm} \psi^{\pm}_\alpha (\vec x) = \epsilon^{\pm}_\alpha
\psi^{\pm}_\alpha (\vec x)
\eqn $$
where
$$
 H^{\pm}= h \pm \left[ g_\omega \omega_0 - g_\rho {\vec\tau \over 2}
                 (\vec\rho_0 + \vec A_0 \gamma_5) \right]
\eqn $$
We just quote the final result. The total energy for a system
with baryon number equal to one has been found to be
$$ E = E_{val} +E_R +E_I +E_{mes}
\eqn $$
where the mesonic contribution
$$
E_{mes} =
     {\mu^2 \over 2} \int d^3 x (\sigma^2 + \vec\pi^2 -f_\pi^2)
    +{m_\rho^2 \over 2}\int d^3 x (\vec \rho_\mu^2 + \vec A_\mu^2)
    +{m_\omega^2 \over 2}\int d^4 x \omega_\mu^2
\eqn $$
the valence contribution
$$
E_{val} = \theta (\epsilon_{val}) \epsilon_{val}^+
\eqn $$
the sea real
$$
E_R = -{N_c \over 2} \sum_\alpha \[
\bar \epsilon_\alpha R(\bar \epsilon_\alpha , \Lambda)-
\epsilon_\alpha^0 R(\epsilon_\alpha^0 , \Lambda) \]
\eqn $$
and the sea imaginary contributions
$$
E_I = -{N_c \over 4} \sum_\alpha {\rm sign} (\bar \epsilon_\alpha)
(\epsilon_\alpha^+ - \epsilon_\alpha^-)
\eqn $$
have been defined. We have also introduced the averaged eigenvalues
$$ \bar \epsilon_\alpha = {1\over 2} (\epsilon_\alpha^+ +
\epsilon_\alpha^-) \eqn $$
where $ \epsilon_\alpha^+ $ and $ \epsilon_\alpha^- $ are the
eigenvalues of the Hamiltonians $H^+ $ and $H^-$ respectively with the
condition that for $g_\omega=0$ they coincide
$ \epsilon_\alpha^+ = \epsilon_\alpha^- $. It
should be mentioned that the expression for the total energy has
been checked asymptotically against the heat kernel expansion
in the pure $\sigma$, $\pi$ and $\omega$ case for fixed meson profiles
(Schueren et al. 1992b). From the former equations it can be seen that
the valence part contributes to the
total energy of the system as  long as the averaged eigenvalue
$\bar\epsilon_{val}$ is non-negative. We will see below that this is in
perfect agreement with the way the valence quarks contribute to the
baryon number. Since the valence part of the energy is discontinuous
for $\bar\epsilon_{val} =0$, the question arises whether the sum of sea
and valence contributions behave continuously. It can be shown
analytically and also numerically (Schueren et al. 1992b) that
continuity is fulfilled due to the contribution of the imaginary part.
The remarkable effect is the strong evidence for a vanishing imaginary
contribution if $\bar \epsilon_{val} >0$.

Let us remind that in the long wavelength limit the imaginary part of
the action reduces to the gauged Wess-Zumino term \qeq{\gaugwz}.
Therefore the valence quarks coupled to all degrees of freedom
seem to be complementary to the gauged Wess-Zumino term as long as
${\bar\epsilon}_{val}$  becomes negative.

Another consistency check
is based on the infinite cutoff limit
$\Lambda \to \infty$. In this case it is possible to perform the whole
calculations also in Minkowski space. The result found coincides
exactly with formula in the infinite cutoff limit.

\subsec{Currents of the B=1 Soliton}

{}From the expression for the total action the corresponding mean field
quark densities may be obtained as functional derivatives of the quark
contribution with respect to the relevant fields. The expectation
value of a bilinear in the soliton background is given by
$$ \eqalign{ <\bar q(x) \Gamma q(x)> &=
\theta(\bar \epsilon_{val})
\bar \psi^+_{val} (\vec x) \Gamma \psi^+_{val} (\vec x)   \cr &
  -{N_c \over 2} \sum_\alpha {\rm sign}(\bar \epsilon_\alpha)
[ \bar \psi_\alpha^+ (\vec x) \Gamma \psi_\alpha^+ (\vec x) -
  \bar \psi_\alpha^- (\vec x) \Gamma \psi_\alpha^- (\vec x) ]   \cr &
-{N_c \over 2} \sum_\alpha [ \RR_2(\bar \epsilon_\alpha, \Lambda) +
\bar \epsilon_\alpha \RR_2'(\bar \epsilon_\alpha, \Lambda)]
{\rm sign}(\bar \epsilon_\alpha)   \cr &
[ \bar \psi_\alpha^+ (\vec x) \Gamma \psi_\alpha^+ (\vec x) +
  \bar \psi_\alpha^- (\vec x) \Gamma \psi_\alpha^- (\vec x) ]  \cr }
\eqn $$
with $\Gamma=1, i\gamma_5 \vec\tau , \gamma_\mu $. Here $\RR_2(\epsilon,
\Lambda)$ and $\RR_2'(\epsilon,\Lambda)$ represent the regularization
function and its derivative as used in sect. 3.3. As we see
there are three distinct contributions to any mean field quark density:
1) a valence one provided the averaged eigenvalue $\bar \epsilon_{val} $
is positive, 2) a real sea contribution which depends on the cutoff
and 3) an imaginary part contribution
which is not regularized. This also true in particular for the baryon
number density. Moreover, if vector fields are switched off the baryon
density coincides with the non-regularized baryon density of chapter 3.

\subsec{Self-Consistent Solitonic Solutions}

The equations of motion for general static configurations can be
obtained by varying the total energy of the soliton with respect to
all the meson fields. This yields
$$ \eqalign{& \sigma(x)= {g\over \mu^2} <\bar q(x) q(x) > \cr
& \vec\pi(x)={g\over \mu^2} <\bar q(x)i\gamma_5 \vec\tau q(x)>\cr
& \vec\rho^\mu (x)={g_\rho \over 2 m_\rho^2 }
<\bar q(x) \gamma^\mu \vec\tau q(x) > \cr
& \vec a^\mu (x)={g_\rho\over 2 m_\rho^2}
<\bar q(x) \gamma^\mu \gamma_5 \vec\tau q(x) > \cr
& \omega^\mu (x) = { g_\omega \over m_\omega^2}
<\bar q(x) \gamma^\mu q(x) >  \cr}
\eqn $$
The first two equations of motion are formally identical to the ones for
the
$\sigma\pi$ case discussed in chapter 3. For vector mesons the
equations of motion reproduce the current field identities, i.e. the
$\omega$, $\rho$ and $a$ fields in the nucleon are proportional to the
expectation values of the baryon, vector and axial currents in the
soliton background respectively (see \queq{\cfi}).

For practical calculations we use the hedgehog ansatz
$$ \eqalign{
    \sigma &= f_\pi \Phi (r) cos\Theta(r),
 \quad \pi_a  = \hat x_a f_\pi \Phi(r) \sin\Theta(r), \quad
 \rho_a^i = 2\epsilon_{ika} \hat x^k \rho(r), \quad \rho_a^4 = 0, \cr
 \omega^i &=0, \quad \omega^4= \omega(r),
            \quad a_a^i =2\delta_a^i A_S (r) +
 2\hat x_a \hat x^i (A_S (r)-{1\over 3} A_T (r)), \quad a_a^4=0  \cr}
\eqn $$
Notice that the chiral circle condition $\sigma^2 (x) + \vec \pi^2 (x) =
f_\pi^2 $ is achieved if the polar field $\Phi(r)$ is taken to be a
constant equal to one. It can be checked that this variational ansatz
is a true solution of the equations of motion provided the Dirac sea
contribution only involves closed shells in grand spin quantum
numbers.

\subsec{Numerical Results}

The mean field equations of motion have been solved iteratively for
the non-linear case (${\Phi(r)=~1}$) by  Schueren et al. (1993) and
in the linear case ($\Phi(r)\neq~1$) by Ruiz Arriola et al. (1993) in
a similar way as it was done in chapter 3.
To simplify the discussion we will refer mainly to the full
$\sigma\pi\omega\rho~a$ system where the parameters have been fixed by
computing the on-shell meson propagators and in the proper-time
regularization. We remind that fitting the meson masses and the pion
decay leaves the constituent quark mass $M$ as the only free parameter.
An interesting result is that vector mesons prevent the soliton to
collapse    if the
chiral circle constraint
$
\Phi(r)=1
$ is relaxed
(Ruiz Arriola et al. 1993).
 This is a consequence of the repulsive
nature of the $\omega$ meson.
For the simpler $\sigma\pi$ case the collapse has been discussed in
sect. 3.5..
The results for various mean field observables both in the non-linear
and in the linear case are shown in \qutab{\tabv1,\tabv2 } for different
values of the constituent quark mass.

As it can be seen, the total soliton mass is around 200 MeV higher
than without vector mesons (see chap. 3). Unfortunately (see
\qutab{\tabv2 }
the isoscalar radius and the axial coupling constant are $25\%$  too
high and $60\%$ too low corresponding to
experimental values respectively being nearly  independent of the
particular constituent quark mass $M$. Actually in the physical region
the computed quantities
do not depend strongly on the chiral circle condition, although the
deviations from unity at the origin are significant in the linear model.
Those become larger if the $\omega$ coupling constant is reduced and
lead
eventually to a collapase of the soliton for values much lower than the
ones nedeed to fit the $\omega$ meson mass. For instance, for $M=340$
MeV a $g_\omega= 2.24 $ is required to fit the $\omega$ meson mass,
whereas collapse takes place below $g_\omega=0.8$. The experimental
isoscalar nucleon mean square radius  can only be obtained for
$g_\omega=1.60
$.
For this last value the averaged quark eigenvalue has been found to be
positive.

For illustration we show in Fig. 7.1 the self-consistent vector meson
fields $\omega, \rho, A_S$ and $A_T $ in the non-linear case for the
particular values $M=340 $MeV, $g_\rho=4.61$ and $g_\omega=2.24$.
In principle, they
resemble the corresponding fields found in other soliton models like
e.g. the Skyrme model (Meissner UG 1988) and the quark-meson model
(Broniowski and Banerjee 1986), and their shapes do not depend
strongly on the regularization scheme employed (Schueren et al. 1993).

As we have discussed in chapter 3 one of the virtues of the
NJL model is that no assumption is made a priori whether the valence
quark picture is correct or not. This is in fact decided upon the
results in the calculation of physical observables.
This means if a calculation reproduces the relevant experimental data
and if then a clearly separated bound single quark state exists in the
energy gap between positive and negative continuum, then the valence
picture is valid. In  SU(2) and SU(3) calculations with non-vector
mesonic couplings all calculations by now yield this picture.

When vector mesons
are present
the validity or invalidity of the valence quark picture depends on the
sign of the averaged quark eigenvalue $\bar \epsilon_{val}$. From
\qutab{\tabv2 } we see that this is positive both in the linear
and in the non-linear model. However in this
context all calculations involving vector mesons performed so far do not
allow to reach final
conclusions because the nucleon observables are not yet reproduced well
enough. In particular
the sign of $\bar \epsilon_{val} $ depends very much on the
particular vector mesons included and the way the parameters are
fixed. Schueren et al. 1993 have shown that in the non-linear model
and as long as the nucleon isoscalar radius is used as nucleon
observables to be reproduced    this question
cannot be decided unambiguously. In other words, it is possible to
adjust the radius but with different signs for $\bar \epsilon_{val}$. In
this sense the calculation of other nucleon
observables is desirable to reach definite conclusions.

\subsec{Alternative Approaches }

For the  $\sigma,\pi,\omega,\rho,A_1$-system there are other approaches
to
the soliton sector of the NJL model with vector mesons. Alkofer et al.
(1993) propose to make the whole numerical calculation in Euclidean
space without rotating the single particle hamiltonian back to Minkowski
space where the problem
is originally formulated. Instead they use a particular operational
prescription
to obtain the mass of the soliton. The parameters are fixed
by means of
a heat kernel expansion although the masses of the vector mesons are not
negligible.
These
authors find solitons for the whole  system. Within the same scheme
Zueckert et al. 1993 have also found more recently that vector mesons
stabilize the soliton of the $\sigma,\pi,\omega,\rho,A_1$-system in
the
linear case. Following  their calculations it is possible to fit the
nucleon radius with a solution having a negative valence quark
eigenvalue.
Watabe and Toki (1992)   made calculations in the
$\sigma,\pi,\omega$-system  without continuation of the omega field into
Euclidean space, which avoids the  appearance of an imaginary part of
the Euclidean action.
 Besides an explicit breaking
of vector gauge symmetry  (which dissappears in the infinite cutoff
limit), their prescription corresponds to have a regularized baryon
density. In addition they release the chiral circle condition.
However  neither bound nor stable solitons exist in this scheme.

As we have mentioned at the begining of this section, the introduction
of vector mesons in the soliton sector is by no means a trivial problem
from a conceptual point of view.
As an unfortunate consequence the various
prescriptions proposed by different authors quoted at the beginning of
sect. 7.3.  lead to different results.
This makes the situation  unsatisfactory. In this
sense it would be highly interesting to study this problem more
deeply and to see which or even if any of the prescriptions suggested is
the correct one.

In particular the problem of regularization has to be addressed since
none of the above authors really starts from a regularized action as
basis for all further developments.

\vfill\eject

%
%

%
\chap{The \NJLM and Other Effective Chiral Models}
%
%
%
\TAB\t81{Various observables (energy mean field energy $E_{MF}$,
isoscalar electric mean square radius $\langle R^2 \rangle_E ^{T=0} $,
moment of inertia $\Theta$,
axial vector coupling $g_A$)
for the selfconsistently solved NJL (valence part and sea part) (1.column),
the selfconsistently
solved NJL, where the sea part of the observables has been approximated
by the 2nd order heat kernel or gradient expansion (valence part and mesonic
approximation of the sea part) (2.column) and the
selfconsistently solved CSM (valence part and mesonic part) (3.column).}

Finally we want to compare shortly the NJL soliton consisting of valence
quarks and the polarized Dirac sea with other effective chiral models which
have been discussed in the literature so far.
Both the Skyrme model (sect.8.1.) and the chiral sigma model of
Gell-Mann and
Levi (sect.8.2.) can be formally related to the NJL model via gradient
and
heat-kernel expansions.
Whereas the Skyrme model contains no explicit valence quark degrees of
freedom but relates the baryon number to the topological winding number of
the Goldstone fields, the valence quarks are present in the Gell-Mann--Levi
model from the very beginning. As we have discussed in chapter 3 both pictures
can   be incorporated in the NJL model and it is therefore interesting
to see how
these models are formally and numerically related.
The renormalized chiral sigma model containing polarized Dirac sea as well as
explicitly the kinetic energy of the mesons shows a translational not
invariant ground state and is described in sect.8.3.
In sect.8.4 we consider recent extension of the NJL model containing dilaton
fields.
General aspects concerning the connection between those various effective
models are given in sect.8.5.

\sect{The Skyrme model}

\subsec{The Skyrme Model Without Vector Mesons}

The $SU(2)$ Skyrme Lagrangian in the scalar sector without vector mesons
reads (Skyrme 1961,1962, Adkins et al.1983,
for a review cf. Zahed and Brown 1986, Holzwarth
and Schwesinger 1987)
$$
\eqalign{
\L_{SKY} &= {1\over 4} f_\pi ^2
\Tr_{\tau} \left [
(\ddmu U^{\dag}) (\dumu U ) \right ]  \cr
&+ {1\over{32 e^2}} \Tr_{\tau}
\left \{
\left [
(\ddmu U^{\dag} \ddnu U )
 (\dumu U^{\dag} \dunu U )
\right ]
-
\left [
(\ddmu U^{\dag} ) ( \dumu U )
\right ]^2
\right \} \cr}
\EQN\e81
$$
where $U (\vec r) = e^{ i \theta (r) \vec\tau {\hat r}}$ with
$\theta (\infty) =0$ denotes the hedgehog chiral field on the chiral circle and
$e$ is a model parameter, which can be chosen to reproduce certain
observables.
The topological winding number $n$ of the chiral field, which is defined by
$\theta (0) = -n \pi$, has to be an integer number in order to obtain a
finite energy $E_{SKY} = \int dt ( - \L_{SKY} )$.
It can be shown that the $4$-divergence of the corresponding
current, which is the topological or anomalous current
(Goldstone and Wilczek 1981, Witten 1983)
$$
j_{top} ^\mu (x) =  {{(-1)}\over{24\pi^2}} \epsilon^{\mu{\nu_1}{\nu_2}{\nu_3}}
\Tr_{\tau} \left [
(\partial_{\nu_1} U )
(\partial_{\nu_2} U )
(\partial_{\nu_3} U )
U^{\dag}
\right ]
\EQN\e82
$$
vanishes and $n$ is identified with the baryon number of the system.
As we have already stated (cf. sect.3.2, app.B), $j_{top} ^\mu (x)$
is obtained as the $4-th$ order gradient expansion of the
$1$-loop (sea) part of the baryon current $\langle b^\mu (x) \rangle$, whereas
the second order vanishes.
If we look at he $4$th order gradient or heat-kernel expansion of the
fermion determinant $S_{eff} ^F (U)$ of the
effective action in the NJL (\qeq{\e217})
itself we find (Aitchison and Frazer 1984,1985a,1985b,
Dhar et al.1985, Ebert and Reinhardt 1986):
$$
\L^{(4)} = \L_{SKY} ^{(4)} + \L_D ^{(4)}
\EQN\e83
$$
with
$$
\eqalign{
\L_{SKY} ^{(4)} = {{N_c}\over{32\pi^2}} &\Gamma  \left ( 2, \left (
{M\over \Lambda} \right )^2 \right )
{1\over 6}
\Tr_{\tau}
\left \{
\left [
(\ddmu U^{\dag} \ddnu U  )
(\dumu U^{\dag} \dunu U  )
\right ]
-
\left [
(\ddmu U^{\dag} ) ( \dumu U )
\right ]^2
\right \} \cr
\L_{D} ^{(4)} = {{N_c}\over{32\pi^2}}
\Biggl \{
&\Gamma  \left ( 1, \left (
{M\over \Lambda} \right )^2 \right )
{1\over 3}
\Tr_{\tau} \left [
( \square U^{\dag} )
( \square U )
\right ]
- \cr
&\Gamma  \left ( 2, \left (
{M\over \Lambda} \right )^2 \right )
{1\over 6}
\Tr_{\tau} \left [
( \ddmu U^{\dag} )
( \dumu U  )
\right ]^2
\Biggr \} \cr}
\EQN\e84
$$
where we have used the proper time scheme for regularizing the fermion
determinant and performed a heat-kernel expansion (cf. sect. 2.5.).
$\Gamma (\alpha ,x)$ denotes the incomplete $\Gamma$-function:
$$
\Gamma(\alpha ,x) = \int_x ^{\infty} dt t^{\alpha -1} e^{-\alpha}
\EQN\e85
$$
Indeed, $\L_{SKY} ^{(4)}$ has the form of the Skyrme stabilizing term in
\qeq{\e81} but in addition a 'destabilizing' term $\L _D ^{(4)}$ appears
(Aitchison et al.1985).
Meissner Th et al.(1990) showed that for small constituent quark masses
$M \leq 700 \MeV$ the total energy in $4$th order heat kernel expansion
is indeed unstable, because $E_D ^{(4)}$ corresponding to $\L _D ^{(4)}$
goes to $-\infty$ for meson profiles with small size $R$
(cf. \qeq{\e37}, \qufig{\f32}).
Due to the behavior of the incomplete $\Gamma$-functions
$E_D ^{(4)}$ changes sign for larger $M$ and the theory in $4$th order
heat kernel expansion gets stable.
Because for very high constituent quark masses the valence quark is
part of the negative spectrum one therefore recovers at least the philosophy
of the Skyrme model for $M\to \infty$.
On the other hand we know from sect.3.3  that the physics of the nucleon
demands $M\approx~400\MeV$, from which we conclude, that the Skyrme
model
in the Goldstone sector is not a good approximation to the NJL
approach.

\subsec{The Skyrme Model With Vector Mesons}

In the mesonic sector the heat kernel approximation of the NJL
Lagrangian has been found to resemble the phenomenological chiral
Lagrangian of Gomm et al.(1984),
where the imaginary part of the effective action of the NJL corresponds
to the gauged Wess-Zumino term in the Skyrme like theory, in which Gomm et al.
(1984)
adjusted the parameters to reproduce the
hadronic processes $\rho\to\pi\pi$ and $A_1 \to\rho\pi$. Using an
on shell mass renormalization (cf. sect. 7.2.)
the NJL prediction for these decays
is rather good for constituent quark masses around
$300$ and $350 MeV$.

Reinhardt and Dang (1989) have
claimed that the NJL model with vector
mesons reduces exactly to the original Skyrme Lagrangian \queq{\e81}
if the vector mesons are integrated out, and the nonlinear
constraint of the chiral circle $\sigma^2 + \piv^2 =f_\pi ^2$ is assumed.
This indeed is true
in a second order heat kernel expansion.
However, in the $4th$ order additional terms appear, whose magnitude
are not known. Hence there arises an ambiguity about the proper way to
compare with the original Skyrme Lagrangian.
It is important to notice that the way vector mesons are presently
described in the \NJL turns out to be in full agreement with the old
massive Yang-Mills idea (Lee and Nieh 1968) that vector mesons are the
gauge bosons of chiral symmetry. Such an ideology has been widely used
within the Skyrme model approach (Meisser UG 1988). Not surprisingly,
the gradient expansion produces a massive Yang-Mills type Lagrangian for
the non-anomalous part and a gauged Wess-Zumino term for the anomalous
part, as it is often used in the Skyrme model
(Zahed and Meissner UG 1986, Meissner UG and Zahed 1987).

Finally we want to state that up to now it remains an open question, if the
solution of the NJL including $\omega$, $\vec\rho$ and $\vec A_1$ mesons as
degrees of freedom supports a bosonized, Skyrme like picture and moreover
if the explicit inclusion of vector mesons in the NJL is reasonable
at all (cf. sect. 7.3.).
In fact it is known that the way they are introduced in the \NJL
through direct vector and axial couplings breaks the proper chiral
anomaly (Wakamatsu 1988, Ruiz Arriola and Salcedo 1993a).

\sect{The Chiral Quark Meson Model without Vector Mesons}

The Chiral Sigma model (CSM) is based on a Lagrangian of Gell-Mann and Levi
(1960) and has been quite successful in the last years for a detailed
description of nucleon and $\Delta$ observables and form factors.
The Lagrangian in $SU(2)$ and for $\sigma$ and $\vec \pi$ fields reads in
the chiral limit ($m_\pi =0$) and on the chiral circle $\sigma^2 + \piv^2 =
f_\pi ^2$:
$$
\L = {\bar{q}} i \sld q - g {\bar{q}} (\sigma + i \vec\tau \piv \gamma_5 ) q
+ \half \left ( \dumu\sigma \ddmu\sigma + \dumu\piv \ddmu\piv
\right )
\EQN\e88
$$
In practical calculations (Birse and Banerjee 1984, Birse 1985,
Goeke et al 1985, Cohen and Broniowski 1986, Fiolhais et al.1987,
Fiolhais et al.1988, Alberto et al.1988, Meissner Th et al.1989, Birse
1990,
Neuber and Goeke 1992) the Dirac sea has been ignored and the $q$ was taken
to be a spinor of the $N_c = 3$ valence quarks. The $\sigma$ and $\piv$ are
first of all classical static fields.
In this approximation the total energy of a soliton reads:
$$
E_{CSM} = N_c \epsilon_{val} + \half \int d^3 x
\left [ (\vec\nabla \sigma )^2 + (\vec\nabla \piv)^2 \right ]
\EQN\e89
$$
where $\epsilon_{val}$ is the valence quark energy. The $\epsilon_{val}$ and
$\sigma (\vec x)$ and $\piv (\vec x)$ are obtained easily from the
equations of motion:
$$
\eqalign{
{{\delta E_{CSM} }\over {\delta {\bar{q}} (\vec x)} } &= 0 \cr
{{\delta E_{CSM} }\over {\delta \sigma (\vec x)} } &= 0 \cr
{{\delta E_{CSM} }\over {\delta \piv (\vec x)} } &= 0 \cr}
\EQN\e810
$$
in the hedgehog approximation.
Solving these equations yields selfconsistent valence quark, $\sigma$ and
$\piv$ fields of the Lagrangian \qeq{\e88}.
Observables can be calculated in the usual way by coupling of $\L_{CSM}$ to
external currents and inserting the selfconsistent fields into the
corresponding expressions, where the quark and meson fields have to be
properly quantized, so that the system carries good spin, isospin
or momentum quantum numbers, respectively.
This quantization has been done in the semiclassical cranking way
(Cohen and Broniowski 1986).
Another way consists in assuming coherent Fock states for the sigma and
pion field and to use Peierls-Yoccoz projection techniques
(Birse 1985, Fiolhais et al. 1987,1988, Alberto et al. 1988, Neuber and
Goeke 1992).

Actually the $E_{CSM}$ of \qeq{\e89} can be directly obtained in the
gradient or heat-kernel expansion of the fermion determinant
$S_{eff} ^F (\sigma,\piv)$ of the effective NJL action
\qeq{\e217} (Meissner Th et al.1988,1990).
If one truncates the expansion after the first nonvanishing order one obtains
directly the kinetic energy
$ \half \int d^3 x
\left [ (\vec\nabla \sigma )^2 + (\vec\nabla \piv)^2 \right ] $ of the
meson fields (cf. sect. 2.5.),
which, by adding $N_c \epsilon_{val}$, yields $E_{CSM}$.
This means that the kinetic terms of the mesons in the CSM have been
generated from the polarized Dirac sea of the NJL by gradient expansion.
A comparison between the two models will be performed now very easily:
We once solve the NJL model in the $1$ quark loop approximation and evaluate
then observables.
This is called NJL (exact) in \qutab{\t81}. Then we perform a gradient
expansion of the energy and several observables up to the first nonvanishing
order. We insert the selfconsistent NJL solutions into those gradient expanded
expressions yielding NJL (grad.).
These numbers are compared with the selfconsistent solutions and the
corresponding observables from the CSM.
Using in all three cases the semiclassical quantization the results of some
relevant observables are compared in \qutab{\t81}, where the contributions
from the Dirac sea and of the meson cloud are treated on a equal footing.
Apparently  the total values of the observables agree within $15 \%$,
while
the separated contributions from valence quarks and mesons (sea quarks)
differ.
The $g_A$ is a clear exception, there the gradient or heat kernel expansion
seems to converge badly. The $g_A$ from the Dirac sea of NJL basically
vanishes, while the mesonic contribution in the CSM is comparable with the
valence quark contribution. Thus a general statement on a comparison between
these models is impossible and can only be done separately for each
observable.

Finally we want to note that also the full linear version of the CSM, which
contains in addition to \queq{\e88} the {\it mexican hat potential} reading
$- {{m_\sigma ^2}\over {8 f_\pi^2}} (\sigma^2 + \piv^2 - f_\pi ^2)^2 $
(if $m_\pi =0$) can be obtained from the NJL by gradient or heat kernel
expansion, where the $\sigma$-mass comes out to be
$m_\sigma ^2 = 4 (g f_\pi)^2$.
In contrast to the NJL, where without constraining the meson fields to the
chiral circle the soliton collapses (cf. sect. 3.5), the mexican hat
potential
arising in second order gradient expansion is highly stabilizing and
therefore the linear CSM shows perfectly stable solitonic solutions.
Moreover it turns out that the nucleon observables calculated from those
solutions in the CSM do not differ very much from the ones which  are
obtained in the nonlinear version.

This fact and the behavior of $g_A$ as described above show that the
{\it gradient expansion} cannot be  regarded as {\it unreliable
in the solitonic sector} as a generally reliable approximation. It has
to be used with care.

\sect{The Chiral Quark Meson Model without Vector Mesons}

The chiral quark meson model with vector mesons proposed by Broniowski
and Banerjee (1985,1986) is based on the massive Yang-Mills Lagrangian
of Lee and Nieh (1968) supplemented with valence quarks coupled to all
mesons ($\sigma\pi\omega\rho~a$). Many practical calculations
(Broniowski and Banerjee 1985,1986; Broniowski and Cohen 1986;
Ruiz Arriola et al. 1990; Alberto et al. 1990b; Ruiz Arriola et al.
1993b)  neglect  the effects of the Dirac sea. Similiarly to the case
without vector mesons, they may be considered to be included in an
approximate way in the kinetic energy of the mesons. It should be also
mentioned that this type of models do not include a gauged Wess-Zumino
term as it is done  in Skyrme type models. From the point of view
of the
\NJL  this is not a problem since the gauged Wess-Zumino terms
correspond to the valence quarks (Schueren et al. 1993)
(see also the discussions in sect. 7.3.)
The dynamical question, whether a chiral quark model with vector mesons
is supported by the NJL model, remains an open problem.

\sect{The Renormalized Chiral Sigma Model - Vacuum Instability}

Some authors (Soni 1987, Ripka and Kahana 1987, Li et al.1989) considered
a renormalized CSM, i.e. they solved the Lagrangian \queq{\e88} in the
$1$-quark loop   approximation and subtracted the divergent parts as
{\it local
counterterms}.
Such an approach does not contain an UV cutoff by construction.
In our language the total energy of the renormalized CSM on the chiral
circle reads:
$$
\eqalign{
E_{CSM}^{ren} &= N_c \eval +
\half \int d^3 r
\left [ (\vec\nabla \sigma)^2 + (\vec\nabla \vec\pi )^2 \right ]
\cr
&+ \lim_{\Lambda\to\infty}  \Bigl \{
E_{sea} (\Lambda) -
(4N_c g^2 I_2 ^\Lambda ) \half \int d^3 r
\left [ (\vec\nabla \sigma)^2 + (\vec\nabla \vec\pi )^2 \right ]
\Bigr \}  \cr}
\EQN\e811
$$
where $I_2 ^{\Lambda} (M)$ is the logarithmically divergent integral
from \qeq{\e227}.
Due to the subtraction of the counterterm
$(4N_c g^2 I_2 ^\Lambda ) \half \int d^3 r
\left [ (\vec\nabla \sigma)^2 + (\vec\nabla \vec\pi )^2 \right ]$ from the
divergent sea energy $E_{sea}$ the whole expression remains finite in the
limit $\Lambda \to \infty$.
In the renormalized CSM the kinetic energy of the mesons
$\half \int d^3 r \left [ (\vec\nabla \sigma)^2
+ (\vec\nabla \vec\pi )^2 \right ]$ is added explicitly rather than
generated by the polarized Dirac sea as in the NJL where the cutoff $\Lambda$
is properly adjusted.
Actually one can easily see, that the total energy
of the NJL and the renormalized CSM formally only differ by the
UV convergent terms. In the NJL those are evaluated with a finite
UV cutoff $\Lambda$ reproducing the pion decay constant, whereas in the
renormalized CSM they are treated with infinite $\Lambda$.
Though this small difference has a tremendous effect.
The authors quoted above showed that the sea energy $E_{sea}$ gets
negative for meson profiles with finite size $R$. This means that the full
vacuum is not translationally invariant.
Moreover it turns out that $E_{sea}$ gets more and more negative,
if one increases the topological winding number $n$ of the pion field and
therefore neither a stable vacuum nor a soliton solution with finite
baryon number $B$ exists.

Recently it has been suggested that this problem does not occur if
dynamical vector mesons  are included in the CSM
(Kahana and Ripka
1992) in a massive Yang-Mills scheme and considering the corresponding
quantum corrections.

\sect{Scale Invariance and Dilaton Fields}

Ripka and Jaminon (1992) extended the NJL to a model which in addition to
the spontaneously broken chiral symmetry also exhibits the anomalous
breaking of scale invariance in QCD.
Following the work of Jain et al. (1987) they introduced to this end a
scalar-isoscalar {\it dilaton} field $\chi$, which is coupled to the
effective action by:
$$
{I^{\prime}} = \half \Sp \propchi {{ds}\over s}
\Bigl ( e^{-s D^{\dag} D } -
      e^{-s {D_V}^{\dag} {D_V} } \Bigr )  +
{{a^2} \over 2}  \int d^4 x \Bigl [ (\sigma^2 + \piv^2)^2
- {\sigma_V} ^4  \Bigr ]  + \int d^4 x \L (\chi )
\EQN\e812
$$
where $a^2$ is a dimensionless parameter.
The first two terms in \qeq{\e812} are scale invariant, the scale breaking
being related to the $\chi$-field Lagrangian $\L (\chi)$.
One chooses $\L (\chi)$ in such a way that the divergence of the dilation
current $s_\mu$ is $\partial_\mu s^\mu = \chi^4$ and therefore the vacuum
value $\chi_V ^4$ can be related to the gluon condensate
$\langle G_{\mu\nu} ^2 \rangle $.
It has been shown (Jain et al.1987, Ripka and Jaminon 1992) that, when
a vacuum value
$\chi_V \approx 350 \MeV$ is used, as deduced from the QCD sum rule
estimate (Shifman et al.1979),
the $\chi$ field stays almost constant throughout
the chiral phase transition, in which the chiral symmetry is restored at
high baryon densities or high temperatures.

Motivated by this fact Meissner Th et al.(1993) and Weiss et al.(1993)
used a constant dilaton field $\chi =\chi_V$ by solving the theory
\queq{\e812} in the solitonic sector.
The scale invariant form of the mesonic interaction in \qeq{\e812}, which
differs from the mesonic mass term in the original NJL \queq{\e217}
$$  {{\mu^2}\over 2} \int d^4 x (\sigma^2 + \piv^2 - \sigma_V ^2) $$
has the effect, that solitonic solutions of \qeq{\e812} exist even in
the case, when both $\sigma$ and $\piv$ are not restricted to the chiral circle
$\sigma^2 + \piv^2 = \sigma_V ^2$ (non-linear model), but both
$\sigma$ and $\piv$ degrees of freedom are fully allowed (linear model).
This was not the case in the original version of the NJL, where the constraint
of the chiral circle turned out to be necessary, because otherwise the soliton
collapses to a configuration with zero size and zero energy but baryon
number $B=1$ (cf. sect. 3.5).
One can easily convince oneself, that, with the notation of sect.3.5, the
meson energy in \qeq{\e812} gets now proportional to $U^4 R^3 \propto
R^{3-4\alpha}$.
The soliton energy can then only fall to zero as $R\to
0$ when $\alpha < {3\over 4}$. However, as it was shown in sect.3.5,
the soliton can
only maintain a baryon number $B=1$ while its total energy tends to zero
if $\alpha > 1$. For this reason the $B=1$ soliton does not collapse when
it is calculated with the scale invariant action \queq{\e812}.

Furthermore Meissner Th et al.(1993) and Weiss et al.(1993)
have shown that for the physical relevant region of the constituent mass
$M\approx 400 \MeV$ the deviation of $\sigma$ and $\pi$ obtained as solitonic
solutions of \queq{\e812} from the chiral circle is numerically very small.
Moreover all calculated observables are nearly identical to the ones which
are obtained from the original version of the NJL, where $\sigma$ and $\pi$
are restricted to the chiral circle from the very beginning.

The nonlinear version of the NJL, as it has always been used in the past
and on which the present review is based on, can therefore be justified by
a model which implements the anomalous breaking of scale invariance in QCD.

\sect{General Aspects}

If one looks from a bird's view on section 8.1 ,8.2 ,6.4. and
7.3., the following
qualitative evaluation can be done:
The simple Skyrme model with a pion field only cannot be compared with NJL.
In the heat kernel expansion a destabilizing term occurs
and it also turns out that the result for semiclassical quantized
observables are different.
The inclusion of vector mesons in the
Skyrme Lagrangian, however, allows a better comparison with the NJL model in
the scalar and pseudoscalar sector.
The observables are similar and
in the collective hamiltonian of $SU(3)$ structurally the same terms appear
and the numerical results are close as well.
Thus it seems that the role of the valence quarks in the NJL is
qualitatively played by the vector mesons in the Skyrme model.
Presently one cannot judge the role of vector mesons in NJL.
Altogether it seems, and this includes the chiral sigma model as well,
that the degrees of freedom relevant for the low energy regime of QCD can
effectively be parameterized in rather different ways still reproducing
the low energy hadronic phenomena with similar accuracy. One, however, is
quite clear: \lb {\it The Goldstone boson fields or Goldstone bosonic
quark-quark interactions are playing the dominant role.}
\vfill\eject

%
%

%
\chap{Summary }

The present review article deals with effective chiral symmetric models
of quarks and non-dynamical mesons
for
the description of mesonic and baryonic ground states.
In general effective theories are
trying to
model  low energy QCD phenomena by identifying the appropriate degrees
of
freedom and the relevant symmetries neccessary for a qualitative and if
possible, quantitative description of low energy hadronic phenomena.
Experience over the last 20 years has cummulated  in the common opinion
that for baryonic ground states
spontaneously broken chiral symmetry and the incorporation of the
corresponding Goldstone-boson fields are the important
ingredients for a proper effective approach.
\lb \indent
The review concentrates on the use of  the \NJLM  and
related approaches in the sector with baryon number $B=1$.
In its basic form
it is defined by being the simplest pure quark theory with local
biquadratic
interactions  of Goldstone  character exhibiting spontaneous breakdown
of chiral symmetry.
One should note that the  \NJLM   plays a central role because it
allows to build conceptual and numerical bridges to other
basic approaches, such as fully bosonized approaches of Skyrme type and
theories, where valence quarks are coupled to dynamical meson fields.
\lb \indent
The \NJLM  is treated explicitly
in the path integral
formalism   performing a suitable bosonization
procedure. Then the
theory is described in terms of quarks and composite, non-dynamical  meson
fields.
Actually  the model is  solved in the zero-boson and
one-quark-loop
approximation, such that it can be uniquely defined by a  so called
fermion determinant.
In this form and neglecting the scalar fluctuations around
the vacuum
it reduces to the solutions of the chiral quark loop and  agrees
with the structure of the Chiral Quark Model.
Performing an expansion of the determinant under the assumption of
slowly varying pionic fields gives the structure of the Skyrme model,
including the
whole anomalous sector being  described by the Wess-Zumino term.
\lb\indent
Actually  in the present model and for time-independent
chiral fields on the chiral circle
the fermion determinant is evaluated exactly and provides  a full
treatment of the polarization of the Dirac sea. The latter is caused by
the presence of a discrete and localized valence level within  the
single-particle spectrum.
The procedure used in the literature consists of three steps: First, a
regularization scheme
is chosen and the parameters of the model are fixed in the mesonic
sector to reproduce PCAC, meson masses and decay constants.
Second
it is checked that the vacuum condensates and the current quark masses
are
reasonable. Third, without changing the parameters solitonic solutions
are selfconsistently obtained
in the sector with baryon number $B=1$.
\lb \indent
In order to describe baryons with the quantum numbers of the spin $1/2$
and spin $3/2$ multiplets, the hedgehog based quark field ansatz has to
be quantized in the collective subspace.
Therefore a
time dependent rotation in the direction of the symmetry of the model is
performed and canonical commutation relations are imposed on the
collective coordinates and the canonical momenta. This results in
proper  commutation relations for the generators of the group, which
in the case of SU(2)
are the components of the total angular momentum and in SU(3) are
complemented by some  generators acting on the spin-baryon
number space.
The hedgehog
ansatz for the SU(2) fields, which  manifests  the relation between
spatial and isospin rotations, results after the quantization in the
constraint that the absolute value of spin always equals isospin.
However the value of the
spin itself can be integer or half integer and therefore the soliton can
be quantized as fermion as well as boson. Only the generalization to
SU(3)
gives a  constraint to  fermionic solitons as long as the number of
colors is odd.
\lb\indent
The numerical results for the solitonic sector of the \NJLM
are basically independent on the regularization scheme used as long as the
cut-off is treated as a parameter of the system,
which is fixed
in the meson sector. The calculations of the  literature, all being based on
hedgehog structure and physical values for the meson sector, can be
summarized as follows (status autumn 1993):

\item{i)} SU(2), $\sigma$ and $\pi$ field, chiral circle: For a
constituent quark
mass $M=420\MeV$ the following observables within $15\%$ are
reproduced.
Isoscalar and isovector charge squared radius of the nucleon, $<r^2>^c_{T=0}$
and $<r^2>^c_{T=1}$, nucleon sigma term, $\Sigma$,  nucleon axial
coupling constant and magnetic moments of proton and neutron (if the
recent $1/N_c$ corrections are included),  $g_A$, $\mu_p$ and $\mu_n$,
the
nucleon-delta splitting, $M_\Delta-M_N$,  the q-dependence of the form
factor
$G_E^p(q^2)$,
$G_M^p(q^2)$,
$G_M^n(q^2)$, $g_A(q^2)$. The nucleon energy comes out at about $800\MeV$ if
rotational and  translational zero modes corrections are included.
The neutron squared radius, being a very sensitive quantity, is twice as large
as the experimental value, the $G_E^n(q^2)$ is generally by a factor
of two too large at finite
$q^2$.

\item{ii)} SU(3), $\sigma$, $\pi$, $K$ and $\eta$ fields. chiral circle
for
sigma and pion, trivial
embedding of SU(2) into SU(3), perturbative treatment of $m_s$ up to second
order
or Yabu-Ando approach: Splitting between and within spin 1/2 and spin
3/2 baryons is reproduced
within few MeV. The splitting within isospin multiplets is reproduced and a
common
value for $m_u-m_d$ is found. If the zero point corrections to translation and
the SU(3)-rotations are included the energies of all octet and decuplet baryons
are too large by a constant shift of approximately 100 MeV. If recent $1/N_c$
corrections
are  included the $g_A^0$, $g_A^3$ and $g_A^8$ of the nucleon are
reproduced slightly outside the experimental errors.
The nucleon $\Sigma$-term is reproduced  with a strangeness content
of $15\%$ and the recently measured Gottfried sum comes out as well.

\item{iii)} SU(2), $\sigma$, $\pi$, $\rho$, $A_1$ and $\omega$ mesons,
chiral circle for
sigma and pion. None of the present approaches has been developed to
an extent that
one  can judge the influences or the necessity of vector mesonic
coulings. The observables
calculated by now are the isoscalar charge radius of the nucleon and $g_A$,
both  deviating noticebly from experiment. There seems to be  also
conceptual problems
with  treating the omega meson. On the other hand the repulsive
character of the
omega  guarantees a stable soliton and one does not require any more the
restriction to the chiral circle.

 The problem encountered  by now can be summarized as following: The
semiclassical
quantization procedure is  not perfectly  understood yet, although
it is common usage.
Its relationship to well-established quantum-mechanical theories like
Peierls-Yoccoz projection approaches is yet unknown.  If one looks at
$g_A$ and
magnetic moments of the nucleon its convergence in the rotational
frequency
$\Omega$ seems to be slow. Furthermore it has been shown that
the PCAC
relation is reproduced within the cranking scheme
if rotational corrections in linear  order of $\Omega$ are
included.
\lb\indent
The quantum corrections due to zero modes of translation and rotation
add up to $30-40\%$ of the resulting baryon mass and hence are by far
too large for a correction. The corresponding corrections
of non-zero modes are of order ${\cal O}(N_c)$ as well  and they  are
not known yet.
There  are indications in the Skyrme model that they are small, but a
clear statement in
the \NJLM is missing.
\lb\indent
The treatment of the $\omega$-meson is not quite settled yet. A
consistent scheme, which
start from the very beginning with an action, the real part   of which
is  regularized
in Euclidean space, is still missing. Furthermore one needs some more
observables to be
evaluated. Hence the question if vector mesons in such a quark
theory are indeed necessary to
be included as quark-quark couplings is not yet settled.
\lb\indent
Altogether one can summarize: Quark models like the \NJLM with a
polarized
Dirac sea and involving Goldstone type couplings (i.e. non-dynamic mesons),
have been by now quite successful in describing the ground states of baryons
in the octet and decuplet.
\lb\noindent
If one ignores vector mesons, whose relevance in the present model is
still under debate, the calculations support a clear picture of the
baryons consisting of three localized valence quarks interacting with a
moderately polarized Dirac sea. This is in contrast to Skyrme-like
models, which are governed by meson fields, exhibiting certain
topological properties, and quark-meson models, where valence quarks are
coupled to dynamical meson fields.
\lb\indent
The problem encountered concern the semiclassical quantization method.
It has certain merits, but it deserves improvement. Last but not least,
the NJL-model lacks confinement which should be necessary for orbitally
excited states of baryons.

{\bf Acknowledgements:} The work was partially supported
by the {\it Alexander von Humboldt Stiftung (Feodor Lynen Program)}
and the {\it US Department of Energy} (under grant DE-FG06-90ER40561) (Th.M.),
by the spanish
{\it DGICYT} under contract PB92-0927 and the {\it Junta de
Andaluc\'{\i}cia} (E.R.A.),
by the {\it Graduiertenstipendium des Landes NRW} (A.B.), the {\it
Bundesministerium f\"ur Forschung und Technologie}(BMFT), the {\it
Deutsche Forschungsgemeinschaft}(DFG) and the {\it COSY} project of the KFA
J\"ulich.

Th.M. and E.R.A. are grateful to the {\it Institut f\"ur theoretische
Physik II} of the {\it Ruhr-Universit\"at Bochum} for hospitality.

Finally we would like to thank
C.V. Christov,
D. Diakonov,
V. Petrov,
P.V. Pobylitsa,
M. Polyakov,
M. Praszalowicz,
M. Wakamatsu,
T. Watabe and
R. W\"unsch
for various discussions along the way and also
J. Berger, W. Ricken and C. Schneider
for intensive organisatoric help.

\vfill\eject

%
%
%
%
\appendix{Minkowski- and Euclidean Space-Time: Notation and Convention}
\FIG\fa1{Contour integration for evaluating static observables of a system
with baryon number $B=0$ (closed lines) and $B=1$ (dashed lines) in
Minkowski and Euclidean space.}
\TAB\ta1{Notation for $4$-vectors $A_\mu$, the metric tensor $g_{\mu\nu}$,
the scalar product of two $4$-vectors and the Dirac $\gamma$-matrices in
Minkowski and Euclidean space.
Hereby Greek indices run from $0-3$ in Minkowski space and from $1-4$ in
Euclidean space, whereas Latin indices generally run from $1-3$.}

\qutab{\ta1} compares the notation for $4$-vectors $A_\mu$, the metric tensor
$g_{\mu\nu}$, the scalar product between two $4$-vectors as well as the
Dirac $\gamma$-matrices in Minkowski and Euclidean space.

Generally a Minkowski $4$-vector is transformed to an Euclidean $4$-vector
by performing a {\it Wick rotation}:
$$
\eqalign{
{A_E}^4 &= (+i) {A_M}^0     \cr
{A_E}_4 &= -{A_E}^4 =(-i) {A_M}^0 =(-i) {A_M}_0    \cr
    {A_E}^i &= {A_M}^i       \cr
    {A_E}_i &= {A_M}_i       \cr    }
\EQN\ea1
$$
Hereby both
${A_E}^4$ and ${A_M}^0$ are assumed to be real numbers, which means that any
function $f({A_M}^0 )$ depending on the real variable
${A_M}^0$ gets analytically continued to the complex plane.
For the $4$-space vector and the $4$-momentum vector
we have especially:
$$
\eqalign{
&{x_M}^{\mu} = (t,\vec r) \qquad {x_E}^{\mu} = (\tau,\vec r)  \cr
&\tau = (+i)\> t \qquad  \int d^4 x_E = (+i) \int d^4 x_M \qquad
{\partial \over {\partial\tau}} = (-i) {\partial \over {\partial t}}
\cr  }
\EQN\ea2
$$
and
$$
\eqalign{
&{p_M}^{\mu} = (\nu,\vec p) \qquad {p_E}^{\mu} = (u,\vec p)  \cr
&u= (+i)\> \nu \qquad  \int d^4 p_E = (+i) \int d^4 p_M \qquad
{\partial \over {\partial u}} = (-i) {\partial \over {\partial\nu}}
\cr  }
\EQN\ea3
$$
with the notations:
\halign { # \hfill \quad & # \hfill \cr
     \noalign{\smallskip}
$t$: Minkowski time & $\tau$: Euclidean time \cr
     \noalign{\smallskip}
$\nu$: Minkowski frequency & $u$: Euclidean frequency \cr
     \noalign{\smallskip} }

Further examples handled in this way
are the cranking frequency $\Omega$ (cf. chap. 3), which
couples like the time component of an isovector vector meson ($\vec\rho$
meson), or
the $\omega$ meson, which is time component of an isoscalar vector meson
(cf. chap. 7).

The transformation of a scalar product reads:
$$
\eqalign{
&(A_M \cdot B_M) =    \cr
&{A_M} ^\mu {B_M}_\mu =
{A_M}^0 {A_M}_0 + {A_M}^i {A_M}_i =
{A_M}^0 {A_M}^0 - {A_M}^i {A_M}^i = \cr
&-{A_E}^4 {A_E}^4 - {A_E}^i {A_E}^i
= {A_E}^4 {A_E}_4 + {A_E}^i {A_E}_i
= {A_E} ^\mu {B_E}_\mu = \cr
&(A_E \cdot B_E) \cr   }
$$
i.e.:
$$
(A_M \cdot B_M) = (A_E \cdot B_E)
\EQN\ea4
$$
Furthermore we notate:
$$
\eqalign{
(A_M)^2 &= (A_M \cdot A_M )   \cr
(A_E)^2 &= {A_E}^\mu {A_E}^\mu = - (A_E \cdot A_E )   \cr }
\eqn
$$
giving:
$$
(A_M)^2 = -(A_E)^2
\EQN\ea5
$$

\qufig{\fa1} shows the contours in the complex plane arising for the
calculation of static observables of the soliton in Minkowski space:
$$
\int {{du}\over{2\pi}} \sum_\lambda {1\over{ u - \epsilon_\lambda}} \dots
\EQN\ea6
$$
and in Euclidean space:
$$
\int {{du}\over{2\pi}} \sum_\lambda {1\over{ iu - \epsilon_\lambda}} \dots
\EQN\ea7
$$
if the one neglects regularization (cf.\qeq{\e312}).

In order to obtain
for a certain expression
the same results in both cases
one has to make the central assumption, that during rotating the contour-line
from Minkowski into Euclidean space no singularity in the complex plane is hit.
This is clearly true for the non-interacting case, but has to be postulated
in general ({\it Euclidean Postulate}).

A system with baryon number $B=1$ can be constructed either by extending
the corresponding contours as shown in \qufig{\fa1} (dashed lines) or by
introducing a thermochemical potential $\mu$, which causes a shift:
$$
\epsl \to \epsl -\mu
\EQN\ea8
$$
in the single particle orbitals and therefore increases the number of
singularities in the closed contour line by $1$,
namely by the valence orbit (cf. sect.3.2).
It is clear that this prescription only works, if
$\mu$ is both in the Minkowski space and the Euclidean space
is a real number and therefore no analytic continuation (Wick rotation),
as it was posted in \qeq{\ea1}, may be performed.
The treatment of the thermochemical potential $\mu$
in Euclidean space is therefore different
from that of the time component of a physical $4$-vector, as it is the
$\omega$ meson.

Also the Dirac $\gamma$-matrices can be formally transformed
into Euclidean space:
$$
{\gamma_E}^4 = (+i) {\gamma_M}^0  \qquad\qquad
{\gamma_E}^i= {\gamma_M}^i
\EQN\ea9
$$
Though
the $\gamma$-matrices are no physical objects, which transform under
Lorentztransformations as a Lorentz $4$-vector.
The \qeq{\ea9} is therefore only a simple redefinition, which means, that
${\gamma_E} ^4$ is antihermitian as it is
$\gamma^i$.
The reason for doing this is just because the rule for transforming
a scalar product \queq{\ea4} can now also be applied to the expression
$\slashchar A$:
$$
{\slashchar A}_M = {\slashchar A}_E
\EQN\ea10
$$
For the matrices
$\beta$ and $\gamma_5$ we will not define
Euclidean values (as it is sometimes done especially for $\gamma_5$),
but always use the same matrices, namely:
$$\beta = {\gamma_M} ^0 = (-i) {\gamma_E} ^4  \EQN\ea11 $$
and
$$\gamma_5 = \gamma^5 =
  (+i) {\gamma_M}^0  {\gamma_M}^1 {\gamma_M}^2  {\gamma_M}^3   =
       {\gamma_E}^1  {\gamma_E}^2 {\gamma_E}^3  {\gamma_E}^4
\EQN\ea12
$$

Generally we will use the Bjorken-Drell notation (1965):
$$
{\gamma_M}^0 = \pmatrix {  {\cal I}_2     &  0        \cr
                            0       &  -{\cal I}_2    \cr}
\qquad
{\gamma_M}^i = \pmatrix {   0       &  \sigma^i \cr
                          -\sigma^i &   0       \cr}
\EQN\ea13
$$
with the Pauli matrices:
$$
{\sigma}^1 = \pmatrix {     0       &  1        \cr
                            1       &  0        \cr}
\qquad
{\sigma}^2 = \pmatrix {     0       &  -i       \cr
                            i       &  0        \cr}
\qquad
{\sigma}^3 = \pmatrix {     1       &  0        \cr
                            0       &  -1       \cr}
\EQN\ea13
$$
and especially:
$$
{\beta}      = \pmatrix {  {\cal I}_2     &  0        \cr
                            0       &  -{\cal I}_2    \cr}
\qquad
{\gamma}_5 =\gamma^5 = \pmatrix {   0       &  {\cal I}_2     \cr
                                    {\cal I}_2    &   0       \cr}
\EQN\ea14
$$
as well as:
$$
\gamma^i = \beta \alpha ^i
\EQN\ea15
$$

\appendix{Gradient Expansion of the Baryon Current}

Because of the importance for our model we show explicitly the gradient
expansion of the baryon current $b_\mu (x)$ and especially the baryon number
$B = \int d^3 x b_0 (x)$, as it has been performed by Goldstone and Wilczek
(1981) and discussed in detail by Witten (1983) in context of the topological
quantization of $B$.

{}From the general treatment given in sect.3.2 we have for the expectation
value
of $b_\mu$:
$$
 \langle b^{\mu} (x) \rangle = {1\over{N_c}} {\bra x} \Tr_{\gamma \lambda c}
\left [ (i D)^{-1} \gamma^\mu \right ] {\ket x}
\, - \, {\rm vac.contr.}
\EQN\eb1
$$
Hereby the $U$ denotes a $SU(n)$ chiral field which is parameterized in terms
of the Goldstone field~$\xi$(x)
$$
U(x) = e^{i \xi (x)}
\EQN\eb2
$$
$\xi (x)$ is hermitian and traceless, i.e.
$$
\xi (x) = \sum_{a = 1} ^ {n^2 -1} \xi^a (x) {{\lambda^a}\over 2}
\EQN\eb3
$$
where the $\lambda^a$ being the generators of $SU (n)$.
$iD$ is the quark Dirac operator with the chiral field $U$:
$$
\eqalign{
iD &= i \sld - M U_5 (x) \cr
(i D)^{-1} &=  \left [ i \sld - MU_5 \right ] ^{-1} =
(-) \left [ (\ddnu\dunu + M^2) + iM \sld {{U_5}} \right ]^{-1}
(i \sld + MU_5^{\dag} )\cr}
\EQN\eb4
$$
with:
$$
U_5 = \half (U + U^{\dag} ) + \half \gamma_5 (U - U^{\dag})
\EQN\eb6
$$
We therefore find:
$$
\eqalign{
&\langle b^{\mu} (x) \rangle - \langle b^{\mu} (x) \rangle _V =
\cr
(-) {1\over{N_c}} \Bigl \{
&{\bra x} \Tr_{\gamma\lambda c} \left [ \gamma^{\mu}
(\ddnu\dunu + M^2 + iM \sld {U_5} )^{-1} (i \sld + MU_5^{\dag} )
\right ] {\ket x}   - \cr
&{\bra x} \Tr_{\gamma\lambda c} \left [ \gamma^{\mu}
(\ddnu\dunu + M^2 )^{-1} (i \sld +M )  \right ] {\ket x}
\Bigr \}   \cr}
\EQN\eb7
$$
In this expression we can now perform the power expansion:
$$
(A+B)^{-1} = A^{-1} - A^{-1} B A^{-1} + \dots = A^{-1} \sum_{m=0}
^\infty  (-)^m (BA^{-1} )^m
\EQN\eb8
$$
with:
$$\eqalign{
A&:= (\ddnu\dunu + M^2) =: G^{-1} \cr
B&:= i \sld {U_5}   \cr}
\EQN\eb9
$$
Because of $\Tr_\lambda \xi =0$ we find that all terms in \qeq{\eb9} up
to order $m=2$ vanish. For $m=3$ we get:
$$
\langle b^{\mu} (x) \rangle - \langle b^{\mu} (x) \rangle _V =
i^3 M^4 \Tr_{\gamma\lambda} {\bra x}  \left [ \gamma^\mu
G (\sld U_5) G (\sld U_5) G (\sld U_5) G {{U_5}^{\dag}} \right ] {\ket x}
+ \dots
\EQN\eb10
$$
The $G$ is diagonal in the momentum space basis $\ket {p_E}$.
In the lowest order gradient expansion
the $\sld U_5$ does not vary with the coordinate $x$ (cf. sect.2.5).
After inserting complete sets of
$\ket {x_E}$ and $\ket {p_E}$, respectively, one obtains:
$$
\eqalign{
&\langle b^{\mu} (x) \rangle - \langle b^{\mu} (x) \rangle _V =
\cr
&i^3 M^4 \id {p_E} \left ({1 \over {p_E ^2 + M^2}} \right) ^4
\Tr_{\gamma\lambda} \left [ \gamma^\mu
(\sld U_5 (x)) (\sld U_5 (x)) (\sld U_5 (x))
{{U_5}^{\dag}} (x) \right ] + \dots   \cr}
\EQN\eb11
$$
and finally after performing the $\gamma$ trace:
$$
\langle b^{\mu} (x) \rangle - \langle b^{\mu} (x) \rangle _V =
 {{(-1)} \over {24 \pi^2}} \epsilon^{\mu \nu_1 \nu_2 \nu_3}
\Tr_{\lambda} \left [ (\partial_{\nu_1} U) (\partial_{\nu_2} U)
(\partial_{\nu_3} U) \kr U \right ]
\EQN\eb12
$$
which is the {\it anomalous baryon current} of Goldstone and Wilczek (1980).

The same result can be obtained, if one starts from the imaginary part of the
Euclidean  effective action and performs a gradient expansion
(cf.sect.2.4),
which leads to the famous Wess-Zumino-Witten term \queq{\e2wzw}.
The algebraic manipulation for doing this is very similar to the one,
which was used
above for gradient expanding the baryon current itself.
'$U(1)$-gauging' the Wess-Zumino-Witten term \queq{\e2wzw}, as it was shown
by Witten (1983), gives also the expression in \qeq{\eb12} as
conserved current.

In the special case of the $SU(2)$-hedgehog form for $U$:
$$
U(\vec r) = \cos \theta (r) + i (\vec\tau \hat r ) \sin \theta (r)
\EQN\eb13
$$
one finds for the time component of \qeq{\eb12}:
$$
\langle b^0 (x) \rangle - \langle b^0 (x) \rangle _V =
{1 \over {2 \pi^2}} {{d\theta}\over{dr}}
{{\sin^2 \theta}\over {r^2}} + \dots
\EQN\eb14
$$
and after integration $\int d^3 r$ and using the fact, that $\theta(\infty)
=0$:
$$
\langle B \rangle =
{{\sin (2\theta (0))} \over{2\pi}} - {{\theta (0)}\over \pi}
\EQN\eb15
$$
For $\theta (0) =- n\pi$
one recognizes that in case of integer $n$ the gradient expanded
baryon number $\langle B \rangle$ coincides with $n$ and therefore the
gradient expansion is valid exactly.
On the other hand for non integer $n$ the gradient expanded expression
\queq{\eb15} is also non integer in contrast to the exactly calculated
$\langle B \rangle$ and therefore the gradient expansion does not converge in
this case.

A necessary condition for the validity of the expansion \queq{\eb8} as well
as the neglegibility of the space dependence of $\sld U_5 $ in \queq{\eb10}
is:
$$
\vert \partial U \vert \ll M
\EQN\eb16
$$
which has to hold at any point $r$ in the radial integral
$\langle B \rangle = \int d^3 r \langle b_0 (\vec r) \rangle $.
If the meson profile $U(\vec r)$ is characterized by some size parameter $R$
(as e.g.in \qufig{\f31,\f32}) one finds therefore:
$$
|n| \pi \ll M \cdot R
\EQN\eb17
$$
which means that the gradient expansion is good either for large profile sizes
or also in case of selfconsistent solutions for high constituent masses $M$.

\appendix{Lorentz Transformation of the Mean Field Soliton Energy}

In this appendix we want to show explicitly that the soliton mean field
energy behaves under Lorentz boost transformations like the time component
of a Lorentz $4$-vector with vanishing space components, and therefore prove
formula \qeq{\e449}, which is the main ingredient in the pushing approach.

Let us to this end consider a Lorentz boost in $x_1$-direction with velocity
$v$:
$$
x^\mu \to {x^{\prime}}^\mu = \Lambda^{\mu^{\prime}} _\nu (\vec\omega) x^\nu
\EQN\ec1
$$
where:
$$\eqalign{
\vec\omega &= {\rm arc tanh} (v) \cdot {\hat v} \cr
{\hat v } &= \left ( \matrix{ 1 \cr 0 \cr 0 \cr } \right ) \cr}
\EQN\ec2
$$
and the contra- and covariant Lorentz transformation read:
$$
\eqalign{
{\Lambda^{\mu^{\prime}}} _\nu (\vec\omega) &=
\pmatrix {
 \cosh \omega & - \sinh \omega & 0 & 0  \cr
-\sinh \omega &   \cosh \omega & 0 & 0  \cr
       0      &         0      & 1 & 0  \cr
       0      &         0      & 0 & 1  \cr } \cr
{\Lambda_{\mu^{\prime}}} ^\nu (\vec\omega) &=
\pmatrix {
 \cosh \omega &   \sinh \omega & 0 & 0  \cr
 \sinh \omega &   \cosh \omega & 0 & 0  \cr
       0      &         0      & 1 & 0  \cr
       0      &         0      & 0 & 1  \cr }
\cr}
\EQN\ec3
$$
respectively.
The generator of this transformation is the unitary matrix:
$$
\B(\vec\omega) = e^{ - \vec\omega\vec K}
\EQN\ec4
$$
where the infinitesimal boost generators $K^i$ are defined by:
$$
K^i = {t\over i} {\partial\over{\partial x^i}} - x^i i
{\partial\over{\partial t}} + i {{\alpha^i}\over 2}
\EQN\ec5
$$
One can easily convince oneself that:
$$
\eqalign{
\B^{-1} (\vec\omega) x^\mu \B (\vec\omega) &= {x^{\prime}}^\mu =
{\Lambda^{\mu^{\prime}}} _\nu (\vec\omega) x^\nu \cr
\B^{-1} (\vec\omega) \partial_\mu \B (\vec\omega) &= {\partial^{\prime}}_\mu =
{\Lambda_{\mu^{\prime}}} ^\nu (\vec\omega) \partial_\nu \cr
\B^{-1} (\vec\omega) \gamma^\mu \B (\vec\omega) &=
{\Lambda^{\mu^{\prime}}} _\nu (\vec\omega) \gamma^\nu \cr}
\EQN\ec6
$$
and therefore derive the action of $\B(\vec\omega)$ on the Dirac operator
$iD$:
$$
\B^{-1} (\vec\omega)
\left [
i \gamma^\mu \partial_\mu - g \left ( \sigma (x)
+ i \vec\tau\piv(x) \gamma_5 \right )
\right ]
\B(\vec\omega)  =
\left [
i \gamma^\mu \partial_\mu - g \left ( \sigma (x^{\prime}) +
i \vec\tau\piv(x^{\prime}) \gamma_5 \right )
\right ]
\EQN\ec7
$$

In the $1$-quark loop approximation the fermionic part of the energy of a
system with meson field configuration
$\{ \sigma (x) ,\piv (x) \}$ is defined as the expectation value of the
$1$-particle hamiltonian
$h = {{\vec\alpha\vec\nabla}\over i} + g\beta
[\sigma (x) + i \vec\tau\piv (x) \gamma_5 ]$ due to \qeq{\e311a}:
$$
\eqalign{
\int d t \langle h (x) \rangle &=
{  {\D q \D {\bar q} \int d^4 x \left [ {\bar q} (x)
\beta h q(x) \right ]
e^{ i \int d^4 x_E {\bar q} (x) (iD[\sigma (x),\piv (x) ] ) q (x)} } \over
  {\D q \D {\bar q} e^{ i\int d^4 x {\bar q} (x) (iD[\sigma (x), \piv (x)])
q (x)} } } \cr
&= i {\partial \over {\partial \Upsilon}} \Sp \ln
[i \partial_t - h(x) - \Upsilon h (x) ]_{\Upsilon =0} \cr}
\EQN\ec8
$$
In case of time independent fields $\sigma (x)$ and $\piv (x)$ this reduces to
$$
\int d t \langle h \rangle _{stat} = (-i \T) \int_{\C_B} {{du}\over{2\pi}}
\sum_{\lambda} {{\epsl}\over{u - \epsl}} = \T \sum_{\lambda \in \C_B} \epsl =
\T E_B
\EQN\ec9
$$
which is of course the same as the unregularized expression obtained in
Euclidean space (\qeq{\e310\e310b}), if the integration contour $\C_B$ is
chosen in such a way that the
appropriate number of orbitals necessary for obtaining
a system with baryon number $B$ gets occupied.
In \qufig{\fa1} this has been indicated for the cases
$B=0$ and $B=1$.

On the other hand in the boosted system ($^\prime$) we have the expectation
value of the boosted hamiltonian
$h_\omega (x) =
{{\vec\alpha\vec\nabla}\over i} + g\beta
[\sigma (x^{\prime} (x)) + i \vec\tau\piv (x^{\prime}(x)) \gamma_5 ]$
$$
\eqalign{
&\int d t \langle h_\omega (x^{\prime}(x)) \rangle = \cr
&= i {\partial \over {\partial \Upsilon}} \Sp \ln
\left [
i \gamma^\mu \partial_\mu - g
(\sigma (x^{\prime}(x) ) + i \vec\tau\piv (x^{\prime} (x)) \gamma_5 )
- \Upsilon \beta h_\omega (x^{\prime} (x) )
\right ]_{\Upsilon =0}  \cr
&= i {\partial \over {\partial \Upsilon}} \Sp \ln \left \{ \B^{-1}
(- \vec\omega)
\left [
i \gamma^\mu \partial_\mu - g
(\sigma (x^{\prime}(x) ) + i \vec\tau\piv (x^{\prime} (x)) \gamma_5 )
- \Upsilon \beta h_\omega (x^{\prime} (x) )
\right ] \B (- \vec\omega) \right \}_{\Upsilon =0}
\cr}
\EQN\ec10
$$
Applying \qeq{\ec6} one finds (Betz and Goldflam 1983):
$$
\eqalign{
&\int d t \langle h_\omega (x^{\prime}(x)) \rangle = \T({\omega}) E({\omega})
= \cr
&\int d t \langle - i \partial_t + h (x) \rangle \cr
&+ (\T \cosh (\omega)) \cdot
\bigl [
\cosh (\omega) \erw{h(x)}
- \sinh (\omega) \erw { (\vec\alpha {\hat\omega})h (x)} \cr
&+ \sinh (\omega) \erw { (\vec p {\hat\omega})h (x)}
- \tanh (\omega) \sinh (\omega) \erw { (\vec\alpha {\hat\omega})
(\vec p {\hat\omega}) } \bigr ] \cr}
\EQN\ec11
$$
For time independent meson field it holds:
$$
\eqalign{
\langle - i \partial_t + h (x) \rangle &= 0 \cr
\langle \vec p \rangle = {i\over 2} \erw{ [ h^2 , \vec x ]} &=0 \cr
\erw{\vec \alpha h } = \langle \vec p \rangle &= 0 \cr}
\EQN\ec12
$$
Furthermore for fixed index $i$ (e.g.$i=1$ in case of \qeq{\ec2}) one has:
$$
\erw{\alpha_i p_i} =
\erw{x_i \cdot
[ g \beta \partial_i (\sigma (x) + i\vec\tau \piv (x) \gamma_5 ) ] }
\EQN\ec12
$$
which can easily be shown to vanish if the meson fields fulfill the classical
mean field equations of motion \queq{\e346} reflecting a virial theorem
(Rafelski 1977).

Therefore we end up finally with:
$$
\int d t \langle h_\omega (x^{\prime}(x)) \rangle =
\T(\omega) E(\omega) =
(\T \cosh (\omega)) (E_{MF} \cosh (\omega) )
\EQN\ec13
$$
Because the time interval $\T$ as itself is boosted to $\T_\omega =
\cosh ({\omega}) \T$, we obtain the desired relation \queq{\e449}
between the energy
in the boosted (moving) system $E_\omega$ and the static
mean field energy $E_{MF}$:
$$
E(\omega) = \cosh (\omega) E_{MF}
\EQN\ec14
$$

\appendix{Quantization condition for $N_c$ and why solitons are
            fermions}

We know from the action of right generators $R_A$ of Sect. 6 on the
baryonic wave-functions $\Psi (A)$, that
(cf. Balachandran et al. 1984)
$$   \exp{i\Theta_a R_a } \Psi (A) = \Psi ( A
      \exp{ (i\Theta_a\half\lambda_a) }  )  \EQN\a1  $$
For the special case of a rotation around the $8th$ axis, we have to
take
into account that $R_8={N_c B/(2\sqrt{3})}$ is constraint. Using the
fact $\exp{(i2\pi\sqrt{3}\lambda_8)}=1$, which corresponds to
$\Theta_8=4\pi\sqrt{3}$ in \qeq{\a1 }, one obtains
\beq  \exp{i4\pi\half B N_c} \Psi (A) = \Psi (A) \EQN\a2 \eeq
so that $B\times N_c$ is integer. The number of colors therefore is an
integer quantity. This corresponds  to the quantization condition for
electric and magnetic charges, when we interpret the Lagrangean
$L_M^{rot}$ as describing the moment of a charged particle in a monopole
gauge field (Jackiw 1983). But we can deduce another quantization
result. Under a rotation of $2\pi$ around the 3th axis, we observe that
\beq \exp{(i\Theta_3\half\lambda_3)} = \exp{(i\Theta_8\half\lambda_8)
    } \EQN\a3 \eeq
for $\Theta_8=2\pi\sqrt{3}$ and $\Theta_3=2\pi$. Using  \qeq{\a1 }
again
one gets on the one hand
\beq   \exp{(i\Theta_8 R_8)} \Psi(A) = \exp{(i\pi)} \Psi(A) = - \Psi(A)
                            \EQN\a4 \eeq
and on the  other hand using \qeq{\a3 } for $N_c=3$
\beq  \exp{(i\Theta_8 R_8)} \Psi(A)
      =    \Psi(A \exp{(i\Theta_3\half\lambda_3)})  \EQN\a5 \eeq
Comparison of \qeqs{\a4,\a5 }
gives  that the wave functions $\Psi(A)$  changes sign under a space
rotation of
$2\pi$. Therefore in the actual case of $N_c=3$, and in fact for an
arbitrary odd value of the number of colors, we have  quantized
the soliton as a fermion (and as a boson in the case of an even number
of colors). It was noted already by Finkelstein and Rubinstein (1968)
and Witten
(1983b), that in the case of SU(2), where
the corresponding Wess-Zumino action - which leads to the constraint for
the eighth generator -
is vanishing, the soliton can be quantized as fermion as well as boson.

\vfill\eject

%
%
%
%
\centerline{\bf REFERENCES }
\bigskip
\noindent
Adjali I Aitchison I R and Zuk J A 1991 {\it Phys.Lett.} B {\bf 256}
497  \lb
Adjali I Aitchison I R and Zuk J A 1992 {\it Nucl.Phys.} A {\bf 537}
457  \lb
Adkins G S Nappi C R and Witten E 1983 {\it Nucl.Phys.} B {\bf 228} 552
\lb
Adkins G S and Nappi C R 1985 {\it Nucl.Phys.} B {\bf 249} 507 \lb
Adler S L 1969 {\it Phys.Rev.} {\bf 177} 2426  \lb
Adler S L and Davies A C 1984 {\it Nucl.Phys.} B {\bf 244} 469  \lb
Adrianov A 1985  {\it Phys. Lett.}{\bf B157} 425  \lb
Ahrens L A et.al. {\it Phys.Lett.} B {\bf 202} 284   \lb
Aitchison J R and Frazer C M 1984 {\it Phys.Lett.} B {\bf 146} 63 \lb
Aitchison J R and Frazer C M 1985a {\it Phys.Lett.} D {\bf 31} 2608 \lb
Aitchison J R and Frazer C M 1985b {\it Phys.Lett.} D {\bf 32} 2190 \lb
Alberto P Ruiz Arriola E Fiolhais M Gruemmer F Urbano J N and Goeke K
1988 {\it Phys.Lett.} B {\bf 208} 75 \lb
Alberto P Ruiz Arriola E  Urbano JN and Goeke K 1990, {\it Phys. Lett.}
{\bf B 247} 210  \lb
Alkofer R and Reinhardt H 1989 {\it Z.Phys.} {\bf C45} 275 \lb
Alkofer R 1990 {\it Phys.Lett.} B {\bf 23} 310 \lb
Alkofer R and Zahed I 1990 {\it Phys.Lett.} B {\bf 223} 149 \lb
Alkofer R  Weigel H and Zueckert U 1993 {\it Phys.Lett.} B
{\bf 298} 132  \lb
Alkofer R and Weigel H  1993, '$1/N_c$   Corrections to
$g_A$ in the Light of PCAC', TU report No. UNITHU-THEP-9/1993 \lb
Balachandran AP Lizzi F  Rodgers VGJ and Stern A 1985, {\it Nucl.
      Phys.}{\bf   B256} 525 \lb
Ball R 1987 {Workshop} on "Skyrmions and Anomalies", Mogilany, Poland,
eds. M. Jezabek and M. Praszalowicz (World Scientific, Singapore) \lb
Ball R 1989 {\it Phys. Rep.} {\bf 182} 1 \lb
Bando M Kugo T Uehara S and Yanagida T 1984 {\it Phys.Rev.Lett.}
{\bf 54} 1215 \lb
Bando M Kugo T and Yamawaki K 1988 {\it Phys.Rep.} {\bf 4} 217  \lb
Bardeen W 1969 {\it Phys.Rev.} {\bf 184} 1848 \lb
Belkov A A Ebert D and Pervushin V N 1987 {\it Phys.Lett.} B
{\bf 193} 315  \lb
Bell J S and Jackiw R 1969 {\it Nuov. Cim.} {\bf 60A} 47 \lb
Berg D Ruiz Arriola E Gruemmer F and Goeke K 1992 {\it J.Phys.} G
{\bf 18} 35 \lb
Berger J 1994, Diploma thesis RUB (unpublished) \lb
Bernard V Brockmann R Schaden M Weise W and Werner E 1984
{\it Nucl.Phys.} A {\bf 412} 349  \lb
Bernard V Meissner U-G and Zahed I 1986 {\it Phys.Rev.Lett.} {\bf 59}
966  \lb
Bernard V 1986 {\it Phys.Rev.} D {\bf 34} 1601 \lb
Bernard V Meissner U-G and Zahed I 1987 {\it Phys.Rev.} D {\bf 36} 819
\lb
Bernard V Jaffe R L and Meissner U-G 1988 {\it Nucl.Phys.} B {\bf 308}
753  \lb
Bernard V Meissner U-G 1988 {\it Nucl.Phys.} A {\bf 489} 647  \lb
Bernard V Kaiser N and Meissner U-G 1990 {\it Phys.Lett.} B {\bf 237}
545      \lb
Bernard V Meissner U-G Blin A H and Hiller B 1991 {\it Phys.Lett.} B
{\bf 253} 443 \lb
Bernard V and Meissner U-G 1991 {\it Phys.Lett.} B {\bf 266} 403 \lb
Bernard V Osipov A A and Meissner U-G 1992 {\it Phys.Lett.} B
{\bf 285} 119  \lb
Bernstein J 1968 {\it Elementary Particles and their Currents} (San
Fransisco: Freeman)  \lb
Betz M and Goldflan R 1983 {\it Phys.Rev.} D {\bf 28} 2848  \lb
Bijnens J Bruno C and de Rafael E 1993, {\it Nucl. Phys.}{\bf B}390 501
\lb
Birse M C and Banerjee M K 1984 {\it Phys.Lett.} B {\bf 136} 284 \lb
Birse M C 1985 {\it Phys.Rev.} D {\bf 31} 118  \lb
Birse M C and Banerjee M K 1985 {\it Phys. Rev.} D {\bf 34} 118 \lb
Birse M C 1990 {\it Prog. Part. Nucl. Phys.}{\bf 26} (A. Faessler ed.,
Pergamon Press) \lb
Bhaduri R K 1988 {\it Models of the Nucleon: From Quarks to Soliton}
(Reading Addison-Wesley)   \lb
Blaizot JP and Ripka G 1988, {\it Phys. Rev.}{\bf D38} 1556 \lb
Blin A H Hiller B and Schaden M 1988 {\it Z.Phys.} A {\bf 331} 75 \lb
Blotz A Doering F Meissner Th and Goeke K 1990 {\it Phys.Lett.}  {\bf B
251} 235 \lb
Blotz A Diakonov D Goeke K Park N W Petrov V and Pobylitsa P V 1992 {\it
Phys.Lett.} B {\bf 287} 29 \lb
Blotz A and Goeke K 1992  {\it Int. J. Mod. Phys. A}, in press \lb
Blotz A Polyakov M and Goeke K 1993a {\it Phys.Lett.}{\bf B302} 151\lb
Blotz A Diakonov D Goeke K Park N W Petrov V and Pobylitsa P V 1993b
{\it Nucl.Phys.}  {\bf A555 }  765   \lb
Blotz A Praszalowicz M and Goeke K 1993c,
        {\it Phys. Lett.}{\bf  B317} 195 \lb
Blotz A Praszalowicz M and Goeke K 1993d,
        Bochum report  No. RUB-TPII-41/1993     \lb
Blotz A Praszalowicz M and Goeke K 1993e, Talk
   given at the {\it International Workshop on the Structure of Baryons}
   in Trento, 4-15. oct. 1993, (to be published) \lb
Blotz A  Christov C V  Goeke K  Pobylitsa Polyakov M
Praszalowicz M and Wakamatsu M  1994, to be published \lb
Braaten E Tse S M and Willcox C 1986 {\it Phys.Rev.Lett.} {\bf 56}
2008  \lb
Braaten E Tse S M and Willcox C 1986 {\it Phys.Rev.} D {\bf 34} 1482 \lb
Brown G E and Rho M 1979 {\it Phys.Lett.} {\bf 823} 177  \lb
Brodsky S Ellis J and Karliner M 1988 {\it Phys.Lett.} B {\bf 206} 309
\lb
Broniowski W and Banerjee M 1985, {\it Phys. Lett.}{\bf B158} 335 \lb
Broniowski W and Banerjee M 1986, {\it Phys. Rev.}{\bf D34} 849 \lb
Broniowski W and Cohen T D 1986, {\it Phys. Lett.}{\bf B177} 141 \lb
Cahill R T and Roberts C 1985 {\it Phys.Rev.} D {\bf 32} 2419 \lb
Cahill R T 1992 {\it Nucl.Phys.} A {\bf 543} 63c \lb
Callan C G  Coleman S Wess J  and Zumino  B 1969 {\it Phys. Rev.}{\bf
177}        2247  \lb
Callan C G Dashen R and Gross D J 1978 {\it Phys.Rev.} D {\bf 17} 2717
\lb
Callan C G Dashen R and Gross D J 1979a {\it Phys.Rev.} D {\bf 19} 1826
\lb
Callan C G Dashen R and Gross D J 1979b {\it Phys.Rev.} D {\bf 20} 3279
\lb
Callan C and Klebanov I 1985 {\it Nucl.Phys.} B {\bf 262} 365 \lb
Callan C Hornbostel K and Klebanov I 1988 {\it Phys.Lett.} B {\bf 202}
296  \lb
Carlitz R D and Craemer D B 1979 {\it Ann. Phys.} {\bf 116} 429 \lb
Caro J and Salcedo LL 1993, {\it Phys. Lett.}{\bf B309} 359 \lb
Chan L H 1985 {\it Phys.Rev.Lett.} {\bf 57} 1199 \lb
Chanfray G Pirner H and Zuk J A 1991 {\it Z.Phys.} A {\bf 339} 503 \lb
Cheng T P and Li L F 1984 {\it Gauge Theory of Elementary Particle
Physics} (Oxford: Clarendon Press)  \lb
Chemtob M 1985 {\it Nucl.Phys.} B {\bf 256} 600 \lb
Chodos A and Thorn C B 1975 {\it Phys.Rev.} D {\bf 12} 2733 \lb
Chon K-C 1961 {\it Sov. Phys.} JETP {\bf 12} 492  \lb
Christov C V Ruiz Arriola E and Goeke K 1990 {\it Nucl.Phys.}A689 \lb
Christov C V Ruiz Arriola E and Goeke K 1990 {\it Phys.Lett.} B {\bf
243} 191  \lb
Christov C V Gorski A Gruemmer F and Goeke K 1993a, in preparation \lb
Christov C V Goeke K Pobylitsa P Petrov V Wakamatsu M and Watabe T
1993b, in preparation
\lb
Cohen T D 1986 {\it Phys.Rev.} D {\bf 34} 2187   \lb
Cohen T D and Bronjowski W 1986 {\it Phys.Rev.} D {\bf 34} 3472 \lb
Coleman S 1985 {\it Aspect of Symmetry} (Cambridge: University Press)\lb
Coleman S and Glashow S L 1961 {\it Phys. Rev. Lett.}{\bf 6} 423 \lb
Coleman S and Glashow S L 1964 {\it Phys.Rev.} B {\bf 134} 671  \lb
Coleman S Wess J and Zumino B 1969 {\it Phys. Rev.}{\bf 177} 2239 \lb
Creutz M 1983 {\it Quarks, Gluons and Lattices} (Cambridge: Cambridge
University Press)  \lb
Dashen R 1969 {\it Phys.Rev.} {\bf 183} 1245 \lb
Dashen R and Weinstein M 1969 {\it Phys.Rev.} {\bf 183} 1261 \lb
DeGrand T Jaffe R L Johnson K and Kiskis J 1975 {\it Phys.Rev.} D {\bf
12} 2060 \lb
Deister S Gari M F Kruempelmann W and Mahlke M 1991 {\it Few-Body Syst.}
{\bf 10} 1-36 \lb
De Swart J J 1963 {\it Rev.Mod.Phys.} {\bf 35} 916  \lb
Dethier J L et.al. 1983 {\it Phys.Rev.} D {\bf 27} 2191  \lb
Dhar A and Wadia R S 1984 {\it Phys.Rev.Lett.} {\bf 52} 959 \lb
Dhar A Shankar R and Wadia R S 1985 {\it Phys.Rev.} D {\bf 31} 3256 \lb
Doering F Blotz A Schueren C Meissner Th Ruiz Arriola E and Goeke K
1992a
{\it Nucl.Phys.} A {\bf 536} 548  \lb
Doering F  Ruiz Arriola E and Goeke K 1992b {\it Z. Phys.}{\bf
A344}159 \lb
Doering F Schueren C Ruiz Arriola E and Goeke K 1993 {\it
Phys. Lett.}{\bf B298} 11 \lb
Donoghue J and Nappi C R 1986 {\it Phys.Lett.} B {\bf 168} 105 \lb
Donoghue J F Golowich E and Holstein B R 1992 {\it Dynamics of the
Standard Model} (Cambridge: Cambridge University Press) \lb
Diakonov D I and Petrov V Yu 1986 {\it Nucl.Phys.} B {\bf 272} 457 \lb
Diakonov D Petrov V and Pobylitsa P 1988 {\it Nucl.Phys.} B {\bf 306}
809 \lb
Diakonov D Petrov V  and Praszalowicz M 1989 {\it
Nucl.Phys.} B {\bf 323} 53 \lb
Ebert D and Reinhardt H 1986 {\it Nucl.Phys.} B {\bf 271} 188 \lb
Eguchi T 1976 {\it Phys.Rev.} D {\bf 14} 2755  \lb
Ellis J 1992 {\it Nucl.Phys.} A {\bf 546} 447c \lb
Ellis J and Jaffe R L 1974 {\it Phys.Rev.} D {\bf 9} 1444 \lb
Ellis J Flores R A and Ritz S 1987 {\it Phys.Lett.} B {\bf 198} 393 \lb
Ellis J and Karliner M 1988 {\it Phys.Lett.} B {\bf 213} 73  \lb
EMC Collaboration, Ashman J et al 1988 {\it Phys.Lett.} B {\bf 206} 364
\lb
EMC Collaboration, Ashman J et al 1989 {\it Nucl.Phys.} B {\bf 328}1 \lb
Esbensen H and Lee T S 1985 {\it Phys.Rev.} C {\bf 32} 1966 \lb
Ferstl P Schaden M and Werner E 1986 {\it Nucl.Phys.} A {\bf 452}
680\lb
Finkelstein D and Rubinstein J 1968 {\it J.Math.Phys.} {\bf 9} 1762  \lb
Fiolhais M Goeke K Gruemmer F and Urbano J N 1988 {\it Nucl.Phys.} A {\bf
481} 727 \lb
Fiolhais M Neuber T Goeke K Alberto P and Urbano J 1991
{\it Phys.Lett.} B {\bf 268} 1  \lb
Fonda L and Ghirardi G 1972 {\it Symmetry principles in Quantum
Physics} ed. A. Barut (New York: Dakkar) \lb
Frazer W R and Fulko J R 1959 {\it Phys.Rev.Lett.} {\bf 2} 365 \lb
Frazer W R and Fulko J R 1960 {\it Phys.Rev.} {\bf 117} 1609 \lb
Gari M and Kruempelmann W 1985 {\it Z.Phys.} A {\bf 322} 689  \lb
Gasiorowicz S and Geffen D A 1969 {\it Rev.Mod.Phys.} {\bf 41} 531 \lb
Gasser J and Leutwyler H 1982 {\it Phys.Rep.} {\bf 87} 77 \lb
Gasser J and Leutwyler H 1984 {\it Ann. of Phys.} {\bf 158} 142 \lb
Gasser J and Leutwyler H 1985a {\it Nucl.Phys.} B {\bf 250} 465 \lb
Gasser J and Leutwyler H 1985b {\it Nucl.Phys.} B {\bf 250} 517 \lb
Gasser J Sainio M E and Svarc A 1988 {\it Nucl.Phys.} B {\bf 307} 779
          \lb
Gasser J Leutwyler H Locher M P and Sainio M E  1988 {\it  Phys.Lett.} B
         {\bf 213} 85 \lb
Gasser J Leutwyler H and Sainio M E 1991 {\it Phys.Lett.} B {\bf 253}
       252, 260  \lb
Gell-Mann M and Zachariasen F 1961 {\it Phys.Rev.} {\bf 124} 953 \lb
Gell-Mann M 1962 {\it Phys.Rev.} {\bf 125} 1067 \lb
Gell-Mann M and Levy M 1960 {\it Nuov. Cim.} {\bf 16} 705  \lb
Gell-Mann M Oakes RJ and Renner B  1968 {\it Phys. Rev.} {\bf 175}
         2195 \lb
Gell-Mann M and Ne'eman Y 1964 {\it The Eightfold Way} (New
          York:Benjamin)     \lb
Goeke K Gorski A Z Gruemmer F Meissner Th Reinhardt H and Wuensch R 1991
{\it Phys.Lett.} B {\bf 256} 321  \lb
Goldberger M L and Treiman S B 1958 {\it Phys.Rev.} {\bf 110} 1178 \lb
Goldstone J 1961 {\it Nuov. Cim. } {\bf 19} 154  \lb
Goldstone J and Wilczek F 1981 {\it Phys.Rev.Lett.} {\bf 47} 986 \lb
Gomm H Kaymakcalan O and Schechter J 1984 {\it Phys.Rev.} D
{\bf 30} 2345 \lb
Gorski A Gruemmer F and Goeke K 1992 {\it Phys.Lett.} B {\bf 278} 24 \lb
Gorski A Christov C V and Goeke K 1993, in preparation \lb
Gourdin M 1974 {\it Phys.Rep.} {\bf 11} 29  \lb
Gross D J and Wilczek F 1973a {\it Phys.Rev.} D {\bf 8} 3633  \lb
Gross D J and Wilczek F 1973b {\it Phys.Rev.Lett.} {\bf 30} 1343 \lb
Guadagnini E 1984 {\it Nucl.Phys.} B {\bf 236} 35  \lb
Hanson T H Prakash M and Zahed I 1990 {\it Nucl.Phys.} B {\bf 335} 67
\lb
Hatsuda T and Kunihiro T 1985 {\it Prog. Theor. Phys.} {\bf 74} 765
\lb
Hatsuda T and Kunihiro T 1987a {\it Phys.Lett.} B {\bf 185} 304 \lb
Hatsuda T and Kunihiro T 1987b {\it Phys.Lett.} B {\bf 198} 126 \lb
Hatsuda T and Kunihiro T 1987c {\it Phys.Lett.} B {\bf 206} 385 \lb
Hatsuda T  1990 {\it Phys.Rev.Lett.}{\bf 65} 543  \lb
Hatsuda T and Kunihiro T 1991 {\it Z.Phys.} {\bf C51} 49  \lb
Hatsuda T and Kunihiro T 1994 {\it Phys. Rep. } to appear  \lb
Heisenberg W 1932 {\it Z. Phys.} {\bf 77} 1 \lb
Holinde K. 1992 {\it Nucl. Phys.}{\bf A 543 } 143c   \lb
Holland D F 1968 {\it J.Math.Phys.} {\bf 10} 531 \lb
Holzwarth G and Schwesinger B 1986 {\it Rep.Prog.Phys.}{\bf 49} 825\lb
Holzwarth G 1992 {\it Phys.Lett.} B {\bf 291} 218 \lb
Huang K 1987 {\it Statistical Mechanics} (New York Wiley) \lb
Itzykson C and Zuber J B 1980 {\it Quantum Field Theory} (New York: Mc
Graw Hill) \lb
Jackiw R 1983  {\it Gauge Theories of the Eighties} eds. R. Raitio and
J. Lindfors
(Berlin: Springer)
\lb
Jaffe R 1989 {\it Phys.Lett.} B {\bf 229} 275 \lb
Jaffe R L and Manohar A 1990 {\it Nucl.Phys.} B {\bf 337} 509 \lb
Jain S and Wadia S R 1985 {\it Nucl.Phys.} B {\bf 258} 713  \lb
Jain P et al. 1987 {\it Phys.Rev.} D {\bf 35} 2230  \lb
Jain P Johnson R Meissner U-G Park N W and Schechter J 1988a {\it
Phys.Rev.} D {\bf 37} 3252  \lb
Jain P Johnson R Park N W Schechter J and Weigel H 1989 {\it Phys.Rev.}
D {\bf 40} 855  \lb
Jain P Johnson R and Schechter J 1988b, {\it Phys. Rev. }{\bf D38} 1571
\lb
Jaminon M and Ripka G 1993, Saclay preprint T93/014 \lb
Jaminon M Ripka G and Stassart P 1989 {\it Nucl.Phys.} A {\bf 504} 733
         \lb
Jaminon M Mendez-Galain R Ripka G and Stassart P 1992 {\it Nucl.Phys.}
        A {\bf 537} 418  \lb
Janssen G Durso JW Holinde K Pearce BC and Speth J 1993,
     {\it Phys. Rev. Lett.}{\bf 71} 1978 \lb
Jenkins E and Manohar A V 1990 {\it Phys.Lett.} B {\bf 255} 558 \lb
Jenkins E and Manohar A V 1991 {\it Phys.Lett.} B {\bf
      259} 353  \lb
Kaiser N Meissner U G and Weise W 1987 {\it Phys.Lett.} B {\bf 198} 319
\lb
Kahana S Ripka G and Soni V 1984 {\it Nucl.Phys.} A {\bf 415} 351 \lb
Kahana S and Ripka G 1984 {\it Nucl.Phys.} A {\bf 429} 462 \lb
Kahana S H and Ripka G 1992 {\it Phys.Lett.} B {\bf 278} 11 \lb
Kato M Bentz W Yazaki K and Tanaka K 1993 {\it Nucl. Phys.}{\bf A551}
541 \lb
Kawarabayashi K and Suzuki M 1966 {\it Phys.Rev.Lett.} {\bf 16} 255 \lb
Kaymakcalan O Rajeev S and Schechter J 1984 {\it Phys.Rev.} D {\bf 30}
594 \lb
Kaymakcalan O and Schechter J 1985 {\it Phys.Rev.} D {\bf 31} 1109 \lb
Kikkawa K 1976 {\it Prog.Theor.Phys.} {\bf 56} 947  \lb
Klebanov I Strangeness in the Skyrme Model, Princeton report PUPT-1158
(unpublished)  \lb
Kleinert H 1976 Erice Summer Institute, Understanding the fundamental
constituents of matter, Plenum Press, NY (1978) A. Zichichi (ed.) p. 289
\lb
Klevansky S P 1992 {\it Rev.Mod.Phys.} {\bf 64} 649 \lb
Klimt S Lutz M Vogl U and Weise W 1990 {\it Nucl.Phys.} A {\bf 516}
429  \lb
Kodaira J Matsuda S Muta T Uematsu T and Sasaki K 1979 {\it Phys.Rev.} D
{\bf 20} 627 \lb
Kodaira J Matsuda S Sasaki K and Uematsu T 1979 {\it Nucl.Phys.} B {\bf
159} 99 \lb
Kodaira J 1979 {\it Nucl.Phys.} B {\bf 165} 129 \lb
Kroll N M Lee T D and Zumino B 1967 {\it Phys.Rev.} {\bf 157} 1376 \lb
Lee B W and Nieh H T 1968 {\it Phys.Rev.} {\bf 166} 1507 \lb
Lee T D {\it Particle Physics and Introduction to Field Theory} (Harwood
Chur: 1981) \lb
Leech R G and Birse M C 1989 {\it Nucl.Phys.} A {\bf 494} 489  \lb
Li M and Perry JP 1988  {\it Phys. Rev.}{\bf D37} 1670 \lb
Li B M Wilets L and Perry R J 1989 {\it J.Comp.Phys.} {\bf 85} 457 \lb
Luebeck E G, M.C. Birse, E.M. Henley and L.Wilets  1986 {\it Phys.Rev.}
D   {\bf  33} 234     \lb
Lutz M et. al. 1992 {\it Nucl. Phys.}{\bf A542} 521  \lb
Lutz M and Weise W 1990, {\it Nucl. Phys.}{\bf A518} 156 \lb
Lutz M and Weise W 1991, {\it Z. Phys.}{\bf A340} 393 \lb
Machleidt R Holinde K and Elster Ch 1987 {\it Phys.Rep.}{\bf 149} 1 \lb
Mazur P Nowak M Praszalowicz M 1984 {\it Phys.Lett.} {\bf B147} 137\lb
McGovern J A and Birse M 1990a {\it Nucl.Phys.} A {\bf 506} 367 \lb
McGovern J A and Birse M 1990b {\it Nucl.Phys.} A {\bf 506} 392 \lb
McKay D W and Munczek H J 1985 {\it Phys.Rev.} D {\bf 32} 266 \lb
McNamee P Chilton S J and Chilton F 1964 {\it Rev.Mod.Phys.} {\bf 36}
1005  \lb
Meissner Th Ruiz Arriola E Gruemmer F Mavromatis H and Goeke K 1988
{\it Phys.Lett.} B {\bf 214} 312 \lb
Meissner Th Gruemmer F and Goeke K 1989 {\it Phys.Lett.} B {\bf 227} 296
\lb
Meissner Th Gruemmer F and Goeke K 1990a {\it Ann.Phys.} {\bf 202} 297
\lb
Meissner Th Ruiz Arriola E and Goeke K 1990b {\it Z.Phys.}{\bf A336}
91  \lb
Meissner Th and Goeke K 1991a {\it Z.Phys.} A {\bf 339} 513 \lb
Meissner Th and Goeke K 1991b {\it Nucl.Phys.} A {\bf 524} 719 \lb
Meissner Th Ripka G Wuensch R Sieber P Gruemmer F and Goeke K 1993 {\it
Phys.Lett.} B {\bf 299} 183 \lb
Meissner U-G 1988 {\it Phys.Rep.} {\bf 161} 213    \lb
Meissner U-G 1989a {\it Phys.Rev.Lett.} {\bf 62} 1013 \lb
Meissner U-G 1989b {\it Phys.Lett.}  {\bf B 220} 1   \lb
Meissner U-G Kaiser N Weigel H and Schechter J 1989 {\it Phys.Rev.} D
{\bf 39} 1956 \lb
Meissner U-G Kaiser N and Weise W 1987 {\it Nucl.Phys.} A {\bf 466} 685
\lb
Meissner U-G and Zahed I 1987,{\it Z. Phys.}{\bf A327} 5 \lb
Meissner U-G 1993 {\it J. Phys.}{\bf G} to appear, {\it Universitaet
Bern preprint BUTP-93/01
} \lb
Moussalam B and Kalafatis D 1991, {\it Phys. Lett.}{\bf B272} 196\lb
Nambu Y 1957 {\it Phys.Rev.} {\bf 106} 1366 \lb
Nambu Y 1960 {\it Phys.Rev.Lett.} {\bf 4} 380  \lb
Nambu Y and Jona-Lasinio G 1961a {\it Phys.Rev.} {\bf 122} 345 \lb
Nambu Y and Jona-Lasinio G 1961b {\it Phys.Rev.} {\bf 124} 246 \lb
Negele J and Orland H 1987 {\it Quantum Many Particle Systems}
(Reading: Addison Wesley) \lb
Nelson T J 1967 {\it J.Math.Phys.} {\bf 8} 857 \lb
Neuber T and Goeke K 1992 {\it Phys.Lett.} B {\bf 281} 202   \lb
NMC Collaboration, Amaudruz P et al. 1991 {\it Phys. Rev. Lett.} {\bf
66}2712 \lb
Okubo S 1962 {\it Progr.Theor.Phys.} {\bf 27} 949  \lb
Pagels H 1975 {\it Phys.Rep.} C {\bf 16} 219 \lb
Park B Y and Rho M 1988 {\it Z.Phys.} A {\bf 331} 151  \lb
Park N W Schechter J and Weigel H 1989a {\it Phys.Lett.} B {\bf 224}
171 \lb
Park N W Schechter J and Weigel H 1989b {\it Phys.Lett.} B {\bf 228}
420 \lb
Park N W Schechter J and Weigel H 1990 {\it Phys.Rev.} D {\bf 41}
2836  \lb
Park N W Schechter J and Weigel H 1991 {\it Phys.Rev.} D {\bf 43}
869  \lb
Park N W and Weigel H 1991 {\it Phys.Lett.} {\bf B268} 155 \lb
Park N W and Weigel H 1992 {\it Nucl. Phys.} {\bf A541} 453 \lb
Pearce BC Holinde K and Speth J 1992 {\it Nucl. Phys.}{\bf A541} 663 \lb
Petersen J L 1985 {\it Act.Phys.Pol.} B {\bf 16} 271  \lb
Pobylitsa P Ruiz Arriola E Meissner Th Gruemmer F Goeke K and
        Broniowski W 1992 {\it J.Phys.} G {\bf 18} 1455  \lb
Politzer H D 1974 {\it Phys.Rep.} {\bf 14C} 129  \lb
Polyakov M V 1990 {\it Sov.J.Nucl.Phys.} {\bf 51} 711 \lb
Praszalowicz M 1985 {\it Phys.Lett.} B {\bf 158} 264  \lb
Praszalowicz M  and Trampetic J 1985 {\it Phys.Lett.} B {\bf 161} 169
\lb
Praszalowicz M 1990 {\it Phys.Rev.} D {\bf 42} 216  \lb
Praszalowicz M Blotz A and Goeke K 1993 {\it Phys.Rev.} D {\bf47} 1127
\lb
da Providencia J Ruivo M A and de Sousa C A 1987 {\it Phys.Rev.} D
{\bf 36} 1882  \lb
Rafelski J 1977 {\it Phys.Rev.} D {\bf 16} 1890  \lb
Rajaraman R 1982 {\it Solitons and Instantons} (North Holland,
Amsterdam) \lb
Reinhardt H and Alkofer R 1988,{\it Phys. Lett.} {\bf B207} 482 \lb
Reinhardt H and Dang B V 1989 {\it Nucl.Phys.} A {\bf 500} 563 \lb
Reinhardt H and Wuensch R 1988a {\it Phys.Lett.} B {\bf 215} 577 \lb
Reinhardt H and Wuensch R 1988b {\it Phys.Lett.} B {\bf 306} 577 \lb
Reinhardt H and Wuensch R 1989 {\it Phys.Lett.} B {\bf 230} 93  \lb
Reinhardt H 1989 {\it Nucl.Phys.} A {\bf 503} 825 \lb
Riazuddin and Fayyazuddin 1967 {\it Phys.Rev.} {\bf 18} 507\lb
Ripka G and Kahana S 1987 {\it Phys.Rev.} D {\bf 36} 1233  \lb
Ripka G and Jaminon M 1992 {\it Ann.Phys.} {\bf 218} 51 \lb
Ring P and Schuck P 1980 {\it The Nuclear Many Body Problem} Springer
Verlag  \lb
Roberts C D Cahill R T and Praschifka J 1988 {\it Ann.Phys.}
{\bf 188} 20 \lb
Ruiz Arriola E  Alberto P  Urbano JN and Goeke K 1989, {\it Z. Phys. }
{\bf A 333} 203 \lb
Ruiz Arriola E Christov Chr V and Goeke K 1990 {\it Phys.Lett.} B {\bf
243} 191  \lb
Ruiz Arriola E 1991a {\it Phys.Lett.} B {\bf 253} 430 \lb
Ruiz Arriola E 1991b {\it Phys.lett.} B {\bf 264} 178 \lb
Ruiz Arriola E  Alberto P  Urbano JN and Goeke K 1990,
                     {\it Phys. Lett.} B {\bf 236} 371 \lb
Ruiz Arriola E and Salcedo LL 1993a, {\it Phys. Lett.}{\bf B}  \lb
Ruiz Arriola E  Alberto P  Urbano JN and Goeke K 1993b, {\it Nucl.
       Phys. }{\bf A} to appear \lb
Ruiz Arriola E Doering F Schueren C and Goeke K 1993c, Bochum preprint
             \lb
Sakurai J J 1960 {\it Ann.Phys.(N.Y.)} {\bf 11} 1 \lb
Sakurai J J 1969 {\it Currents and Mesons} (University of Chicago Press,
Chicago) \lb
Salam A and Strathdee J 1982 {\it Ann. Phys.}{\bf 141} 316  \lb
Schaden M Reinhardt H Amundsen P A and Lavelle M J 1990
{\it Nucl.Phys.} B {\bf 339} 595 \lb
Schechter J and Weigel H 1991 The breathing mode in the SU(3) Skyrme
model, SU-4228-466, IPNO/TH90-83, {\it Phys.Rev.} D to be published;
{\it Phys.Lett.} B {\bf 261} 235 \lb
Schechter J Soni V Subbaraman A and Weigel H 1990 {\it Phys. Rev. Lett.}
{\bf 64} 1495, {\it Phys. Rev. Lett.}  {\bf 65} 2955 \lb
Schlienz J Weigel H Reinhardt H and Alkofer R 1993, {\it Phys. Lett.}
   {\bf B315} 6 \lb
Schneider C 1994, Diploma thesis RUB (unpublished) \lb
Schueren C Ruiz Arriola E and Goeke K 1992a {\it Nucl.Phys.} A
{\bf 547} 612  \lb
Schueren C Ruiz Arriola E and Goeke K 1992b {\it Phys. Lett. } B
{\bf 287} 283   \lb
Schueren C Doering F Ruiz Arriola E and Goeke K, to be published (1993)
\lb
Schwesinger B Weigel H Holzwarth G and Hayashi A 1989 {\it Phys.Rep.}
{\bf 173} 173  \lb
Schwesinger B and Weigel H 1987 {\it Nucl.Phys.} A {\bf 465} 733 \lb
Shifman M A Vainshtein A I and Zakharov V I 1979 {\it Nucl.Phys.} B {\bf
147} 385 448 519 \lb
Shore G M and Veneziano G 1990 {\it Phys.Lett.} B {\bf 244} 75 \lb
Shuryak E V 1984 {\it Phys.Rep.} {\bf 115} 151 \lb
Sieber P Meissner Th Gruemmer F and Goeke K 1992 {\it Nucl.Phys.} A
{\bf 547} 459 \lb
Skyrme T H R 1961 {\it Proc.R.Soc.} A {\bf 260} 127  \lb
Skyrme T H R 1962 {\it Nucl.Phys.} {\bf 31} 556 \lb
SMC Collaboration, Adeva B et al. 1993 {\it Phys. Lett. }{\bf B302} 533
\lb
Soni V 1987 {\it Phys.Lett.} B {\bf 183} 91 \lb
Stern J and Clement G 1989 {\it Nucl.Phys.} A {\bf 504} 621 \lb
Takizawa M Tshushima K Kohyama Y and Kubodera K 1990 {\it Nucl.Phys.}
A {\bf 507} 611 \lb
Takizawa M Kubodera K and Myhrer F 1991 {\it Phys.Lett.} B {\bf 261} 221
\lb
Theberge S Thomas A W and Miller G A  1980,
{\it Phys.Rev.} D
{\bf
22}
2838
\lb
t'Hooft G 1973 {\it Nucl.Phys.} B {\bf 62} 444 \lb
t'Hooft G 1976a {\it Phys.Rev.Lett.} {\bf 37} 8 \lb
t'Hooft G 1976b {\it Phys.Rev.} D {\bf 14} 3432 \lb
t'Hooft G 1986 {\it Phys.Rep.} {\bf 142}   357 \lb
Thomas A W 1983 {\it Adv.  Nucl.Phys.} {\bf 13} 1 \lb
Toyota N 1987 {\it Prog.Theor.Phys.} {\bf 77} 688 \lb
Veneziano G 1979 {\it Nucl. Phys.}{\bf 159} 213 \lb
Vogl U Lutz M Klimt S and Weise W 1990 {\it Nucl.Phys.} A {\bf 516} 469
\lb
Vogl U and Weise W 1991 {\it Prog. Part. Nucl. Phys. }
{\bf 27} 195  \lb
Wakamatsu M 1989 {\it Ann.Phys.} {\bf 193} 287  \lb
Wakamatsu M 1990a {\it Phys.Lett.} B {\bf 234} 223 \lb
Wakamatsu M 1990b {\it Phys.Rev.} D {\bf 42} 2427  \lb
Wakamatsu M 1992a {\it Phys.Lett.} B {\bf 280} 97  \lb
Wakamatsu M and Yoshiki H 1991 {\it Nucl.Phys.} A {\bf 524} 561  \lb
Wakamatsu M 1992b {\it Phys.Rev.} D {\bf 46} 3762  \lb
Wakamatsu M 1993 {\it Phys.Lett.} B {\bf 300} 152  \lb
Wakamatsu M and Watabe T 1993, {\it Phys. Lett.}{\bf B312} 184 \lb
Wakamatsu M and Weise W 1988 {\it Z.Phys.} A {\bf 331} 173 \lb
Watabe T and Toki H 1992 {\it Prog.Theor.Phys.} {\bf 87} 651  \lb
Wegner F J 1971 {\it J. Math. Phys.} {\bf 12} 2259  \lb
Weigel H Schwesinger B and Hayashi A 1987 {\it Phys.Lett.} B {\bf 197}
11      \lb
Weigel H 1988 {\it Phys.Lett.} B {\bf 215} 24  \lb
Weigel H Schechter J Park NW and Meissner U-G 1990 {\it Phys.Rev.}
D {\bf 42} 3177 \lb
Weigel H Alkofer R and Reinhardt H 1992a {\it Phys.Lett.} B {\bf 284}
296 \lb
Weigel H Alkofer R and Reinhardt H 1992b {\it Nucl. Phys.} B {\bf 387}
638  \lb
Weigel H Reinhardt and Alkofer R 1993, {\it Phys. Lett.}{\bf B313} 377
        \lb
Weiss C Alkofer R and Weigel H, 1993 {\it Mod.Phys.Lett.} A {\bf
8}    79 \lb
Weinberg S 1967 {\it Phys.Rev. Lett.} {\bf 18} 188 \lb
Weinberg S 1968 {\it Phys.Rev.} {\bf 166} 1568 \lb
Weinberg S 1979 {\it Physica} {\bf 96A} 327  \lb
Wess J 1972 {\it Acta Phys.Austr.} {\bf 10} 494 \lb
Wess J and Zumino B 1971 {\it Phys.Lett.} B {\bf 37} 95 \lb
Wilson K 1974 {\it Phys.Rev.} D {\bf 10} 2445  \lb
Witten E 1979a{\it Nucl.Phys.} B {\bf 156} 269 \lb
Witten E 1979b{\it Nucl.Phys.} B {\bf 160} 57  \lb
Witten E 1983a{\it Nucl.Phys.} B {\bf 223} 422 \lb
Witten E 1983b{\it Nucl.Phys.} B {\bf 223} 433 \lb
Yabu H and Ando K 1988 {\it Nucl.Phys.} B {\bf 301} 601 \lb
Yabu H 1989 {\it Phys.Lett.} B {\bf 218} 124 \lb
Zahed I and Brown GE 1986 {\it Phys. Rep. }{\bf 142} 1 \lb
Zahed I and Meissner UG 1986, {\it Phys. Rev. Lett.}{\bf 56} 1035\lb
Zueckert U Alkofer R Reinhardt H and Weigel H 1993, preprint HEP-PH
             9303271 \lb
Zuk J A and Adjali I 1992 {\it Int.J.Mod.Phys.} A {\bf 7} 3549  \lb

\vfill\eject

\figout
\tabout

\end